\documentstyle[12pt,epsfig]{article}
 1
\def\as{\alpha_{\rm S}}

\def\citenum#1{{\def\@cite##1##2{##1}\cite{#1}}}
\def\citea#1{\@cite{#1}{}}

\def\as{\alpha_{\rm S}}

\def\g{\gamma}

\def\ra{\rightarrow}

\def\s{\sigma}

\def\({\left(}
\def\){\right)}

\def\citenum#1{{\def\@cite##1##2{##1}\cite{#1}}}
\def\citea#1{\@cite{#1}{}}

\def \VEV#1{\left\langle #1\right\rangle}

\def\beq{\begin{equation}}
\def\eeq{\end{equation}}
\def\bea{\begin{eqnarray}}
\def\eea{\end{eqnarray}}

\def\eq#1{{Eq.~(\ref{#1})}}

\def\bbbz{{\mathchoice {\hbox{$\sf\textstyle Z\kern-0.4em Z$}}
{\hbox{$\sf\textstyle Z\kern-0.4em Z$}}
{\hbox{$\sf\scriptstyle Z\kern-0.3em Z$}}
{\hbox{$\sf\scriptscriptstyle Z\kern-0.2em Z$}}}}
\relax

\begin{document}
\newcounter{savefig}
\newcommand{\alphfig}{\addtocounter{figure}{1}%
\setcounter{savefig}{\value{figure}}%
\setcounter{figure}{0}%
\renewcommand{\thefigure}{\mbox{\arabic{savefig}-\alph{figure}}}}
\newcommand{\resetfig}{\setcounter{figure}{\value{savefig}}%
\renewcommand{\thefigure}{\arabic{figure}}}

\newcounter{saveeqn}
\newcommand{\alpheqn}{\setcounter{saveeqn}{\value{equation}}%
\setcounter{equation}{0}%
\renewcommand{\theequation}{\mbox{\arabic{saveeqn}-\alph{equation}}}}
\newcommand{\reseteqn}{\setcounter{equation}{\value{saveeqn}}%
\renewcommand{\theequation}{\arabic{equation}}}

\begin{titlepage}
\begin{flushright}
TAUP 2465 - 97\\
DESY 97 - 213\\
October 1997\\
{\tt hep - ph 9710546}
\end{flushright}

\centerline{\Huge \bf EVERYTHING ABOUT REGGEONS }
\centerline{}
\centerline{}
\centerline{\Large \bf Part I : REGGEONS IN  ``SOFT" INTERACTION. }
\centerline{}
\centerline{} 
\centerline{\Large \bf  E u g e n e\,\,\,  L e v i n}
\centerline{}
\centerline{}
\centerline{\it School of Physics and Astronomy, Tel Aviv
University}
\centerline{\it  Ramat Aviv, 69978, ISRAEL}
\centerline{\it  and}
\centerline{\it  DESY Theory, Notkrstr. 85, D - 22607, Hamburg, GERMANY}
\centerline{\tt  leving@ccsg.tau.ac.il; \,\,levin@mail.desy.de;}
\begin{flushright}
Academic Training Program\\
DESY, September 16 - 18, 1997\\
{\tt First revised version}
\end{flushright}
\centerline{}
\centerline{}
\centerline{}
{\bf Abstract:}
This is the first part of my lectures on the Pomeron structure which I am
going to read during  this academic year at the Tel Aviv university. The
main goal
of these  lectures is to remind  young theorists as well as  young 
experimentalists of  what are the Reggeons that  have  re-appeared
in
the high energy phenomenology to describe the HERA and the Tevatron  data.
Here, I show   how and why the Reggeons  appeared in the
theory, what theoretical problems they have solved and what they have
failed to solve. I describe in details what we know about Reggeons and
what we do not. The major part of these lectures is devoted to the Pomeron
structure, to the answer to the questions:  what is the 
so called Pomeron; why it is so different from other Reggeons and why we
have to introduce it.
 In short, I hope that this
lectures will be the shortest
way to learn everything about Reggeons from the beginning to the current
understanding. I  concentrate on the problem of Reggeons in this
lectures while the Reggeon interactions or the shadowing corrections I
plan  to discuss in the second part of my lectures. I include here only
a short lecture with an  outline of the main properties of the shadowing
corrections to discuss a correct ( from the point of the Reggeon approach
) strategy of the experimental investigation of the Pomeron structure.
\end{titlepage} 
\thispagestyle{empty}
\parbox{10cm}{~ \epsfig{file=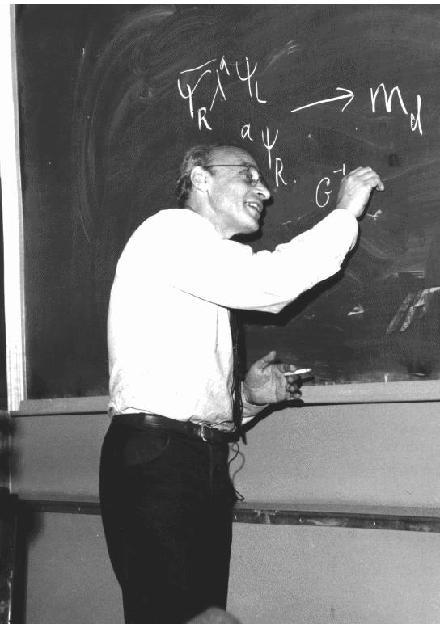,width=90mm}}
\parbox{5cm}{{\it I dedicate these lectures to the memory of Prof.
Vladimir Gribov (1930 -~
1997) who was my teacher and one of the most outstanding physicists in  
this century. I grew up as a physicist under his strong influence and for
the rest of my life I will look at physics with his eyes. Being a father
of Regge approach to high energy asymptotic, he certainly stood behind
my shoulders, when I was preparing these lectures. I hope that I will be
able to share with you all our hopes, ideas, results and disappointments
that we discussed, solved and accepted under his leadership in 
seventies in our  Leningrad Nuclear Physics Institute. }} \hfill

{\it In the next page I put the obituary of the Russian Academy of
Science. This document gives a good description of the best Gribov's
results which will be really with us in these lectures.
You need to know a bit about our life in the Soviet Union to understand
this document and to try to answer such questions like why he had to teach
at the night school, why he was not elected as a full member of the
Academy of Science or why the fact that he was the  leader and the
head of the theoretical physics department  at the Leningrad Nuclear
Physics
Institute during his best twenty years was not mentioned in the document.
This is a good illustration that you need to know a little bit to
understand. I hope, that my lectures will give you this ``a little bit"
for the Reggeon approach.

fter putting these lectures in the hep-ph archive, I got a lot of
comments, critisism and remarks. In particular, the author of the obituary
 sent to me two pieces of his text that were cut in the Russian Academy of
Science. Here they are:

1.  "... after graduating Leningrad University in
politically troubled year 1952, he shared the fate of many a talented
Jewish students who were nort allowed to pursue science, and the only job
available to him was a teacher...." .

2. "Already in early 60's, great Landau considered Vladimir Naumovich his
heir apparent in field theory and particle physics. When the Leningrad
Nuclear Physcis Institute (LIYaF, situated in Gatchina) has separated
from the Ioffe Institute in 1971, Vladimir Naumovich has become the
Head of the Theory Division. While still at the Ioffe Institute in 60's,
and later in Gatchina in 70's, Gribov's department has become a Mecca of
theoretical high energy physics. Gribov's influence transcended the
departmental and national borders, the latter was striking indeed
considering that for more that 20 years Vladimir Naumovich has been
consistently denied a right to attend international conferences abroad to
which he has been invited as a plenary speaker and rapporteur. It is proper
to recall that for more than dozen years Gribov's papers toped the list
of most cited publications by Sov'iet physicists. The annual LIYaF
Winter schools of physics, where one of focal points were post-dinner
discussions chaired by Gribov which extended to small hours, have always
been the major event of the year and the indispensable source of
inspiration for participants. Many a leading high energy theorists in
Leningrad, Moscow, Novosibirsk, Armenia, Georgia, Khar'kov grew out of
these schools, and quite a few renowned theorists in USA and Europe do
rightfully belong to the list of his pupil."

I think, no comments are needed.}
 \newpage
\thispagestyle{empty}

\parbox{10cm}{~ \epsfig{file=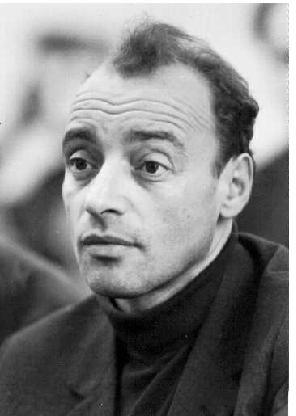,width=90mm}}
\parbox{5cm}{
 The  eminent  Russian  theorist  Corresponding Member of Russian   
 Academy of Sciences Vladimir Naumovich  Gribov  passed  away  on
 August 13, 1997, in Budapest, after short and grave illness.

 V.N.Gribov  was  born  on  March  25,  1930 in Leningrad. Having
 graduated Leningrad University in 1952 he  was  forced  to  work
 as  a
 teacher  at  the  night  school  for working people. In 1954, he
 joined the Ioffe Physico-Technical Institute in  Leningrad,  and
 before long became the leading member of the Theory Division and
 one   of   unquestionable   leaders  in  Soviet  and  the  world   
 theoretical high energy physics. Many a theorists in Russia  and
 FSU  are his pupil.

 Gribov's theory of threshold multiparticle reactions, the Gribov
 -  Froissar  expansion,  Gribov's  factorization,  the  Gribov -
 Pomeranchuk shrinkage of the diffraction}
\noindent
 cone, Gribov -  McDowel
 symmetry,   Gribov   -  Volkov  conspiracy,  Gribov  -  Morrison   
 selection  rules,  Glauber  -  Gribov  theory   of   diffraction
 scattering  on  nuclei,  Gribov  - Pomeranchuk - Ter-Martirosian
 reggeon  unitarity,  Gribov's  reggeon  calculus,  Abramovski  -
 Gribov   -   Kancheli  (AGK)  cutting  rules,  Gribov  -  Ioffe-
 Pomeranchuk growth of spatial  scales  in  strong  intreactions,
 Gribov's extended vector dominance, Gribov - Pontecorvo analysis
 of neutrino oscillations, Gribov's theorem for Bremsstrahlung at
 high  energies,  Gribov  - Lipatov evolution equations, Gribov -
 Lipatov reciprocity, Gribov's vacuum copies, Frolov - Gorshkov -
 Gribov - Lipatov theory of Regge processes in gauge theories are   
 jewels in the crown of modern theoretical physics.
 
 Gribov  was  the  foreign member of the American Academy of Arts
 and Sciences in Boston. His distinctions include the  L.D.Landau
 medal  of  the  Academy of Sciences of USSR, of which he was the
 first ever recipient, J.J.Sakurai prize of the American Physical
 Society, Alexander von Humboldt prize and the  Badge  of  Honor.
 Since  1980 he has been a member of the L.D.Landau Institute for
 Theoretical Physics (Moscow) and the last decade also a Professor
 at  the
 R.Etvoesh University in Budapest.
 
 Exceptionally warm  and  cheerful  personality,  who  generously
 shared  his  ideas and fantastic erudition with his students and
 collegaues,  indefatigable  discussion   leader   V.Gribov   was
 universaly  revered  in  his  country and elsewhere. His family,
 friends,  colleagues  and  the  whole  of  high  energy  physics
 community shall miss him dearly.

\newpage

\centerline{ \Large {\bf Several words about these lectures:}}

~
\centerline{ \Large {\bf goals, structure and style.}}

~

Let me start saying several words about my personal activity in the
Reggeon approach. I started to be involved in the Pomeron problem in early
70's not because I felt that this was an interesting problem for me but
rather because everybody around worked on  this problem.

 It was a heroic time in our department
when
we worked as one team under the leadership of Prof. Gribov. He was
 for us, young guys,  not
only a dominant leader but a respected  teacher. I took his
words seriously: `` Genya, it seems to me that you are a smart guy.  I am
sure, 
you will be able to do something more reasonable than this quark stuff."
So I decided to try and, frankly speaking,  there was
another reason behind my decision. I felt that I could not develop my
calculation skill,
doing the additive quark model. However, I must admit,  very soon I
took  a deep interest in the Pomeron problem, so deep  that I decided to  
present here all my ups and downs in the attempts to attack this
problem. 

 I hope, that these lectures  show the
development of the main ideas in their historical perspective from the
first enthusiastic attempts to find a simple solution to understanding
of
the complexity and difficulty of the problem.
 In other words, these are lectures of a man
 who wanted to understand everything in high  
energy interactions, who did his best but who is still at the beginning,
 but who has not lost his temper and considers the Pomeron structure as
a  beautiful and difficult problem, which deserves his time and
efforts to be solved.

 Based on this emotional background, I will try to 
give a selected  information on the problem, to give not only a
panorama of ideas and problems but to teach you to perform all calculations
within the Reggeon approach. I  tried to collect here the full information
that you need to start doing calculations. It is interesting to notice
that that the particular contents of my lectures will cover the most
important of Gribov results which we can find in the obituary of the
Russian Academy of Science attached to these lectures. However, you will
find more than that.  
 If you will not find something,
do not think that I forgot to mention it. It means that I do think that
you do not need this information. The criteria of my selection is very
simple. I discuss only those  topics which you, in my opinion, should take
with you in future. As an example, everybody has  heard that the Regge
pole
approach is closely related to the
analytical properties of the scattering amplitude in angular moment
representation. You will not find  a single word about this here
because, I think, it was only the historical way of how  Regge poles
were introduced.
But, in fact, it can be   done without  directly  mentioning the
angular
moment representation. I hope, that you will understand  an idea, that my
attempt to
give you a full knowledge will certainly not be the  standard way.
 
 In this particular part I restrict
myself to the 
consideration of Reggeons ( Regge poles)  in  ``soft" interaction while
discussions of
the shadowing corrections, the high energy asymptotic in perturbative
QCD, the 
so called  BFKL Pomeron and the shadowing corrections at short
distances ( high parton density QCD )
will be given later in the following parts of these lectures and will be
published elsewhere. However, to give you some  feeling on what
kind of experiment will be useful from the point of view of the Reggeon
approach I will outline some of the  main properties of the shadowing
corrections
in the ``soft" processes in the last lecture. I would like also to
mention that I am going to read at the Tel Aviv university this year the
complete course of lectures:
``Everything about Reggeons"   and it will consist of  four parts:

1. Reggeons in the ``soft" interaction;

2. Shadowing corrections and Pomeron interactions in ``soft" processes;

3. The QCD Pomeron;

4. The high parton density QCD.
\thispagestyle{empty}
 
The `` soft" processes at high energy were, are  and
will be the area of the Regge phenomenology which has been useful for 25
years. Unfortunately, we, theorists, could not suggest something better.
This is the fact which itself gives enough motivation to learn what is the
Reggeon approach. The second motivation for you is, of course, new data
from HERA which can be absorbed and can be discussed only with some
knowledge about Regge phenomenology. What you need to know is what is
Regge phenomenology, where can we apply it and what is the real 
outcome for the future theory.   

Let me make some  remarks on the  style of the presentation. It is not a
text book or paper. It is a  lecture. It means that it is a living
creature
which we together can change  and make it better and better. I will be
very
thankful
for any questions and suggestions. I am writing these lectures after
reading them at DESY. I got a lot of good questions which I answer here.

Concluding my introductory remarks I would like to mention  Gribov's
words:`` Physics first". Have a good journey.

\centerline{\bf Comments to the first revised version.}
After putting my first version of these lectures in the hep-ph archive, I
recieved a lot of comments, remarks, both critical and supportive,
advises,
recommendations and questions. I am very grateful to everybody who
interacted with me . This is my   first attempt to take into account 
everything that I recieved. This is the first revised
version in which I tried only to    improve the presentation without
adding anything. However, I realized that lectures have to be enlarged
 and several items should be added. In particular, it turns out that my
readers need the minimal information about the formal approach based on
the analyticity of the scattering amplitudes in the angular moments
plane. I am planning to work out a new version which will include several
new subjects.

Therefore, I am keeping my promise to continue writing of these lectures
in an interactive regime. My special  thanks to  G. Wolf and S.
Bondarenko,
 who sent me their copies of these lectures with  corrections in
every page. I cannot express in words how it was useful for me.

 \thispagestyle{empty}

\newpage
\tableofcontents
\section{Instead of introduction}

\subsection{ Our strategy.}

My first remark is that in the 70's  finding  the  asymptotic
behaviour of the scattering amplitude at high energies 
was
a highly priority job. During the last five years I have traveled a lot
around the globe and I have found that it is very difficult to explain
to  young physicist why it was so. However, 25 years ago the common
believe was that the analytical property of the scattering amplitude
together with its asymptotic behaviour would  give us the complete and
selfconsistent theory of  the strong interaction. The formula of our
hope was very simple:
\begin{center}
\begin{tabular} {l c l c l}
{\it Analyticity}& & & &\\
{\it Unitarity}  & {\it +} &{\it  Asymptotic} &{\it =}&{ \it Theory \,\,of
\,\,Everything.}\\
{\it Crossing} & & & & \\
\end{tabular}
\end{center}
Roughly speaking we needed the  asymptotic behaviour  at high energy to
specify how many
substractions were necessary in  dispersion relations to calculate the
 scattering amplitude.

 Now the situation is quite different:  we have good microscopic theory (
QCD )
although  we have a lot of problems in QCD which  have to be solved. High
 energy asymptotic behaviour is only one of many. I think it is time to
ask yourselves
why we spend our time and brain trying  to find the high energy
asymptotic behaviour  in QCD. My lectures will be an answer to this
question, but I would
like to start by  recalling you the main theorems for the high energy
behaviour
of the scattering amplitude which follow directly from the general
properties  of
analyticity and crossing symmetry and should be fulfilled in any microscopic
theory including QCD. Certainly, I have to start by  reminding you  what
is
analyticity and unitarity for the scattering amplitude and how we can use
these general properties   in our theoretical analysis of the experimental
data.
\subsection{Unitarity and main definitions.}
Let us start with the case where  we have only {\bf  elastic scattering}.
\begin{figure}[htbp]
\centerline{\epsfig{file=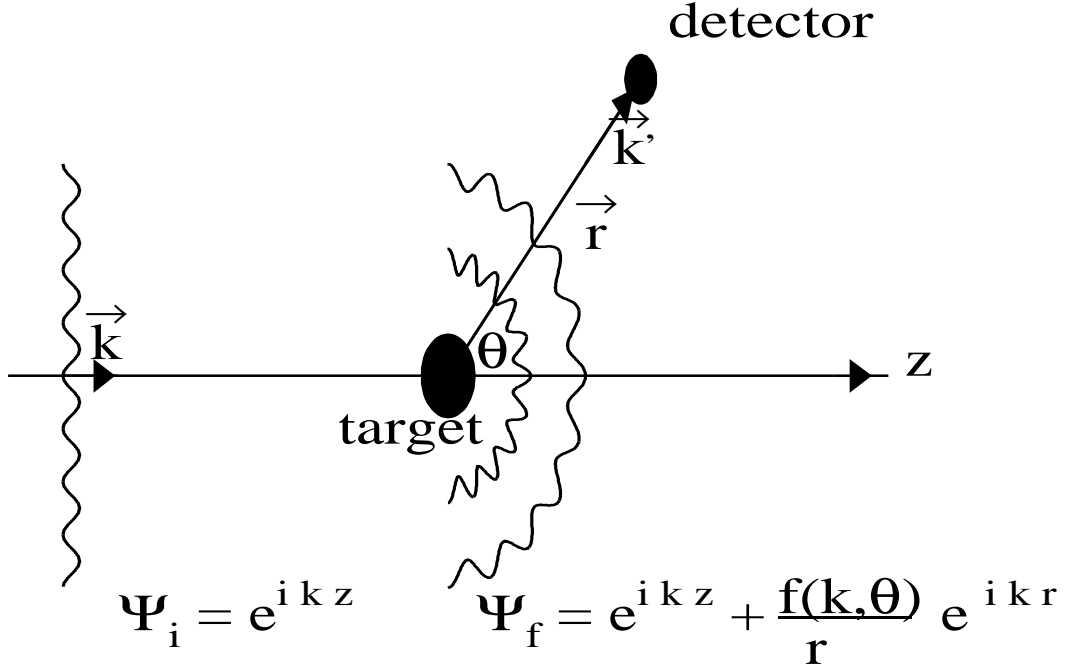, width=160mm,height=100mm}}
\caption{{\it The elastic scattering.}}
\label{fig1}
\end{figure}
The wave function of the initial state ( before collision ) is a plane
wave moving  say along the  z-direction, namely:
$$\Psi^{i}\,\,=\,\,e^{i k z}\,\,=\,\,e^{i \vec{k} \cdot \vec{r}}$$
The wave function of the final state ( after collision) is more
complicated
and  has two terms: the plane wave and an  outgoing spherical wave. The
scattering amplitude is the amplitude of this spherical wave.
$$
\Psi^{f}\,\,=\,\,e^{i \vec{k} \cdot \vec{r}}\,\,\,+\,\,\,\frac{f(\theta,
k)}{r}\,e^{i k r}\,\,,
$$
where $r \,=\,\sqrt{x^2 + y^2 + z^2}$, \,$\vec{k} \cdot \vec{r}\,=\,
k r cos \theta$  and $k$
is the value of the
momentum. The scattering amplitude in this form is very useful since the
cross section is equal to
\beq
d \s\,\,=\,\,\frac{v | \Psi^f|^2 d S}{v}\,\,,
\eeq
where $v$ is the velocity of the incoming particle and $d S $ is the
element of the sphere $d S \,=\,2 \pi\, r^2\, sin \theta d \theta$.
Finally,
\beq 
\frac{d \s}{d \Omega}\,\,=\,\,2 \pi | f(k,\theta)|^2\,\,,
\eeq
where $\Omega \,=\,2 \pi sin \theta \,d\,\theta$.
{\bf Assumption:}The  incoming wave is restricted by some aperture ($L$)
$$
r\,\,\gg\,\,L\,\,\gg\,\,a$$ where $a$ is the typical size of the forces
(potential ) of interaction. Then one  can neglect the incoming wave as $r
\,\rightarrow\,\infty$.

To derive the unitarity constraint we have to consider a more general case
of scattering, namely, the incoming wave is a package of plane waves
coming from different angles not  directed only  along z-direction. 

\begin{figure}[htbp]
\centerline{\epsfig{file=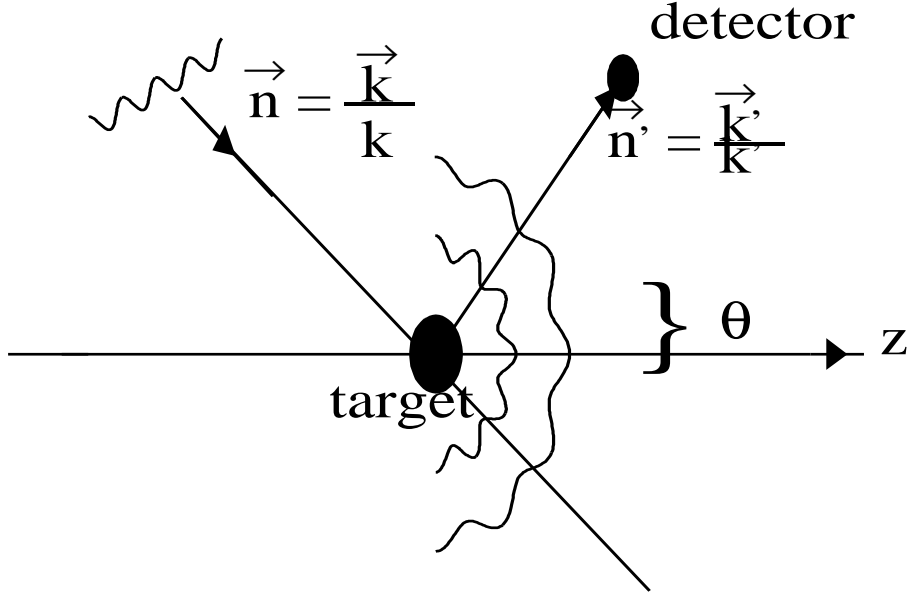, width=160mm,height=100mm}}
\caption{{\it The elastic scattering ( general case).}}
\label{fig2}
\end{figure}

It means that
\beq
\Psi^i\,\,=\,\,\int d \Omega_n \,F(\vec{n})e^{i k r \vec{n} \cdot
\vec{n'}}\,\,
\eeq
where we introduce two unit vectors $\vec{n}\,=\,\frac{\vec{k}}{k}$ and
$\vec{n'}\,=\,\frac{\vec{k'}}{k'}$ (see Fig.2). $F(\vec{n})$ is an  
arbitrary
function.

After scattering we have
\beq
\Psi^f\,\,=\,\,\int d \Omega_n \,F(\vec{n})e^{i k r \vec{n} \cdot
\vec{n'}}\,\,+\,\,\frac{e^{i k r}}{r}\,\int d \Omega_n
\,F(\vec{n})\,f(k,\vec{n},\vec{n'})\,\,.
\eeq

The first term can be explicitly calculated. Indeed, in vicinity of $ z =
cos
\theta = -1$ we have

\parbox{6cm}{~\epsfig{file=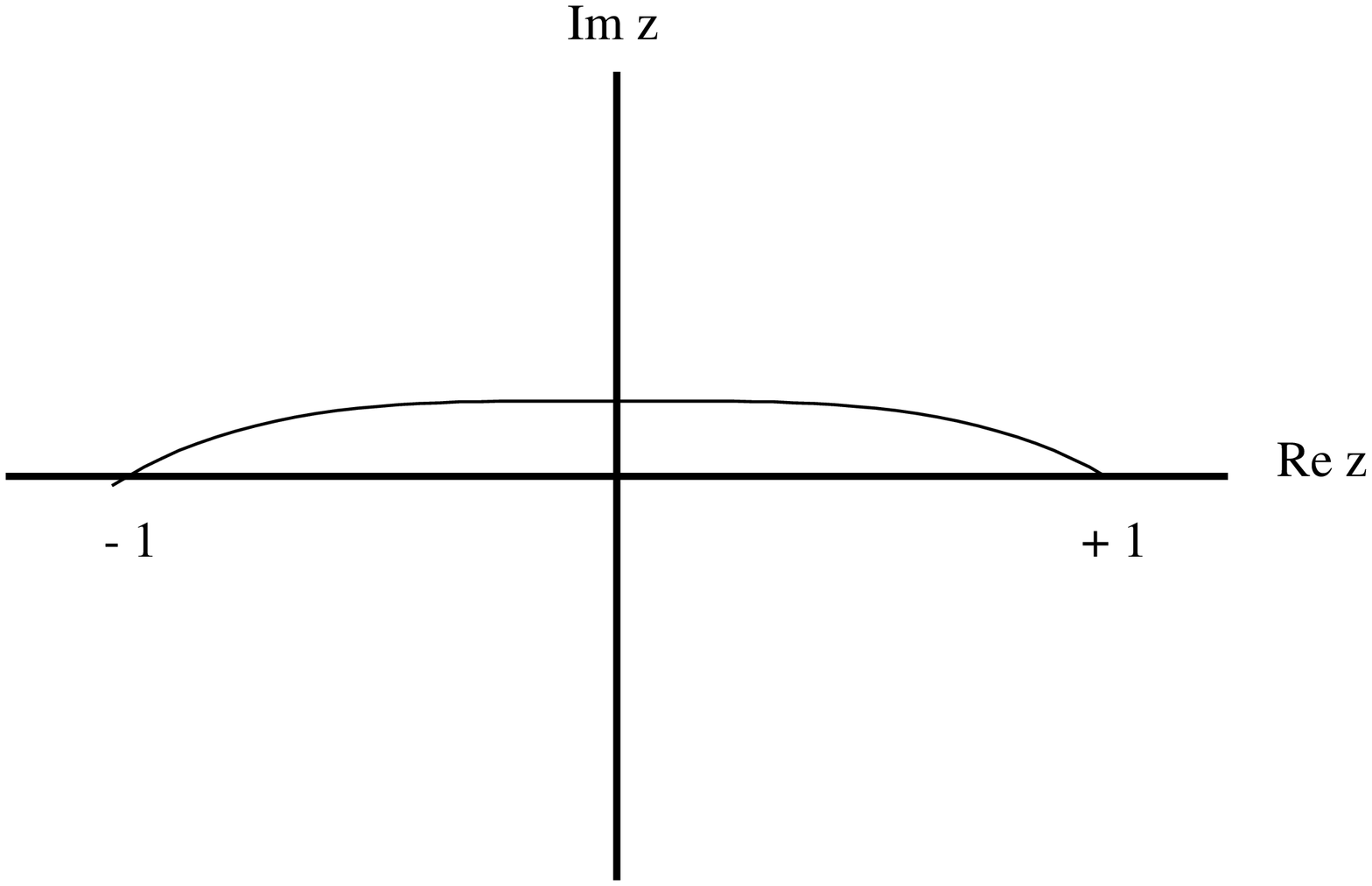,width=65mm}}
\parbox{10cm}{
$$\int d \Omega_n \,F(\vec{n})e^{i k r \vec{n} \cdot
\vec{n'}}\,\,=\,\,2 \pi \int^{-1 + i z_0}_{-1} d z F(z) e^{i k r z}\,\,=
$$
$$ 
2 \pi F( -\vec{n})
\frac{1}{i kr}\,\{\,e^{ - kr z_0}\,\,-\,\,e^{ - i k r}\,\}\,\rightarrow
\frac{2 \pi i}{kr}\,e^{- i k r}\,F( -\vec{n})\,.$$ }\hfill

Finally,
 $$
\int d \Omega_n \,F(\vec{n})e^{i k r \vec{n} \cdot
\vec{n'}}\,\,=\,\,2 \pi i \,\,\{\,F( - \vec{n}) \,\frac{e^{ - i k r}}{k
r}\,\,-\,\,F( \vec{n}) \,\frac{e^{ i k r}}{k r}\,\}$$
The resulting wave function for the final state is equal
$$
\Psi^f\,\,=\,\,\frac{2 \pi i}{k}\,\{\,F( - \vec{n}) \,\frac{e^{ - i k 
r}}{r}\,\,-\,\,F( \vec{n}) \,\frac{e^{ i k r}}{r}\,\}\,\,+\,\,\frac{e^{i k
r}}{r}\,\int f(k,\vec{n},\vec{n'} )\,\,F(\vec{n})\,\, d\Omega_n
\,\,,
$$   
   while the initial wave function is
$$
\Psi^i\,\,=\,\,\frac{2 \pi i}{k}\,\{\,F( - \vec{n}) \,\frac{e^{ - i k
r}}{r}\,\,-\,\,F( \vec{n}) \,\frac{e^{ i k r}}{r}\,\}\,\,.
$$
The conservation  of probability means that ($V$ is the volume):
\begin{large}
\beq
\int \,|\Psi^i|^2\,d\,V\,\,\,=\,\,\,\int\,|\Psi^f|^2\,d\,V\,\,.
\eeq
\end{large}
Taking into account that 
$$\int \frac{e^{-2i k r}}{r^2} r^2 d r d \Omega\,\,=\,\,0 $$
we obtain ( see Fig. 3 ):
\begin{large}
\beq
Im\,f( \vec{n}, \vec{n'}, k )\,\,=\,\,\frac{k}{4 \pi} \,\int
f(\vec{n}, \vec{n"}, k)\,\,f^*(\vec{n"}, \vec{n'}, k)\,d \,\Omega_{n"}
\eeq
or
\beq
Im f(k, \theta)\,\,\,=\,\,\frac{k}{4 \pi}\,\int\,f(k,\theta_1) \,f^*(k,
\theta_2) \,\,d \Omega_1\,\,,
\eeq
\end{large}
where $cos \theta\,=\,cos \theta_1 cos\theta_2\,-\,\,sin \theta_1
sin\theta_2\,cos \phi$.
\begin{figure}[hbtp]
\centerline{\epsfig{file=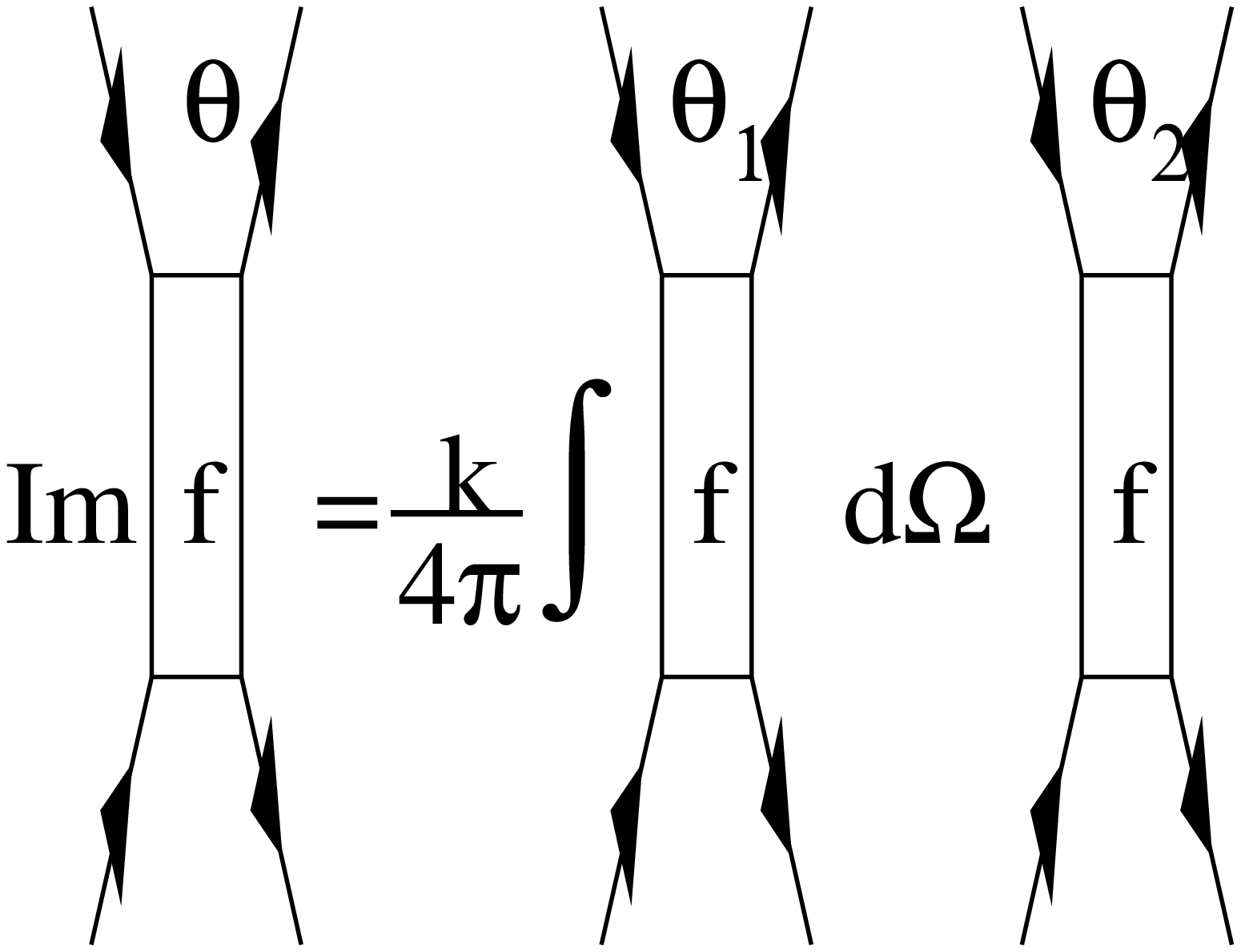, width=100mm,height= 40mm}}
\caption{{\it The s - channel unitarity.}}
\label{fig3}
\end{figure}
Now, let us introduce the partial waves, namely, let us expand the elastic
amplitude in terms  of Legendre  polynomials:
\beq
f(k,\theta)\,\,=\,\,\sum^{\infty}_{l = 0}\,\,f_l(k)\,P_l( cos
\theta)\,(\,2l +
1 )\,\,.
\eeq
Using two properties of Legendre polynomials, namely,

{\bf 1.}
$$
\int^{1}_{-1}\,d z P_l( z) P_{l'}(z)\,\,=\,\,0
\,\,\,(\,\,if\,\,l\,\neq\,l')\,\,\,\,=\frac{2}{2 l + 1}\,\,\,(\,\,
if\,\,l\,=\,l'\,\,)\,\,;
$$
~

{\bf 2.}
$$
P_l(cos\theta)\,\,=\,\,P_l (cos \theta_1)\,P_l(cos\theta_2)\,\,+
\,\,2\,\sum^l_{m = 1}\,\frac{(l - m)!}{(l + m )!}P^m_l
(cos\theta_1)\,P^m_l(cos\theta_2)\,cos m\phi\,\,;
$$
one can easily get that the unitarity constraint looks very simple for
the partial amplitude:
\begin{large}
\beq
Im f_l (k)\,\,\,=\,\,k\,| f_l (k) |^2
\eeq
\end{large}

$\bullet$ {\it What is $k$?}

\begin{figure}[htbp]
\centerline{\epsfig{file=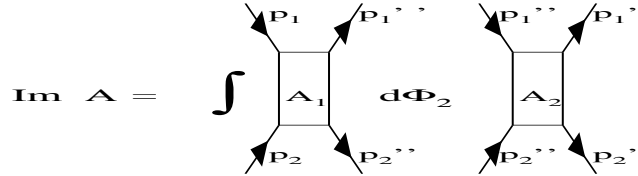, width=100mm,height =40mm}}
\caption{{\it The s - channel unitarity ( relativistic normalization).}}
\label{fig4}
\end{figure}

We define  $s = (p_1 + p_2 )^2$ and the unitarity constraint can be
written
as (see Fig.4):

$$
Im A \,=\,\int A_1 A^*_2 \frac{d^3 {p"}_1}{2 {p"}_{10}}\,\frac{d^3
{p"}_2}{2
{p"}_{20}}\,\delta^{(4)}( p_1 - p_2 - {p"}_1 - {p"}_2 )\,\frac{(2
\pi)^4}{(2 \pi)^6}\,\,;
$$
Note, that ${p"}_1$ and ${p"}_2$ are momenta of the intermediate state in
the unitarity constraint ( see Fig.4).
In c.m.f. $\sqrt{s}\,\,=\,\,2 \,\sqrt{m^2 + k^2}$ or
$$
k\,\,=\,\,\frac{\sqrt{s \,-\,4m^2}}{2}\,\,
$$ 
and 
$$ d^3 {p"}_1 \,\,=\,\,{p"}^2_1 d {p"}_1 d
\Omega\,\,=\,\,\frac{1}{2}{p"}_1 d {p"}^2_{10}$$
Therefore, the unitarity condition can be rewritten in the form for $f =
\frac{A}{\sqrt{s}}$ in the usual way.  The physical meaning
of $k$ in the unitarity constraint is, that $k$ is equal to  the phase
space of two colliding
particles.

\subsection{General solution to the elastic unitarity.}

It is easy to see that the elastic unitarity has a general solution,
namely,
\beq \label{GSU}
f_l\,\,=\,\,\frac{1}{g_l(k^2)\,\,-\,\,i k}\,\,,
\eeq
where $g_l(k^2)$ is real, but otherwise  arbitrary function.

{ \Large  \bf  Two applications of elastic unitarity:}

{\bf 1. The scattering length approximation and the deutron structure.}

Let us assume that
$$
k a \,\,\ll\,\,1$$
where $a$ is the size of the target or more generally, let us assume that
we consider our scattering at such small values of $k$ that
$$
g_{l = 0} (k^2)\,\,=\,\,\frac{1}{a}\,\,+\,\,k^2\,r_0\,\,+\,\,...
$$
and $r_0 k \,\,\ll\,\,1$. At such energies 
$$
f_0 (k) \,\,=\,\,\frac{a}{1\,\,-\,\,i k a}\,\,,
$$
where $ k \,=\,\frac{1}{2} \sqrt{s - 4m^2} $.
One can see that $f_0$ could have a pole at $ 1 - i k a =0$ !?.
However, first we have to define how we continue $k$ into  the complex
plane.  I want to stress that our analytical continuation is defined by
our first formula ( definition of the scattering amplitude ) and from the
condition that the bound state should have a wave function which decreases 
at large distance. Actually, $\sqrt{s - 4 m^2}$ at $s  < 4 m^2$ could
be $ \pm \kappa = \pm  \sqrt{4 m^2 - s}$. Therefore the wave function for
$s  < 4 m^2$  has the  form:
$$
\Psi \,\rightarrow\,f(\kappa, \theta))\,\frac{e^{\pm \kappa r}}{r}
$$
As I have  mentioned we have to choose $\int |\Psi|^2 d v \,<\,\infty$
and this condition tells  us that
$$
- i k \,\,\rightarrow\,\,\kappa\,\,.
$$
Therefore at $s < 4 m^2 $ our scattering amplitude has the  form
\beq
f_0 (s) \,\,=\,\,\frac{a}{1 + a \sqrt{4 m^2 - s}}\,\,.
\eeq
One can see that at $a < 0 $ we have a pole:
$$
1 + a \sqrt{4 m^2 - s}\,\,=\,\,0\,\,;
$$
$$
s = 4m^2 \,\,-\,\,\frac{1}{a^2}\,\,=\,\,m^2_D\,\,.
$$ 
Introducing the binding energy ($\epsilon_D$) $ m_D \,=\,2 m +\epsilon_D
$ we calculate $ a \,=\,\frac{1}{\sqrt{m_D |\epsilon_D|}}$. For small
values of the binding energy the value of $a$ is big and therefore  we can
use our
approximation. Such a situation, for example, occurs in proton - neutron
scattering
where the scattering length is large  and negative. This is the reason for
the deutron state in this reaction. It is interesting to calculate the
wave function of the deutron by just using the definition  ( for $l = 0$
)
$$
\Psi_D( t, r)\,\,=\,\,\int d E \,e^{- i Et } f_0(k)\frac{e^{-\kappa r}}{r}
\,\,.
$$
One can check that taking this integral one finds  the deutron wave
function
which can be found in any text book on quantum mechanics.

{\bf 2. The Breit - Wigner resonance.}

Let us assume that $g_l (s = M^2) \,=\,0$ at some value of $M$ which is
not particularly close to $s = 4 m^2$. In this case  $k = k_M
=
\frac{1}{2}\sqrt{M^2 - 4m^2}$ and
$$
g_l(s) \,\,=\,\,{g'}_l (M^2 - s) \,\,+\,\,{g''}_l ( M^2 - s)^2\,\,+\,\,..
$$
The partial amplitude at $s\,\rightarrow \,M^2 $ reduces to
\beq
f_l\,\,=\,\,\frac{1}{g'_l\,( M^2 - s ) \,\,-\,\,i k_m}\,\,=\,\,
\frac{\g}{M^2 -  s - ik_M \g}\,\,.
\eeq 
I hope that everybody recognizes the Breit - Wigner formula. If we
introduce
a new  notation: $ \Gamma\,\,=\,\,k_M \,\g$ then
\beq \label{BWE}
f_l\,\,=\,\,\frac{\Gamma}{k_M\,(\,M^2\,\,-\,\,s\,\,-i\Gamma\,)}\,\,.
\eeq
This formula can be obtained in a little bit different form. For
$s
\,\rightarrow\,M$ $ s - M^2\,= (\sqrt{s} - M) (\sqrt{s} + M ) \,\approx\,
2\, M ( E\, -\, M)$, \eq{BWE} can be rewritten as
\beq \label{BWENR}
f_l\,\,=\,\,\frac{\tilde \Gamma}{k_M\,(\,E\,-\,M\,-\,i\,\tilde \Gamma\,)}
\eeq
It is very instructive to check that this formula as well as \eq{BWE}
satisfies the unitarity constraint.

Substituting this Breit - Wigner  form for the  general formula we obtain

the following wave function
$$
\Psi\,\,=\,\,\int d E  \frac{e^{- iE t} \tilde \Gamma}{k_M \,(\,E - M -
i\,\tilde \Gamma)}\,\frac{e^{ikr}}{r}\\,=
$$
$$
=\,\,e^{-i M t} \,e^{ - \tilde \Gamma t}\,\frac{\tilde \Gamma}{k_M}
\frac{e^{i k_M r}}{r}\,\,.
$$
Therefore the physical meaning of $\tilde \Gamma$ is that $\tau
\,=\,\frac{1}{\tilde \Gamma}$ is the lifetime of the resonance.
 
The contribution of the resonance to the total cross section at $s = M^2$
 (maximum) is equal to
$$
\s_l\,\,=\,\,\,4 \pi (\,2 l \,+\,1\,)\,|f_l(k)|^2\,\,=\,\,\frac{4 \pi^2 (
2 l + 1 )}{k^2_M}\,\,=\,\,\s^{max}_l\,\,.
$$

\subsection{Inelastic channels in unitarity.}
As  discussed for the elastic scattering we have 
$$
\Psi^i_{el} ( t < t_{interaction})\,\,=\,\,e^{i k z}
$$
and
$$
\Psi^f_{el} ( t > t_{interaction})\,\,=\,\,e^{i k
z}\,\,+\,\,\frac{f(k,\theta)}{r} e^{i k r}
$$
\begin{figure}[htbp]
\begin{tabular}{ c}
\centerline{\epsfig{file=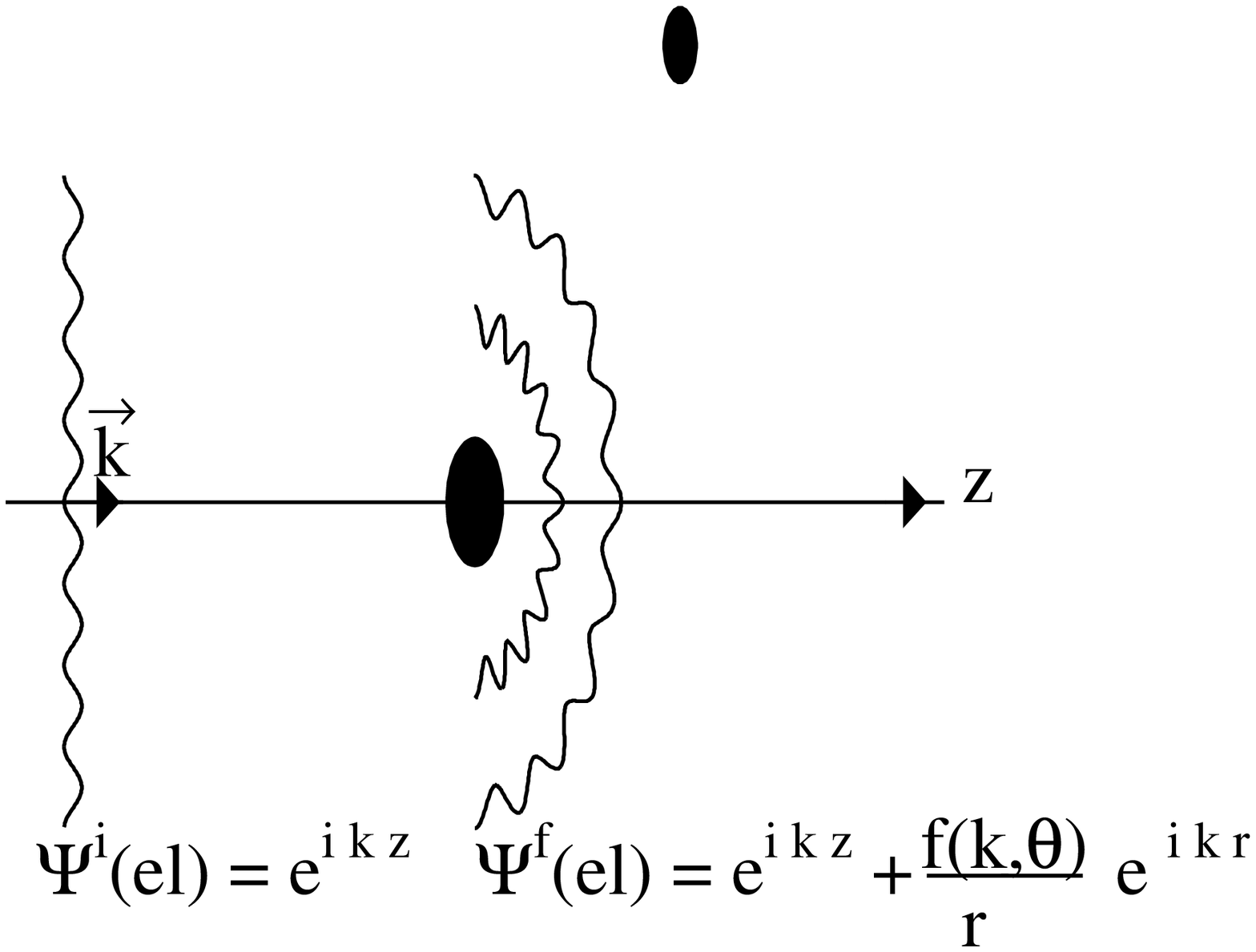, width=100mm}}\\
\centerline{\epsfig{file=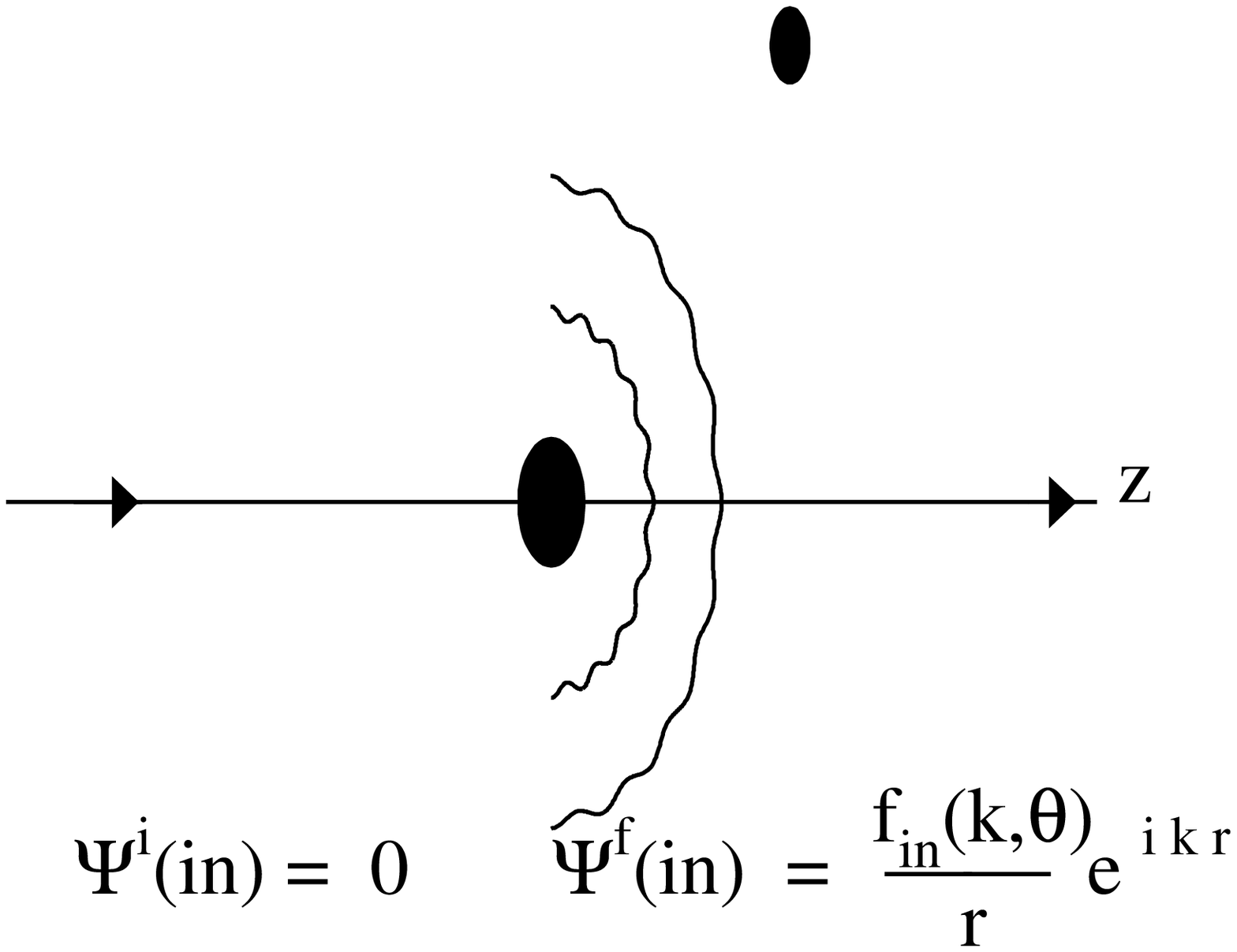, width=100mm}}\\
\end{tabular}
\caption{{\it Elastic and inelastic  scattering.}}
\label{fig5}
\end{figure}

For  inelastic ( production ) processes we have a totally different
situation:
$$
\Psi^i_{in} ( t < t_{interaction})\,\,=\,\,0
$$
  after the  collision there is a wave function, namely, a spherical
outgoing wave:
$$
\Psi^i_{in} ( t > t_{interaction})\,\,=\,\,\frac{f_{in}
(k,\theta)}{r}\,\,e^{i k r}\,\,.
$$
It is obvious that we have to add the inelastic channel to  the r.h.s. of
the unitarity constraint which reads 
\beq \label{UN}
Im f^{el}_l\,\,=\,\,k \,| f^{el}_l|^2\,\,+\,\,\Phi_i
f^{in}_{l;i}f^{*\,in}_{l;i}\,\,
\eeq 
\begin{figure}[htbp]
\centerline{\epsfig{file=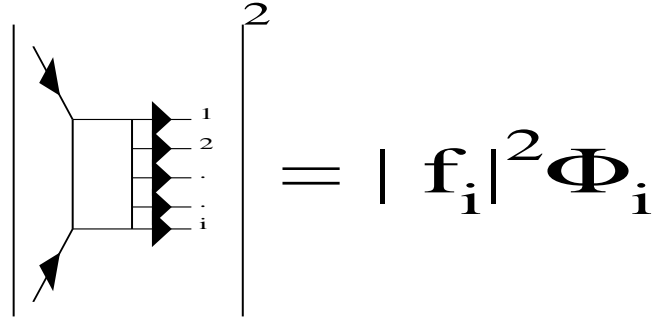, width=100mm,height= 60mm}}  
\caption{{\it The  inelastic contribution to the unitarity constraint.}}
\label{fig6}
\end{figure}

For the Breit - Wigner resonance the full unitarity constraint of
\eq{UN} has a simple graphical form ( see Fig.7).
\begin{figure}[h]
\centerline{\epsfig{file=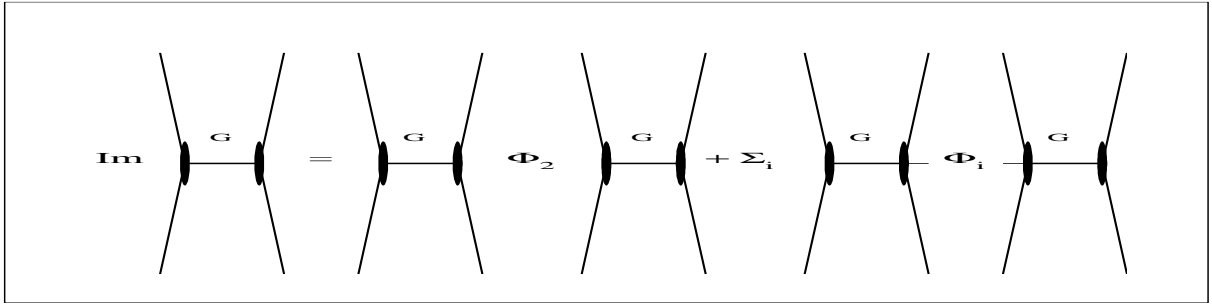, width=160mm, height= 40mm}}  
\caption{{\it The   unitarity constraint for  Breit - Wigner
resonances.}}
\label{fig7}
\end{figure}

Introducing the  notation $g_i g_i = \g_i$ and using the Breit - Wigner 
form for the propagator $G$ 
$$
G\, \,=\,\,\frac{1}{k_M \,(s \,\,- \,\,M^2 \,\,-\,\,i \,\Gamma_{tot}\,)}
$$
we can  rewrite the unitarity constraint in the form:
\beq \label{UNBW}
\Gamma_{tot}\,\,=\,\,k_M \,\g_{el} \,\,+\,\,\sum \gamma_i \Phi_i\,\,=\,\,
\Gamma_{el}\,\,+\,\,\sum \Gamma_i\,\,.
\eeq
\subsection{Unitarity at high energies.}

$\bullet$ {\bf  Elastic scattering.}

High energy means that
$$ k a \,\,\gg\,\,1$$
where $a$ is the typical size of the interaction.
\begin{figure}[htbp]
\centerline{\epsfig{file=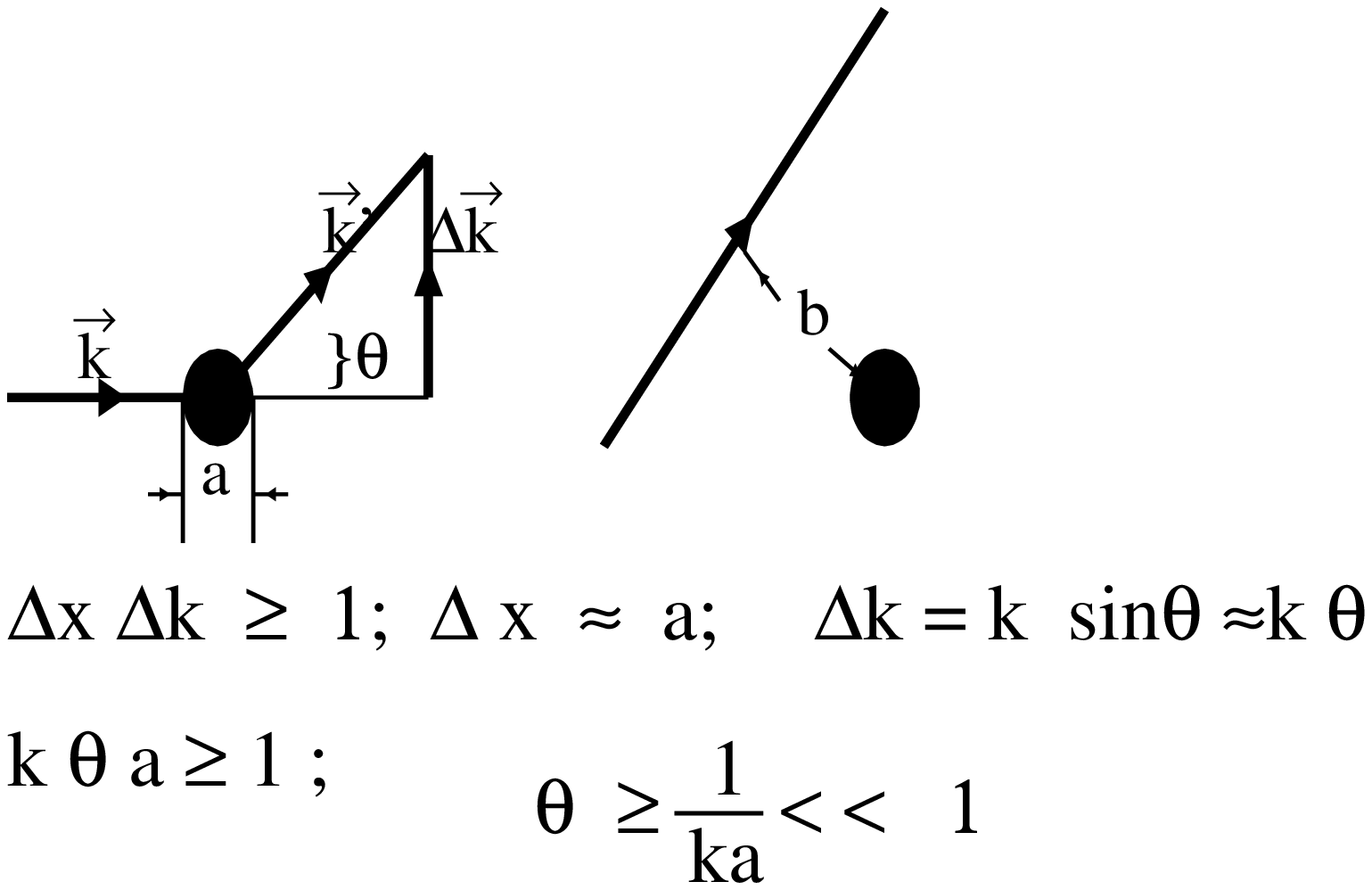, width=140mm}}  
\caption{{\it Useful kinematic variables and  relations for high energy
scattering.}}
\label{fig8}
\end{figure}
It is very useful to introduce the impact parameter $b$ ( very often 
we denote it as $b_t$) by  $l = k b$. The meaning of $b$ is quite clear
from the classical formula for the angular moment $l$ and from Fig.8.
The  new variable  $b$ will be of the order of
$a$  ( $b\,\approx\,a$ ) while $l$ itself will be very large at high
energy. Following this expectation, we  rewrite the scattering amplitude
in the form:
\beq \label{HEFL}
f\,\,=\,\,\sum^{\infty}_{l = 0} f_l(k) ( 2l + 1) P_l( cos \theta)\,\,=\,\,
\eeq
$$
=\,\,\sum^{\infty}_{l = 0} f_l(k) ( 2l + 1) J_0(l \theta)\,\,\rightarrow\,
2 \int l dl f_l J_0 ( l \theta)\,\,.
$$
Here we made  use of  the following property of $P_l$:
$$
P_l (cos \frac{\theta l}{l})\,\,\rightarrow \,\,J_0 ( \theta l )\,\,,
$$
for  $\theta l \,\,\approx\,\,1$. $J_0(z)$ is the Bessel function of the
zero order.
 Then  $$\theta l\,=\,k b \theta
\,\,\approx\,\,\frac{k b}{k a} \,\,=\,\,\frac{b}{a}\,\,.
$$

Now let us introduce a new notation:

1.$ t$\,\,=\,\,$ ( p_1 - p'_1)^2$ = $( p_{10} - {p'}_{10} )^2 \,-
\,2\,p^2_1 (
1 - cos \theta )\,\,\rightarrow\,\, - k^2 \theta^2$ at high energy;

2.  $ - t \,=\,q^2_{\perp}$\,;

3.  $A\,\,=\,\,\frac{f}{k}$ , this is a new definition for the scattering
amplitude

4. Instead of the partial amplitude $f_l (k)$  we introduce $a(b,k)\,=\,
2 f_l k$\,;

5. $ \int d^2 b \,=\,\int b d b d \phi \,=\,\int 2\pi b db$\,.

In the new notation the scattering amplitude can be written in the form:
\beq \label{BR}
A(s,t)\,\,=\,\,\frac{1}{2 \pi} \int \,\,a(b,s)\,d^2 b \,e^{ i \vec{b}
\cdot
\vec{q}_{\perp}}\,\,,
\eeq
where $ \vec{q}_{\perp} \,=\,\vec{p_1}\,-\,\vec{p'_1} $.

\subsection{ Unitarity constraint in impact parameter ( b ).}
The {\bf elastic unitarity} for the new amplitude looks as follows:
\begin{large}
\beq \label{ELUNB}
2 \,Im \,a_{el}(s,b)\,\,=\,\,| a_{el}(s,b)|^2
\,\,.
\eeq
\end{large}
The generalization of \eq{ELUNB}  for the case of inelastic interactions
is obvious and it leads to
\begin{Large}
\beq \label{UNB}
2 \,Im \,a_{el}(s,b)\,\,=\,\,| a_{el}(s,b)|^2
\,\,+\,\,G_{in}(s,b)\,\,,
\eeq
\end{Large}   
where $G_{in}$ is the sum of all inelastic channels. This equation is our
master equation which we will use frequently   during these lectures.
 Let us discuss once more  the normalization of our amplitude.
 
 The scattering amplitude in $ b $-space is defined as
 \begin{equation}
 a_{el}(s,b) = \frac{1}{2 \pi} \int d^2 q\,\,
e^{-i \vec{q_{\perp}}\cdot\vec{ b}}\,\,
  A(s,t)
\end{equation}
where $ t= - q^{2}_{\perp} $ .
The inverse transform  for $A(s,t)$ is
\beq
A(s,t)\,\,=\,\,\frac{1}{2 \pi} \int a_{el}(s,b)\,\,d^2 b\,\,
e^{i\vec{q}_{\perp}\cdot\vec{b}}\,\,.
\eeq
 In this representation
 \begin{equation}
 \sigma_{tot} = 2 \int d^2 b \,\, Im a_{el}(s,b)\,\,;
\end{equation}
 \begin{equation}
 \sigma_{el} = \int d^2 b\,\, \vert a_{el}(s,b) \vert^{2}\,\,.
\end{equation}
\subsection{A general solution to the unitarity constraint.}
Our master equation ( see \eq{UNB} ) has the general solution 
\beq \label{SLTN}
G_{in} (s, b)\,\,=\,\,1\,\,-\,\,e^{- \,\Omega(s,b)}\,\,;
\eeq
$$
a_{el}\,\,=\,\,i\,\{\,1 \,\,-\,\,e^{ - \,\frac{\Omega(s,b)}{2}\,+\,i
\,\chi(s,b)}\,\}\,\,;
$$
where opacity $\Omega$ and phase $\chi = 2 \delta (s,b)$ are real
arbitrary functions.

The algebraic operations, which help  to check that \eq{SLTN} gives the
solution, are:
$$
Im\, a_{el}\,\,=\,\, 1 \,\,-\,\,e^{-\,\frac{\Omega}{2}} \,cos \chi\,\,;
$$
$$
\vert a_{el} \vert^2\,\,=
\,\,( \, 1 \,\,-\,\,e^{-\,\frac{\Omega}{2} \,+\,i\,\chi} \,)\cdot
( \, 1 \,\,-\,\,e^{-\,\frac{\Omega}{2} \,-\,i\,\chi} \,)\,\,=\,\,
1 \,\,-\,2\,e^{-\,\frac{\Omega}{2}}\,cos \chi\,\,+\,\,e^{ - \Omega}\,\,.
$$
The opacity $\Omega$ has a clear physical meaning, namely  $e^{-
\Omega}$ is the probability to have no inelastic interactions with the
target.

One can check that \eq{UNB} in the limit, when $\Omega$ is  small and the
inelastic processes  can be neglected, describes the well known solution
for the elastic scattering: phase analysis. For high energies the most
reasonable assumption  just opposite, namely,  the real part of the
elastic amplitude to be  very small. It means that $\chi\,\rightarrow\,0$
and  the general solution is of   the form:
\beq \label{GSHE} 
G_{in} (s, b)\,\,=\,\,1\,\,-\,\,e^{- \,\Omega(s,b)}\,\,;
\eeq
$$
a_{el}\,\,=\,\,i\,\{\,1 \,\,-\,\,e^{ - \,\frac{\Omega(s,b)}{2}}\,\}\,\,.
$$
We will use this solution to  the end of our lectures. At the moment, I do
not want to discuss the theoretical reason why the real part should be
small at high energy . I prefer to claim that this is a general feature
of all experimental data at high energy.
 \section{ The great theorems.}
Now, I would like to show you how one can prove the great theorems just
from
the unitarity and analyticity. Unfortunately, I have no time to prove all
three theorems but I hope that you will do it yourselves following  my
comments. 

\subsection { Optical Theorem.}
 \par 
 The optical theorem gives us the relationship between the
behaviour
of the imaginary part of the scattering amplitude at zero scattering angle and
the total cross section that can be measured experimentally.
It follows directly from \eq{UNB}, after  integration  over $b$.
Indeed,
\beq \label{OT}
4\pi  Im A(s,t =0)\,\,=\,\,\int 2 Im a_{el} (s,b) \,d^2 b\,\,=
\eeq
$$
\,\,\int
\,d^2 b \{\,| a_{el}(s,b)|^2
\,+\,G_{in}(s,b)\,\,=\,\,\s^{el} \,+\,\s^{in}\,\,=\,\,\s_{tot}\,\,.
$$

\subsection { The Froissart boundary.}

We call the Froissart boundary the following limit of the energy growth of
the total cross section:
\beq
\sigma_{tot} \,\,\,\leq\,\,\,C\,\,ln^2 s
\eeq
where $s$ is the total c.m. energy  squared of our elastic reaction:
$a(p_a ) + b (p_b)
 \rightarrow a + b$, namely $s \,=\, (p_a \,+\,p_b )^2$.
The coefficient $C$ has been calculated but we do not need to know  its
exact
value.
What is really important is the fact that $ C \,\propto \frac{1}{k^2_t}$, where
$k_t$ is the minimum  transverse momentum for the reaction under study.
Since
 the minimum  mass in the hadron spectrum is the pion mass the
Froissart
theorem predicts  that $ C \,\propto \,\frac{1}{m^2_{\pi}}$. The exact
calculation
 gives $C \,= 60 mb$.

I think, it is very instructive to discuss a proof of the Froissart
theorem.
 As we have mentioned the total cross section can be expressed
through the opacity $\Omega$ due to the unitarity constraint. Indeed,
\beq \label{FT1}
\s_{tot}\,\,=\,\,2\,\int \,d^2 b \,\{\,1\,\,-\,\,e^{ -
\frac{\Omega(s,b)}{2}}\,\}\,\,.
\eeq
For small $\Omega$ \eq{FT1} gives
$$\s_{tot}\,\,\rightarrow\,\,\int \,d^2 b\,\Omega(s,b)\,\,.$$

It turns out that we know the $b$ - behaviour of $\Omega$ at large $b$.
Let us consider first the simple example  : the exchange of a  vector
particle (see Fig.9). 
\begin{figure}
\centerline{\epsfig{file=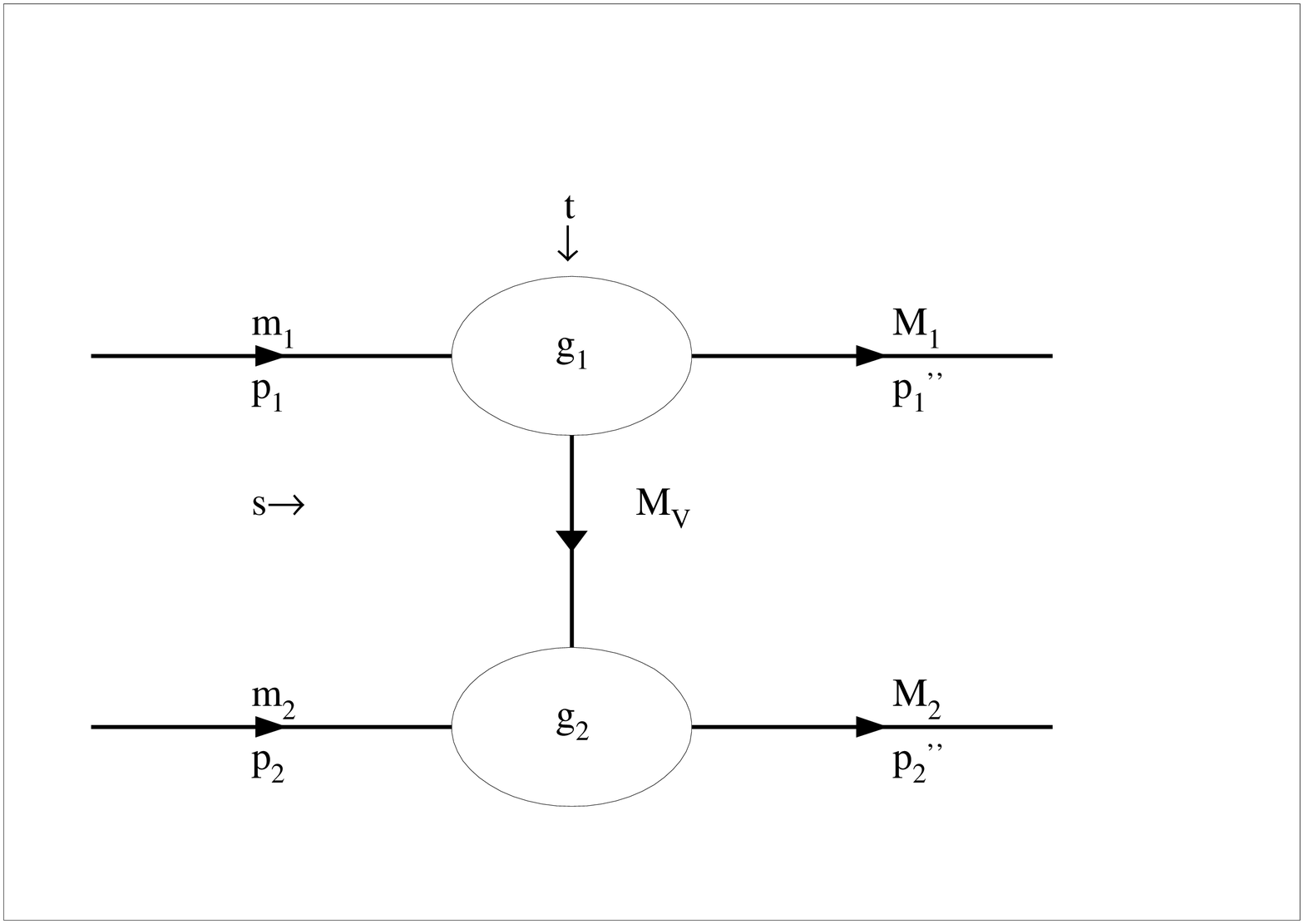, width=100mm}}
\caption{{\it Exchange of particle with mass $M_V$.}}
\label{fig9}
\end{figure} 

Introducing Sudakov variables we can expand each vector in the following
form:

$
q\,\,=\,\,\alpha_q\,p'_1\,\,+\,\,\beta_q\,p'_2\,\,+\,\,q_t
$

where
$p'^2_1\,=\,p'^2_2\,=\,0$

and

$
p_1\,\,=\,\,p'_1\,\,+\,\,\beta_1\,p'_2\,\,\,\,;\,\,\,\,
$

$
p_2\,\,=\,\,\alpha_2\,p'_1\,\,+\,\,\,p'_2\,\,\,\,;
$

It is easy to find that

$$
\beta_1\,\,=\,\,\frac{m^2_1}{s}\,\,\,\,;\,\,\,\,\alpha_1\,\,=\,\,\frac{m^2_2}{s}
$$

since

$
p^2_1\,\,=\,\,m^2_1\,\,=\,\,2\,\beta_1 p'_1 \cdot p'_2\,\,=\,\,\beta_1 \,s   
$  
and $
s = ( p_1\,+\,p_2 )^2 =  (\, ( 1 + \alpha_2 )\, p'_{1 \mu}\, +\,( 1 +
 \beta_1 )\,p'_{2\mu}\,)^2\,\approx\,2\, p'_{1\mu}\,p'_{2 \mu}\,\, ;$

From the  equations

$  
p"^2_1\,\,=\,\,M^2_1\,\,\,\,; \,\,\,\,p"^2_2\,\,=\,\,M^2_2
$

we can find the values of $\alpha_q$ and $\beta_q$:
$$
p"^2_1\,\,=\,\,( p_1 - q )^2\,=\,(\, ( 1 - \alpha_q) p'_1\,+\,(\beta_1 -
\beta_q)
p'_2\,)^2\,=\,(1 - \alpha_q ) (\beta_1 - \beta_q )s\,-\,q^2_t\,=\,M^2_1
$$

Therefore
\beq
\alpha_q\,=\,\frac{M^2_2 - m^2_2}{s}\,\,\,;\,\,\,\beta_q\,=\,
\frac{m^2_1 - M^2_1}{s}
\eeq
 The above equations   lead to
\beq
|q^2|\,\,=\,\,|\alpha_q \beta_q s - q^2_t |\,=\,
\frac{( M^2_2 - m^2_2) (M^2_1 - m^2_1 )}{s}\,+\,q^2_t\,\,\rightarrow |_{s
\ra \infty}\,\,q^2_t
\eeq
Now, we are ready  to write the  expression for the diagram of Fig.9:
\beq
A(s,t)\,\,=\,\,g_1\,g_2 \frac{( p_1 + p'_1 )_{\mu} ( p_2 +p'_2 )_{\mu}}
{ q^2_t + m^2_V}\,=\,g_1\,g_2 \,\frac{4 s}{q^2_t + m^2_V}
\eeq
where $m_V$ is the mass of the vector meson.

Now, we are able to calculate the amplitude as a function of $b$:
\beq \label{BEX}
a_{el} (s,b)\,\,=\,\,g_1\,\, g_2\,\, 4\, s\,\, \int \frac{J_0 (b q_t) q_t
d
q_t}{q^2_t\,\,+\,\,m^2_V}\,\,=
\eeq
$$
=\,\,g_1 g_2 4s K_0 (b m_V)\,\,\rightarrow\,\vert_{b
\,\rightarrow \,\infty}\,\,g_1\,g_2\,4s\,\frac{e^{ - m_V\,b}}{\sqrt{m_V
b}}
$$

This   calculation can be used in more general form  based on  the
dispersion relation for the amplitude A(s,t)
$$
A(s,t)\,\,=\,\,\frac{1}{\pi} \int^{\infty}_{4 \mu^2} \,\,\frac{Im_t
A(s,t')}{ t'\,\,-\,\,t}\,\,d t\,\,,
$$
where $\mu$ is the mass of the lightest hadron (pion). Using $t\,=\,-
q^2_t$ we calculate $a_{el} (s,b)$
\beq \label{BGEN}
a_{el}(s,b)\,\,=\,\,\frac{1}{\pi}\, \int^{\infty}_{4 \mu^2}
\,\,d t' \int q_t d q_t  J_0 (b q_t)\frac{Im_t
A(s,t')}{ t'\,\,+\,\,q^2_t}\,\,\rightarrow \,\vert_{b
\,\ll\,\mu}\,\,e^{- 2 \mu b} 
\eeq

{\bf Assumption:} $ a_{el}(s,b)\,\,<\,\,s^N e^{- 2\mu b}$ at large values
of $b$.

As we have discussed if $a_{el}(s,b)\,\ll\,1$ opacity $\Omega\,=\,a_{el}$.
Using the above assumption we can estimate the value of $b_0$ such that
$\Omega \,\ll\,1 $ for $b \,>\,b_0$.
For $b_0$ we have
$$
s^N e^{- 2 \mu b_0}\,\,=\,\,1
$$
with the solution for $b_0$
\beq \label{B0}
b_0\,\,=\,\,\frac{N}{2 \mu} \,\ln s\,\,.
\eeq

Eq.(29) can be  integrated  over $b$ by  dividing the integral
into  two parts $b >b_0$ and $b < b_0$. Neglecting  the second part of the
integral where $\Omega$ is very small yields 
$$
\s_{tot}\,\,=\,\,4\,\pi \int^{b_0}_0 \,b db \,
[\,1\,\,-\,\,e^{-\frac{\Omega(s,b)}{2}}\,]\,\,+\,\,4\,\pi
\int^{\infty}_{b_0}
\,b db \,
[\,1\,\,-\,\,e^{-\frac{\Omega(s,b)}{2}}\,]\,\,<
$$

$$
<\,\,4 \pi \int^{b_0}_0 b d b\,\,=\,\,2\,\pi \,b^2_0\,\,=\,\,\frac{2 \pi
N^2}{4 \,\mu^2}\,\ln^2 s\,\,.
$$ 
This is the Froissart boundary. Actually, the value of $N$ can be
calculated but it is not important. Indeed, inserting all
numbers we have for the Froissart bound 
$$
\s_{tot}\,\,<\,\,N^2 \,30 \,mb\,\ln^2 s\,\,.
$$ 
Because of the large coefficient in front of $\ln^2 s$ this bound has
no practical application. What is really important for  understanding of
the
high energy behaviour of the total cross section is the logic of
derivation and the way how the unitarity constraint has been used.

\subsection { The Pomeranchuk Theorem.}

The Pomeranchuk theorem is the manifestation of the { \bf crossing symmetry},
 which
can be formulated in the following form: one analytic function of two
variables $s$ and $t$ describes the scattering amplitude of
 two different reactions $ a + b \,\rightarrow
a + b$  at $s > 0 $ and $t < 0 $ as well as $ \bar a + b \,\rightarrow \,
\bar a + b $ at $ s < 0$\, $  (\,u = (p_{\bar a} + p_b )^2 > 0\,) $ and $
t < 0$.

The Pomeranchuk theorem says that the total cross sections of the above two
 reactions should be equal to each other at high energy
if the real part of the amplitude
is smaller than  imaginary  part.

To prove the Pomeranchuk theorem we need to use the dispersion relation
for the elastic amplitude at $t = 0$. 
$$
A(s,t=0)\,\,=\,\,\frac{1}{\pi}\,\,\{\,\int \,\,\frac{Im_s
A(s',t=0)}{s'\,\,-\,\,s}\,\,+\,\,\int \,\,\frac{Im_u
A(u',t=0)}{u'\,\,-\,\,u}\,\}\,\,.
$$
Using the optical theorem we have:
$$
Im_s A(s,t = 0)\,\,=\,\,\frac{s}{4 \pi} \s_{tot}( a + b )\,\,;
$$

$$
Im_u A(s,t = 0)\,\,=\,\,\frac{u}{4 \pi} \s_{tot}( \bar a + b )\,\,.
$$
Recalling that at large $s$\,\, $ u \,\rightarrow \,-\,s$ we have
$$
A(s,t = 0)\,\,=\,\,\frac{1}{4 \pi^2} \,\int\,s'\,d s'\,\{\,\frac{\s_{tot}
(a +
b)}{s'\,-\,s}\,\,+\,\,\frac{\s_{tot} (\bar a + b)}{s'\,+\,s}\,\}\,\,.
$$
Since  $\s_{tot}$ can rise as $\ln^2 s$ we have to make one substraction
in the dispersion relation. Finally, we obtain
\beq \label{DRPO}
A(s,t = 0)\,\,=\,\,\frac{s}{4 \pi^2} \,\int\,s'\,d s'\,\{\,\frac{\s_{tot}
(a +    
b)}{s'(\,s'\,-\,s\,)}\,\,-\,\,\frac{\s_{tot} (\bar a +
b)}{s'(\,s'\,+\,s\,)}\,\}\,\,.
\eeq
To illustrate the final step in our proof let us assume that at high
 $s'$\,\, $\s_{tot}( a + b )\,\rightarrow \,\s_0(a + b ) \ln^{\g}s$ and
$\s_{tot}(\bar  a + b )\,\rightarrow \,\s_0(\bar a + b ) \ln^{\g}s$.
Substituting these expressions into  \eq{DRPO}, we obtain
$$
Re A (s, t = 0 )\,\,\rightarrow\,\,\frac{\s_0(a + b )\,-\,\s_0(\bar a + b
)}{\g\,+\,1}\,\ln^{\g + 1}s\,\,\gg\,\,Im A(s,t = 0)\,\,.
$$
This is in contradiction with the unitarity constraint which has no
solution   if $Re A $ increases with energy and  is bigger than $Im A$.
Therefore, the only way out of this contradiction is to assume that
\beq \label{POM}
\s_{tot} ( a\,+\,b )\,\,=\,\,\s_{tot} (\bar  a\,+\,b
)\,\,\,\,at\,\,\,\,s\,\rightarrow\,\,\infty\,\,.
\eeq

\subsection{ An instructive model:  the `` black disc " approach.}

Let us assume that 
$$
\Omega\,\,=\,\,\infty\,\,\,\,at\,\,\,\,b\,\,<\,\,R(s)\,\,;
$$
$$
\Omega\,\,=\,\,0\,\,\,\,at\,\,\,\,b\,\,>\,\,R(s)\,\,.
$$
The radius $R$ can be a function of energy and it can rise as
$R\,\,\approx\,\,\ln s$ due to the Froissart theorem. 

As you may  guess from the assumptions, this ``black" disc model is just a
rough realization of what we expect in the Froissart limit at ultra high
energies. 

It is easy to see that the general solution of the unitarity constraints 
simplifies  to
$$
a(s,b)_{el}\,\,=\,\,i\,\Theta( \,R\,-\,b\,)\,\,;
$$
$$
G_{in} (s,b)\,\,=\,\,\Theta( \,R\,-\,b\,)\,\,;
$$
which leads to
$$
\s_{in} \,\,=\,\,\int d^2 b G_{in}\,\,=\,\,\pi\,R^2\,\,;
$$
$$
\s_{el} \,\,=\,\,\int d^2 b \,\vert a_{el} \vert^2 \,\,=\,\,\pi \,R^2\,\,;
$$
$$
\s_{tot}\,\,=\,\,\s_{el}\,\,+\,\,\s_{in}\,\,=\,\,2\,\pi\,R^2\,\,;
$$
$$
A(s,t)\,\,=\,\,\frac{i}{2 \pi}\,\int\,\,d^2 b \Theta( \,R\,-\,b\,)
\,e^{i \vec{q_t}\cdot\vec{b}}\,\,=\,\,i \int^R_0 b d b J_0 (b
q_t)\,\,=\,\,i \frac{R}{q_t}\,J_1 (q_t R)\,\,;
$$
$$
\frac{d \s}{d t}\,\,=\,\,\pi \vert A \vert^2 \,=\,\pi \frac{J^2_1(  R
\sqrt{|t|})}{|t|}\,\,.
$$

{\bf EXPERIMENT:}
The experimentalists used to present their data on $t$ - dependence in the
exponential form, namely
$$
Ra \,=\,\frac{\frac{d \s}{d t}}{\frac{d \s}{d t}\,\vert_{t =
0}}\,\,=\,\,e^{- B
|t|}
$$
Comparing this parameterization of the experimental data with the
behaviour of the  ``black
disc" model  at small $t$, namely,
$\frac{d \s}{d t}\,=\,\frac{\pi R^2}{4}\,(\,1\,-\,\frac{R^2}{2} \,|t|\,)$,
 leads to  $B\,=\,R^2/2$.

From experiment we have for proton - proton collision $B \,=\,12 GeV^{-2}$
and, therefore, $R = 5  GeV^{-1} =  1  fm$. The  $t$ - distribution
for  this
radius in the `` black disc " model is given in Fig.10.
The value of the total cross section which corresponds to this value of 
 the radius is
$80
mb$ while the experimental one is one half of this value , $\s_{tot}(exp)
= 40 mb$.
In Figs. 11 and 12 $b$,$\s_{el}, \s_{tot}$ and $\frac{d \s}{d t}$
are given as function of energy. One can see that experimentally
$\s_{el}/\s_{tot}\,\approx\,0.1 - 0.15$ while the ``black disc" model
predicts 0.5. Experimentally, one finds  a diffractive structure in the 
$t$
distribution but the ratio of the cross sections in the second and  the
first maximum is about $10^{-2}$ in  the model and much smaller
($\approx\,10^{-4}$) experimentally.

{\it Conclusions:} In spite of the fact that  the ``black disc" model 
predicts all qualitative features of the experimental data the quantative
comparison with the experimental data shows that the story is much more
complicated than this simple model. Nevertheless, it is useful to have
this model in front of your eyes in all further attempts to discuss the
 asymptotic behaviour at high energies.  Nevertheless  the
``black" disc
model produces  all qualitative features of the experimental high energy
cross
sections, furthermore  the errors  in the numerical evaluation is only
200 -
300 \%.  This model certainly is not  good  but  it is not so  bad as
it could be. Therefore, in spite of the fact that the Froissart limit will
be never reached, the physics of ultra  high energies is not
too far away from the experimental data. Let us remember this for future
and move on  to understand number of  puzzling problems.
\begin{figure}[p]
\centerline{\epsfig{file=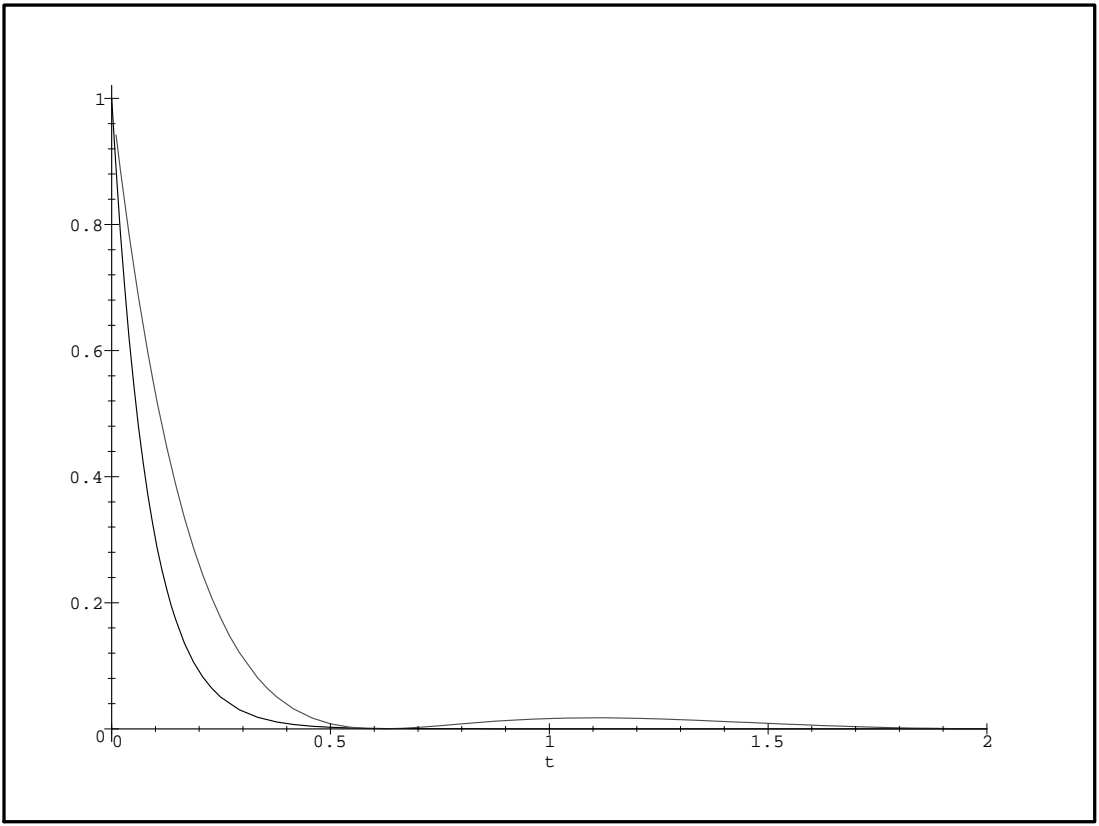,angle=-90, width=160mm}}  
\caption{{\it The $t$ - dependence of the  ratio $Ra $ for the ``black"
disc
model ( dashed line ) and for the exponential parameterization ( solid
line ). }}
\label{fig10}
\end{figure}
\newpage

\alphfig
\begin{figure}[p]
\centerline{\epsfig{file=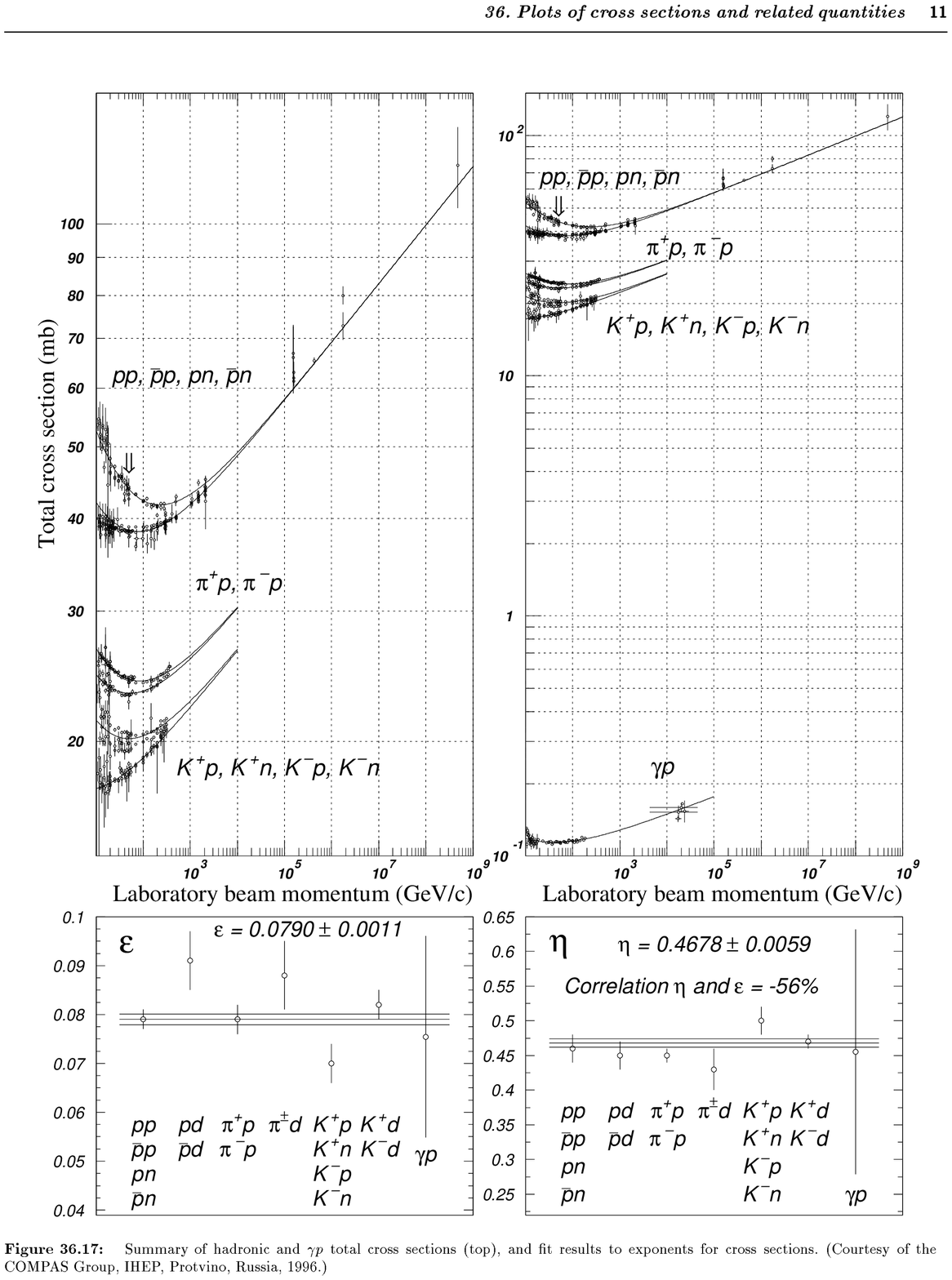, width=160mm}}
\caption{{\it The experimental data on the total cross
section ( Particle Data Group 1996 ) .}}
\label{fig11a}
\end{figure}
\begin{figure}[p]
\centerline{\epsfig{file=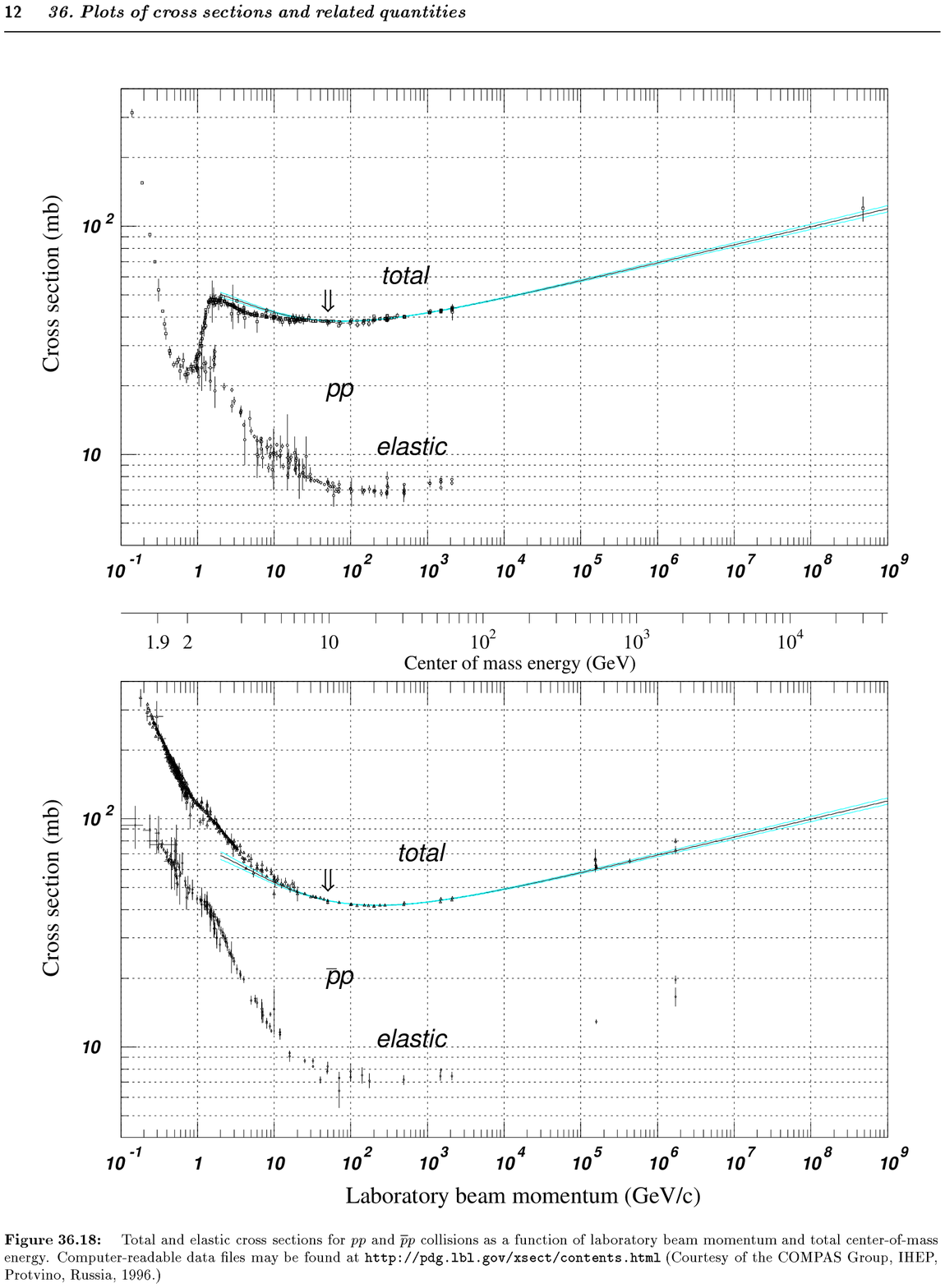, width=160mm}}
\caption{{\it The experimental data on total and elastic
cross sections ( Particle Data Group 1996 ).}}  
\label{fig11b}
\end{figure}
\resetfig

\newpage
\begin{figure}[htbp]
\centerline{\epsfig{file=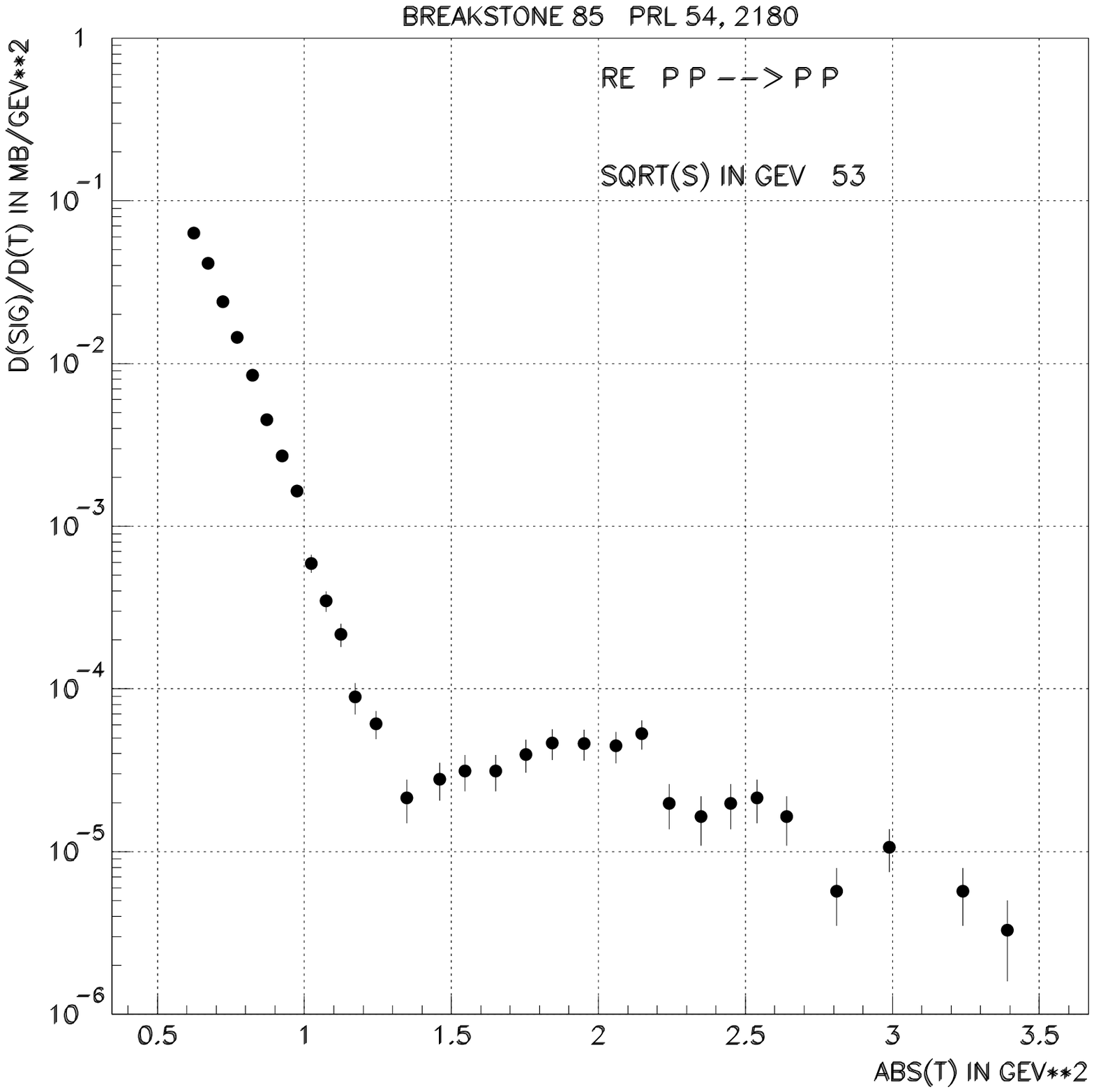, width=160mm}}  
\caption{{\it Differential elastic cross section.}}
\label{fig12}
\end{figure}
\newpage
\section{ The first puzzle and Reggeons.}
\subsection{The first puzzle.}
The first puzzle can be formulated as a question:{\it What happens in
t- channel exchange of
 resonances for which  the  spin  is bigger than 1 ?"}
On the one hand such resonances have been observed experimentally, on the
other
hand the exchange of the resonances with spin $j$ leads to the scattering
 amplitudes to be  proportional to $s^j$ where $s$ is the energy of two
colliding hadrons. Such a behaviour contradicts the Froissart bound.
It means that we have to find the theoretical solution to this problem.

 We have considered the exchange of a vector particle and saw that
it gives an  amplitude which grows as $s$.
Actually, using this  example, it is easy to  show that the exchange of
 a
resonance
with spin $j$ gives rise to a  scattering amplitude of the form
\beq \label{RE}
A^{res}( s, t= - q^2_t )\,=\,g_1\,g_2\,\frac{ ( 4s)^j}{ q^2_t +
m^2_R}\,\,.
\eeq
Indeed, the amplitude for a  resonance in the  $t$-channel (for the
reaction 
with energy $\sqrt{t} $ ) is equal to
\beq \label{REX}
A^{res} (s,t)\,\,=\,\, (2 j + 1) g_1(t) g_2(t) \frac{P_j (z)}{ t - M^2_R -
i\Gamma}\,\,,  
\eeq
As we know $g_i$ has an additional kinematic factor $k^j$ where $k$ is
momentum of decay  particles measured in the frame with c.m. energy 
$\sqrt{t}$. $z$ is the  cosine
of the scattering angle in the same frame. To find the high energy
asymptotic behaviour we need to continue this expression into  the
scattering
kinematic region where $t < 0$ and $z\,\rightarrow\,\frac{s}{2
k^2}\,\gg\,\,1$. This  is easy to do, if we notice that $P_j
(z)\,\rightarrow
\,z^j$ at $z \,\gg\,1$ and all non- analytical factors ( of $k^j$ - type ) 
cancel in \eq{REX}. Recalling that $t\,\rightarrow\, - q^2_t$ at high
energy we obtain \eq{RE}, neglecting the width of the resonance in 
first approximation. In terms of  $b$ \eq{REX} looks as follows:
$$ A\,\,\propto\,\, s^j \, exp ( - m_R b )$$ at
large values of $b$ and $s$.

Of course, the exchange of a single  resonance gives a real amplitude
and,
therefore, does not contribute to the cross section since the total cross 
section is proportional to the imaginary part of the amlitude accordingly 
to the optical theorem. To calculate
the imaginary part of the scattering amplitude we can use the unitarity
constraint, namely
\beq \label{UNEX}
Im A(s,t = 0)\,\,=\,\,\int \,d^2 q \delta({p"}^2 - M^2_1) \delta( {p"}^2_2
 - M^2_2)\,\,\vert A^{res} (s,q^2) \vert^2\,\,.
\eeq
Recalling that the phase space factor is $d^4 q \,\,=\,\,\frac{|s|}{2} d
\alpha_q
\beta_q d^2 q_t$ where the  values of $\beta_q $ and $\alpha_q$ given in
Eq.(30) ( all other notations are clear from Fig.9),
we can perform  the integration  in \eq{UNEX}. It is clear that at high
energy
$$
Im A(s,t = 0)\,\,\propto\,\,s^{2 j\,-\,1}\,\,.
$$
This equation can be easily generalize for the case of the contribution to 
  the total cross section  due to exchange of two particles
with spin $j_1 $ and $j_2$, namely
\beq \label{CREX}
\s_{tot}\,\,\propto\,\,s^{j_1\,+\,j_2\,-\,2}\,\,.
\eeq
This formula is very instructive when we will discuss the contribution of
different exchange at high energies. From \eq{CREX} it is seen that the
exchange of the resonance with spin bigger than 1 leads to a  definite
violation of the Froissart theorem.

\subsection{Reggeons - solutions to the first puzzle.}
The solution to the first puzzle is as follows. It turns out that
when considering the exchange of a resonance with spin $j$ one has to
include also all exitation with spin $j + 2$, $j + 4$, ...  (keeping all
other quantum numbers the same). These particles lie on a Regge trajectory
$\alpha_R(t) $ with $\alpha_R( t  = M^2_j ) = j $. The contribution to the
scattering amplitude of
  the exchange
of all resonances can be described as an exchange of
the new object - Reggeon and its contribution to the scattering
amplitude is given by the simple function:
\beq \label{RGEX}
A_R (s,t)\,\,=\,\,g_1(m_1,M_1, t) \,g_2( m_2,M_2,t) \cdot \frac{ s^{\alpha(t)}
\,\pm \,( - s )^{\alpha (t)}}{ sin \pi \alpha(t)}
\eeq
$\alpha (t)$ is a function of the momentum transfer which we call the Reggeon
 trajectory.
\begin{figure}[htbp]
\centerline{\epsfig{file=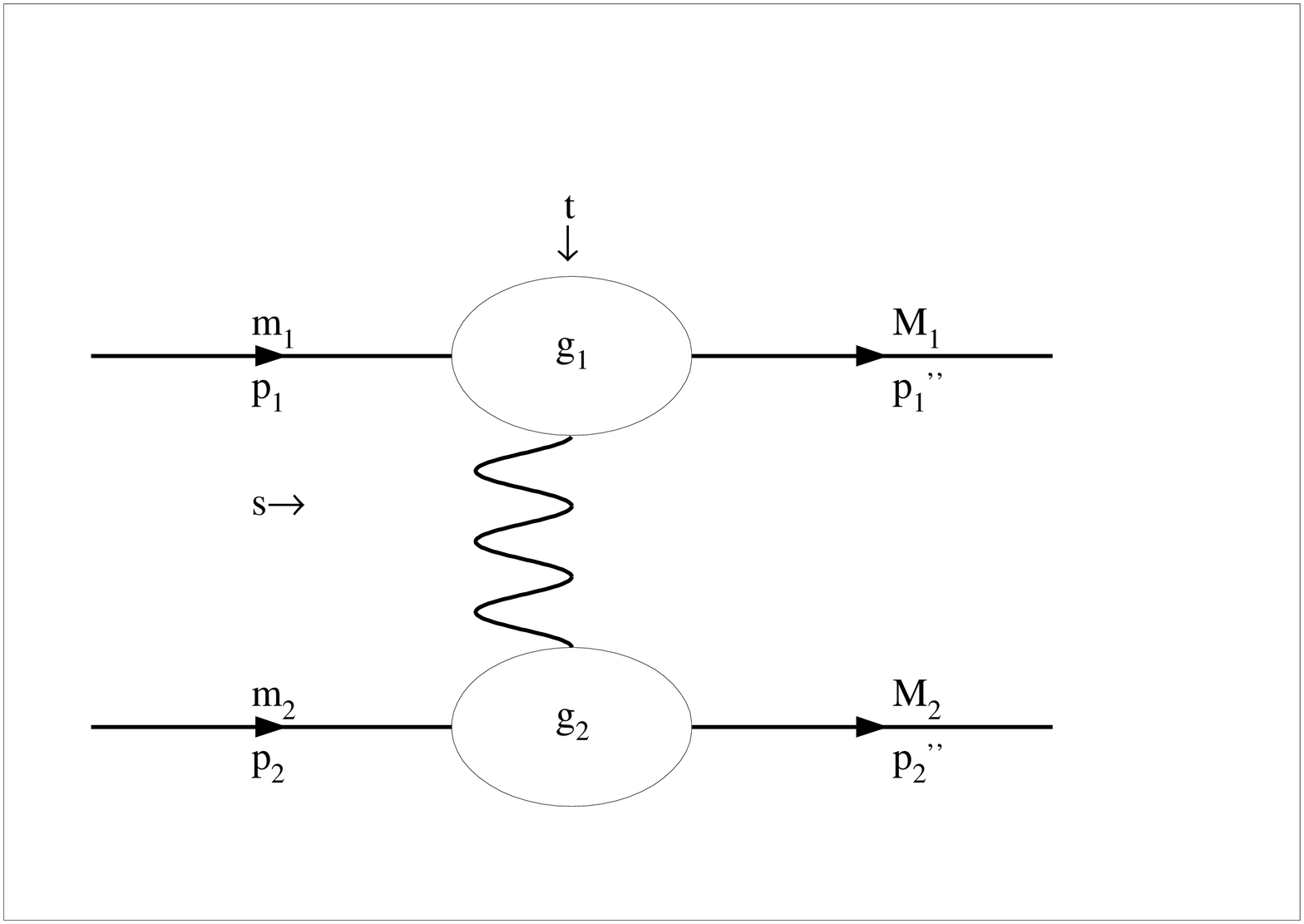,width=80mm}}
\caption{{\it The Reggeon exchange.}}
\label{fig13}
\end{figure}

 The name of the new object as well as the form of the amplitude
came from the analysis of the properties of the scattering amplitude in
the $t$
 channel using the  angular momentum representation. For the following
one does not need to follow the full historical development. We need only 
to understand  the main properties of the above function which
plays
 a crucial role in the theory and phenomenology of  high energy
 interactions.

\section{The main properties of the Reggeon exchange.}
\subsection{Analyticity.}
First, let me recall that function $( - s )^{\alpha(t)}$  is an analytical
 function of the complex variable $s$ with the cut starting from $s$ = 0 
and going to infinity ( $ + \infty $ ) along the $x$ - axis.
One can calculate the discontinuaty along this cut. To do this you have to
take two values of $s$: $ s + i \epsilon$ and  $ s - i \epsilon$
and calculate
$$
\Delta ( - s )^{\alpha(t)}\,\,=\,\,2 i  Im ( - s )^{\alpha(t)} \,\,=\,\,
s^{\alpha(t)}\,\{ e^{i \alpha(t) \pi( 1 - \frac{\epsilon}{s} )} \,\,-\,\,
 e^{- \,i s^{\alpha(t)} \pi ( 1 - \frac{\epsilon}{s}}\,\,=\,\,2 i
sin(\,\pi \,\alpha(t)\,)\,s^{\alpha(t)}\,\,.
$$

After these remarks
it is obvious that the Reggeon exchange is the analytic function in $s$,
which
 in the $s$ - channel has the imaginary part 
$$
\pm \,g_1\,g_2 \,s^{\alpha(t)}
$$
 and  in the $u$ - channel the imaginary part
$$
g_1\,g_2 \,s^{\alpha(t)}
$$
i ( for $ \bar a + b \rightarrow \bar a + b$ reaction ).
For different signs in \eq{RGEX} the function has different properties
with
 respect to crossing symmetry. For plus ( positive signature) the function is
symmetric while for minus ( negative signature )  it is antisymmetric.
\subsection{ s - channel unitarity}
To satisfy the $s$ - channel unitarity we have to assume that the  
trajectory
 $\alpha (t)  \leq 1 $ in the scattering kinematic region ( $t < 0 $ ).
In this way the exchange of  Reggeons can solve our first puzzle.
\subsection{Resonances.}
Let us consider the same function but in the resonance kinematic region at $t
> M^2_0$. Here $\alpha (t)$ is a complex function. If $ t \rightarrow
t_0$  then  $\alpha (t_0) \,=\,j\,=\,
2 k $ ( or $\alpha_0(t)\, =\,j\,=\,2\,k + 1$,it depends on the sign in
Eq.(42))  where $k = 1,2,3,..$.
The Reggeon
 exchange for  positive signature has a form:
\beq
A_R(s,t)\,\,\rightarrow_{t \rightarrow t_0} g_1\,g_2 \cdot \frac{s^{2k}}{
\alpha'( t_0) \,( t - t_0 ) \,-\,i Im \alpha(t_0)}
\eeq
Since in this kinematic region the amplitude $A_R$ describes the reaction
$ \bar a + a \rightarrow \bar b + b $
$$
s\,\,=\,\,p^2 sin \theta
$$
where $p \,=\,\frac{\sqrt{t_0}}{2}$, the amplitude has the  form
\beq
A_R\,\,=\,\,g_1\,g_2\cdot\frac{s^{j}}{\alpha'( t_0) ( t - t_0 ) \,-
\,i Im \alpha( t_0 ) }\,\,=\,\, \frac{g_1 g_2 }{\alpha' (t_0)}
\cdot \frac{p^{2j} sin^j \theta}{t - t_0 - i \Gamma}
\eeq
where the resonance width $\Gamma = \frac{Im \alpha (t_0)}{ \alpha' (t_0)}$.

Therefore the Reggeon gives the Breit - Wigner  amplitude of the resonance
 contribution
at $t > 4 M^2_0$. It is easy to show that the Reggeon exchange with the
negative signature describes the contribution  of a resonance with odd
spin $j = 2k + 1 $.
\subsection{Trajectories.}
We can rephrase the previous observation in different words saying that
a  Reggeon describes the family of resonances that lies on the same
trajectory
$\alpha(t)$. It gives us a new approach to  classification of the
resonances,
which is quite different from usual $SU_3$ classification.
\begin{figure}[htbp]
\centerline{\epsfig{file=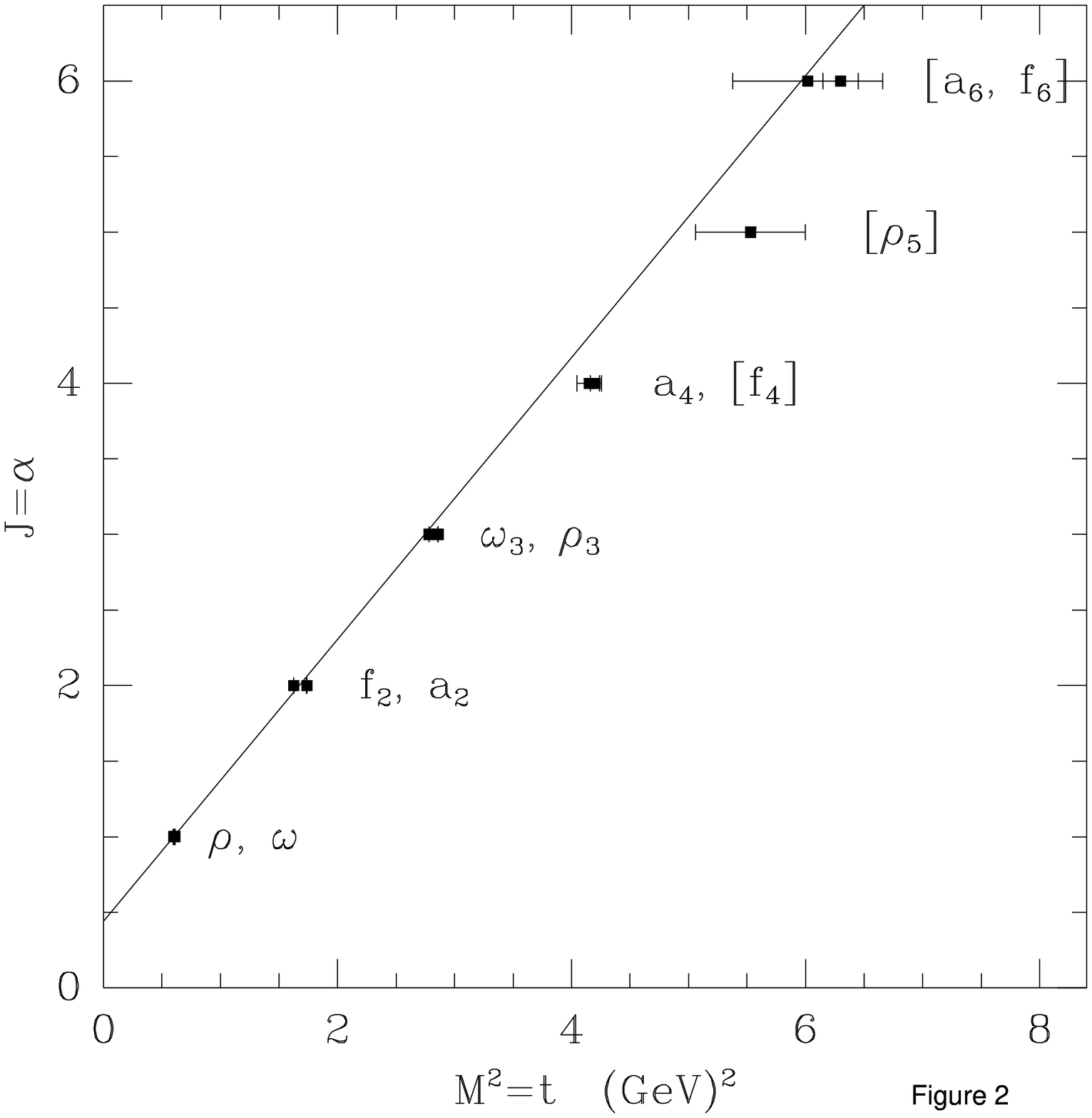,width=100mm}}
\caption{{\it The experimental $\rho$, $f$ and $a$ trajectories .}}
\label{fig14}
\end{figure} 
 Fig.14 shows
the
bosonic resonances classified  by  Reggeon trajectories.
The surprising experimental fact is that all trajectories seem to be
approximately   straight
 lines
\beq
\alpha( t ) \,\,=\,\,\alpha (0) \,\,+\,\,\alpha'\,t
\eeq
with the similar  slope $\alpha' \simeq 1 GeV^{- 2}$.

We would like to draw your attention to the fact that this simple linear form
comes from two experimental facts: 1) the width of resonances are much
smaller
than their mass ($ \Gamma_i \,\ll \, M_{R_i}$) and 2) the slope of the
trajectories which is responsible for the shrinkage of the diffraction peak
turns out to be the same from the experiments in the scattering kinematic
 region.

 $SU_3$ symmetry requires that
$$\alpha_{\rho} (0)\,=\,
\alpha_{\omega} (0)\,=\,
\alpha_{\phi} (0)$$
and the slopes to have the same value. The simple picture drawn in Fig.14 
shows the intercept $\alpha(0) \,\simeq 0.5$. Therefore the exchange of
Reggeons leads to  the cross section which  falls  as a function of
the energy and therefore do not violate 
the Froissart theorem.

\subsection{ Definite phase.}
The Reggeon amplitude of \eq{RGEX} can be rewritten in the form:
\beq
A_R\,\,=\,\,g_1 g_2 \eta_{\pm}  s^{\alpha(t)}
\eeq
where $\eta$ is the signature factor
$$
\eta_{+}\,\,=\,\,ctg\frac{\pi \alpha(t)}{2} \,\,+\,\,i
$$
$$
\eta_{-}\,\,=\,\,tg \frac{\pi \alpha(t)}{2}\,\,-\,\,i\,\,.
$$
The exchange of a Reggeon defines also the  phase
 of the scattering amplitude. This fact is very important especially for
 the description of the interaction with a polarized target.
\subsection{Factorization.}
The amplitude of \eq{RGEX} has a simple factorized form in which all
  dependences on the particular properties of colliding hadrons are
concentrated
in the vertex functions $g_1$ and $g_2$. To make this clear let us rewrite
this
factorization property in an explicit way:
\beq
A_R\,\,=\,\,g_1 (m_1,M_1,q^2_t) \,g_2 (m_2,M_2,q^2_t) \cdot \eta_{\pm} \cdot
s^{\alpha( q^2_t)}
\eeq
For example, it  means that if we try to
 describe  the total cross section of the  deep inelastic scattering
 of virtual photon  on a
 target
 through the Reggeon
 exchange, only the vertex function should depend on the value of the
 virtuality of photon ($Q^2$) while the energy dependence  ( or the
intercept of the Reggeon )
does not depend on
$Q^2$.

It should be stressed that these factorization properties are the direct
consequences of the Breit - Wigner formula in the resonance kinematic
region ( $t \,\geq\,4m^2_{\pi}$).

\subsection{The Reggeon exchange in $b$.}

It is easy to show that  Reggeon exchange has the following form in
$b$:
\beq \label{RINB}
a_R(s,b)\,\,=\,\,g_1(0)\,g_2(0) s^{\alpha(0)} \cdot \frac{1}{4\pi
 ( R^2_1 + R^2_2 + \alpha' \ln s)} \cdot e^{- \,\frac{b^2}{4( R^2_1
+ R^2_2 + \alpha' \ln s)}}
\eeq
 if we assume the simple exponential parameterization for the
vertices:
$$
g_1 (q^2_t) = g_1(0) \,e^{- R^2_1 \,q^2_t}\,\,;
$$
$$
g_2(q^2_t) = g_2(0) \,e^{- R^2_2 \,q^2_t}\,\,.
$$
To do this we need to consider the following  integral ( see eq.(21) ):
$$
a_R(s,b)\,\,=\,\,g_1(0)\,g_2(0) s^{\alpha(0)}\,\frac{1}{2}\int d q^2_t
J_0(q_t b) \,\,e^{ - [\,R^2_1\,+\,R^2_2\,+\,\alpha'\,\ln s\,]\,q^2_t}\,\,,
$$
which leads to  \eq{RINB}.  From \eq{RINB} one can see that  Reggeon
exchange leads to a radius of interaction, which is proportional to 
$\sqrt{ R^2_1\,+\,R^2_2\,+\,\alpha'\,\ln s}\,\,\rightarrow
\,\,\sqrt{\alpha' \ln s}$ at very high energy. We  recall  that in the
``black
disc "  ( Froissart ) limit  the radius of interaction 
increases proportional to $\ln s $ only. Therefore, we see that 
Reggeon exchange gives a picture which is quite different from the ``
black disk" one.

 \subsection{Shrinkage of the diffraction peak.}
Using the linear trajectory for  Reggeons it is easy to see that the
elastic cross section due to the exchange of a  Reggeon can be written
 in the form:
\beq
\frac{d \sigma_{el}}{d t}\,\,=\,\,g^2_1(q^2_t)\cdot g^2_2( q^2_t)
\cdot s^{2( \alpha(0) - 1)} \cdot e^{- 2 \alpha'(0)\,\ln s\,q^2_t}
\eeq
The last exponent reflects the phenomena which is known  as the
shrinkage of
 the diffraction peak. Indeed, at very high energy  the elastic cross section
is concentrated at values of $q^2_t < \frac{1}{\alpha' \ln s}$. It means that
the diffraction peak becomes narrower at higher  energies.
\section{Analyticity + Reggeons.}
\subsection{Duality and Veneziano model.}
Now we can come back to the main idea of the approach and try to construct the
amplitude from the analytic properties and the Reggeon asymptotic
behaviour at high
 energy. Veneziano  suggested the scattering  amplitude  is
the sum of all resonance contributions  in $s$ - channel with zero width
and, simultaneously,
the same amplitude is the sum of the $t$ - channel exchanges of all
possible Reggeons.
\begin{figure}[htbp]
\centerline{\epsfig{file=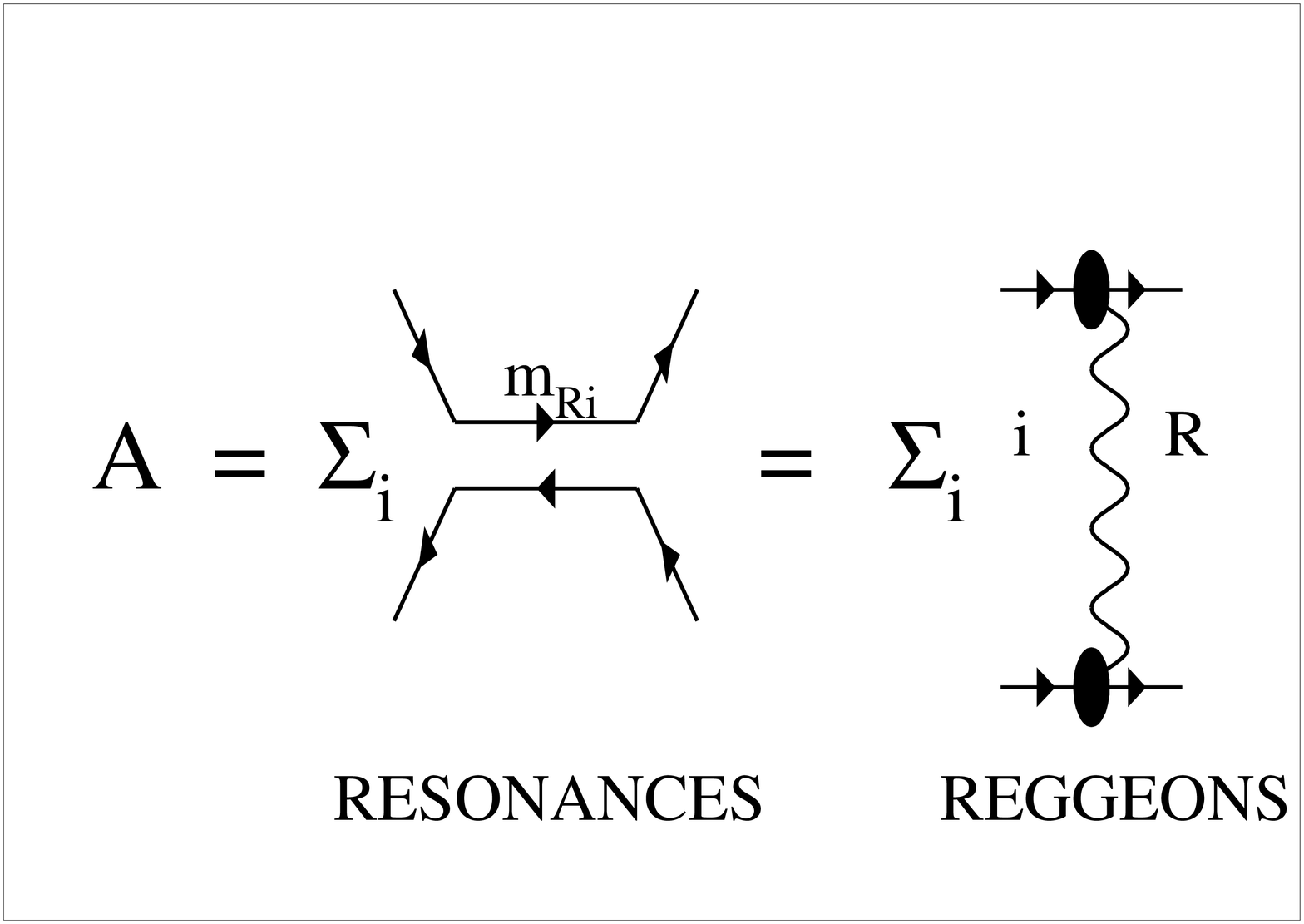,width=140mm}}
\caption{{\it  Duality between the resonances in  the $s$-channel
and  Reggeon exchange in the $t$- channel.}}
\label{fig15}
\end{figure}

Taking a  simplest case  the scattering of a scalar particle, the
Veneziano
 amplitude looks as follows:
\beq
A\,\,=\,\,g\,[\,V(s,t)\,+\,V(u,t)\,+\,V(s,u)\,]
\eeq
where
\beq
V(s,t)\,\,=\,\,\frac{\Gamma( 1 - \alpha(t) ) \Gamma( 1 - \alpha (s ) )}{\Gamma
( 1 - \alpha(t) - \alpha(s) )}
\eeq
$\Gamma(z)$ is the Euler gamma function defined as
$\Gamma(z)\,=\,\int^{\infty}{0}\,e^{- t}\,t^{z - 1}\,dt$ for $Re z
\,>\,0$. What we need to know about this function is the following:

1.$\Gamma(z)$ is the analytical  function of $z$ with the simple poles in
$z\,=\, -n \,( n = 0,1,2..)$;

2. At $ z \,\rightarrow \,- n$ $\Gamma (z) \,\rightarrow \,\frac{ ( - 1
)^n}{( z + n ) n!}$;

3.$\Gamma(z + 1 )\,\,=\,\,z\,\Gamma (z)$, $\Gamma(1) = 1$;

4. At large $z$ $\Gamma(z)\,\rightarrow\,z^{z - \frac{1}{2}}\,e^{-z}
\,\sqrt{2 \pi}(\, 1\, + \,O(\frac{1}{z}) \, )$;

5. $lim|_{|z| \rightarrow \infty} \frac{\Gamma( z + a )}{ \Gamma( z
)}\,e^{- a \ln z}\,=\,1$.

Taking these properties in mind one
  can see that the Veneziano amplitude has resonances at $\alpha(s) = n +
1$
where $n = 1,2,3...$, since $\Gamma (z) \rightarrow \frac{1}{z + n}$ at
 $z \rightarrow - n $. At the same time
$$
A\,\,\rightarrow_{s \rightarrow \infty}\,\,\Gamma( 1 - \alpha(t) ) [ (
- \alpha(s))^{\alpha(t)} \,\,+\,\, (- \alpha(u) )^{\alpha(t)} ]
$$
which  reproduces  Reggeon exchange at high energies.

This simple model was the triumph of our general ideas
 showing us how we can construct the theory using analyticity and asymptotic.
The idea was to use the Veneziano model as the first approximation or in other
word as a  Born term in the theory and to try to build the new theory
 starting with the new Born Approximation. The coupling constant $g$ is 
dimensionless and smaller than unity. This fact certainly also encouraged
the theoreticians in 70's to explore this new approach.

\subsection{Quark structure of the Reggeons.}
In this section, I would like to recall  a consequence of the duality
approach, described in Fig.15, namely, any Reggeon can be viewed as the
exchange of a  quark - antiquark pair in the  $t$ - channel. Indeed, the
whole
spectrum of  experimentally observed  resonances can be described as
different kinds of excitation of either a  quark - antiquark pair for
mesons
or of  three quarks for baryons.  This fact is well known and  is one of
the main  experimental results that led to   QCD.
In Fig.16 we draw the quark diagrams reflecting the selection rules for
meson - meson and meson - baryon scattering. One can notice that all  
 diagrams shown have quark - antiquark exchange in the $t$ - channel.
Therefore, the Reggeon - resonance duality leads to the quark - antiquark
structure of the Reggeons. I would like to mention that I do not touch
here the issue of so called baryonic Reggeons which contribute to   meson
- baryon scattering at small values of $u$, since our main goal to discuss 
the Pomeron here, which is responsible for the total cross section or for
the scattering amplitude at small $t$.
 
\begin{figure}[htbp] \begin{tabular}{c} 
\centerline{\epsfig{file=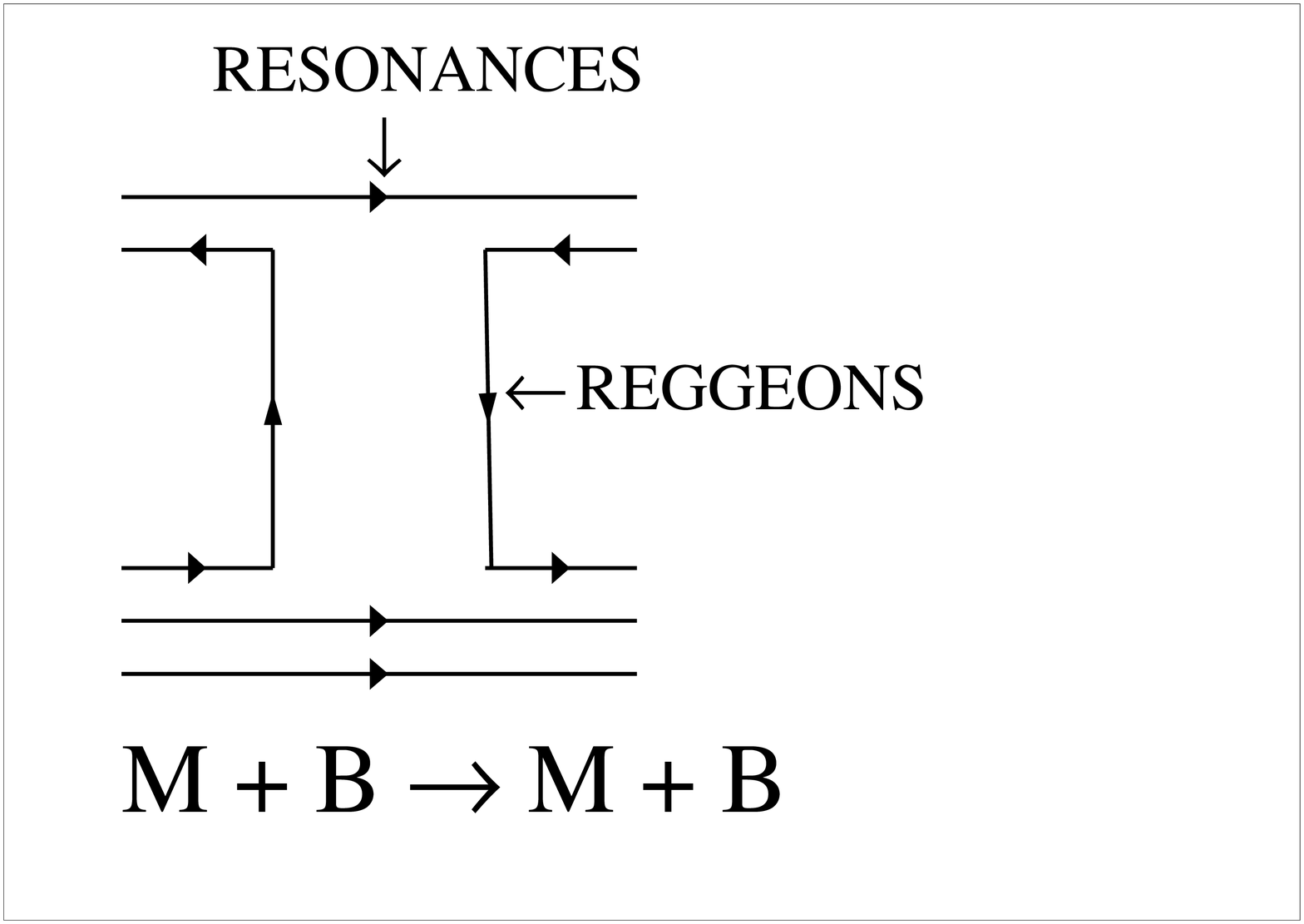,width=100mm}}\\
\centerline{\epsfig{file=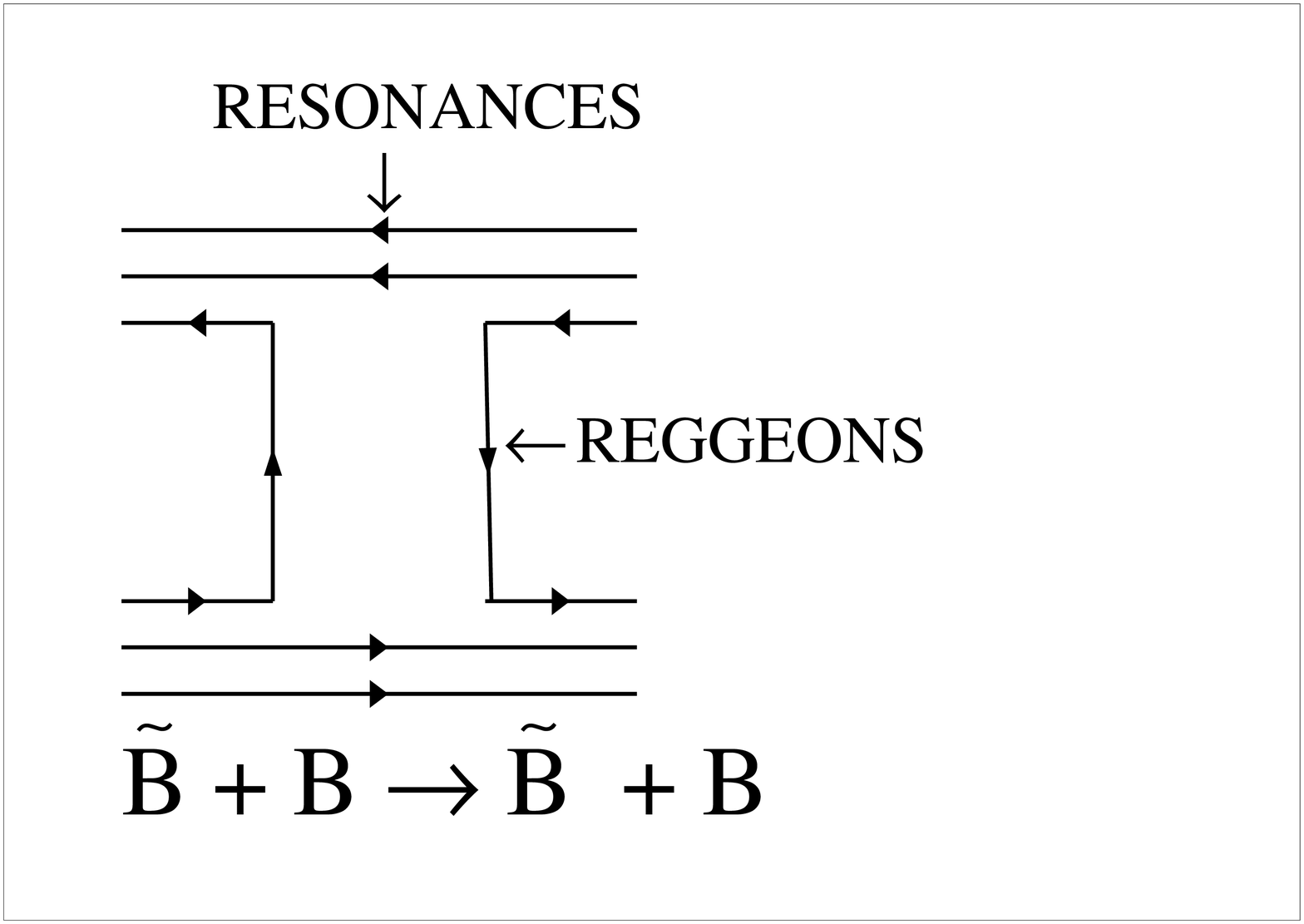,width=100mm}}
\end{tabular}
\label{fig16}
\caption{{\it  Quark diagrams for meson - meson and meson -
baryon scattering.}}
\end{figure}

A  clear manifestation of these rules is the fact that reggeons do not
contribute to the total cross sections of either proton - proton
 or  to $K^{+} p$ collisions. For
both of these processes we cannot draw the diagram with quark - antiquark
exchange in the $t$ - channel and /or with three quarks ( quark -
antiquark
pair) in $s$ - channel (see Fig.17).
\begin{figure}[htbp]
\begin{tabular}{c}
\centerline{\epsfig{file=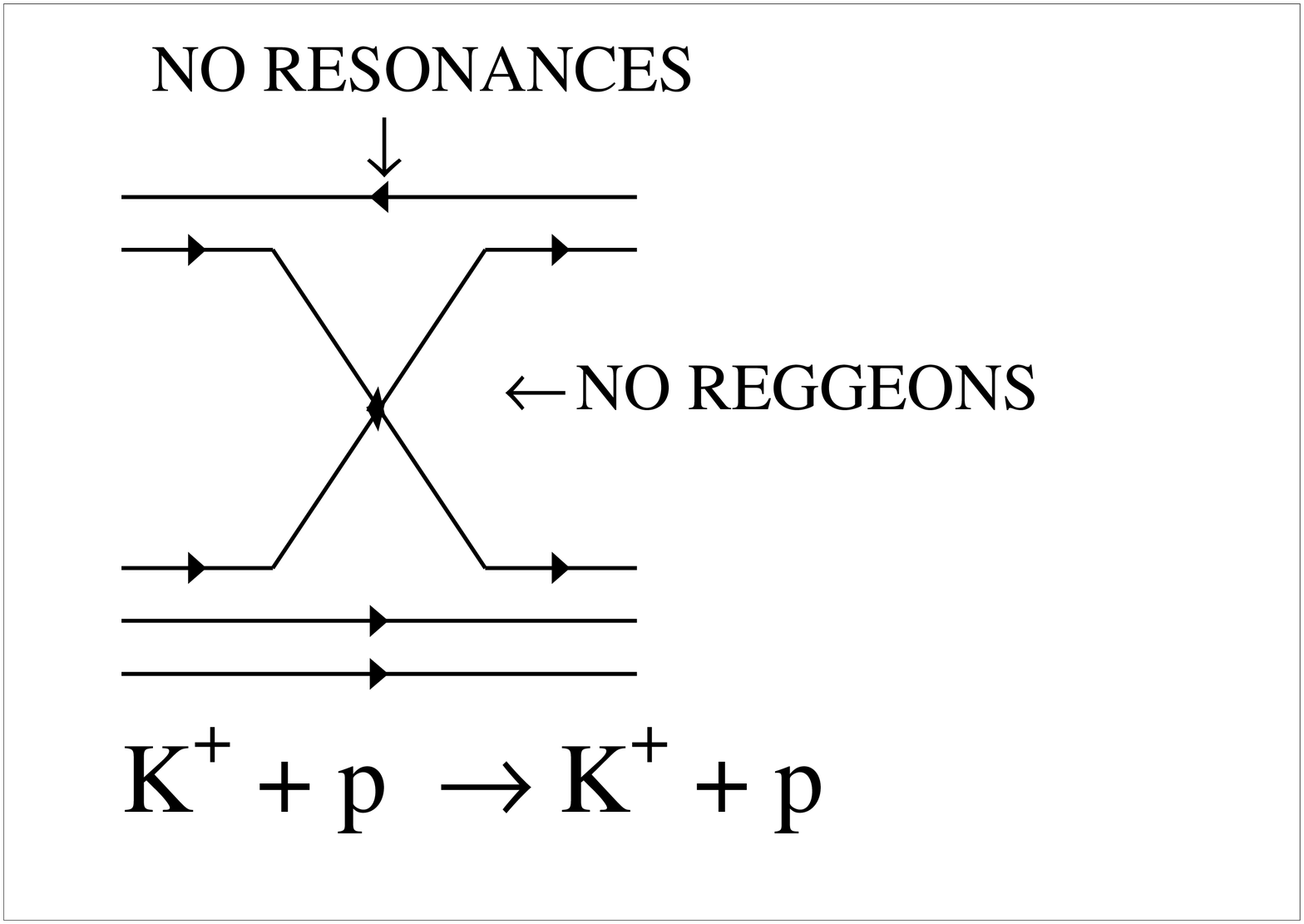,width=100mm}}\\
\centerline{\epsfig{file=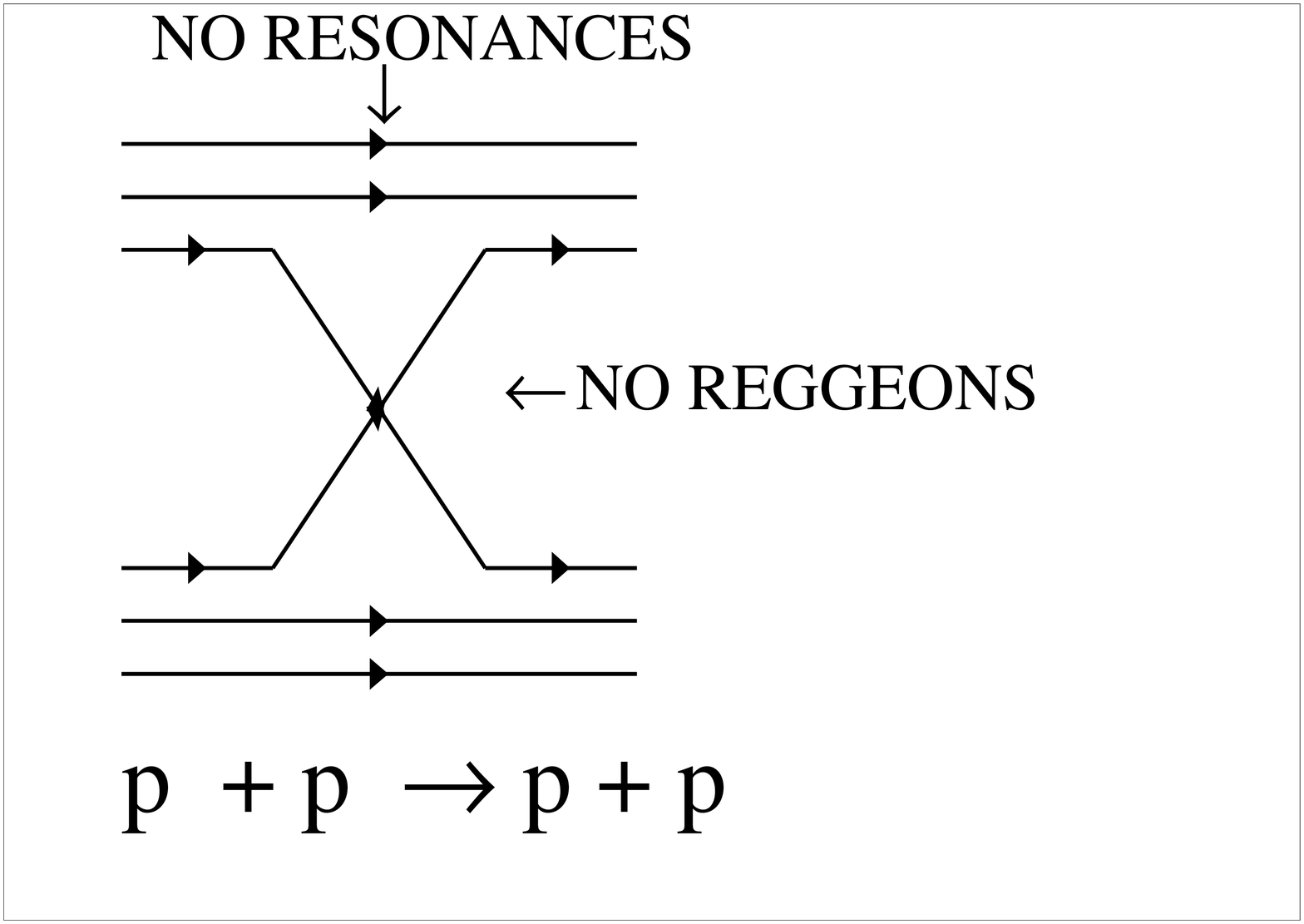,width=100mm}}
\end{tabular}
\label{fig17}
\caption{{\it  Quark diagrams for elastic proton - proton and $K^{+} p $
scattering.}} 
\end{figure}

In Reggeon language the quark - antiquark structure of Reggeons
leads to so called signature degeneracy.  All Reggeons with positive and
negative signature have the same trajectory. Actually, this  was shown in
Fig.14. 
   
Topologically, the duality quark diagrams are  planar diagrams.
It means that they can be drawn in one sheet of paper without any two 
lines would be crossed. It is worthwhile
mentioning, that all more complicated diagrams, like the exchange of many
gluons inside of  planar diagrams, lead to the contributions which
contain the maximal power of $N_c$ where $N_c$ is the number of colours.
Therefore, in the limit where $\as \,\ll\,1$ but $\as N_c\,\approx\,1 $ 
only planar diagrams  give the leading contribution.   
\begin{figure}[htbp]
\centerline{\epsfig{file=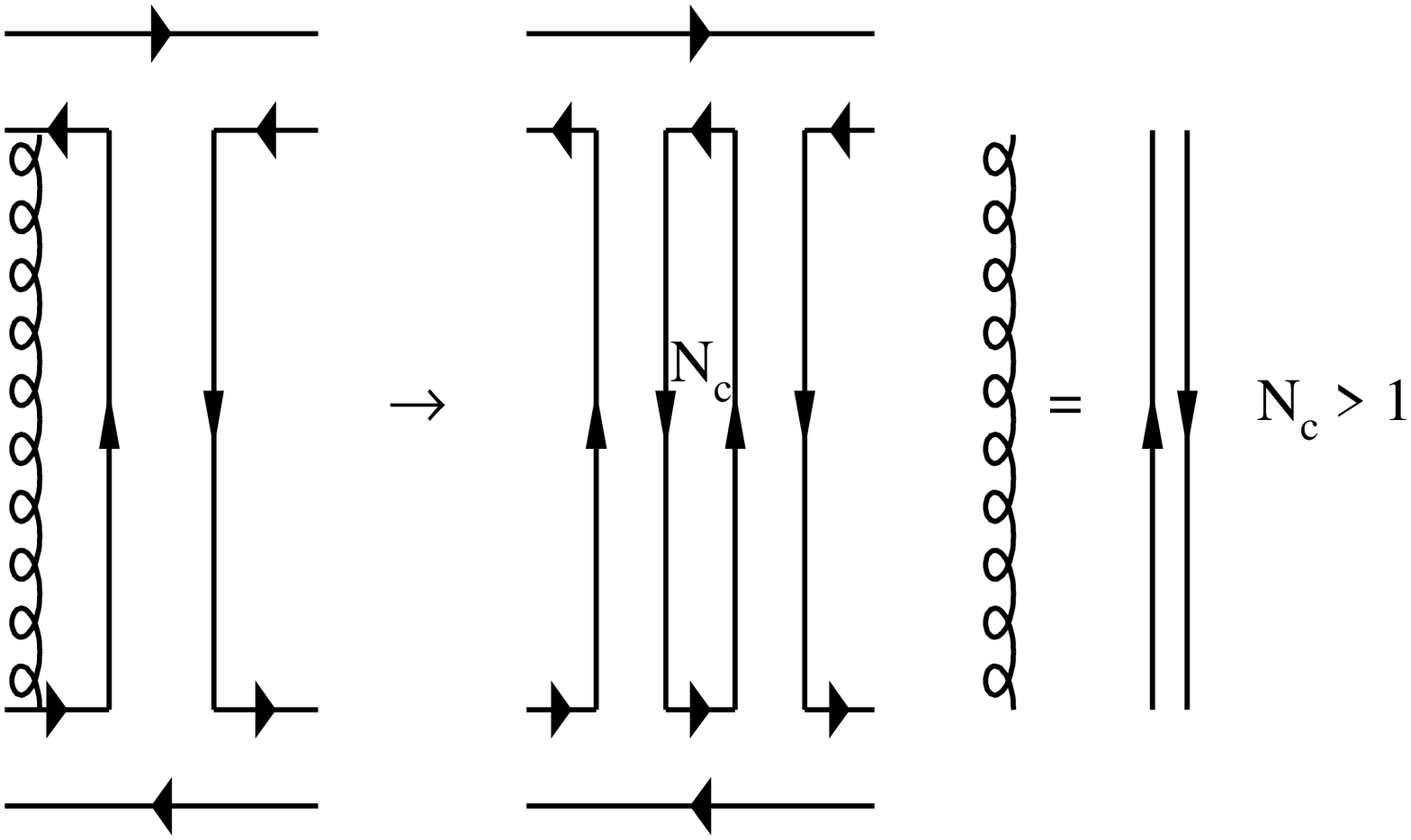,width=160mm}}
\label{fig18}
\caption{{\it   An example of a  planar diagram in the 
$N_c\,>\,1$ approximation.}}
\end{figure}
    
\section{The second and the third puzzles: the Pomeron?!}
\subsection{Why we need the new Reggeon - the Pomeron?}
The experiment shows that:

{\bf 1.} {\it There are  no particles ( resonances) on a Reggeon
trajectory
with
an  intercept  close to unity ( $\alpha(0) \rightarrow
 1$).} As mentioned before  the typical highest has an  intercept
of  $\alpha(0) \sim 0.5$ which generates a  cross section 
$\sigma_{tot} \propto s^{\alpha(0) - 1} \propto s^{-\frac{1}{2}}$.
Hence, we introduced Reggeons to solve the first puzzle and  found
to our surprise that these Reggeons give us a total cross section which
 decreases   rapidly  at high energy.

{\bf 2.}{\it The measured  total cross sections are  approximately
constant at high energy.}
We see that fighting against the rise of the cross section due to
exchange of  high
spin resonances  got us into
 the  problem to describe the  basic experimental fact that the total
cross hadron - hadron 
sections  do not  decrease with energy.

We call the above two statement the second and the third puzzles
that have to
 be solved by  theory. 
The ad-hoc solution to this problem  was the  assumption  that a
Reggeon with the intercept close to 1 exists. One than had to understand
how
 and why this Reggeon is  different from other Reggeons, in particular,
why there
 is no particle on this trajectory.  Now let us ask ourselves why we
introduce one more Reggeon? And we have to answer: only because of the
lack of imagination. We can also  say that we did not want to
multiply the essentials and even find some philosopher to refer to.
However, it does not make any difference and one cannot hide the fact that
the
new Reggeon is just the simplest attempt to understand the constant total
cross section at high energy.

 As far as I know the first who introduce
such a new Reggeon was V.N. Gribov who liked very much  this
hypothesis since it could explain  two facts: the constant cross section
and the shrinkage of the diffraction peak in high energy proton - protron
collision . By the way, high energy was $s  \,\approx\, 10 - 20 GeV^2$ !!!
However, I have to remind you that at that time ( in early sixties )
the popular working model for high energy scattering was the ``black
disck" approach with the radius which does not depend on energy. 
Certainly, such a model was much worse than a Reggeon and could not be
consistent with our general properties of  analyticity and unitarity
as was shown by. V.N. Gribov. Of course,  on this background the Reggeon
hypothesis was a relief since this new Reggeon could exist in a theory.
I hope to give you enough illustration of the above statement during the
course of the lectures.

Now, let me introduce for the first time the word Pomeron. The first
 definition of the Pomeron:

{\bf The Pomeron is the Reggeon with $\alpha(0) - 1 \,\equiv \,\Delta\,\ll
\,1$}

The name  Pomeron was introduced in honour of the  Russian physicist
Pomeranchuk who
did a lot to understand this funny  object. By the way,  the general name
of
the Reggeon was given after the Italian physicist Regge, who gave a beautiful
theoretical argument why such objects can exist in  quantum mechanics.

The good news  about the Pomeron hypothesis is that it  leads to a large 
number of predictions since the Pomeron should possess 
all properties of a  Reggeon exchange ( see the previous section).
And in fact, all  Reggeon properties have been
established  experimentally for the Pomeron.

\subsection{Donnachie - Landshoff Pomeron.}

The phenomenology, based on the Pomeron hypothesis turned out to be very
successful. It  survived, at least   two decades  and  even if you  do
not like
the hypothesis you have to learn about it nowdays  to cope with the large
amount of  experimental
information on the high energy behaviour of strong interactions.

 Let
me summarize what we know about the Pomeron from  experiment:

1. $\Delta \,\simeq\,0.08$

2. $\alpha'(0)\,\simeq\,0.25 GeV^{-2} $

Donnachie and Landshoff gave an elegant description of almost all existing
experimental data using the hypothesis of the Pomeron with the above parameters
of its trajectory. The fit to  the data is  good. Let us use this
DL Pomeron as an example to which we are going to apply everything that we
have learned. 

The first regretful property of the DL Pomeron is the fact that it
violates
unitarity since $\Delta \,>\,1$.  However, the Froissart limit requires  
$\s_{tot} \,<\,30 mb \,N^2 \ln^2 s/s_0$. Taking $ N = 1$
and $s_0 = 100 GeV^2$ we have $\s_{tot} \,<\,300\,mb$ for $\sqrt{s} = 1800
GeV$ (  Tevatron energy ). Therefore, at first sight, the theoretical
problem with
unitarity 
exists but we are far away from this problem in all experiments  in the
near future.

However, we have to be more careful with such statements, since the
unitarity constraint is much richer than  the Froissart limit. Indeed,
from the
general unitarity constraint we can derive the  so called ``weak
unitarity",
namely

\beq \label{WU}
Im a_{el}(s,b)\,\,\leq\,\,1\,\,,
\eeq
which  follows directly from the general solution for the  unitarity
constraint,

\beq \label{WEEKU}
 Im a_{el}(s,b)\,\,=\,\,1\,\,-\,\,e^{-\frac{\Omega}{2}}\,\,\leq\,\,1\,\,.
\eeq

For the DL Pomeron we can calculate the amplitude for  Pomeron exchange
in impact parameter representation ( see \eq{RINB} ). The result is
\beq \label{DLINB}
a^{DL}_{el}\,\,=\,\,\frac{\s(s=s_0)}{4\, \pi\, [\, 2 R^2_p(s = s_0)
\,+\,\alpha'\,\ln (s/s_0)\,]} \,\,(
\frac{s}{s_0})^{\Delta}\,\,e^{-\,\frac{b^2}{4[\, 2 R^2_p(s = s_0) 
\,+\,\alpha'\,\ln (s/s_0)\,]}}\,\,=
\eeq
$$
\,\,\frac{\s(s=s_0)}{4\,
\pi\,B_{el}}\,\,( \frac{s}{s_0} )^{\Delta}
\,\,e^{ - \frac{b^2}{2 B_{el}}}\,\,,
$$
where $B_{el} $ is the slope for the elastic cross section ($\frac {d
\s}{d t} = \frac{d \s}{d t}\vert_{t = 0}\,e^{ - B_{el} | t |}$).
One can see, taking the  numbers from Fig.11, that $a^{DL}_{el}(s,b=0)
\,\sim\,1/2$ at $\sqrt{s} \,\sim\,100 \,GeV$ . For higher energies the DL
Pomeron violates  the ``weak unitarity". 

Therefore, in the range of energies between the fixed target FNAL energies
and the Tevatron energy the DL Pomeron cannot be considered as a good
approach from the theoretical point of view in spite of the good
description of the experimental data. We will discuss this problem later
 in more details.

I would like to draw your attention to a new parameter $s_0$ that has
appeared in our estimates ( see Eq.(54) for example). Practically, it
enters the master formula of Eq. (47) in the following way
$$
A_R\,\,=\,\,g_1\,g_2\,\eta_{\pm}\,\,(\frac{s}{s_0})^{\alpha( q^2_t)}\,\,
$$
and gives the normalization for the vertices $g_1$ and $g_2$. As you can
see in Veneziano  approach ( see Eq.(50) )
$s_0\,=\,\frac{1}{\alpha'_R}\,\approx\,1\,GeV^2$. However, in spite of the
fact that the value of $s_0$ does not affect any physical result since the
value of vertices $g_i$ we can get only from fitting the
experimental data the choice of the value of $s_0$ reflect our believe
what energy are large. The Reggeon approach is asymptotic one and it 
can be applied only for large energies $s \,\>\,s_0$. Therefore, $s_0$ is
the energy starting from which we believe that we can use the Reggeon
approach.

\section{The Pomeron structure in the parton model.}
\subsection{The Pomeron in the Veneziano model ( Duality approach).}
As has been mentioned the Pomeron does not appear in the new Born term of our
approach. Therefore, the first  idea was to  attempt to
calculate
the next to the  Born approximation in the Veneziano model.
 The basic equation that we want to use is graphically pictured
 in Fig.19, which is nothing more than the optical theorem.

\thispagestyle{empty}
\begin{figure}[htbp]
\begin{tabular}{c}
\centerline{\epsfig{file=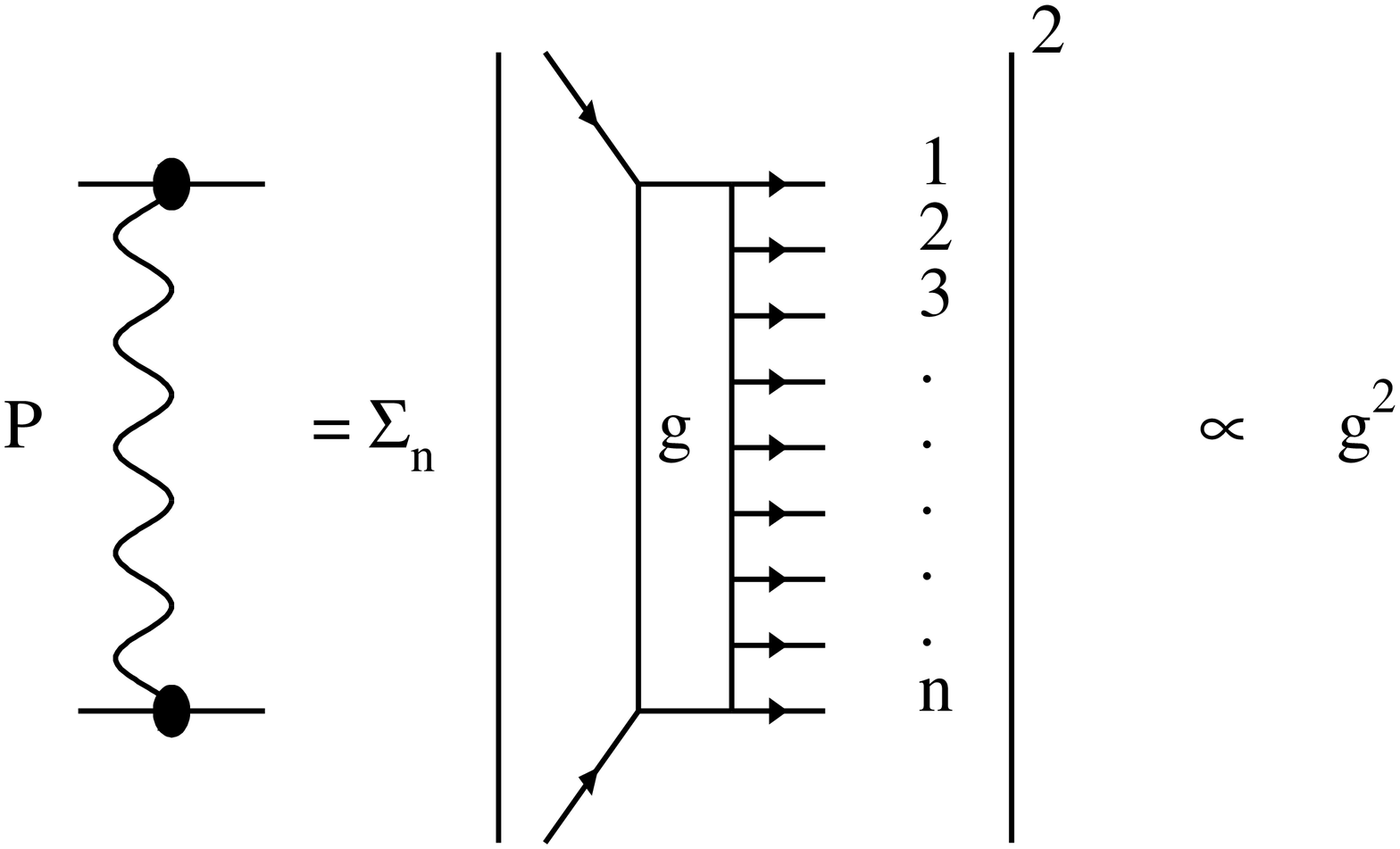,width=160mm}}\\
\centerline{\epsfig{file=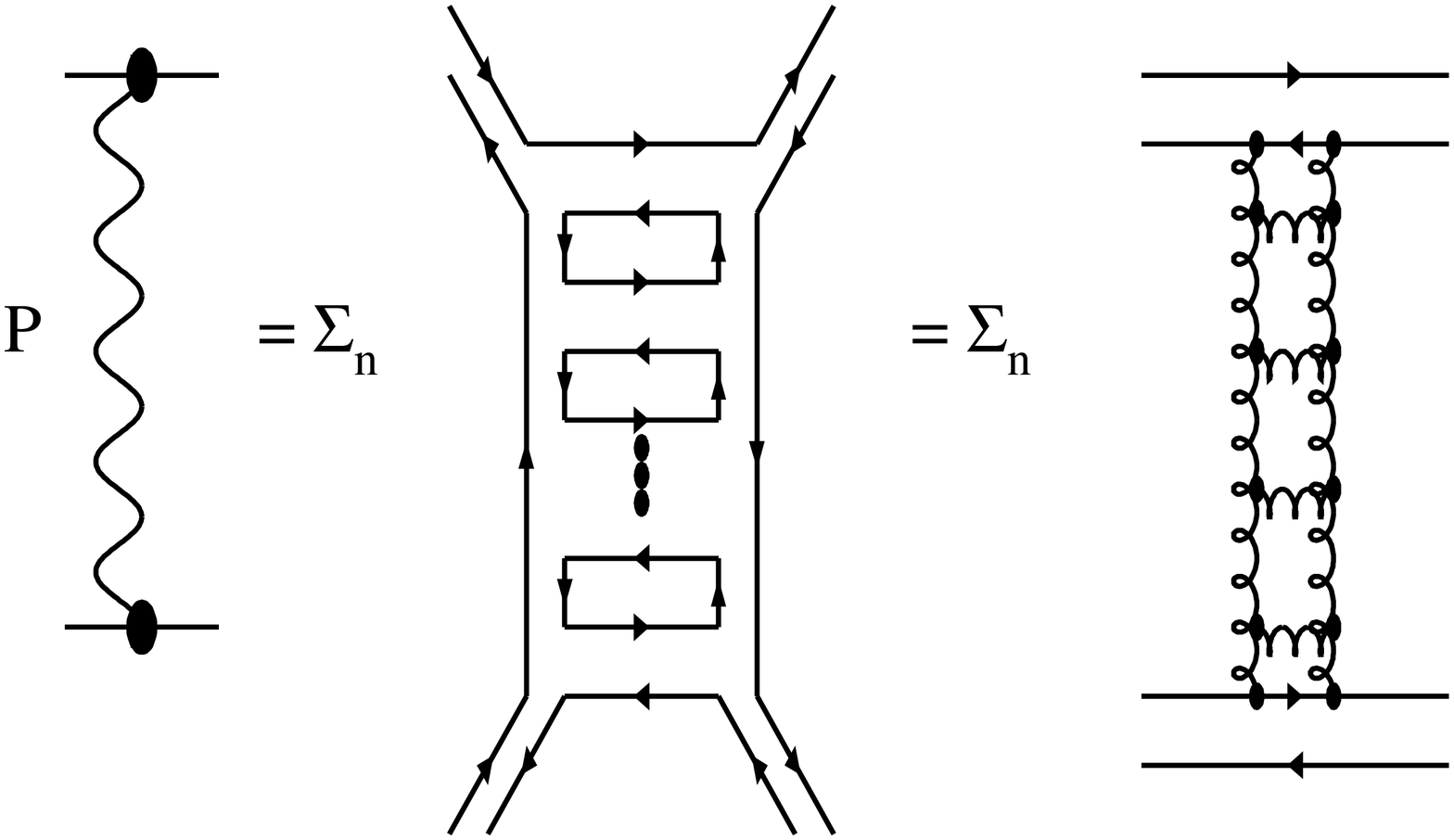,width=160mm}}
\end{tabular}
\label{fig19}
\caption{{\it  The Pomeron structure in the duality approach.}}
\end{figure}

We need  to
 know the Born approximation for the amplitude of production of $n$
particles.
Our hope was that we would need to know only a general features of this
production amplitude to reach an understanding of the Pomeron structure.

To illustrate the main  properties and problems which can arise in
this
approach let us calculate the contribution in equation of Fig. 19   of the
first
 two particle state ( $\s^{(2)}_{tot}$). This contribution is equal to:
\beq
\sigma^{(2)}_{tot}\,\,=\,\,s^{2(\alpha_R(0) - 1 )} \,\,\int d^2 q'_t
\Gamma^2
(1-\alpha(- q'^2_t)) \cdot e^{ - 2\alpha' q'^2_t \ln s}\,\,.
\eeq
Since
$$
\Gamma( 1 - \alpha(- q'^2_t))\,\propto\,
( - \,\frac{\alpha(- q'^2_t)}{e})^{ - \,2 \alpha(- q'^2_t)}\,\,\rightarrow
\,\,e^{\alpha'  q^2_t \ln(\alpha' q^2_t)}
$$
one can see that the essential value of $q'$ in the integral is rather big,
namely 
 of the order of $q'^2_t \,\sim s $. This  means that we have to believe
in the
Veneziano amplitude at  larges values of the momentum transfer. Of course
nobody believed in the Veneziano model as the  correct theory at small
distances  neither 25 years ago nor now. 
 However we have learned  a  lesson from this exercise ( the
lesson!),namely

{\it To understand the Pomeron structure we have to understand better the
structure of the scattering amplitude at large values of the momentum
 transfer or in other words we must   know the interaction at small
distances.}
\subsection{ The topological structure of the Pomeron\\ in  the  duality
approach.}
The topological structure of the quark diagrams for the Pomeron is more
complicated. Actually, it corresponds to the two sheet configuration . It
means that we can draw the Pomeron duality diagram, without any two lines
being crossed,   not in the sheet of
paper but on  the surface of a cylinder.  For us even more important is 
that
the counting the power of $N_c$ shows that the Pomeron leads to  one power of
$N_c$ less than the planar diagram. In Fig.19 you can see the reduction of
the duality quark diagram to QCD gluon exchange (gluon ``ladder" diagram).
Counting of the $N_c$ factors shows that the amplitude for emission
of
$n$-gluon  is   $\propto \,\,\as^{2 + n}\, N^{1 + n}_c$. Therefore, the
approach 
becomes a bit messy. Strictly speaking there is no Pomeron in  leading
order of $N_c$, but in spite of the fact that it appears in the next
order with respect to $N_c$ the $s$-dependence can be so strong that it
will compensate this smallness. The estimate for ratio yields:
$$
\frac{Pomeron}{Reggeon}\,\,\propto\,\,\frac{\as}{N_c}
(\frac{s}{s_0})^{\Delta - \alpha_R(0) + 1 }\,\,\approx \,\,
\frac{\as}{N_c}
(\frac{s}{s_0})^{0.51}\,\,.
$$

\subsection{The general origin of $\ln^n s$ contributions.}
We show here that the $\ln^n s$ contribution results  from  phase
space and, because of this,  such logs appear in any theory (at least in
all
theories that I know).
The main contribution to  equation of Fig.19  comes from a  specific
region of integration. Indeed, the total cross section can be written as
 follows:
\beq \label{CRD}
\sigma_{tot}\,\,=\,\,\Sigma_n \int |M^2_n ( x_i p, k_{ti})|
 \Pi_i \frac{ d x_i}{x_i}\,\, d^2 k_{ti}
\eeq
where $x_i$ is the fraction of energy that is  carried by the $i$-th
particle.
Let's call all secondary particle partons.
One finds  that the biggest contribution in the above equation comes from
 the region of integration with strong ordering in $x_i$ for all produced
 partons,namely
\beq
x_1\,\gg \,x_2\,\gg .....\gg \,x_i\,\gg\, \,x_{i + 1} \,\gg .....\gg \,x_n\,=\,
\frac{m^2}{s}
\eeq
Integration over   this kinematic region allows to  put all $x_i= 0 $ into 
the amplitude
$M_n$. Finally,
\beq \label{CRD1}
\sigma_{tot}\,\,=\,\,\Sigma_n \int \Pi_i d^2 k_{ti} |M^2_n (k_{ti})| \,
\int^1_{\frac{m^2}{s}} \,\frac{d x_1}{x_1} .....\int^{x_{i -
1}}_{\frac{m^2}{s}}
\,\frac{d x_i}{ x_i}\,...\int^{x_{n - 1}}_{\frac{m^2}{s}} \,\frac{d x_n}{ x_n}
\eeq
$$=\,\,\Sigma_n \int \Pi_i d^2 k_{ti} |M^2_n (k_{ti})| \,\cdot\,
\frac{1}{n!}\,\ln^n s
$$
This equation shows one very general property of   high energy
interactions,
 namely the longitudinal coordinates ( $x_i$) and the transverse ones
($k_{ti}$)
are  separated and should be treated differently. The
integration over longitudinal coordinates gives the  $\ln^n s $ - term.
It 
does not depend on a specific theory ,  while  the
transverse momenta integration depends on the  theory and  is a rather
complicated problem  to be solved in general.
 In some sense the
above equation reduced the problem of the  high energy behaviour of the
total
cross section to the calculation of the amplitude $M_n$ which depends  
only on
transverse coordinates. Assuming, for example, that $\int \Pi d^2 k_{ti}
|M^2_n( k_{ti})|^2\,\propto \frac{1}{m^2} g^n$ we can derive from the
eq.(58)
\beq
\sigma_{tot}\,=\,\frac{1}{m^2}\,\Sigma_n\,\frac{g^n}{n!} \,\ln^n s\,=\,
\frac{1}{m^2}\cdot s^{g}
\eeq
which looks just as  Pomeron - like behaviour. This example
shows the way
how the Pomeron can be derived in the theory.

Let us consider a more sophisticated example, the  so called $ g \phi^3$ -
theory. This theory has a nice property, namely, the coupling constant is
dimension and all integrals over transverse momenta are  convergent. Such
a 
theory is one of the theoretical realizations of the  Feynman - Gribov
parton
model, in which  was assumed that the mean transverse momentum of the
secondary
particles ( partons) does not depend on the  energy ($k_{it}\,=\,Const
(s)$). 
 The parton  model is  the simplest model for the scattering amplitude at
small distances which reproduces the main experimental result  in deep
inelastic scattering. For us it seems natural to try this model for the
structure of the Pomeron.
To do this, let us first formulate the approximation in which we are going
to work: the leading log$ s$ approximation (LLA). In the LLA we sum in
each order of perturbation theory, say $g^n$, only contributions of the
order $(g \ln s)^n$, which are big at high energies. As we have discussed
the 
$\ln s^n$  term  comes  from the phase space integration  at high energy.
To have a
contribution
of  order $(g \ln s)^n$ we have to consider a  so called ladder diagram
(see
Fig.20). These ladder diagrams give a  sufficiently simple two dimensional 
theory for $M_n(k_{it})$, namely the product of bubble as shown in Fig.20.
\begin{figure}[htbp]
\centerline{\epsfig{file=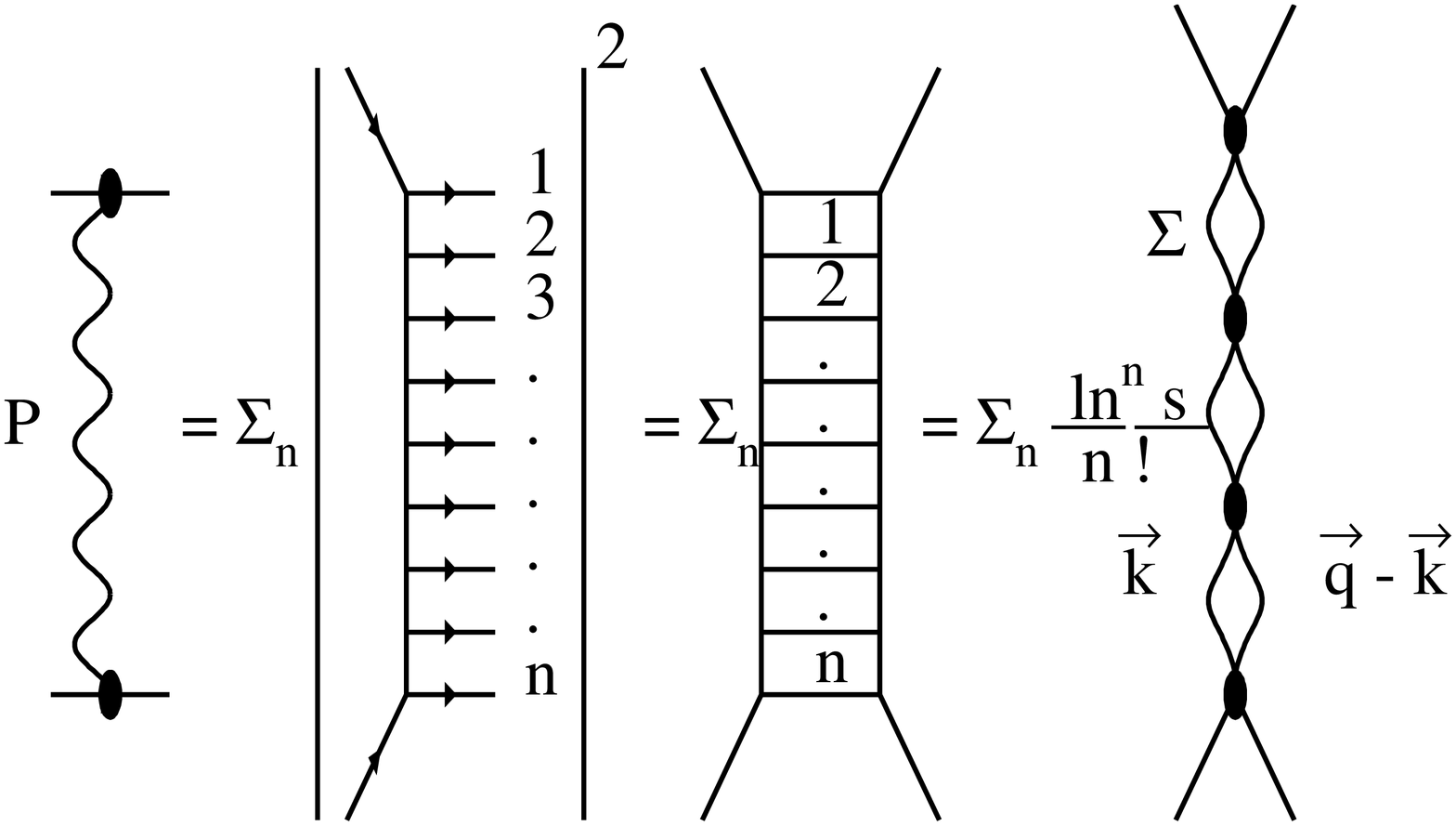,width=160mm}}
\label{fig20}
\caption{{\it  The Pomeron in the  LLA for  $g\phi^3$ - theory}}
\end{figure}
 
One can see that the cross section  of emission  for the  $n$-partons is
equal to
\footnote{Here we introduce $\sigma^n(q^2)$ from \eq{CRD}.}:
\beq \label{CRINGFC}
\s^n\,\,=\,\,\frac{1 }{s}\frac{\ln^n( s/\mu^2)}{n!}\,\alpha
\,\Sigma(q^2)
\,\,(\Sigma(q^2))^n
\,\,
\eeq
where $\alpha = \frac{g^2}{4 \pi}$ and
$$
\Sigma(q^2)\,\,=\,\,\alpha\,\,\int\,\frac{d^2 k_t}{(2 \pi)^2} \,\frac{1}{
(\,k^2_t\,\,+\,\,\mu^2\,)\,(( k - q )^2_t\,\,+\,\,\mu^2\,)}
$$
with $\mu$ being the  mass of a parton.

For the total cross section we have:
\beq \label{TOTCR}
\s_{tot} \,\,=\,\,\sum^{\infty}_{n=0}\,\s^n\,\,=\,\,\frac{1}{s}\,\alpha
\,\Sigma(q^2)( \frac{s}{\mu^2})^{\Sigma(q^2)}\,\,.
\eeq
Therefore, one can see that we reproduce the Reggeon in this theory with
trajectory $ \Sigma(q^2)$. To justify that this Reggeon is a Pomeron
we
have to show that the intercept   of this Reggeon is equal to $1 + \Delta$
 ( $\Sigma (0) = 1 +  \Delta $ )  as we have discussed. In the LLA of   $g
\phi^3$ - theory we can fix the value of the coupling constant $g$ (
$\alpha$), namely, $\Sigma(0) \,=\, \frac{\alpha}{4 \pi  
\mu^2}\,=\,1\,+\,\Delta$ or $\alpha \,=\,( 1 + \Delta)  4 \,\pi\,\mu^2$.
It gives you the parameter of the perturbation approach in $ g \phi^3$ -
theory, $\frac{\alpha}{ 2 \pi \mu^2} \,\approx 2 $. This sufficiently
large value indicates that the   LLA can be used onlt as a qualitative 
attempt to understand the physics of the high energy scattering in this
theory but not for serious quantative estimates. Actually, this is the
main reason why we are doing all calculation in the LLA of $g \phi^3$ -
theory but after that we call them parton model approach, expressing our
belief that such calculations performing beyond the LLA will reproduce the
main qualitative features of the LLA.
 
 \subsection{Random walk in $ b$.}
The simple parton picture reproduces also the shrinkage of the diffraction
peak.
Indeed, due to the uncertainty principle
\beq
\Delta b_{i} \,\,k_{ti}\,\sim \,1
\eeq
or in a different form
$$
\Delta b_{i}\,\sim \frac{1}{ < k_t >}
$$
Therefore, after each emission the position of the parton will be shifted
by an amount  $\Delta b$ which is the same on average.
\alphfig
\begin{figure}[htbp]
\centerline{\epsfig{file=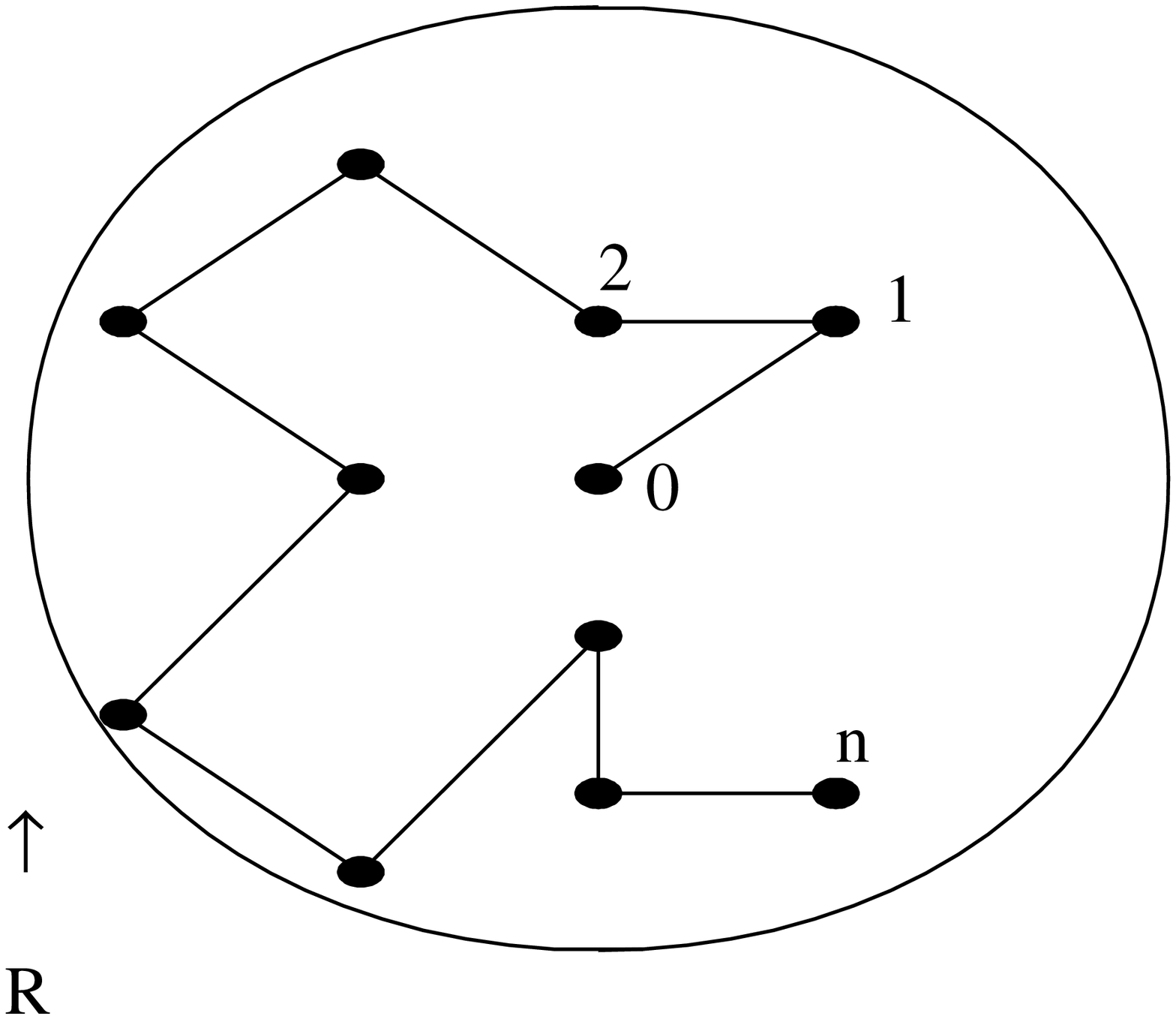,width=100mm}}
\caption{{\it Parton random walk in the transverse plane.}}
\label{fig21a}
\end{figure}
 After $n$ emissions  we
 have the picture given in Fig.21a, namely the total shift in $b$ is
equal to
\beq
b^2_{n}\,\,=\,\,\frac{1}{< k_t>^2} \cdot n
\eeq
which is the typical answer for a  random walk in two dimensions ( see
Fig.21a).
The value of the average number of emissions  $n$ can be estimated from
the
expression for the total cross section (see Eqs.(60)- (61)), since 
$$
\sigma_{tot}\,=\,\frac{\alpha \Sigma(0)}{s}\,\Sigma_n\,\frac{\Sigma^n 
(0)}{n!} \,\ln^n \frac{s}{\mu^2}\,=\,
\sigma_0\,\Sigma_n \frac{ < n >^n}{n!}
$$
which leads to  $< n > \simeq \Sigma(0) \ln s$.
If we substitute this value for  $< n >$ in the eq.(63 ) we get the radius
of
 interaction
$$
R^2\,\,=\,\,<b^2_{n}>\,\,=\,\,\frac{\Sigma(0)}{<k_t>^2} \cdot \ln s\,\,.
$$

Taking into account the  $b$ - profile for the Reggeon ( Pomeron)
exchange ( see  Eq.( 48 ) ) one can calculate the mean radius of
interaction, namely
$$
R^2\,\,=\,\,\frac{\int d^2 b\,b^2\, \,Im\, a_R (s, b)}{ \int \,d^2 b\,Im
\,a(s,b)}\,\,= 4\,\alpha'\,\ln s\,\,.
$$
Comparing these two equations we get
 $$\alpha'\,\,=\,\,\frac{\Sigma(0)}{ 4\,<k_t>^2}\,\,.$$

Therefore, in $g \phi^3$ theory we obtain the  typical properties  of
Pomeron exchange, but the value of the Pomeron intercept is still an open
question which is crucially correlated with the microscopic theory.

In spite of the  primitive level of calculations, especially if you
compare
them with typical QCD calculations in DIS, this model was a good guide for
the
 Pomeron structure for years and, I must admit, it is still the model
where we can see everything that we assign to the Pomeron. Therefore, we 
can formulate the second definition for what  the Pomeron is, ( for
completeness I repeat the first definition once more):
\centerline{}
\centerline{}
\centerline{}
\centerline{{\large \bf What is Pomeron?}}
\centerline{}
\centerline{}
\centerline{}
{ \large \bf A1: The  Pomeron  is the  Reggeon with $ \alpha_P(0)\, - \,1
=\,
\Delta
\,\ll\,\,1$.}
\centerline{}
\centerline{}
\centerline{}
{\large \bf A2: The Pomeron is a  ``ladder" diagram for a  
superconvergent} 

~
{\large \bf\,\,\,\, theory like {$ \mathbf g \phi^3$.}}
\centerline{}
\centerline{}
\centerline{}

The second definition turns out to be extremely useful for practical
purposes, namely, for development of the Pomeron phenomenology, describing 
a variety of different processes at high energy. This subject  will be
discussed  in the next section. However, let us first describe the very
simple picture of the Pomeron structure that results  from the second
definition.

\subsection{Feynman gas approach to multiparticle production.}
\centerline{}
\subsubsection{ Rapidity distribution.}
\centerline{}
\eq{CRD1} can be rewritten in the new variables transverse momenta and
rapidities ofthe  produced particles. Let us start from the  definition of
rapidity
($y$).  For particle with energy $E$ and  transverse momentum $k_t$
the rapidity $y $ is equal to  :
\beq \label{DEFY}
y\,\,=\,\,\frac{1}{2}\,\ln\frac{ E
\,+\,p_L}{E\,-\,p_L}\,\,\rightarrow\,|_{E\,\gg\,max( k_t,m)}\,\,
\ln\frac{2\,E}{m_t}\,\,,
\eeq
where $m^2_t\,=\,m^2\,+\,k^2_t $.

It is easy to see  that the phase space factor in \eq{CRD1} can be written
in terms of 
rapidities in the following way:
\beq \label{CRY}
\sigma_{tot}\,\,=\,\,\sum_{n}\,\,\int \prod_i d^2 k_{ti} d y_i \,\vert
M^2_n(k_{ti} \vert\,\,,
\eeq
with strong ordering in rapidities:
$$
y_1\,>\,y_2\,>\,...\,>\,y_{n - 1}\,>\,y_n\,\,.
$$

This  means that \eq{TOTCR} can be interpreted as the sum over the partial
cross
sections $\sigma^n$, each of them is the cross section for the  production
of
$n$ particles   uniformly distributed in rapidity ( see Fig.21b).
\begin{figure}[htbp]
\centerline{\epsfig{file=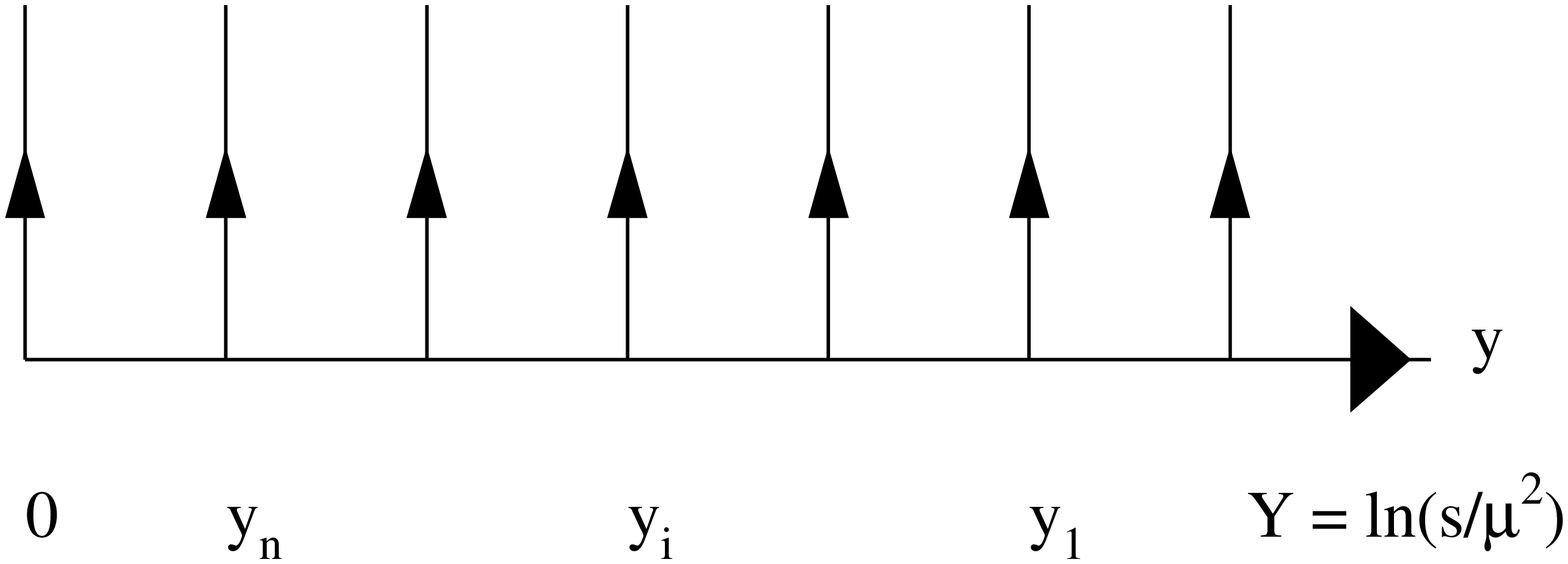,width=100mm}}
\caption{{\it Uniform rapidity distribution of the produced particles in 
the parton model}}
\label{fig20b}
\end{figure} 
\centerline{}
\subsubsection{ Multiplicity distribution.}
\centerline{}
\eq{TOTCR} can be rewritten in the form:
\beq \label{MULTI}
\sigma_{tot}\,\,=\,\,\sigma_0\,\sum_n
\,\sigma^n\,\,=\,\,\sigma_0 \,\sum_n\,
\frac{< n >^n}{n!}\,\,=\,\,\sigma_0\,e^{ <n>}\,\,,
 \eeq
where $< n> $ is the average multiplicity of the  produced partons (
hadrons).

From \eq{MULTI} one can calculate
$$
\frac{\sigma^n}{\sigma_{tot}}\,\,=\,\,\frac{< n >^n}{n!}\,e^{ - < n
>}\,\,.
$$
This is nothing more than the Poisson distribution. The physical meaning of
this distribution is that the produced particles can be considered as 
a system of free particles without any correlation between them.
\centerline{}
\subsubsection{ Feynman gas.}
\centerline{}
 It is clear ( from Fig.20, for example) that the transverse momentum
distribution of one produced particle does not depend how many other
particles have  been produced. Now, we can collect everything that has
been
discussed about particle production in the parton model and draw the
simple picture for the Pomeron structure:
 \centerline{}
\centerline{}
{\bf The Pomeron exchange has the following simple structure:}

{\bf 1. The dominant contribution comes from the production of a large
number of particle, namely $\VEV{ n }\,\,{ \mathbf \propto\,\,\ln
(s/\mu^2)}$;}

{\bf 2. The produced particles are uniformly distributed in rapidity;}

{\bf 3. The correlations between the produced particles are small and can
be
neglected in  first approximation, therefore, the final state for 
Pomeron exchange can be viewed as the perfect gas ( Feynman gas ) of
partons ( hadrons)
 in
the cylinderical phase space with the coordinates: rapidity $y$ and
transverse
momentum ($k_t$) ( see Fig.21c).}

This simple picture generalizes  our experience with the parton
model
and with the experimental data. I am certain that this picture is our
reference point in all our discussions of the Pomeron structure. I think
that the real Pomeron is much more complex  but everybody should
know these  approximations since only this simple picture leads to
the Reggeon with such  specific properties as factorization, shrinkage of
the
diffraction peak and so on.
\begin{figure}[htbp]
\centerline{\epsfig{file=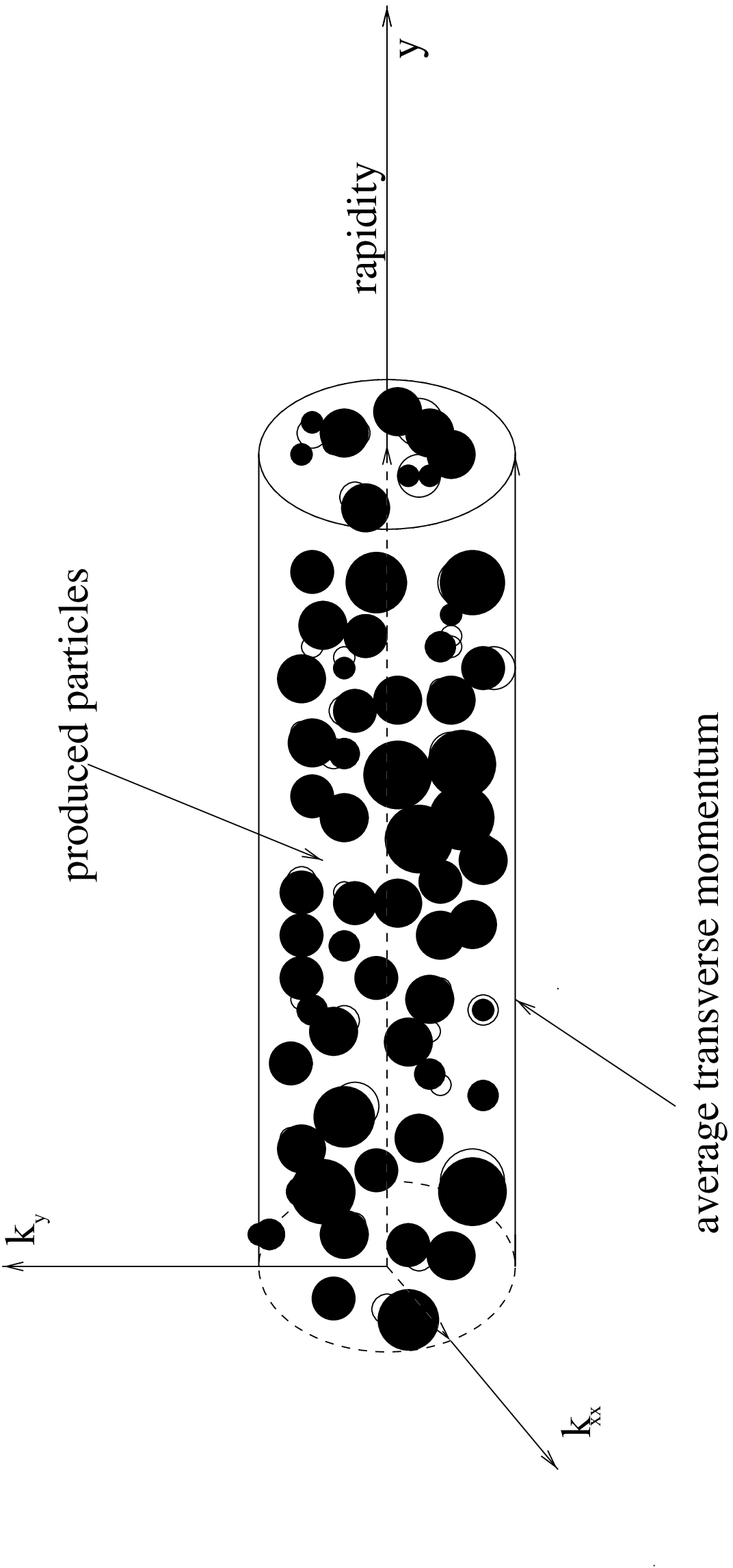,width=100mm}}
\caption{{\it  The cylinderical  phase space for the Feynman gas
approximation.}}
\label{fig21c}
\end{figure}

\section{ Space - time picture of interactions in the parton model.}
\centerline{}
\centerline{}
\subsection{Collision in the  space - time representation.}
\centerline{}
To understand the space - time picture of the collision process in the
parton model is  instructive to rewrite 
\eq{CRD1} in the mixed representation,namely, in  the variables:
transverse
momentum $k_t$ and two space - time coordinates $ x_+\,=\,ct\, +\, z $ and
$x_-\,=\,ct\,-\,z$, where $z$ is the beam direction, for each parton
$``i"$. Since the  amplitude $M_n (k_{t i})$ does not depend on energy and
longitudinal momentum, it is easy to write \eq{CRD1} in such a
representation. Indeed, the time - space structure of the $n$ - th term in
\eq{CRD1} reduces to the simple integral ( see Fig.21d for all
notations)\footnote{To obtain \eq{STCR1} we first did the Fourier
transform to space - time coordinates for the  amplitude of $n$ -
particles
production.
After that we squared this amplitude and integrated it over energies and
longitudinal momenta.The result is written in \eq{STCR1} and it looks so
simple only because all our operations  did not touch the transverse
momenta integral which stands in front of \eq{STRC1}.}:
 \beq \label{STCR1}
\int \prod_i d t_i d z_i d t'_i d z'_i\int^{E}_{\mu} \,\frac{d E_1}{E_1}\,
e^{i (\, E_1 (t_1\,+\,t'_1)\,\,-\,\,p_{L1} (z_1\,+\,z'_1)\,)}...
\eeq
$$
\int^{E_{i
- 1}}_{\mu}\,\frac{d E_i}{E_i}\,e^{i (\, E_i (t_i\,+\,t'_i)\,\,-\,\,p_{Li}
(z_i\,+\,z'_i)\,)}...
$$
$$
\int^{E_{n - 1}}_{\mu}\,\frac{d E_n}{E_n}\,e^{i (\,
E_n (t_n\,+\,t'_n)\,\,-\,\,p_{Ln} (z_n\,+\,z'_n)\,)}\,\,.
$$
\begin{figure}[htbp]
\centerline{\epsfig{file=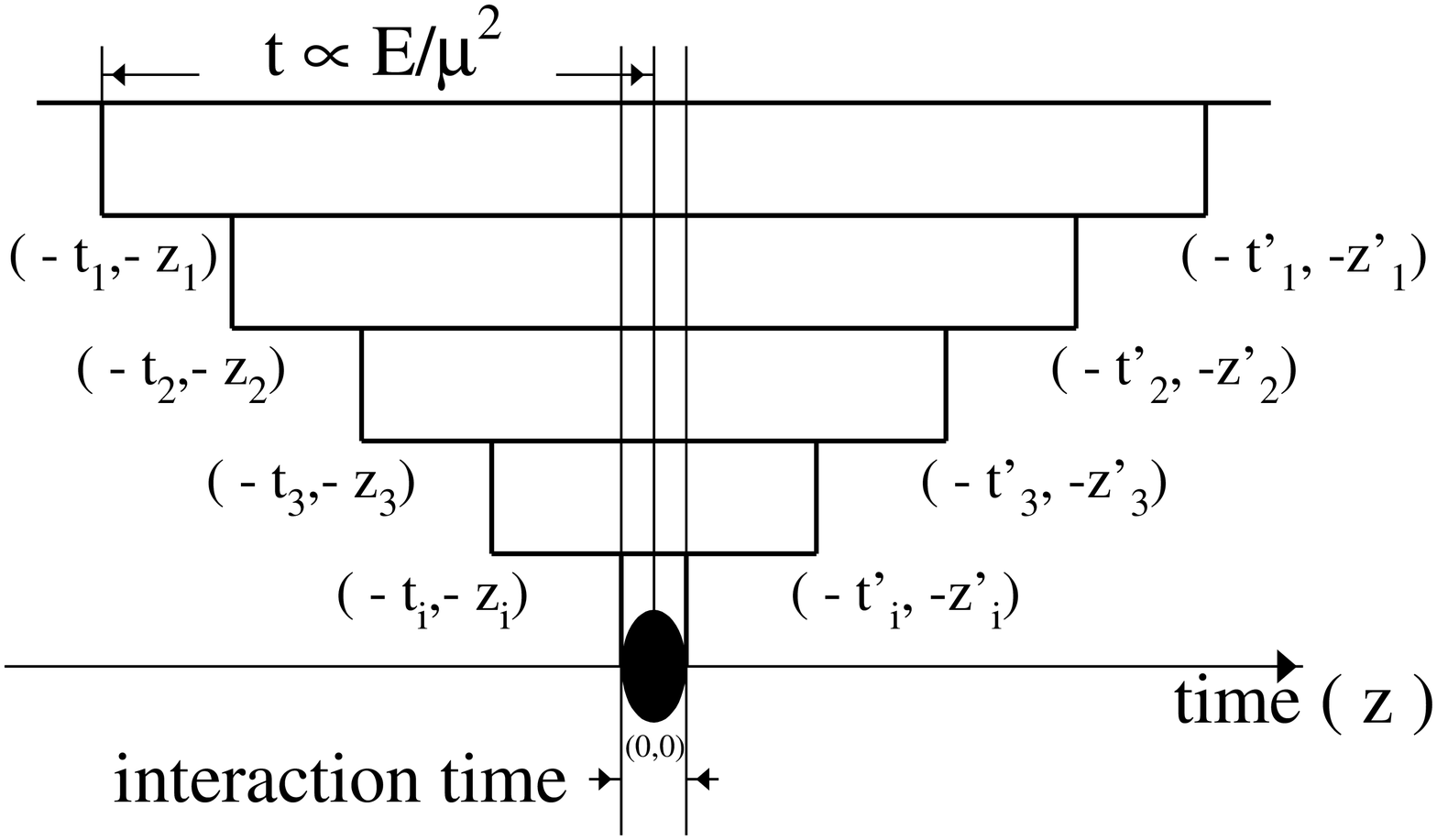,width=160mm}}
\caption{{\it Space - time picture for high energy interactions in the
parton model.}}
\label{fig21d}
\end{figure} 
For a high energy parton ($E_i\,\gg\,\mu$ ) the longitudinal momentum $
p_{Li}\,=\,E_i\,-\,\frac{m^2{ti}}{2 E_i}$, where $m^2_{ti}\,=\,\mu^2
\,+\,k^2_{ti}$. Therefore, 
$$
E_i(t_i\,+\,t'_i)\,\,-\,\,p_{Li}(z_i\,+\,z'_i)\,\,
=\,\,p_{+,i}\,(\,x_{-,i}
\,-\,x'_{-,i}\,)\,\,-\,\,p_{-,i}\,(\,x_{+,i} \,-\,x'_{+,i}\,)\,\,,
$$
where $p_{+,i}\,=\,\frac{E_i\,+\,p_{Li}}{2}$ and
$p_{-,i}\,=\,\frac{E_i\,-\,p_{Li}}{2}$.

For fast partons we have 
$$
E_i(t_i\,+\,t'_i)\,\,-\,\,p_{Li}(z_i\,+\,z'_i)\,\,\rightarrow\,\,
 E_i\, ( c ( t_i \,+\,t'_i) \,-\,(z_i + z'_i)) \,-\,\frac{m^2{ti}}{4 E_i}
\,(  c ( t_i \,+\,t'_i) \,+\,(z_i + z'_i))\,\,.
$$
Therefore,   $ c ( t_i \,+\,t'_i) \,=\,(z_i + z'_i)
\,+\,O(1/E_i) $ at high energy. One can also see that the  integrand in
\eq{STCR1} does not depend on the variables $ t_i - t'_i$ and $z_i -
z'_i$.
Since the  amplitude does not depend on energy and longitudinal momenta,
it contains  $\delta$ - functions  with respect each of these 
variables. Integrating these two $\delta$ - functions yeilds  $t_i =
t'_i$ and $z_i = z'_i$. After taking the integral over $ c ( t_i
\,+\,t'_i)
\,-\,(z_i + z'_i)\,$, which leads to an  extra power of $E_i$ in the
dominator, we   are  left with the following integrations:
\beq \label{STCRFI}
\int dz_{1} \int^{E}_{\mu} \,\frac{d E_1}{E^2_1}\,e^{-
\frac{m^2_{t1}}{E_1}\,z_1}...\int d z_{i} \,\int^{E_{i - 1}}_{\mu}
\,\frac{ d E_i}{ E^2_i}\,e^{-\frac{m^2_{ti}}{E_i}\,z_i}\,...
\int \,d z_{n}\,\int^{E_{n -1}}_{\mu}\,\frac{d E_n}{E^2_n}\,
e^{-\frac{m^2_{tn}}{E_n}\,z_n}\,\,.
\eeq
Due to the factor $E^2_i$ in the dominator     the main contribution in
the integral over $E_i$ comes from the region of  small $E_i$
 but $E_i
\,>\,\frac{z_i}{m^2_{ti}}$ since the integral is suppressed for smaller
energies due to oscillations of the exponent. 

Finally, \eq{CRD1} in space - time coordinates can be written in a form
which is very similar to the form of \eq{CRD1}, 
\beq \label{STFIN}
\sigma_{tot}\,\,=\,\,\Sigma_n \int \Pi_i d^2 k_{ti} |{\tilde M}^2_n
(k_{ti})| \,
\int^{\frac{s}{m^3_{t1}}}_{\frac{1}{\mu}} \,\frac{d z_1}{z_1}
.....\int^{z_{i - 1}}_{\frac{1}{\mu}}
\,\frac{d z_i}{ z_i}\,...\int^{z_{n - 1}}_{\frac{1}{\mu}} \,\frac{d z_n}{
z_n}
\eeq
$$
=\,\,\Sigma_n \int \Pi_i d^2 k_{ti} |{\tilde M}^2_n (k_{ti})| \,\cdot\,
\frac{1}{n!}\,\ln^n s\,\,.
$$
Fig. 21d    illustrates  the high energy interactions accordingly   to
\eq{STFIN} in the lab. frame,
where the target is at  the rest. 
A long time  ( $t_1 \,\propto\,\frac{s}{2m m^2_{t1}}$, where $m$ is the
target mass while $m^2_{t1}$ is the transverse mass of the first parton)
before the  collision with the target the fast hadron decays into a  
system
of partons which to    first approximation  can be considered as non -
interacting ones. The target interacts only with those  partons which have
energy
and /or longitudinal momentum of the order of the target size, which is
$\approx\,1/\mu$. It takes a short  time ( of the order of $\frac{1}{\mu}$
)
in comparison with the long  time of the whole parton cascade which is of
the order of $E/\mu^2$ where $E$ is the energy of the incoming hadron. 
In the parton model we neglect the time of interaction (see Fig.21d) in
comparison with the time of the whole parton fluctuation. Therefore,
we can rewrite \eq{STFIN} in the  general form:
\beq \label{GENST}
\sigma\,\,=\,\,\sum_n \vert \Psi^{partons}_n(t) \vert^2
\,\sigma(\,parton\,\, +\,\,
target\,)  (t)\,\,,
\eeq
where $\Psi^{partons}_n(t)$ is the partonic wave function ( wave function
of $n$ - partons)
 at the moment of
interaction $t$ ( $t$ = 0 in Fig.21d ). From \eq{GENST} one can see that
the target affects only a  small number of partons with sufficiently low
energies while  most properties of the high energy interaction
are absorbed into  the partonic wave function which is the same for any
target. One can recognize the  factorization properties of the Pomeron in
this picture.
\centerline{}
\centerline{}
\subsection{Probabilistic interpretation.}
\centerline{}
\centerline{} 
One can see from \eq{GENST} that the physical observable ( cross section )
can be written as sum of $\vert \Psi_n \vert^2$. $\vert \Psi_n \vert^2$
has obviously a  physical meaning: it is the  probability to find $n$
partons in the parton cascade ( in the fast hadron). Therefore, when  
discussing
the Pomeron structure in the parton model one  can forget 
interference
diagrams and all other specific features of quantum mechanics. In some
sense we can develop a Monte Carlo  simulation for the Pomeron content.
This is so  because the Pomeron is  mostly an inelastic
object and one  can neglect the contribution of the elastic amplitude to
the
unitarity constraint. 

\begin{figure}[htbp]
\centerline{\epsfig{file=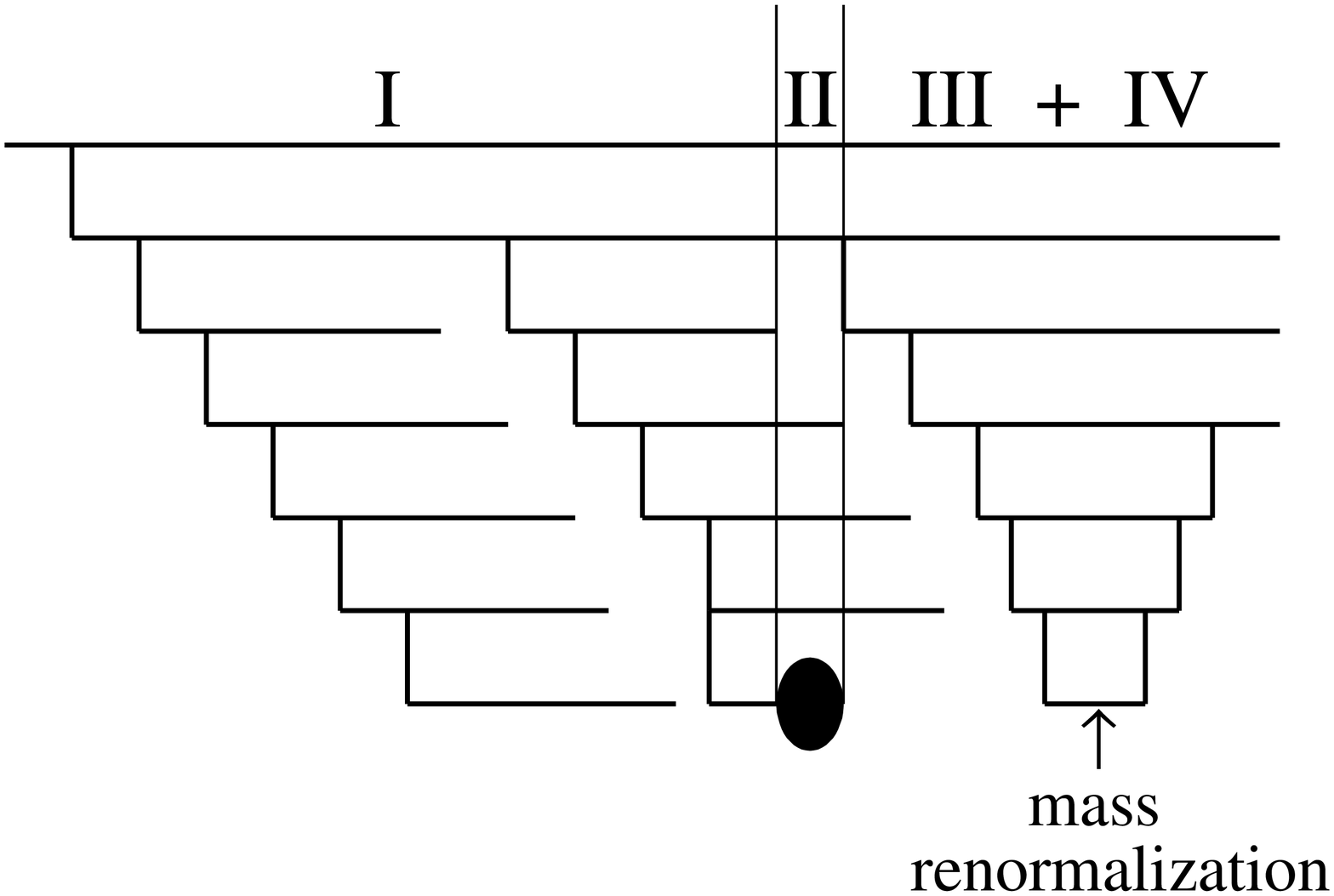,width=160mm}}
\caption{{\it Four stages of high energy interactions in the parton
model.}}
\label{fig21e}
\end{figure}

The second important observation is that for each term in the  sum of
\eq{GENST}, the transverse and longitudinal degrees of freedom are
completely separated and can  be treated independently namely a
simple uniform distribution in rapidity and a rather complicated field
theory for the transverse degrees of freedom.
\centerline{}
\centerline{}
\subsection{ ``Wee" partons  and hadronization.} 
 \centerline{}
\centerline{}
Taking this simple probabilistic ( parton ) interpretation we are able to
explain all features of  high energy interactions  without explicit
calculations. As we have mentioned only very slow parton can interact with
the target in the lab. frame. Such active partons are  called  ``wee"
partons ( Feynman 1969 ).
\begin{figure}[htbp]
\centerline{\epsfig{file=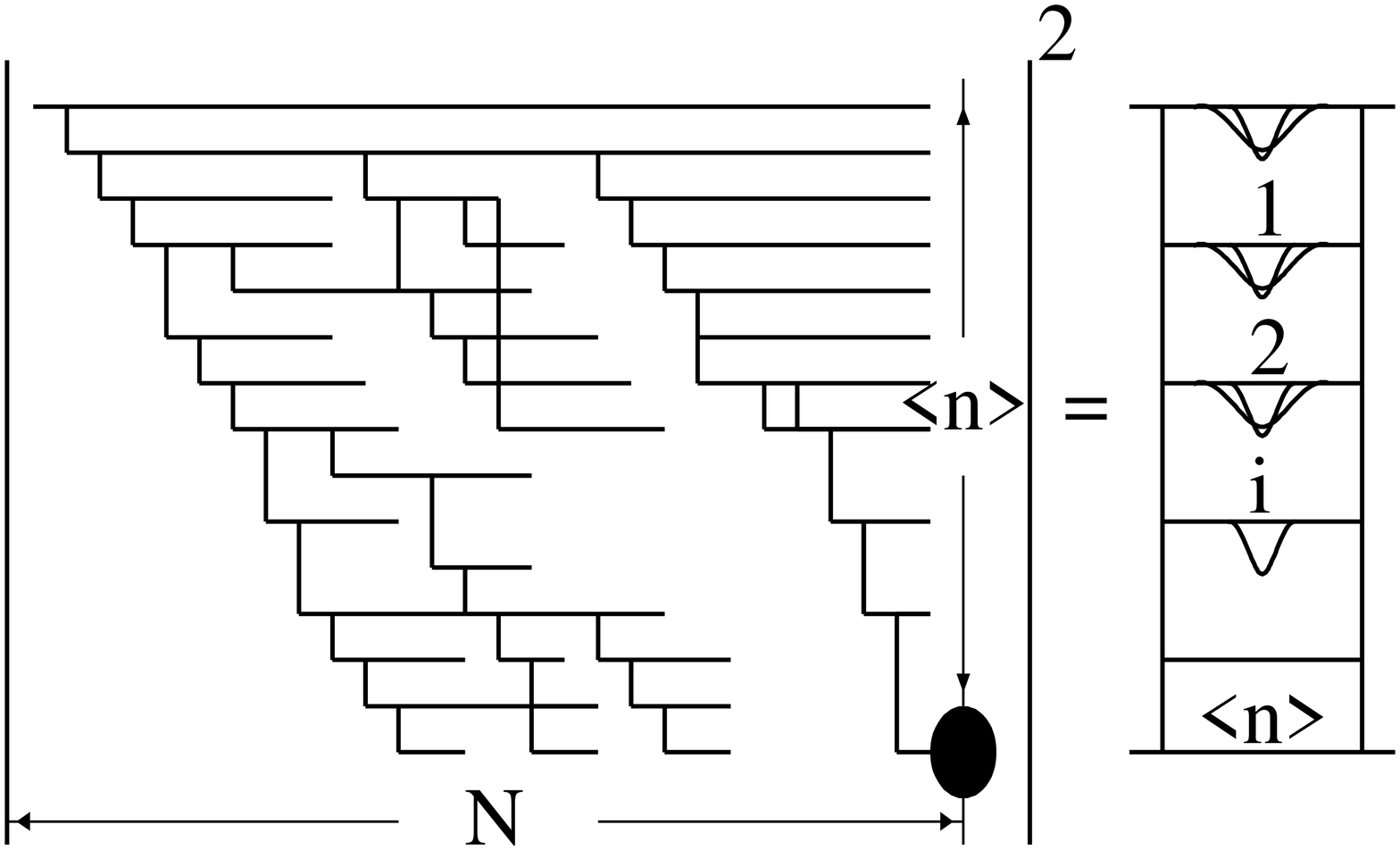,width=140mm}}
\caption{{\it ``Wee" partons and hadrons.}}
\label{fig21f}
\end{figure}

Therefore, in the parton model a  high energy interaction proceeds  four
clearly 
separated stages (see Fig.21e):

{\bf Stage I :} Time ($ t \,\propto\,\frac{E}{\mu^2}$)  before
interaction
of the ``wee" parton with the
target. The fast hadron can be considered as a state of many point-like
particles ( partons) in a  coherent state which we
describe by the means of a  wave function.

{\bf Stage II :} Short time of interaction of the ``wee" parton with the
target. The cross section of this  interaction depends on the target. 
The most important effect of the interaction is 
it destroys the coherence of the partonic wave function.

{\bf Stage III :} Free partons in the final state which, to first
approximation, can be described
as
Feynman gas.

{\bf Stage IV :} Hadronization or in other words the creation of
the observed
hadrons from free partons. There is the  wide spread but false belief  
that
 the simple parton model has no hadronization stage . We will show
in a short while that this  is just wrong.

The total  cross section of the interaction is equal to
\beq \label{PARTON}
\sigma_{tot}(s)\,\,=\,\,N(``wee" \,\,partons) \,\sigma( \,``wee"
parton\,+\,target\,)\,\,.
\eeq
Actually, we can write \eq{PARTON} in a more detailed form by  including
the
integration over the rapidity of the ``wee" parton, namely:
\beq \label{PART1}
\sigma_{tot}(s)\,\,=\,\,\int d y_{wee} \,N( y_{wee} ) \sigma
(y_{wee})\,\,,
\eeq
 and we assume that the cross section for ``wee" parton - target
interaction decreases  as a function of $y_{wee}$ much faster that
$y_{wee}$ - dependence in $N$.

We can estimate the dependence of  $N$  on the  energy. Indeed,  a
 glance on
Fig.21f shows that the number of ``wee" partons should be very large
because each parton can decay into  its own chain of partons. Let us
assume
that we know the multiplicity of partons in one chain ( let us denote it
by $<n>$ ). One can show that $N \,\,\propto\,\,e^{ <n>}$. Therefore,
$$
\sigma_{tot}\,\,\propto\,\,\sigma_0 \,e^{< n >}\,\,,
$$
where $\sigma_0$ stands for the  ``wee" parton - target interaction cross
section. 

We have calculated $ < n > \,\,\approx\,\,\Sigma(o)\, \ln s$  ( subsection
7.4
where we discussed the random walk in the transverse plane ). Graphically,
we show this calculation in Fig.21f. 

An  obvious  question is why the multiplicity in one chain coincides
with the multiplicity of produced particles as it follows from our
calculation of $<n>$ ( see also Fig. 21f). The difference  is in the
hadronization stage. Indeed, a ``wee" parton interacts with the target. We
have $N$ `` wee" partons but only one of them interacts.
Of course, since it 
can  be any, the  cross section is proportional to the total
number of  ``wee" partons. However, if one of the ``wee" partons hits
the target all other pass the target without interaction. They gather
together and contribute to the renormalization of the mass in our field
theory ( see Figs. 21e and 21f). Therefore,in the simple partonic picture  
the number of produced
``hadrons" is the same as the number of partons in one parton chain. Of
course, this is an oversimplified picture of hadronization, but it should
be stressed that even in the simplest field theory such as $g \phi^3$ -
theory we have a hadronization stage which reduces the number of partons
in the parton cascade from $N \,\propto\,e^{<n>}$ to $<n>$.
\centerline{}
\centerline{}
\subsection{ Diffraction Dissociation.}
\centerline{}
\centerline{}
As we have discussed the typical final state in the parton model is an 
inelastic event with large multiplicity and with uniform distribution  of
the
produced hadrons in rapidity. All events with small multiplicity, such as
 resulting from diffraction dissociation, can be considered as a
correction to the
parton model which  should be small. Indeed, diffraction dissociation
events correspond to such interactions of the ``wee" parton with the
target which do not destroy the coherence of the partonic wave function
for  most of partons belonging to it. This process is shown in Fig.21g. In
Fig.21g one can see that the interaction of a ``wee" parton with the
target
does not change the wave function for all partons  with rapidities
$y_i\,<\,y_M$ but  destroys  the coherence completely for partons with
$y_i\,>\,y_M$. In the next section we show how to describe such
processes. 

\begin{figure}[htbp]
\centerline{\epsfig{file=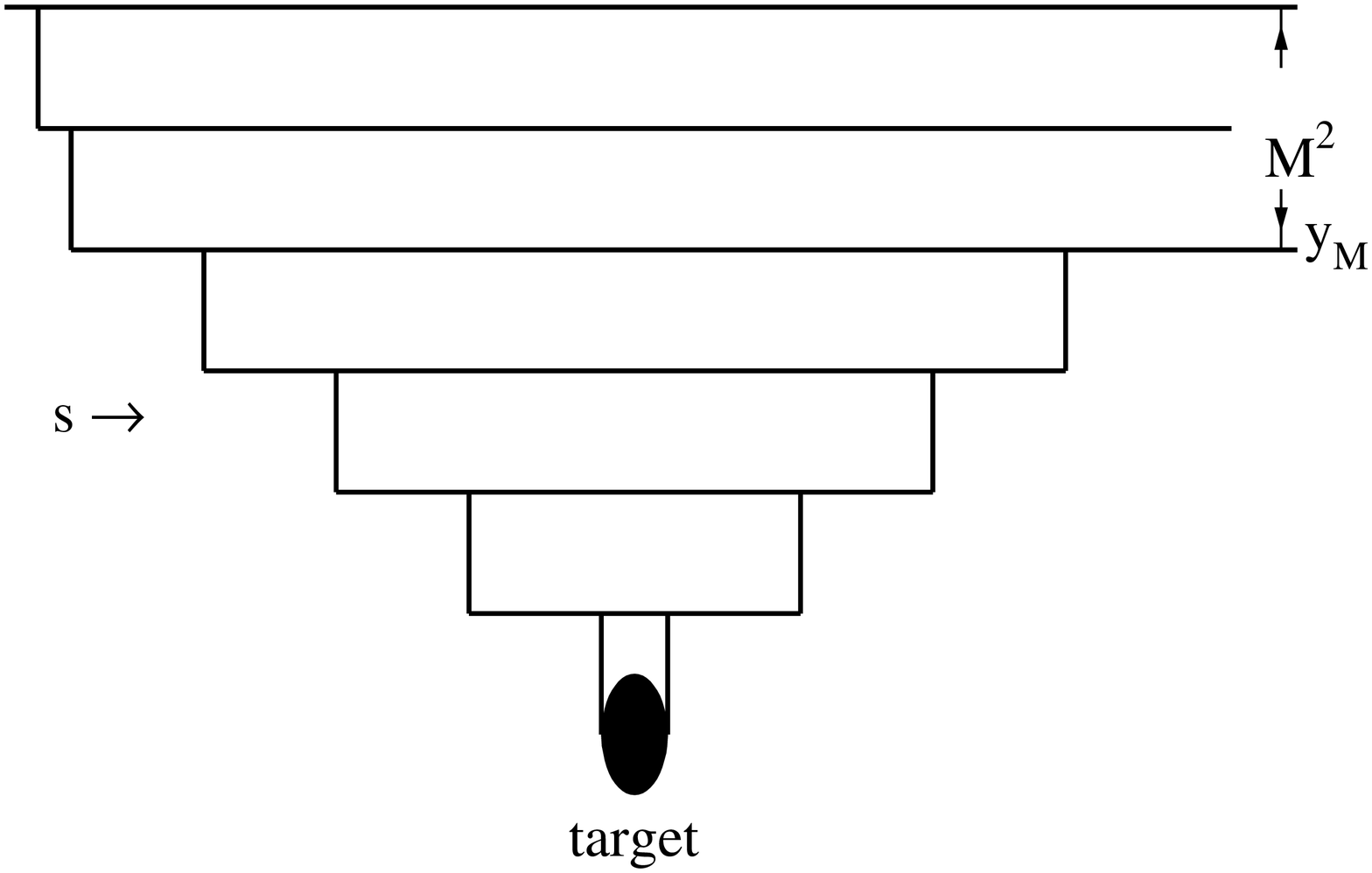,width=120mm}}
\caption{{\it Diffractive dissociation in the parton model.}}
\label{fig21g}
\end{figure} 
\resetfig

\section{ Different processes in the Reggeon
Approach.}

\centerline{}
 \subsection{Mueller technique.}
Using the second definition of the Pomeron, namely:`` 
 The Pomeron is a  ``ladder" diagram for a
superconvergent  theory like {$g \phi^3$ one"( better: using  the second
definition
as a guide) we can easily understand a very powerful technique suggested
by
Al Mueller in 1970. The first observation is  that the optical theorem in
LLA
looks very simple as shown in Fig.22.  The Mueller technique can be
understood
in a very simple way:{ \it for every process try to draw ladder diagrams
and
use
 the optical theorem in the form of Fig.22}.

\begin{figure}[htbp]
\begin{tabular}{c}
\centerline{\epsfig{file=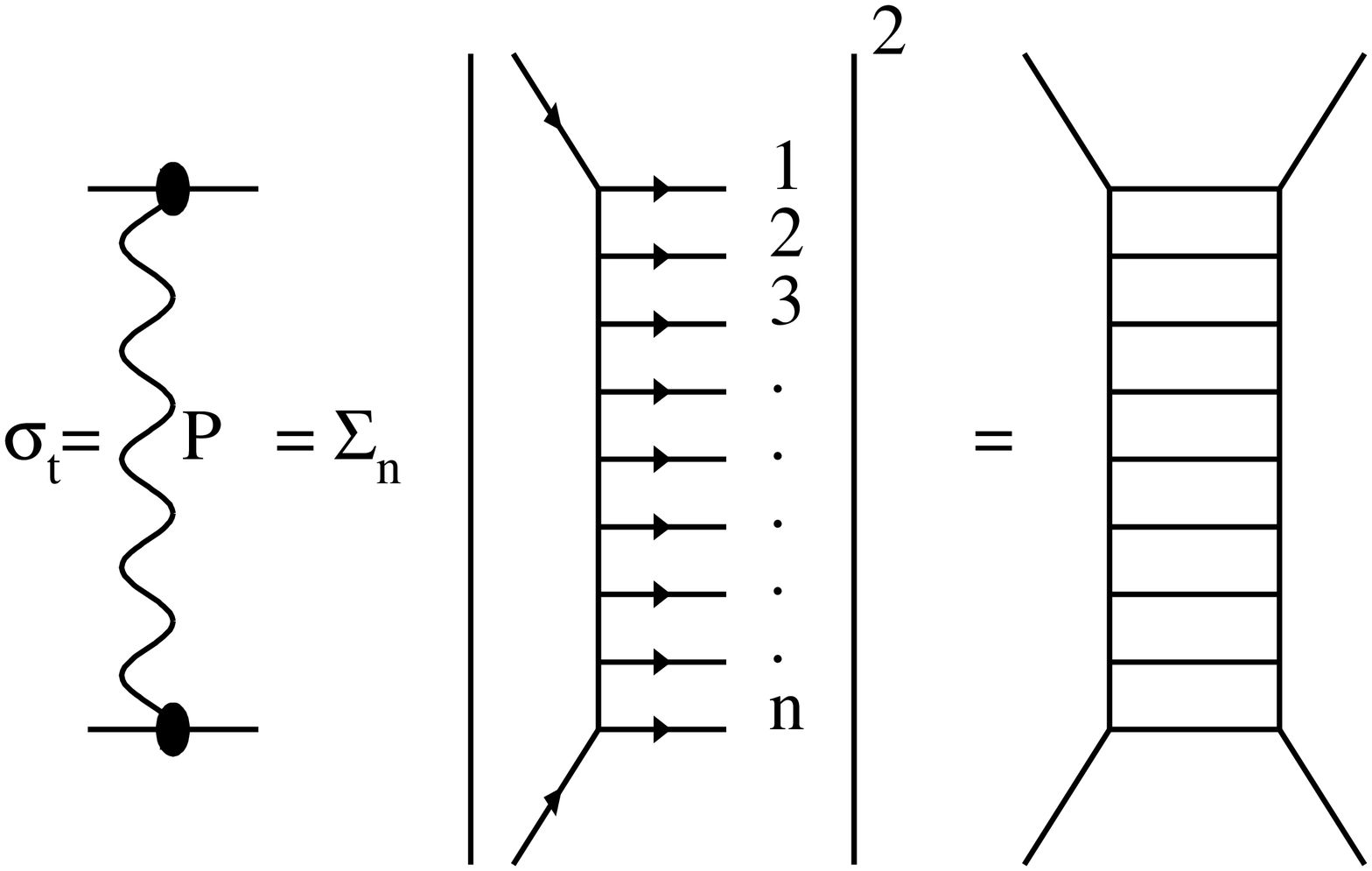,width=120mm}}\\
\centerline{\epsfig{file=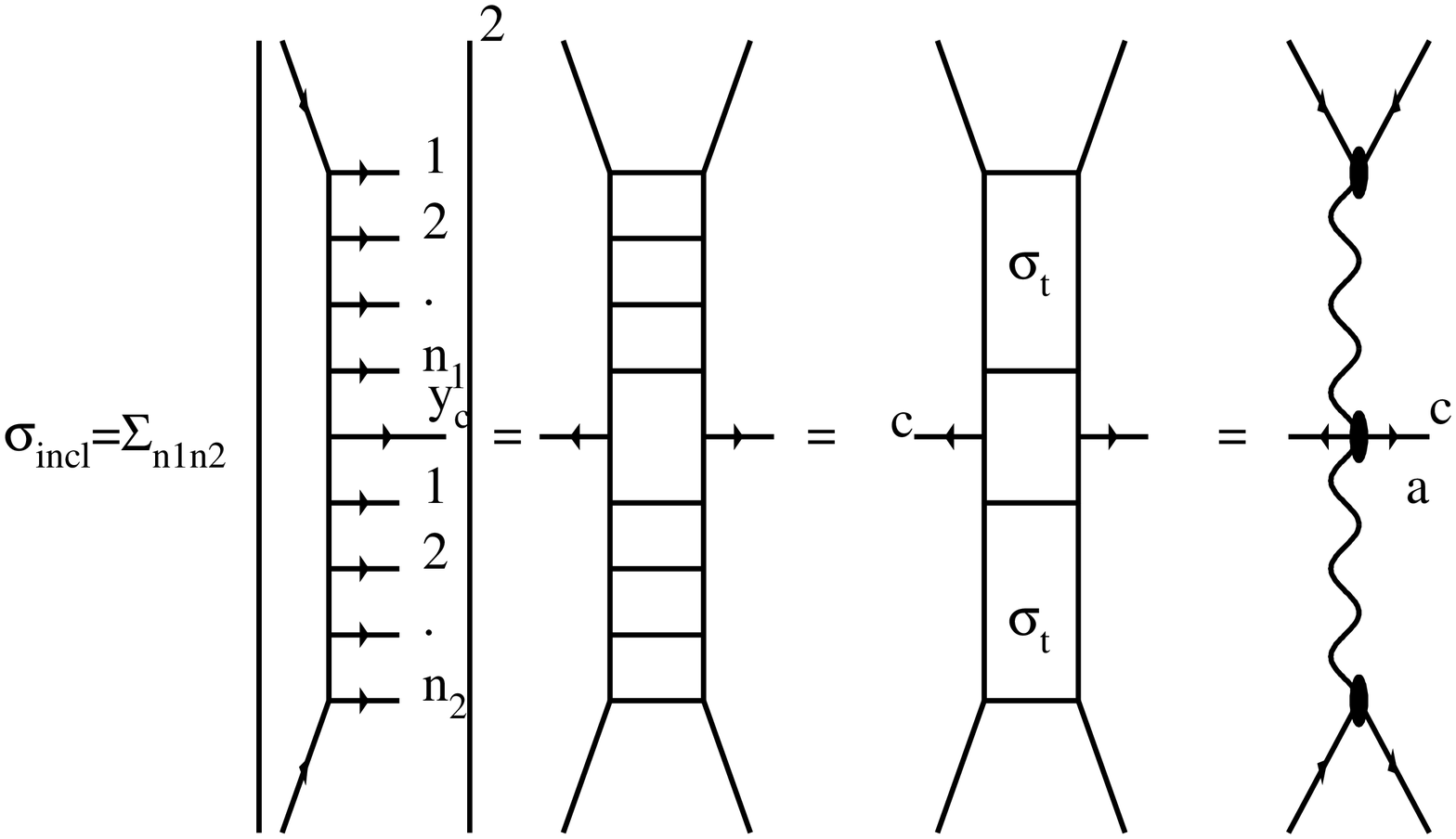,width=120mm}}
\end{tabular}
\caption{{\it Optical theorem and inclusive production in the LLA for $g 
\phi^3$-theory ( parton model).}}
\label{fig22}
\end{figure}

Let us illustrate this technique by considering the single inclusive cross
section for production  of a  hadron $c$ with rapidity $y_c$ integrated
over its transverse
momentum (see Fig.22). Summing over hadrons $n_1$ and $n_2$ with
rapidities
more or less than $y_c$ and using the optical theorem we can rewrite this
cross section as a product of two cross sections with rapidities (
energies)
$Y - y_c$ and $y_c$. Using  Pomeron exchange for the total cross section
and introducing a general vertex $a$ we obtain a Mueller diagram for the
single inclusive cross section.  In spite of the fact that this technique
looks very simple it is a powerful tool to establish a unique description 
of
 exclusive and inclusive processes in the Reggeon Approach. Here
we want to write down several examples of  processes which  can be
treated on the same footing in the Reggeon approach.
\subsection{Total cross section.}
 In the  one Pomeron exchange approximation the total
cross
section  is given by the following expression  (see Fig.23a):
\beq
\sigma_{tot}\,\,=\,\,4 \pi g_1(0) \,g_2(0)  (\frac{s}{s_0} )^{\Delta}\,\,=
\,\,\sigma (s = s_o) ( \frac{s}{s_0})^{\Delta}
\eeq
Remember that  multi-Pomeron exchange is very
essential for describing 
  the total cross section but we postpone this  discussion of their
 contribution to the second part of our lectures.

The energy behaviour of the total cross section in the region of not too
high energy depends on the contribution of the secondary Reggeons. 
It has become customary to consider  only one
secondary Reggeon. It is certainly 
not correct and we have to be  careful. For example, for the  $p - p$
total
cross section we there is a  secondary Reggeon with positive signature
($P'$
or $f$ Reggeon) and two Reggeons with negative signature  ($\omega$ and
$\rho$ ). The first one is responsible for the energy dependence of the
total cross section and  gives the same contribution to $p - p $
 and $\bar p - p $  collisions. The  $\omega$- Reggeon contributes with
opposite
signs   to these two reactions and  is responsible for the value and energy
dependence of the difference $\s_{tot} (\bar p p) \,\,-\,\,\s_{tot} (p
p)\,
\propto\,\, (\,\frac{s}{s_0}\,)^{\alpha_{\omega}(0) - 1}\,\,\approx\,\,
(\,\frac{s}{s_0)}\,)^{- 0.5}$.

Finally, the total cross section can be written in the general form
\beq \label{GENCR}
\s_{tot} \,\,=\,\,4 \pi g^P_1(0) \,g_2^P(0)\,\,  (\frac{s}{s_0})^{\Delta}\,\,
+\,\,\sum_{R_i}( - 1)^{S} 4 \pi g^{R_i}_1(0) \,g^{R_i}_2(0)\,\,
(\frac{s}{s_0})^{\alpha_{R_i}(0) - 1}\,\,,
\eeq
where $S = 0$ for positive signature and $S= 1$ for negative signature
Reggeons  for
particle - particle scattering ( for antiparticle - particle scattering
all contributions are positive).
\alphfig
\begin{figure}[htbp]
\centerline{\epsfig{file=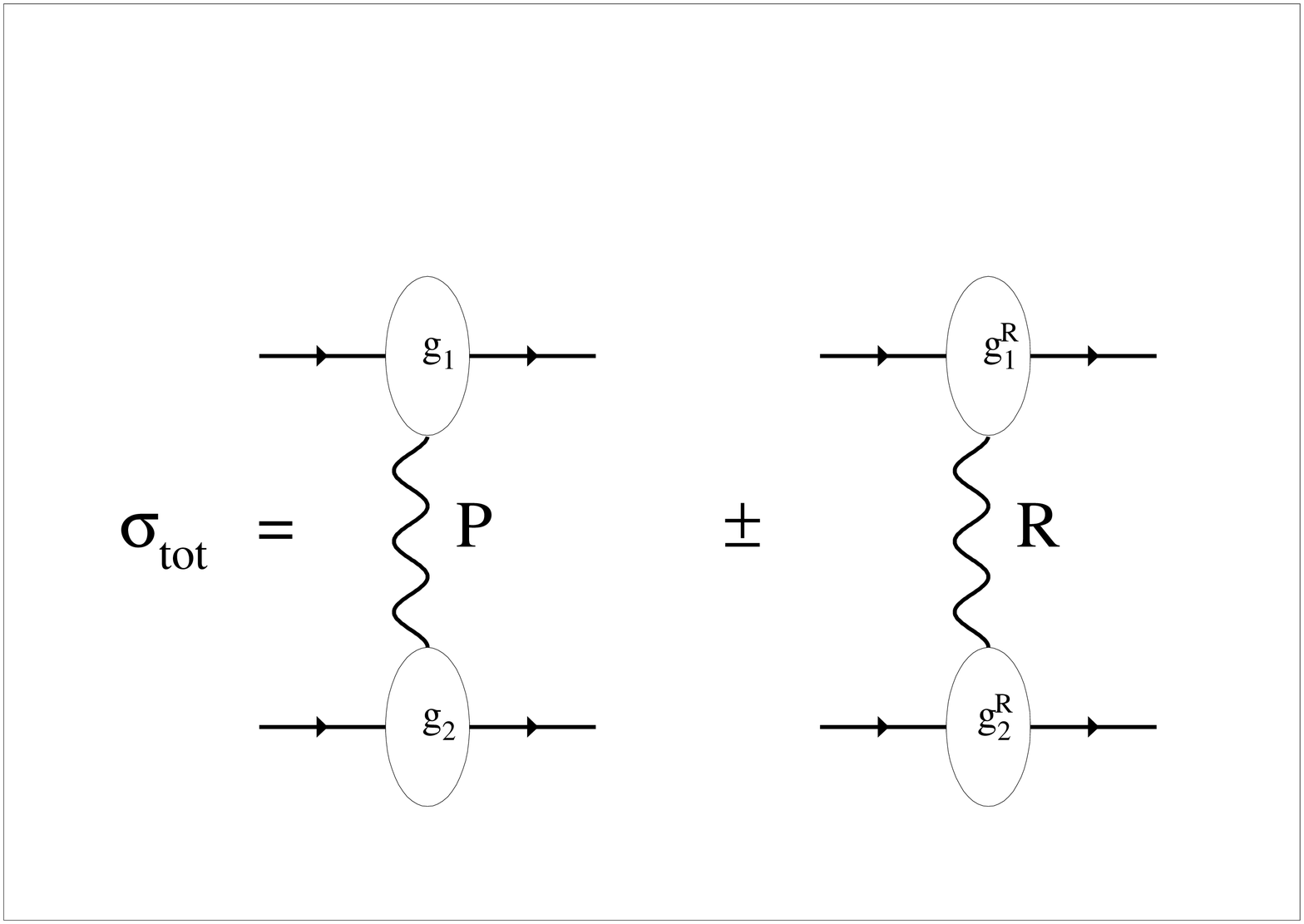,width=120mm}}
\caption{{\it Total cross section in the Mueller technique.}}
\label{fig23a}
\end{figure}

 \subsection{Elastic cross section.}
Collecting everything  we have learned on Reggeon ( Pomeron )
exchange we can show that for one Pomeron exchange the total elastic
cross
 section is equal to (see Fig.23b )
$$
\sigma_{el}\,\,=\,\,\frac{\sigma^2_{tot}}{16 \pi B_{el}}
$$ where
 $$B_{el} \,\,=\,\,2 R^2_{01}\,+\,2 R^2_{02} \,+\, 2 \alpha'_P \ln s/s_0$$
in the exponential parameterization of the vertices.

It is interesting to note that this formula is written in the
approximation
where  we neglected the real part of the amplitude. The correct formula is
$$
\sigma_{el}\,\,=\,\,\frac{\sigma^2_{tot}}{16 \pi B_{el}} \,(
\,1\,\,+\,\,\rho^2\,)\,\,,
$$
where $\rho = \frac{Re A(s,t)}{Im A (s,t)}$. For the Pomeron the quantity   
$\Delta $ is
small (see first definition) and 
$$
\rho\,\,=\,\,\frac{\pi}{2} \,\frac{1}{Im A(s,t)} \,\frac{d Im A(s,t)}{d
\ln
s}\,=\,\frac{\pi}{2}  \Delta\,\,.
$$
For the secondary trajectories we have to take into account the signature
factors ( $\eta_+ $ or $\eta_-$ ) and calculate the real part of the
amplitude. 
\begin{figure}[htbp]
\centerline{\epsfig{file=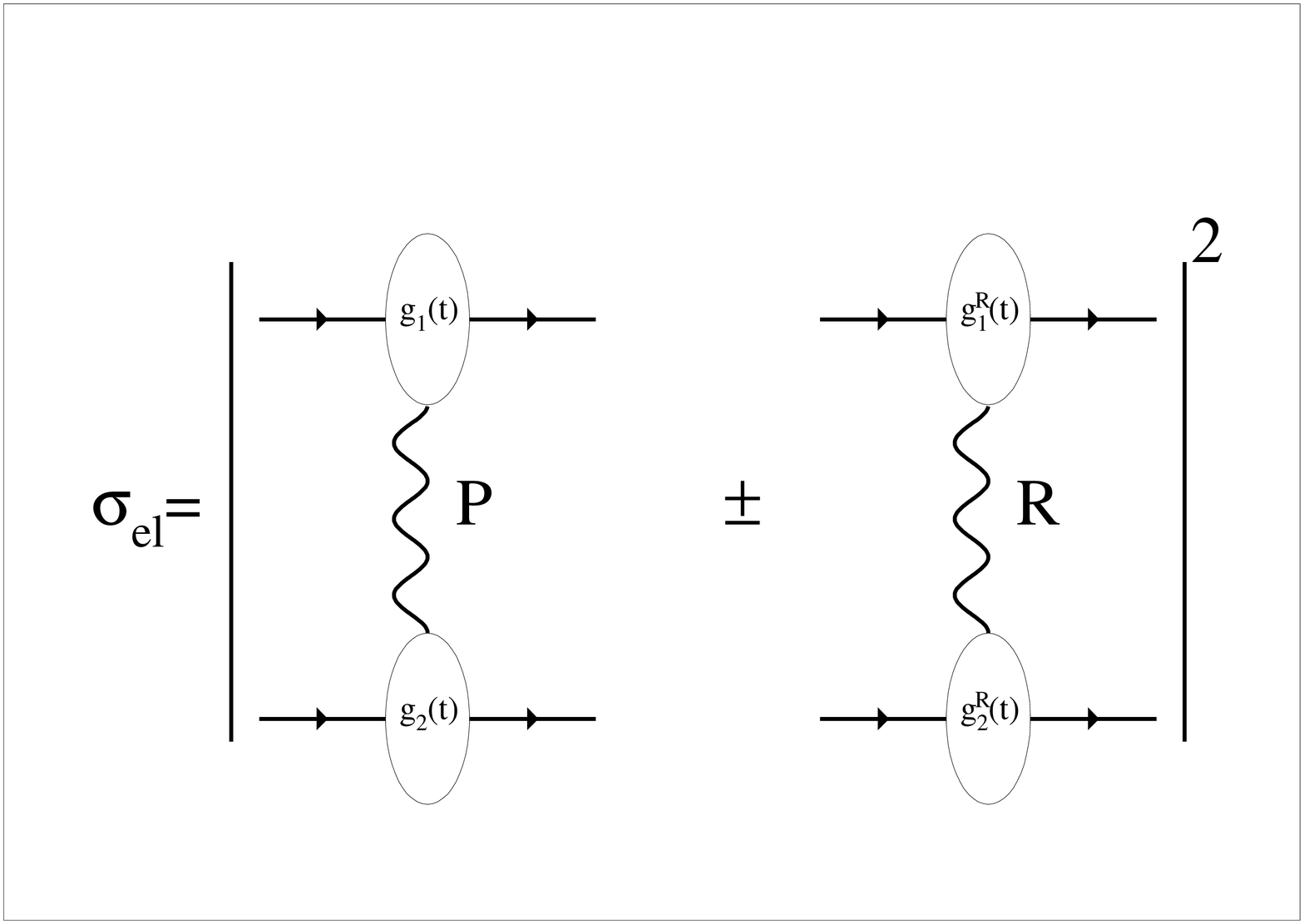,width=120mm}}
\caption{{\it Elastic  cross section in the Mueller technique.}}
\label{fig23b}
\end{figure}

\subsection{Single diffraction dissociation}
The cross section for  single diffraction (see Fig.23c) has the
following
form when  $M^2$ is large):
\beq \label{SDR}
\frac{M^2 d \sigma_{SD}}{d M^2\, d t}\,\,=\,\,\frac{g^2_p(t)}{4}
(\frac{S}{M^2})^{2 \Delta}\,\s_{tot}(2 + P)\,\,,
\eeq
where $\s_{tot}$ is the total cross section ofor  Pomeron + hadron 2
scattering. 
\begin{figure}[htbp]
\centerline{\epsfig{file=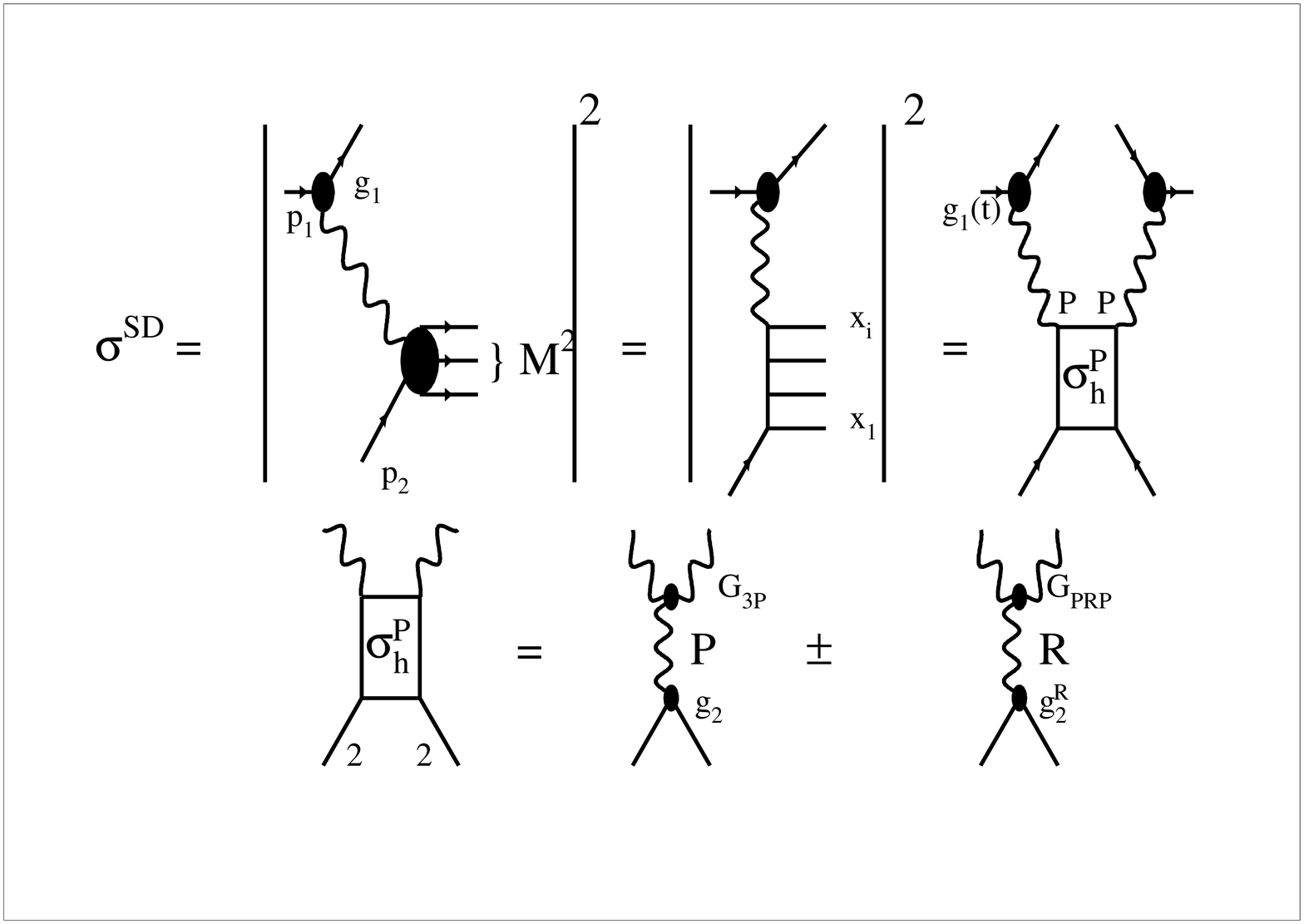,width=120mm}}
\caption{{\it Single Diffraction  in the Mueller technique.}}
\label{fig23c}
\end{figure}

 At first sight, it seems  strange that Pomeron exchange depends on
the 
ratio $\frac{s}{M^2}$. In  LLA for the  $g \phi^2$ - theory ( parton
model )
this process is shown in Fig.23. Here    Pomeron exchange certainly
depends on the energy $s' = x_i p_2 m_1 $ in the rest frame of hadron 1 (
in lab.
frame). In  LLA $$x_1\,\gg\,x_2\,\gg\,...\gg\,x_i$$ and 
$$M^2\,\,=\,\,
( \sum^{i}_{l = 1} p_{0l})^2 - (\sum^{i}_{l = 1} p_{Ll})^2
\,\,\approx\,\,\frac{p^2_{i t}}{x_i}\,\,. $$
Therefore,  in the framework of the parton model where $ < p^2_t >$ does
not depends on energy, $$x_i \,\,\approx\,\, \frac{< p^2_t>}{M^2}$$ and
$$
s'\,\,=\,\,2 x_i p_2 m_1 \,\,=\,\,< p^2_t>\,\cdot\,\frac{s}{M^2}\,\,.$$
We absorb some of the  factors in the definition of the cross section in
\eq{SDR}.

The next question is how well can we determine  the normalization of the
Pomeron -
hadron cross section. The natural condition for the  normalization is 
that the cross section, defined by  \eq{SDR},  coincides  with the
cross section of the interaction of hadron 2 with the resonance (R) at $t
= m^2_R$.  Of course, when $m_R$ is small all formulae give the same
answer. However, let us consider the  $\rho$ - trajectory and compare the
exchange of the $\rho$ - Reggeon at $t = 0$ in two different models: the
Veneziano model and the simple formula of \eq{RGEX}. Taking
$\alpha_{\rho} (t
= 0)$ = $0.5$ we see that cross section in the Veneziano model is $\pi$
times bigger that in \eq{RGEX}.  Therefore,  even for
Reggeons we have a problem with the  normalization. For the
Pomeron the 
situation is even worse since we  know there is even  one
resonance on the Pomeron - trajectory. My personal opinion is that we have
to
make a common agreement what we will expect for the cross section but we
should be very
careful in comparison the value of this  cross section with that for
ordinary   hadron
- hadron scattering.

In the region where   $M^2$ is large  we can  apply for   Pomeron -
hadron
cross section the same expansion with respect to  Pomeron + Reggeons
exchange ( see  \eq{GENCR}
 \beq \label{SD}
\frac{M^2 d \sigma_{SD}}{d M^2}\,\,=\,\, \frac{\sigma_0}{2\pi
R^2(\frac{s}{M^2})}
 \,\cdot\,( \frac{s}{M^2})^{2 \Delta} \,\cdot\,[\,G_{3P}
(0)
\,\cdot\,
(\frac{M^2}{s_0})^{\Delta} \,+\,G_{PPR}(0)
 (\frac{M^2}{s_0} )^{ \alpha_R (0) - 1}\,]
\eeq
Let me summarize what we learned experimentally about  the triple Pomeron
vertex:

1. The value of $G_{3P}$  is smaller than the value of $g_i$ ( the hadron
-
Pomeron vertex):  $G_{3P}\,\,\approx 0.1\,g_i$;

2. The dependence  of the triple Pomeron vertex on $t$ can be
characterized by 
the radius of the triple Pomeron vertex $r_0$ ($G_{3P} (t)
\,\approx\,e^{ - 2 \,r^2_0 |t|}$), which turns out to be
rather
small, namely, $r^2_0 \,\leq\,1\,GeV^{-2}$ .

It is easy to  show that in  the exponential parameterization of the  
vertices
 (see \eq{RINB})
$$
R^2 (\frac{s}{M^2})\,\,=\,\,2 R^2_{01} \,+\,2 r^2_0  \,+\,2 \alpha'_P \ln
(s/M^2)\,\,.
$$ 
\subsection{Non-diagonal contributions.}
Eq.(76) gives the correct descriptions for  the energy and mass behaviour
of
the SD cross section  in the Reggeon phenomenology but only for 
$\frac{s}{M^2}\,\,\gg\,\,1$. For
$\frac{s}{M^2}\,\approx\,\,1$ we have to modify Eq.(76) by  including
secondary Reggeons. It means that we have to substitute in Eq.(76)
$$
(\,\frac{s}{M^2}\,)^{\Delta}\,\,\rightarrow\,\,(\,\frac{s}{M^2}\,)^{\Delta}
\,\,+\,\,\sum_i\,A_{R_i}\,(\,\frac{s}{M^2}\,)^{\alpha_{R_i}(0) - 1}\,\,,
$$
where $i$ denotes the Reggeon with trajectory $\alpha_{R_i} (t)$.
As can be  seen  the $s$ - dependence of \eq{SD} is only governed by the
Pomeron ( $ M^2 d \s_{SD}/d M^2\,\,\propto\,\, s^{2\Delta}$)  while
\eq{SDG} leads to a  more general energy dependence, namely,
\alpheqn
\beq \label{SDE}
\frac{M^2 d \s_{SD}}{d M^2}\,\,\propto\,\,\{\,
(\,\frac{s}{M^2}\,)^{\Delta}\,\,
\pm\,\,A_R \,(\,\frac{s}{M^2}\,)^{\alpha_R(0) - 1}\,\}^2\,\,.
\eeq

 Therefore the general equation has the  form:
\beq \label{SDG}
\frac{M^2 d \s_{SD}}{d M^2\,d t}\,\,
=\,\,\sum_{i,j,k}\,\,g_i\,g_j\,g_k\,G_{ijk}\,
(\,\frac{s}{M^2}\,)^{\Delta_i}\,(\,\frac{M^2}{s_0}\,)^{\Delta_j}\,
(\,\frac{s}{M^2}\,)^{\Delta_k}\,\,,
\eeq
where $i$,$j$ and $k$ denote all Reggeons including the Pomeron and 
$\Delta_l \,\,=\,\,\alpha_l (0)\, -\, 1$ with $l$ = $i$,$j$ and $k$.
Note, that factor $( \frac{M^2}{s_0} )^{\Delta_j}$ comes from the Pomeron
- hadron cross section as has been  discussed above ( see \eq{SD} ). 
 
  We can see from \eq{SDG} as well as from \eq{SDE} that the first
correction to the Pomeron induced energy behaviour  comes from the  so
called
interference or non-diagonal (N-D)  term, which can be written in the 
general form:
\beq \label{INSD}
\frac{M^2 d \s^{N-D}_{SD}}{d M^2\, d t}\,\,=\,\,
\s_{P + p \rightarrow R +p}( M^2 )  
\,g_P\, g_R
\,(\,\frac{s}{M^2}\,)^{\Delta_P}\,(\,\frac{s}{M^2}\,)^{\Delta_R}\,\,,
\eeq
where $P$ stands for the Pomeron and $R$ for  any secondary Reggeon. 
$\s_{P + p \rightarrow R +p}( M^2 )$ can be written as a sum of the
Reggeon contributions as
\beq \label{CSSD}
\s_{P + p \rightarrow R +p}( M^2 ) \,\,=\,
\,g_P \,G_{PPR}\,(\,\frac{M^2}{s_0}\,)^{\Delta_P}
\,\,\pm\,\, 
\,g_R \,G_{PPR}\,(\,\frac{M^2}{s_0}\,)^{\Delta_R}\,\,.
\eeq
It is clear that the non-diagonal term gives the first and the most
important correction which has  to be taken into account in any
phenomenological approach to provide a correct determination of the
diffractive dissociation contribution. By definition, we call diffractive
dissociation only the contribution which survives at high energy or, i.e. 
it corresponds to the Pomeron exchange term in  the Reggeon
phenomenology. Unfortunately, we know almost nothing about this
non-diagonal term. We cannot guarantee even the sign of the contribution.
This lack of our knowledge I have tried to express by  putting $\pm$ into
the
above
equations.

 Let me discuss here what I mean by  ``almost nothing".

First, we can derive an inequality for $\s_{P + p \rightarrow R +p}( M^2)$
using the generalized optical theorem that we discussed in section 9.1.
By definition 
$$
\s_{P + p \rightarrow R +p}( M^2 ) \,\,=\,\,\frac{1}{2}\sum_{n,q}
\{\, g_P(M^2,n,q)\cdot g^*_R(M^2,n,q)\,\,+\,\, g_P(M^2,n,q)\cdot
g^*_R(M^2,n,q)\,\}\,\,,
$$
where $n$ is the number of produced particles with mass $M$ and $q$
denotes 
all other quantum numbers of the final state. Notice that the  factor
$1/2$ in
front is due to the definition in \eq{SDG} where we sum separately over
$PR$ and $ RP$ contributions.

 The
inequality 
$$
\sum_{n,q}|\, g_P(M^2,n,q)\, -\,z\, g_R(M^2,n,q)\,|^2\,\,\geq\,\,0
$$
implies for  any value of $z$, therefore
\beq \label{SB}
\s^2_{P + p \rightarrow R +p}( M^2 )
\,\,\leq\,\s_{P + p \rightarrow P +p}( M^2 )\cdot \s_{R + p
\rightarrow R +p}( M^2 )
\eeq
Using \eq{CSSD} and the  analogous expressions  for the  $p + p$ and $R +
p$ cross
sections,  \eq{SB} leads to
\beq \label{SBV}
 G^2_{PPR}\,\,\leq\,\,G_{PPP}\cdot G_{RPR}\,\,;\,\,\,\,\,\,\,\,
G^2_{PRR}\,\,\leq\,\,G_{PRP}\cdot G_{RRR}\,\,.
\eeq

Actually, \eq{SBV} is the only reliable knowledge  we have on the
interference (
non-diagonal ) term. However, in the simple parton model or/and in the LLA
of $g\phi^3$-theory we know that the sign of the non-diagonal vertices is
positive and that the inequality of \eq{SBV} is saturated:
\beq \label{SBVR}
 G^2_{PPR}\,\,=\,\,G_{PPP}\cdot G_{RPR}\,\,;\,\,\,\,\,\,\,\,
G^2_{PRR}\,\,=\,\,G_{PRP}\cdot G_{RRR}\,\,.
\eeq
The same result  can be obtained  in other models where   the
Pomeron scale is assumed to be  larger than the typical hadron size, as
for instance, in the
additive quark model. However, in general we have no proof for \eq{SBVR}
and, therefore,  can recommend only to introduce a new parameter
$\delta$ and use the following form:
\beq \label{SBVF}
 G_{PPR}\,\,=\,\,sin \delta_P\cdot \sqrt{G_{PPP}\cdot
G_{RPR}}\,\,;\,\,\,\,\,\,\,\,
G_{PRR}\,\,=\,\,sin\delta_R \cdot \sqrt{G_{PRP}\cdot G_{RRR}}\,\,.
\eeq
\reseteqn
 This section is a good illustration how badly we need a theory
for the high energy scattering.
  
\subsection{Double diffraction dissociation.}
  
  From Fig.23d one can see that the cross section for  double
diffraction 
 dissociation ( DD ) process  is equal to:
\beq \label{DD}
\frac{M^2_1 M^2_2 d^2 \sigma_{DD}}{d M^2_1 d M^2_2}\,\,=\,\,
\frac{\sigma_0}{ 2 \pi R^2 (\frac{s s_0}{M^2_1 M^2_2} )} \cdot
G^2_{3P} (0) \cdot ( \frac{s s_0}{M^2_1 M^2_2} )^{2 \Delta} \cdot
( \frac{M^2_1}{s_0})^{\Delta}
  \cdot
( \frac{M^2_2}{s_0})^{\Delta}
\eeq
in the region of large values of produced masses ( $M_1$ and $M_2$ ).

It is important to note that the energy dependence is contained in the
variable
$$
s"\,\,\propto\,\,\frac{s s_0}{M^2_1 M^2_2}\,\,.
$$
Repeating the  calculation of the previous subsection we  see
that (see Fig.23d):
\beq \label{ARG}
x^1_i\,\,=\,\,\frac{< p^2_t >}{ M^2_1}\,\,;
\eeq
$$
x^2_{i'}\,\,=\,\,\frac{< p^2_t >}{ M^2_2}\,\,;
$$
and
$$
s"\,\,=\,\,2 \,x^1_i\,x^2_{i'}\,p_{1 \mu}\,p_{2 \mu}\,\,=\,\,( < p^2_t
>)^2 \,\,\frac{s}{M^2_1\,M^2_2}\,\,\approx\,\,s_0\,\frac{s
\,s_0}{M^2_1\,M^2_2}\,\,.
$$

Note also, that, in the exponential parameterization of the vertices ( see
\eq{RINB} )
  $R^2 ( \frac{s \,s_0}{M^2_1\,M^2_2})$ in \eq{DD} is equal to
$$
R^2 ( \frac{s \,s_0}{M^2_1\,M^2_2})\,\,=\,\,2\, r^2_0 \,\,+\,\, 2
\alpha'_P \ln ( s s_0/M^2_1 M^2_2)\,\,.
$$

\begin{figure}[htbp]
\centerline{\epsfig{file=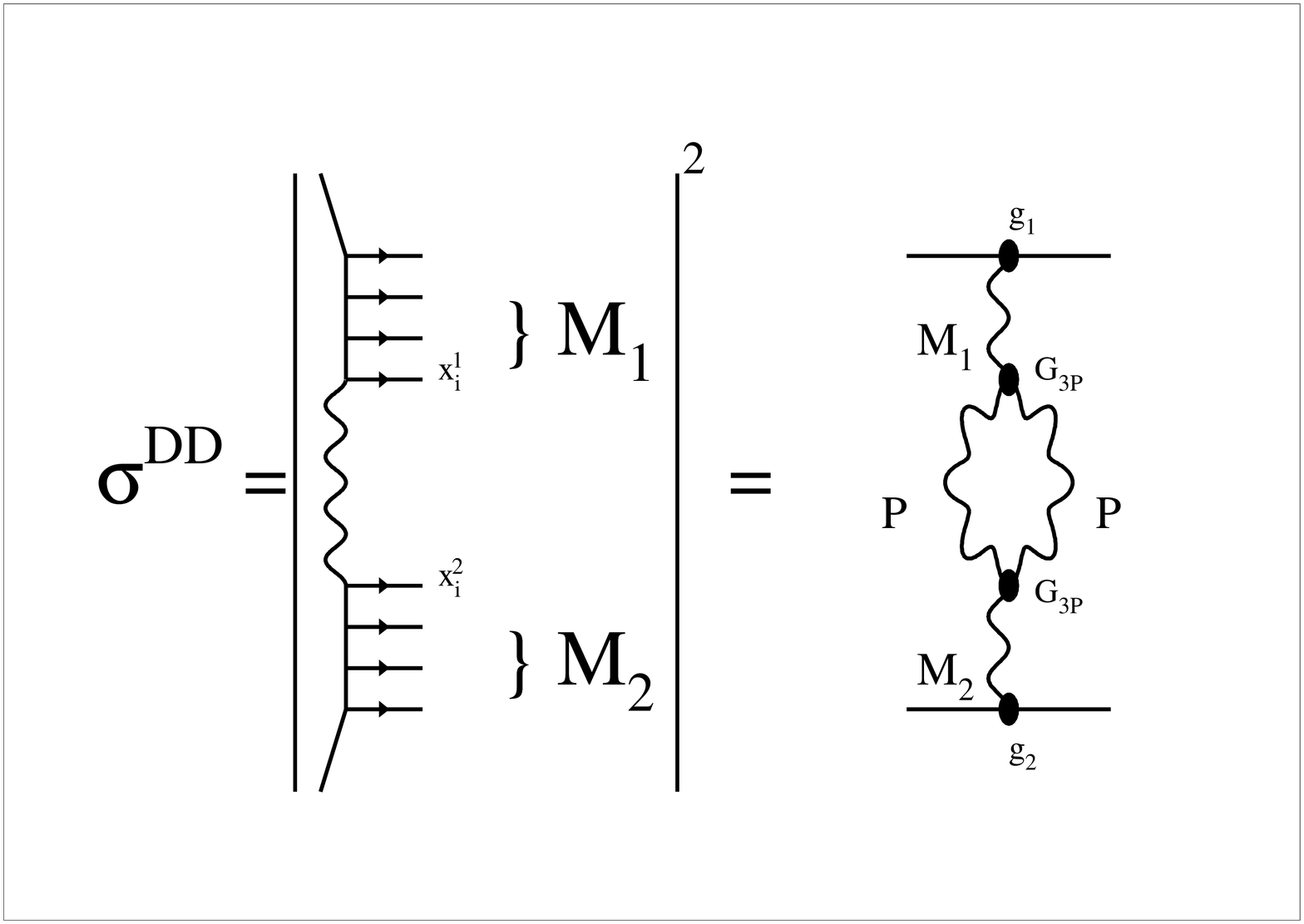,width=120mm}}
\caption{{\it Double Diffraction  in the Mueller technique.}}
\label{fig23d}
\end{figure}

\subsection{Factorization for diffractive processes.}

Comparing the cross sections for elastic,  double and  single diffraction
 the following factorization
relation can be derived:

\beq
\frac{M^2_1 M^2_2 d^2 \sigma_{DD}}{d M^2_1 d M^2_2}\,\,=\,\,   
\frac{\frac{M^2_1 d \sigma_{SD}}{d M^2_1} \frac{M^2_2 d \sigma_{SD} }{d
M^2_2}}
{ \sigma_{el}}\,\cdot\,\frac{R^2 (\frac{s}{M^2_1}) R^2
(\frac{s}{M^2_2})}
{R^2 (\frac{s s_0}{M^2_1 M^2_2}) R^2_{el} (s)}
\eeq
where
$$R^2_{el} = 2 R^2_{01} + 2 R^2_{02} + 2 \alpha'_P \ln (s/s_0) $$ 
The  event structure for double diffraction is
sketched 
in Fig.24 in a  lego -  plot. No particles are produced with
rapidities $\Delta y = \ln \frac{s s_0}{M^2_1 M^2_2}$.
\subsection{ Central Diffraction.}
This process leads to  production of   particles in
the central
 rapidity region, while there are no particles in other regions
of  rapidity.

The cross section for the production of particles of   mass
$M$ in proton - proton collision can be written in the form:
\beq \label{CD}
M^2\,\frac{d \s}{d M^2}\,\,=\,\,\s_{PP} \,\s^2_{pp}\,\frac{1}{
R^2_{0p}\,+\,\alpha'_P(0)\,\ln(s/M^2)}\,\frac{1}{2 \alpha'_P
(0)}\,\ln(\frac{R^2_{0p} \,+\,\alpha'_P(0) \ln(s/M^2)}{R^2_{0p}}\,\,.
\eeq
The last factor in \eq{CD} arises from the integration over rapidity $y_1$
 of the factor 
$1/( R^2_{0p} + \alpha'(0)_P (Y - y_1) )( R^2_{0p} + \alpha'(0)_P (Y -
y_1) )$ which arises  from the integrations over $t_1$ and $t_2$ (see
Fig.23e).
\begin{figure}[htbp]
\centerline{\epsfig{file=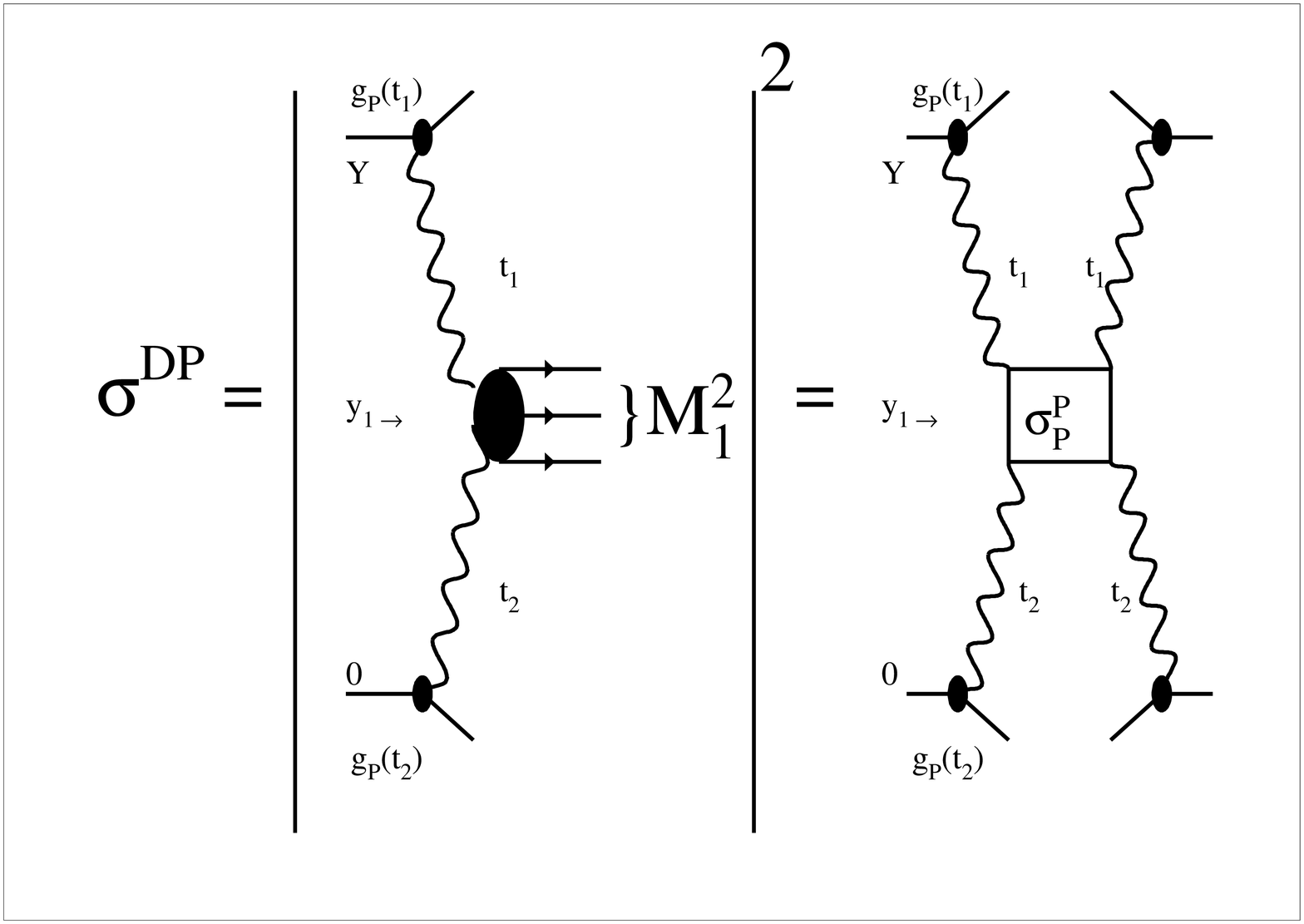,width=120mm}}
\caption{{\it Central Diffraction in the Mueller technique.}}
\label{fig23e}
\end{figure}

\subsection{Inclusive cross section.}
 As  discussed before the inclusive cross section according to the
Mueller  theorem can be described by the diagram of Fig.22.
For reaction $$
a\,\, +\,\, b \,\,\rightarrow\,\,c(y)\,\,+\,\,anything 
$$
  the inclusive cross section can be written
 in the simple form:
\beq
\frac{d \sigma}{d y_c} \,\,=\,\, a \cdot \sigma_{tot}
\eeq

where $a$ is the new vertex for the emission of the particle $c$ which
you can see in 
Fig.22.
\subsection{Two particle rapidity correlations.}
The Reggeon approach  can be used  for  estimating  of the two particle   
 rapidity correlation function, which is defined as:
\beq
R \,\,=\,\,\frac{\frac{d^2 \sigma( y_1, y_2 )}{\sigma_{tot} d y_1 d y_2}}{
\frac{d \sigma(y_1)}{\s_{tot}\, d y_1} \frac{d \sigma (y_2)}{\s_{tot} d
y_2}}\,\,-1
\eeq
where  $ \frac{d^2 \sigma}{d y_1 dy_2}$ is the double inclusive
cross section for  the reaction:
$$
a\,\,+\,\,b\,\,\rightarrow \,\,1 (y_1)\,\,+\,\,2 (y_2)\,\,+ \,\,anything
$$
\par

The double inclusive cross section can be described in the Reggeon
exchange approach by two diagrams ( see Fig.23f). Indeed, at high energies
$Y - y_1\,\gg\,1$ and $y_2 - 0 \,\gg\,1$ and we can restrict ourselves by
considering only Pomerons exchanges in these rapitidy intervals while for
the rapidity interval $\Delta y = |y_1 \,-\,y_2|$ we take onto account the
exchange by the Pomeron and the Reggeon since this interval can be rather
small in the correlation function. 
 The contribution from 
the first diagram drops out  in the definition of the correlation function
but the second one survives and leads to  the correlation function:
\beq \label{CF}
R (\Delta y = |y_1 - y_2|)\,\,=\,\,\frac{a^2_{PR}}{a^2_{PP}}\,\cdot\,
e^{ ( 1 - \alpha_R (0) ) \,\Delta y
}\,\,.
\eeq
All notations are clear from Fig.23f.

It should be stressed that the Reggeon approach gives an estimate for
 the correlation length ( $ L{cor}$).
 One  can rewrite the above formula in the
form:
$$
R \,\,=\,\, SR \cdot e^{- \frac{\Delta y}{L_{cor}}}\,\,,
$$  
The  Reggeon formula yields  $L_{cor} \simeq 2 $. 
Note  that this formula
describes the correction to the Feynman gas model.  One can see that these 
corrections vanish  at large differences in rapidity.

We would like to stress that the correlation function depends strongly on
the value of shadowing corrections ( many Reggeon exchange).  This
problem  will be  discussed  later.
\begin{figure}[htbp]
\centerline{\epsfig{file=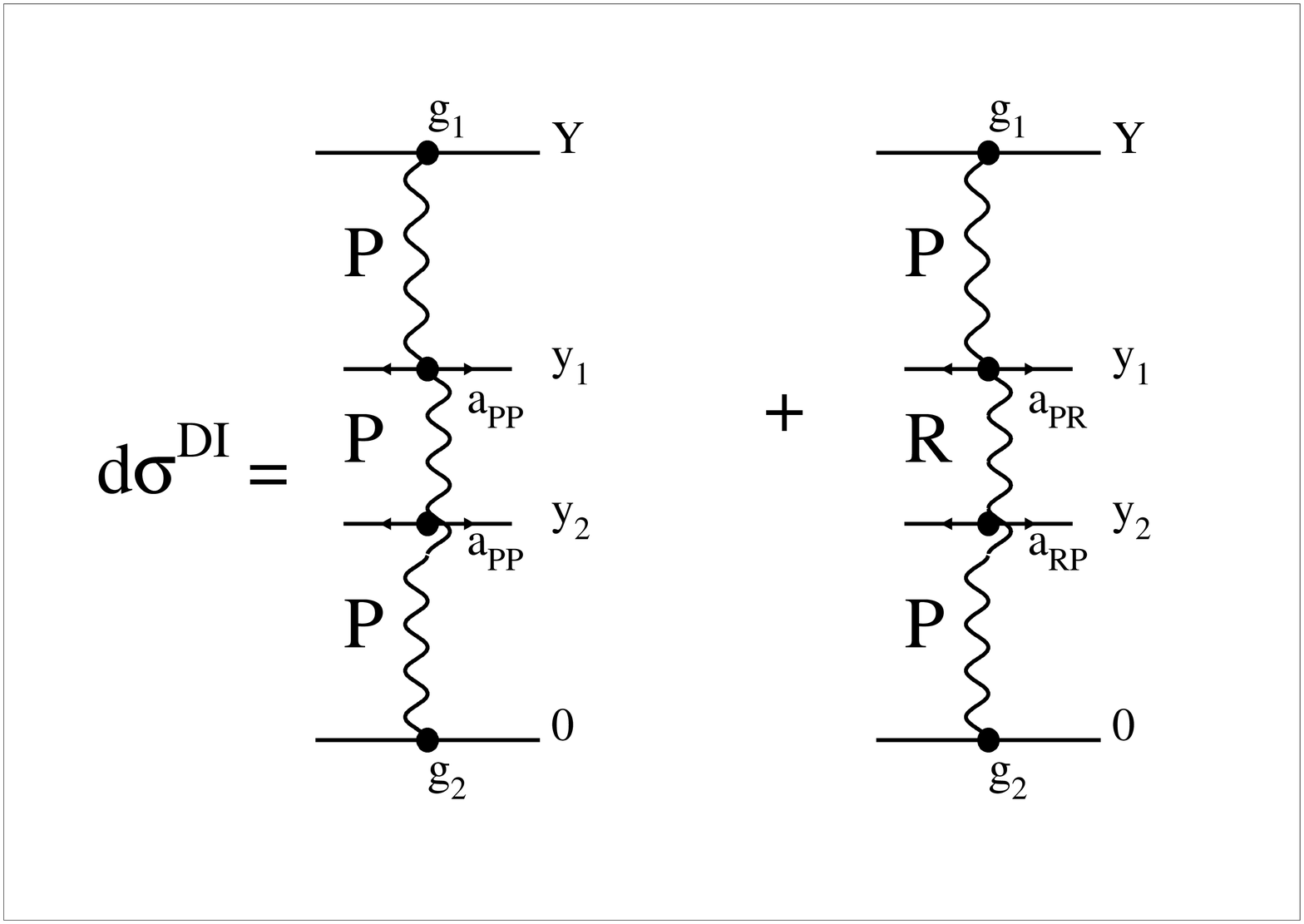,width=120mm}}
\caption{{\it Double inclusive production  in the Mueller technique.}}
\label{fig23g}
\end{figure}
\resetfig

\begin{figure}[htbp]
\begin{tabular}{c}
\centerline{\epsfig{file=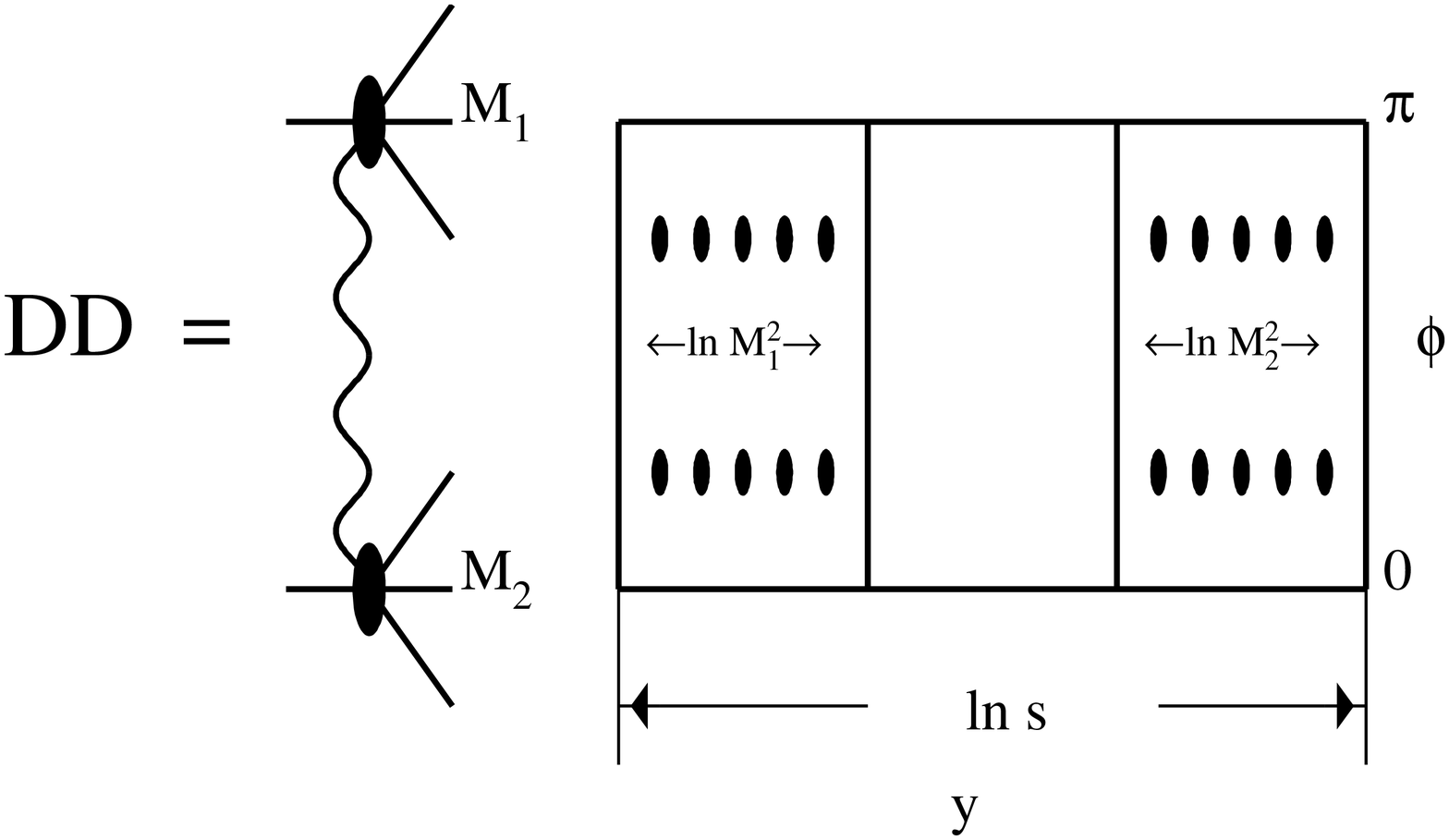,width=120mm,height=50mm}}\\
\centerline{\epsfig{file=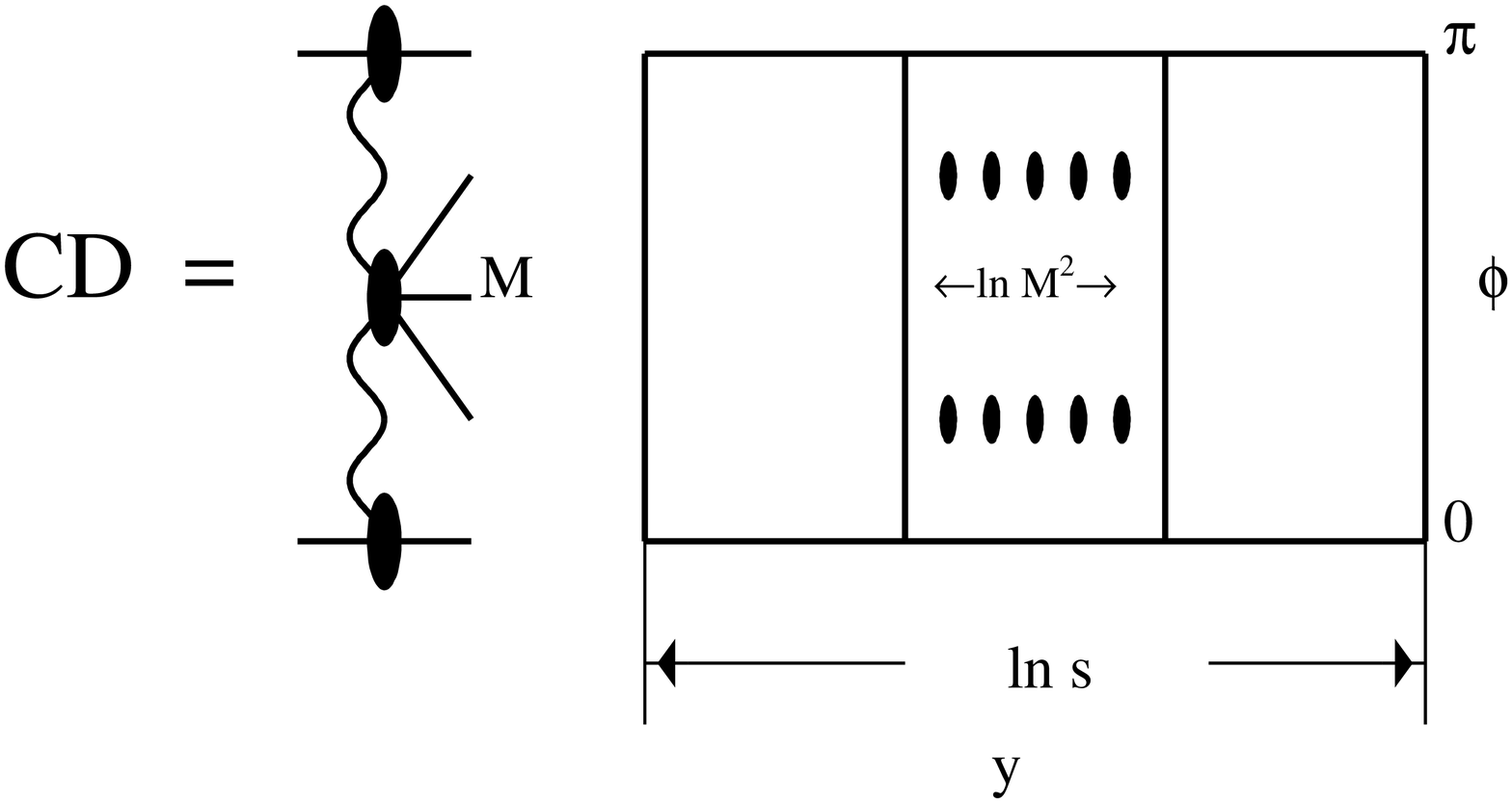,width=120mm,height=50mm}}
\end{tabular}
\caption{{\it Lego - plot for Double and Single Diffraction.}}
\label{fig24}
\end{figure}

\section{Why the Donnachie - Landshoff Pomeron cannot  be  the correct  
one.}
We have discussed in detail that the DL Pomeron gives the amplitude 
$a(s,b)$ which reaches at energies almost accessible now , namely, at
energies 
slightly higher than the Tevatron energies (see Fig.25).
 Let us discuss in more detail what kind of information on the hadron -
hadron collisions we can obtain from the value of $a(s,b = 0)$. At 
Tevatron energies $Im \,a(s,b = 0 )\,=\,1 - \,\exp( - \frac{\Omega}{2})\,
\approx\,0.95$. Therefore, $\Omega\,\approx\,6$, which is a rather large
value. Indeed, we can calculate the value of $G_{in}(s,b=0)\,=\,1 \,-\,
\exp( - \Omega )$ which is equal to 0.9975 and  is close to the
unitarity limit ($G_{in}\,<\,1$ ). Using the unitarity constraint we see
that
$$
Im\, a ( W = 1800\,GeV,b = 0)\,\,=\,\,45\%\,
(\,elastic\,)\,\,\,+\,\,\,50\%\,(\,inrlastic\,)
$$
This fact certainly contradicts the parton model structure of the Pomeron,
in which the Pomeron results predominantly from inelastic
multiparticle production processes.  Nevertheless, we have to admit that
the above observation cannot rule out the first definition of the Pomeron
as the Reggeon with the intercept close to 1.

However, there are experimental data which disagree with the D-L Pomeron,
namely, the data on single diffraction in  proton - proton (
antiproton) collisions. In the Pomeron exchange approach, the cross
section for  single diffraction dissociation as well as elastic
scattering cross section behaves as
$$
\sigma^{SD}\,\,\propto\,\,\,\sigma^{el}\,\,\propto\,\,
(\,\frac{s}{s_0}\,)^{2\Delta} \,\,,
$$
where $\Delta\,\,=\,\,\alpha_P(0)\,-\,1$. Recalling that $
\sigma_{tot}\,\,
\propto\,\,(\,\frac{s}{s_0}\,)^{\Delta}$, we see that we expect
\beq \label{EXSD}
\frac{\sigma^{el}}{\sigma_{tot}}\,\,\propto\,\,\frac{\sigma^{SD}}{\sigma_{tot}}
\,\,;
\eeq
$$
\frac{\sigma^{SD}( t = 0 )}{\sigma^{el} ( t = 0 )}\,\,=\,\,Const (s )\,\,.
$$
One can see from Fig.26 that the experimental data behave  completely
differently than predicted by the D-L Pomeron..

Therefore, we  conclude although  the D-L
Pomeron describes a large amount  of experimental data it cannot be
considered as a
serious candidate for  Pomeron exchange. The problem how to correct the
D-L Pomeron is related to the problem of shadowing corrections which I
 am not able to touch in these lectures because of lack of time.
\alphfig
\begin{figure}[htbp]
\centerline{\epsfig{file=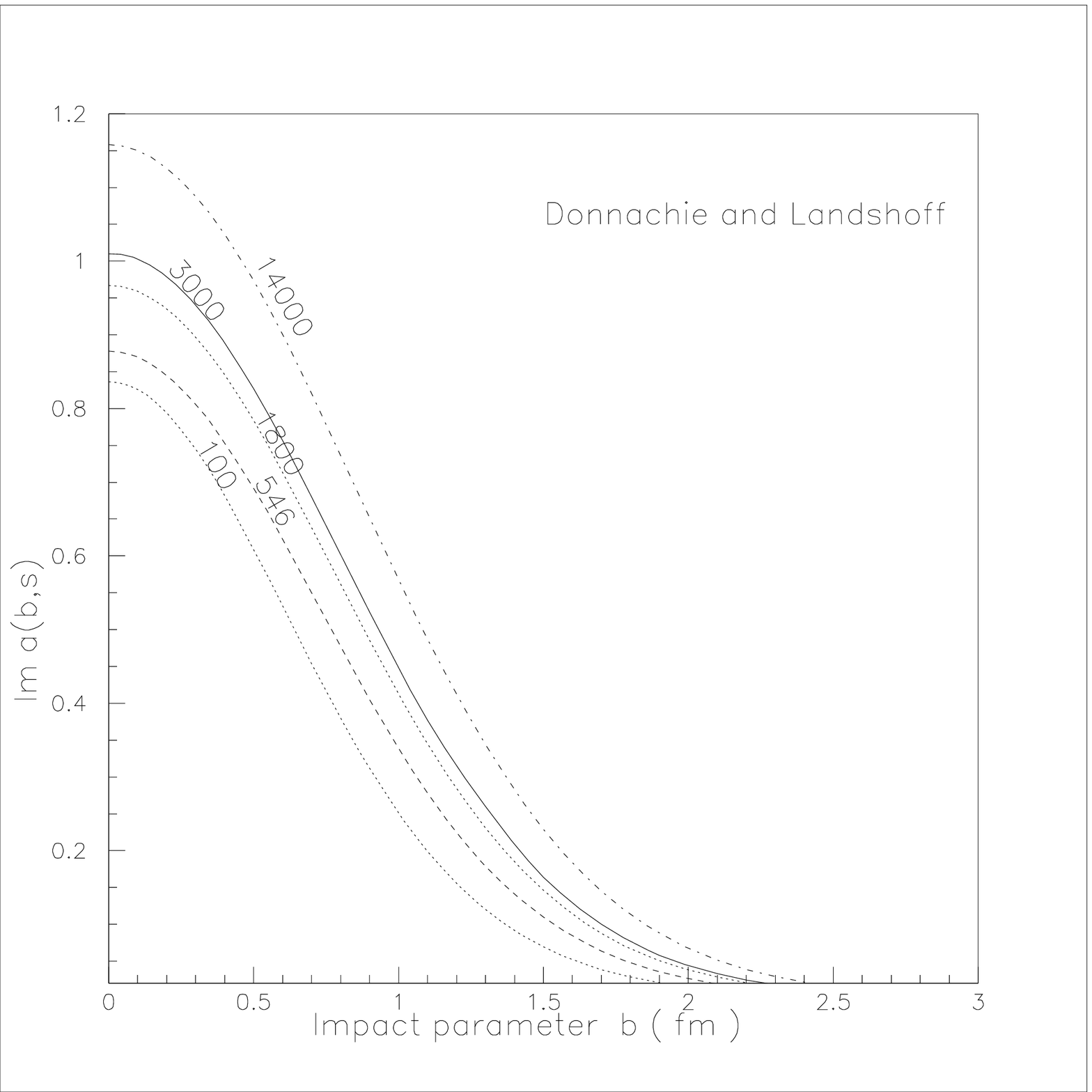,width=120mm}}  
\caption{{\it Impact parameter profile for the  Donnachie - Landshoff 
Pomeron at different values of $W = \sqrt{s}$.  I am very grateful to 
E.Gotsman, who did and gave  me   this picture. }}
\label{fig25a}
\end{figure}

\begin{figure}[htbp]
\centerline{\epsfig{file=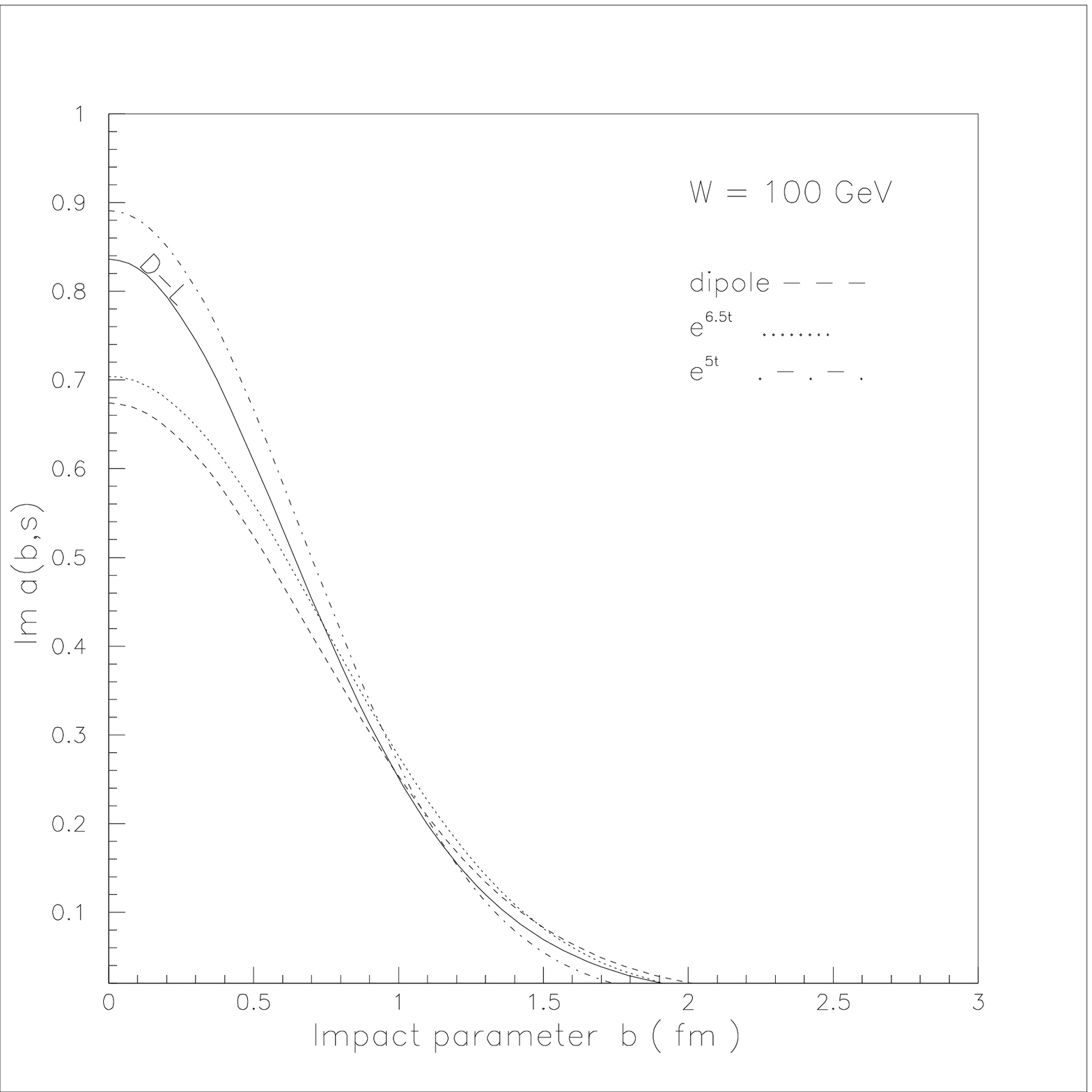,width=120mm}}
\caption{{\it Impact parameter profile for the  Donnachie - Landshoff
Pomeron and other models for the Pomeron  at $W =
\sqrt{s}= 100 GeV$. I am very grateful to
E.Gotsman, who prepared    this picture. }}
\label{fig25b}
\end{figure}
\resetfig

\begin{figure}[htbp]
\centerline{\epsfig{file=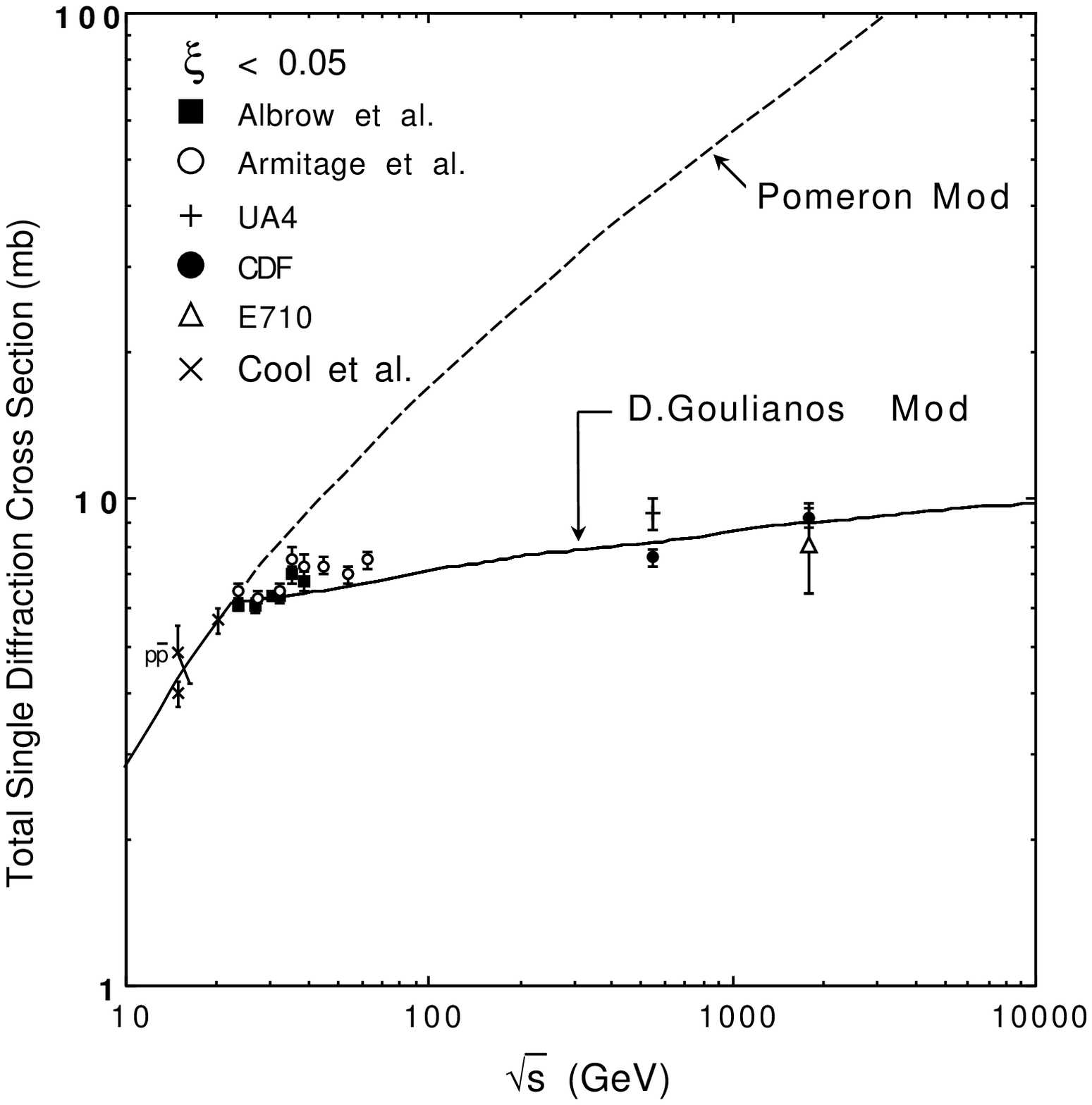,width=150mm}}  
\caption{{\it Experimental data on $\sigma_{SD}$ as a function of center
mass energy. The dashed curve is the triple-Pomeron prediction while the
solid one is one of the model which fits the data. Picture is taken from 
one of the  numerous talks of D. Goulianos.}}
 \label{fig26} 
\end{figure}

 \section{Pomeron in photon - proton interaction.}
The photon - hadron interaction has two aspects which makes it quite
different from the hadron - hadron interaction ( at least at first sight
).  First, the total cross section is of the order of $\alpha_{em}\,\,
\approx\,\,1/137\,\,\ll\,\,1$ and, therefore, we can neglect the elastic
cross section which is an  extra $\alpha_{em} $ times smaller.
We need to understand   the unitarity constraint here. I
Recall that for hadron - hadron collisions  the square of scattering
amplitude enters the unitarity constraint. Second, we have a unique way to
study the dependence of the cross section on the mass of the  incoming
particle by 
changing the virtuality of photon. As is well known at large
virtualities we are dealing with deep inelastic scattering ( DIS ) which
has
a simple partonic interpretation. I would like to remind you that DIS will
be a subject of the third and fourth  parts of my lectures, if any.  Here,
let us try
to understand how to incorporate the Reggeon exchange for photon - proton
interaction at high energies.

\begin{figure}[htbp]
\centerline{\epsfig{file=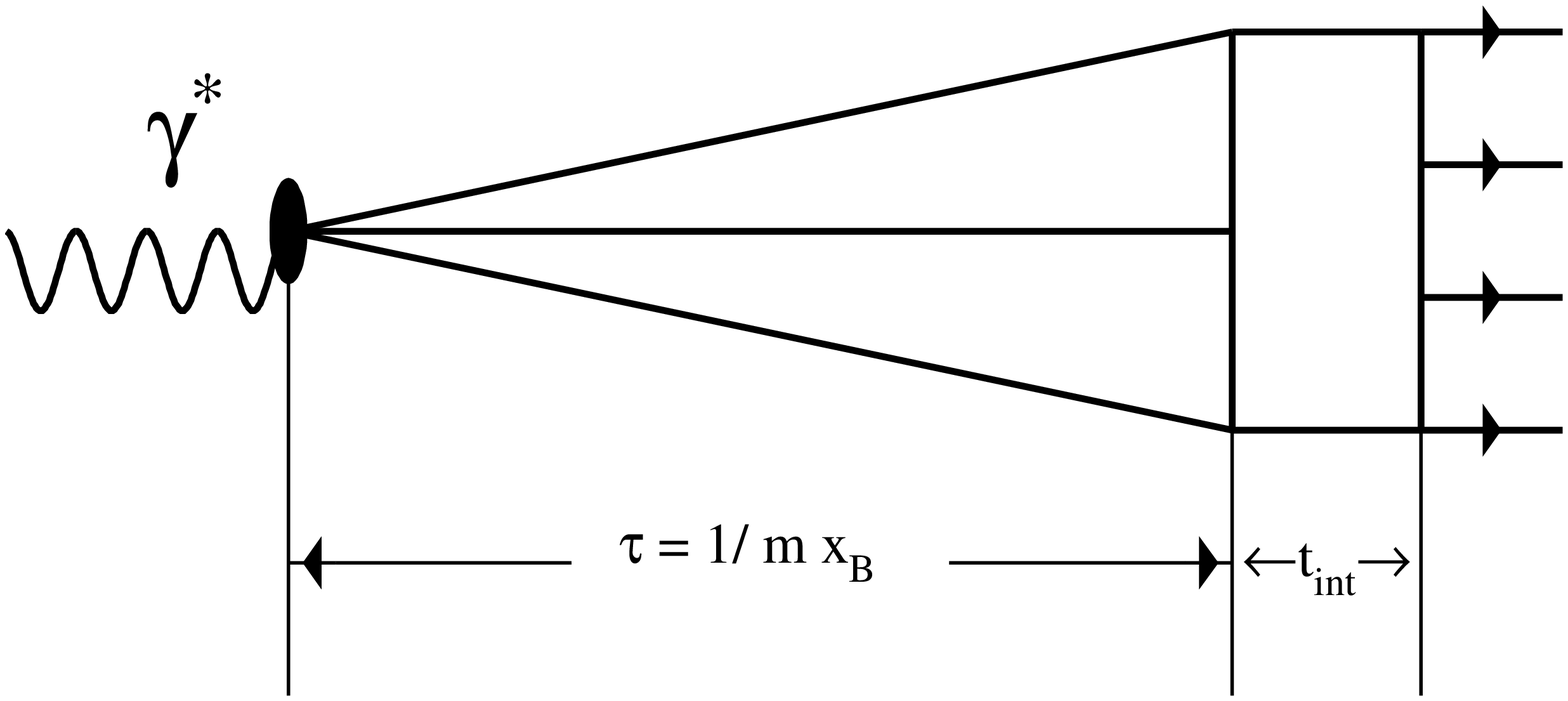,width=140mm}}
\caption{{\it  Two stages of photon - hadron interaction at
high energy.}}
\label{fig27}
\end{figure}

 \subsection{Total cross section.}

Gribov was the first one to  observe  that a photon ( even virtual one)
fluctuates into a  hadron system with  life time ( coherence length )
$\tau\,\,=\,\,l_c\,\,=\,\,\frac{1}{m x_B}$ where $x_B = \frac{Q^2}{s}$,
$Q^2$ is the
photon virtuality and $m $ is the mass of the target. This life time is
much larger at high energy than the size of the target and therefore,
we can consider the photon - proton interaction as a processes which
proceeds  in
two stages: 

(1) Transition  $\gamma^* \,\rightarrow \,hadrons $  which is
not affected by the target and, therefore, looks similar to  electron -
positron annihilation; 

and 

(2) $hadron \,\,-\,\,target $ interaction,
which can be treated as standard hadron - hadron interaction, for example,
in the Pomeron ( Reggeon ) exchange approach ( see Fig.27).

These two separate stages of the photon - hadron interaction   allow
us to  use a
 dispersion relation with respect to the masses $M$ and $M'$ ( Gribov,
1970)
 to describe  the photon - hadron interaction  ( see Fig.28a) for
 notation ), as  the  correlation length $l_c =
\frac{1}{m x}\,\gg\,R_N$,  the target size. Based on this idea
we can write a  general formula for the  photon - hadron interaction,
     
\beq \label{GENGF}
\s(\g^* N )\,\,=\,\,
\frac{\alpha_{em}}{3\,\pi}\,\int \frac{\Gamma(M^2)\,\,d\,M^2}{\,
Q^2\,+\,M^2\,}
\,\,\s(M^2,M'^2,s)\,\,
\frac{\Gamma(M'^2)\,\,d\,M'^2}{\, Q^2\,+\,M'^2\,} \,\,.
\eeq
where $\Gamma(M^2) = R(M^2)$ (see \eq{RATIO} ) and $\sigma(M^2,M'^2,s)$ is
proportional to the imaginary part of the forward amplitude  for $ V\, +\, 
p \,\rightarrow\,V' \,+\,p$ where $V$ and $V'$ are the vector states with
masses $M$ and $M'$. For the case of the diagonal transition ( $M = M'$)
$\sigma(M^2,s) $ is the total cross section for $V- p$ interaction.
Experimentally, it is known that a diagonal coupling of the Pomeron is
stronger than an off-diagonal coupling. Therefore, in first approximation
we can neglect the off-diagonal transition  and substitute in \eq{GENGF}
$\sigma(M^2,M'^2,s)\,=\,\sigma)M^2.s)\,M^2 \,\,\delta(\,M^2\,-\,M'^2)$. 
\alphfig
\begin{figure}
\centerline{\psfig{file= 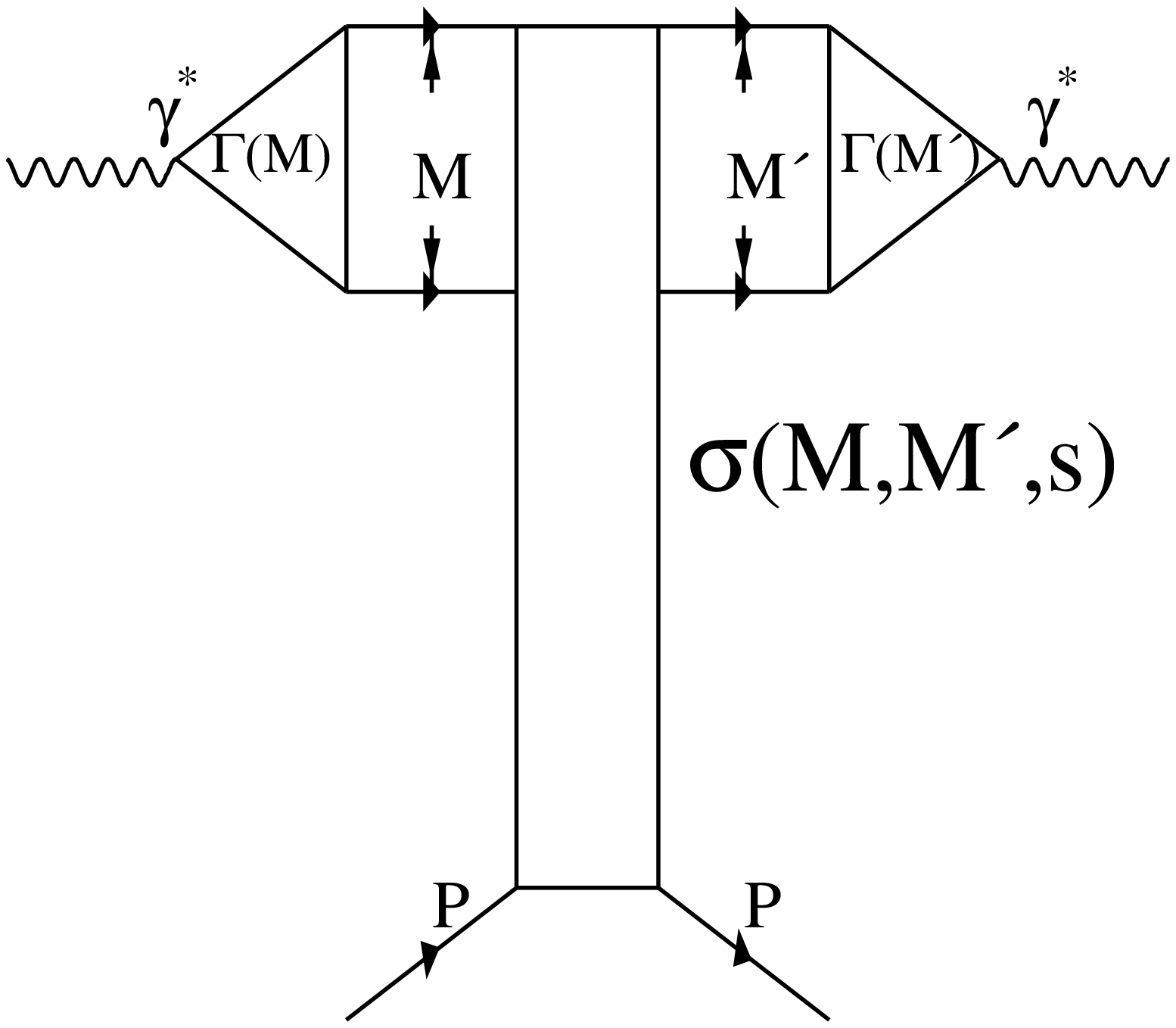,width=100mm}}
\caption{The generalized Gribov's formula for DIS.}
 \end{figure}
The resulting photon - nucleon  cross section can be written written  as:
\beq \label{GRF}
\s(\g^* N )\,\,=\,\,\frac{\alpha_{em}}{3 \,\pi}\,\int\frac{R(M^2)\,M^2\,d
\,M^2}{(\, Q^2\,+\,M^2\,)^2}\,\s_{M^2N}(s)\,\,,
\eeq
where $R(M^2)$  is defined as the ratio
\beq \label{RATIO}
R(M^2)\,\,=\,\,\frac{\s(e^{+} e^{-} \,\rightarrow\,hadrons)}{\s(e^{+}
e^{-}\,\rightarrow\,\mu^{+} \mu^{-} )}\,\,.
\eeq
The  notation is illustrated  in Fig.28 where $M^2$ is the mass squared of
the  hadronic system, $\Gamma^2(M^2)\,=\,R(M^2)$  and  $\s_{M^2
N}(s)$
 is the cross section for the hadronic system to scatter off the
nucleonic target.

\begin{figure}[htbp]
\centerline{\epsfig{file=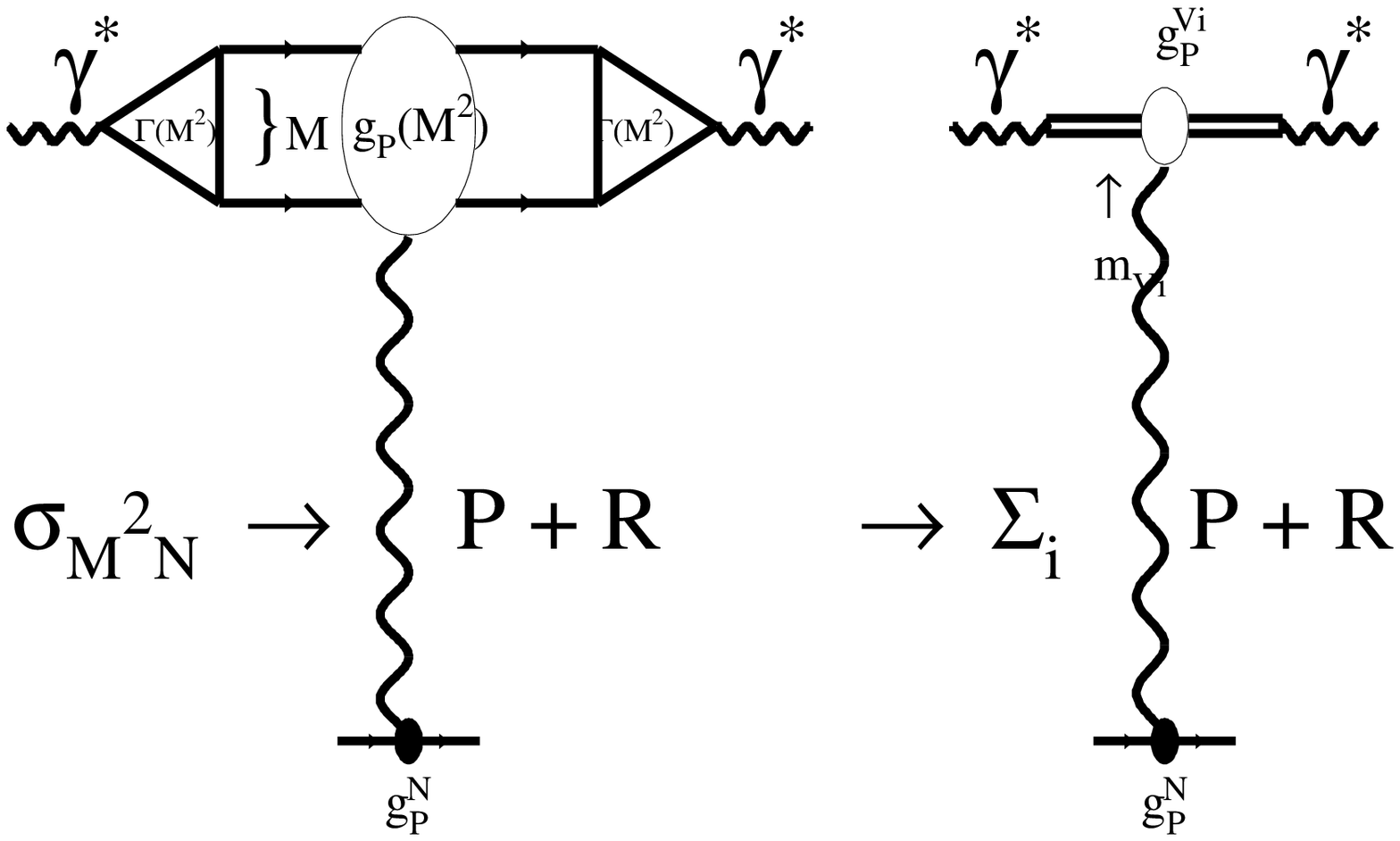,width=170mm}}
\caption{{\it  Gribov's formula for DIS.}}
\label{fig28}
\end{figure}

In Fig.29 one can see the experimental data for  $R(M^2 = s )$. It is
clear
that $R(M^2)$ can be described in a  two component picture: the
contribution
of resonances such as $\rho, \omega, \phi$, J/$\Psi$, $\Psi'$ and so on
and
the contributions from  quarks which give a  more or  less constant term
but it
changes abruptly with every  new open quark - antiquark channel, $R(M^2)\,
\approx\,3\,\sum_q \,e^2_q$, whetre $e_q$ is the charge of the quark and
the summation  is done over  all active quark team.

\begin{figure}[htbp]
\centerline{\epsfig{file=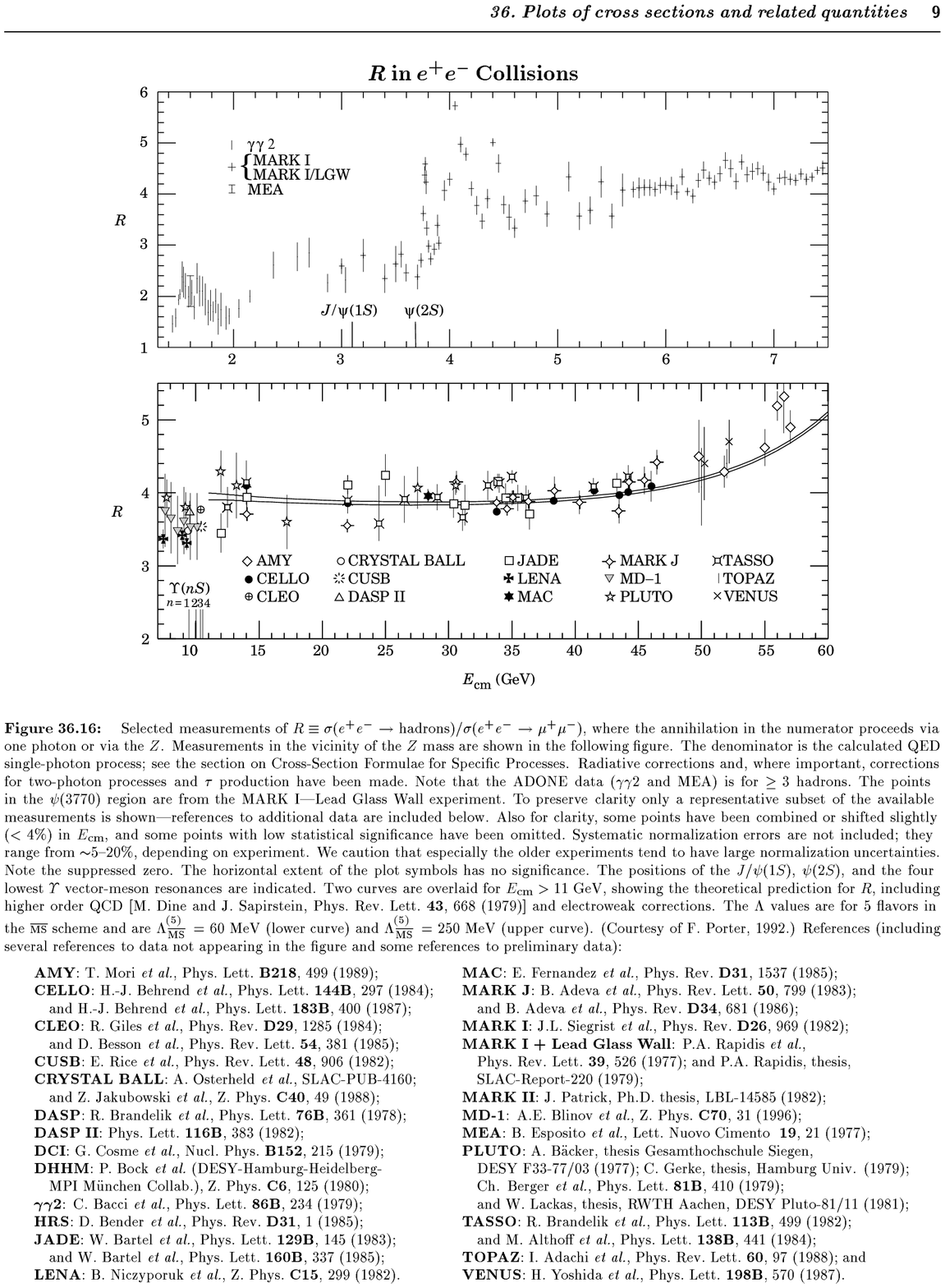,width=160mm}}
\caption{{\it  Experimental behaviour of $R(M^2 = E^2_{cm} )$ (
Particle Data Group 1996 ).}}
\label{fig29}
\end{figure}
\resetfig
For $\s_{M^2 N}$ we can use the Reggeon exchange approach  which gives
for the total cross section \eq{GENCR} where only the  vertices $g_P$ and
$g_R$ depend on $M^2$. 

If we take into account only the contribution of the $\rho$ - meson,
$\omega$ - meson ,  $\phi$  and J/$\Psi$ resonances 
   in $R(M^2)$  we obtain the so called vector dominance model
(VDM) (  J.J. Sakurai 1969)  which gives for the total cross section the
following formula:
\beq \label{VDM}
\s_{tot}(\gamma^*\,+\,p\,)\,\,=\,\,\frac{\alpha_{em}}{3 
\pi}\,\,\s_{tot}( \rho + p )\sum_i\,\,R(M^2=m^2_{V_i}) 
\{\,
\frac{m^2_{V_i}}{Q^2\,\,+\,\,m^2_{V_i}}\,\}^2\,\,
\eeq
where $m_{V_i}$ is the mass of vector meson, $Q^2$ is the value of the
photon virtuality and $ R(M^2 =m^2_{V_i})$ is the value of the $R$ in the
mass of the vector meson which can be rewritten through the ratio of the
electron - positron decay width to the total width of the resonance. 
Of course, the summation in \eq{VDM} can be extended to all vector
resonances ( Sakurai and Schildknecht 1972) or even we can return to a
general approach of \eq{GENGF} and write a model for the off-diagonal
 transition between the vector meson resonances with different masses
( Schildknecht et al. 1975).  We have two problems with all generalization
of the VDM: (i)  a number of unknown experimentally values
such as masses and electromagnetic width of the vector resonances with
higher masses than those that have been included in VDM  and  (ii) a lack
of   theoretical constraints on all mentioned observables. These two
reasons 
 give so much
freedom in fitting of the
experimental data on photon - hadron interaction that it becomes
uninteresting and non - instructive.   

One can see from \eq{VDM}, that if $Q^2\,\gg\,m^2_V$  VDM predicts a 
$\frac{1}{Q^4}$
behaviour of the total cross section. Such a behaviour is in clear
contradiction with the experimental data which show an  approximate
$\frac{1}{Q^2}$ dependence at large values of $Q^2$ ( i.e. scaling ). It
should be stressed
that the background contribution to \eq{GRF} can explain   the 
experimental $Q^2$ -
dependence. Indeed, for $M^2 \,\,\gg\,\,m^2_{\rho}$ 
$R(M^2) \,\,\approx\,\,2$, (see Fig.29).
 Assuming that $\s_{M^2 N}$ does not
depend on $M^2$, we have
\beq \label{BGC}
\s_{tot}(\gamma^*\,+\,p\,)\,\,=\,\,\frac{\alpha_{em}}{3
\pi}\,\,\s_{tot}( \rho + p )\,\,2 \,\,\ln
(\frac{Q^2 \,+\,M^2_0}{Q^2\,+\,m^2_{\rho}})\,\,,
\eeq
where $M_0$ is the largest mass up to which we can consider that  $\s_{M^2
N}$ is independent of $M^2$. Directly from \eq{BGC} one can see
that
 $$
\s_{tot}(\gamma^*\,+\,p\,)\,\,=\,\,\frac{2\,\alpha_{em}}{3
\pi}\,\,\s_{tot}( \rho + p ) \,\,\frac{M^2_0 \,-\,m^2_{\rho}}{Q^2}\,\,.
$$

The value of $M_0$ is the new scale of our approach which separates the
``soft" ( long distance ) processes from the ``hard"  ( short  distance )
ones. Actually, any discussion of this problem is out of the schedule of
these lectures and I want only to point out that  QCD leads to the
cross section which decreases  as $\frac{1}{M^2}$ at large $M^2$.

In order to write explicitly the Pomeron ( Reggeon ) contribution to
 the photon -
nucleon cross section we have to specify the energy variable. Assume, that
the
energy
variable $s/M^2$ will be familiar to the readers.

The contribution of the Pomeron as well as of any Reggeon
to photon - nucleon scattering has the following form:
\beq \label{POMINPN}
\s(\g^* N )\,\,=\,\,\frac{\alpha_{em}}{3 \,\pi}\,\int\frac{R(M^2)\,M^2\,d
\,M^2}{(\,
Q^2\,+\,M^2\,)^2}\,g_P(M^2)g^N_{P}\,\,(\,\frac{s}{M^2})^{\alpha_P (0)
\,\,-\,\,1}\,\,.
 \eeq
A  very important
feature of \eq{POMINPN}is the fact that  the mass dependence is
concentrated only
in the dependence of vertex
$g_P(M^2)$ and in the energy variable $s/M^2$. It is interesting to note
that the integral over $M^2$ converges even if $g_P(M^2)$  is independent
of
 $M^2$. Note also,  that at large $M^2$ we are dealing
with short  distance physics where the powerful methods of perturbative
QCD have to be applied.

\subsection{Diffractive dissociation}
\centerline{}
\centerline{}
 Diffractive dissociation is the process  which has 
 a large rapidity gap in the final state (see Fig.30). This process can be
described
using the same approach as in \eq{GRF}, namely, the dispersion relation  
with respect to $M^2$ in Fig.30. However, it is useful to discuss
separately two cases: the exclusive final state with a restricted number
of
hadrons and inclusive diffraction dissociation without any selection in   
the final state. To understand the difference between these two cases, 
 recall, that, according to the second definition of the Pomeron, the
typical final state in  diffractive processes is still the parton
``ladder", namely, the final state hadrons are uniformly distributed in
rapidity. In other words for a fixed but large mass $M$ of the final state
the typical multiplicity of the produced hadrons $N \,\propto\,a \ln M^2$,
where $a$ is the particle  density in rapidity ( the number of particles
per 
 unit rapidity interval). Therefore, the final state with a fixed number
of hadrons ( partons ) has a suppression which is proportional to
$\exp( - a\,\ln M^2 )$. Hence  for the exclusive process with fixed
number of  hadrons in the final state  one  has to find a new mechanism
different from the normal partonic  one. This mechanism is rather obvious
since
the large value of $M$ can  be due to a  large value of the parton (
hadron ) transverse momenta ( $k_{it}$ in the final state which are of the
order of $M$ ( $ k^2_{it}\,\approx\,M^2$ ). Concluding this introduction
to the DD processes in Reggeon approach, we  emphasize that

(i) for exclusive final states with a  fixed number of hadrons (partons)
the
large diffractive mass  originates from  the parton ( hadron ) interaction
at small distances;

(ii) while the inclusive DD in the region of large mass looks like a
normal inclusive process in the parton model with a  uniform
distribution
of
the secondary hadrons in rapidity and with a limited  hadron transverse
momentum which does not depend on $M$.

\subsection{ Exclusive ( with restricted number of hadrons) final state.}
\centerline{}
\centerline{}
For this process we can use  Gribov's formula ( see Fig.30 ). Futhermore, 
for large values of the mass  one  can calculate the transition from the 
virtual photon to a  quark - antiquark pair in perturbative QCD. So, here,
for the first time in my lecture I recall   our  microscopic theory
- QCD, but, actually, we will discuss here only a very simple idea of
QCD,
namely, the fact that the correct degrees of freedom are quark and gluons.

\begin{figure}[htbp]
\centerline{\epsfig{file=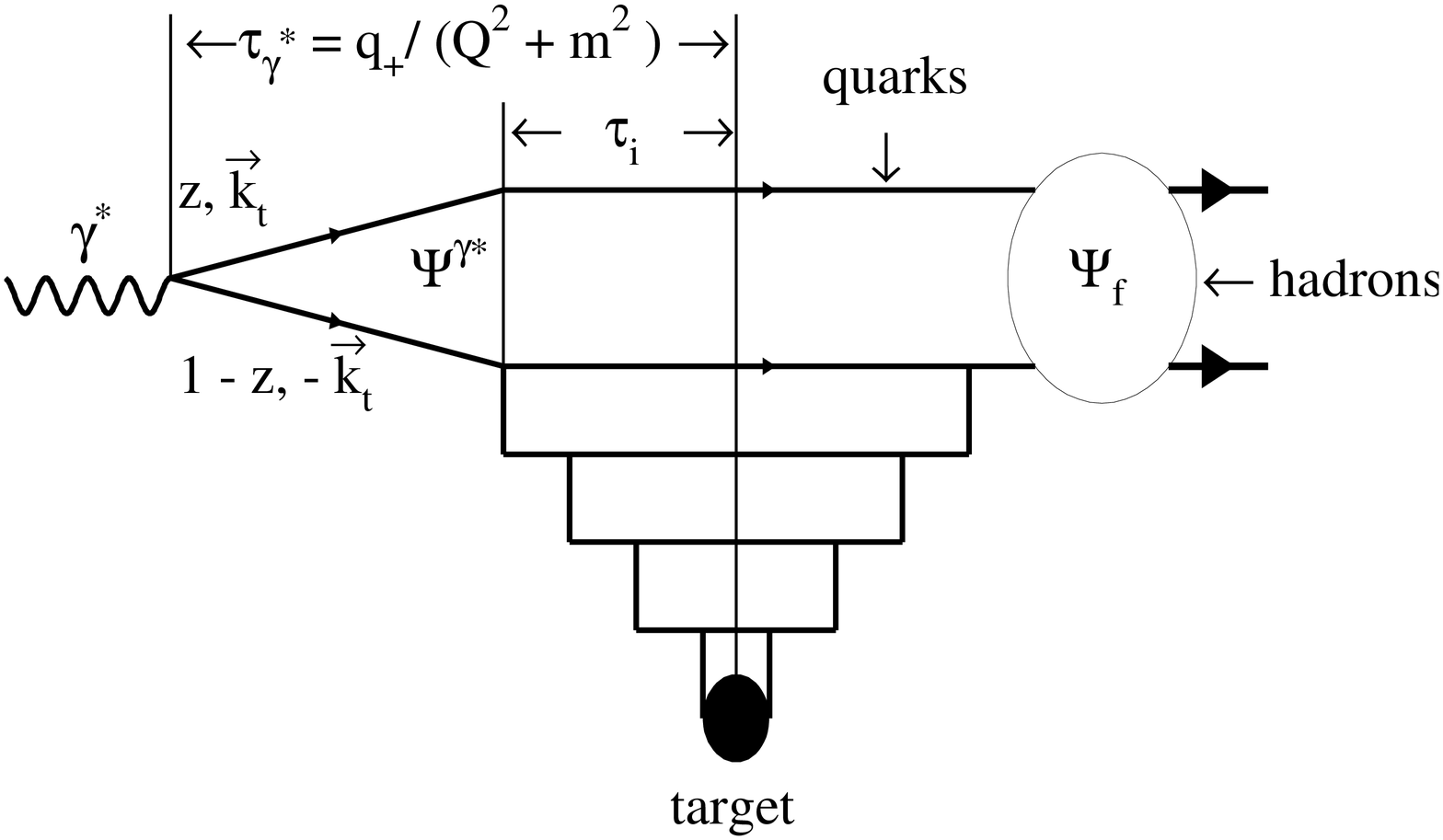,width=160mm}}
\caption{{\it Gribov's formula for diffractive dissociation
 (DD) in DIS.}}
\label{fig30}
\end{figure}

 Let us start with the definition of the  variables:

1. The Bjorken $x$ variable is given by 
$$
x_B\,\,=\,\,x\,\,=\,\,\frac{Q^2}{s}\,\,,
$$  
where $s$ is the square of the total c.m. energy of the virtual   photon -
proton system and $Q^2$ is the photon virtuality;

2. As usual, we introduce 
$$
x_P\,\,=\,\,\frac{ Q^2\,\,+\,\,M^2}{Q^2\,\, +\,\,s}\,\,,
$$
which is the  fraction of the proton energy carried by the
Pomeron;

3.  $
\beta\,\,=\,\,\frac{Q^2}{Q^2\,\,+\,\,M^2}\,\,=\,\,\frac{x_B}{x_P}\,\,,$
where $\beta$ is the fraction of the Pomereon energy carried by the struck
parton in the Pomeron;

4. The transverse momenta of the out going quark and antiquark are denoted
by  $\pm\,\,\vec{k}_t$, and those of the exchanged partons ( gluons)
 by $\pm\,\,\vec{l}_t$;

5. It is convenient to use light -  cone perturbation theory and to
express
the particle four momenta in the form:
$$ k_{\mu}\,\,=\,\,(\,k_{+},k_{-},\vec{k}_t\,)\,\,,$$
where $k_{\pm}\,\,=\,\,k_0\,\,\pm\,\,k_3$;

6. In the frame in which the target is essentially at rest we have
$$
q_{\mu}\,\,=\,\,(\,q_{+}, -\,\frac{Q^2}{q_{+}},\vec{0}_t\,)\,\,;
$$ 
$$
k_{\mu}\,\,=\,\,(\,k_{+}, -\,\frac{m^2_t}{k_{+}},\vec{k}_t\,)\,\,;
$$
$$
l_{\mu}\,\,=\,\,(\,l_{+}, -\,\frac{l^2_t}{l_{+}},\vec{l}_t\,)\,\,;
$$
with $m^2_t\,\,=\,\,m^2_Q\,\,+\,\,k^2_t$ where $m$ is the quark mass.
$q_{\mu}$ is the four momentum of the photon ($ q^2\,\,=\,\,- \,Q^2$).

The Gribov factorization ( Gribov's formula ) follows from the fact that 
the time of
quark - antiquark fluctuation $\tau_{\gamma^*}$ is much longer than the
time of interaction with partons ($\tau_i$ ). It is instructive to
illustrate why this is so. According to the uncertainty principle the
fluctuation time is equal to:
\beq \label{FLUCTTIME}
\tau_{\gamma^*}\,\,\sim\,\,\vert\, \frac{1}{q_{-}\,\,-\,\,k_{1
-}\,\,-\,\,k_{2 -}}\,\vert\,\,=\,\,\frac{q_{+}}{Q^2\,\,+\,\,M^2}\,\,,
\eeq
where $k_1$ and $k_2$ are four momenta of the quarks of mass $m$, and
$$
M^2\,\,=\,\,\frac{m^2\,\,+\,\,k^2_t}{z(1 \,-\,z)}\,\,.
$$
To estimate the interaction time we calculate the typical time of the
emission of a     parton with momentum $l_{\mu}$ from quark $k_1$, which
is
\beq \label{INTTIME}
\tau_i\,\,=\,\,\vert\,\frac{1}{k_{1
-}\,\,-\,\,k'_{1-}\,\,-\,\,l_{-}}\,\vert
\,\,=\,\,\vert\,\frac{q_{+}}{\frac{m^2_t}{z}\,\,-\
,\,\frac{m^2_t}{z'}\,\,-\,\,\frac{l^2_t}{\alpha}}\,\vert\,\,,
\eeq
where $\alpha\,=\,\frac{l_{+}}{q_{+}}$ and $z'\,=\,z\,-\,\alpha$.
In leading log  $s$ approximation in which we  have obtained the Pomeron,
we
have $\alpha\,\,\ll\,\,z$ and hence
$$
\tau_i\,\,=\,\,\frac{\alpha q_{+}}{l^2_t}\,\,\ll\,\,\tau_{\gamma^*}\,\,.
$$
Therefore, we can write  Gribov's formula as it is given in Fig.30,
namely:
\beq \label{GFSM}
M(\,\,\gamma^*\,+\,p\,\,\,\rightarrow\,\,\,M^2\,+ \,\,[\,
LRG\,]\,\,+\,\,p\,\,)\,\,=
\eeq
$$
\,\,\int \,d^2 k_t\,\int^1_0\,d z
\,\Psi^{\gamma^*}
(z, k_t,Q)\,A_P(M^2,s)\,\,\Psi_{final}(\bar q q \rightarrow hadrons)\,\,,
$$
where LRG stands for ``large rapidity gap".

The wave function of the virtual photon can be written in the following
way:
\beq \label{WFP}
\Psi^{\gamma^*}(z,k_t,Q)\,\,\,=\,\,\,- e Z_f \,\frac{\bar
u(k_t,z)\,\gamma\cdot\epsilon\,v( - k_t, 1 -
z)}{\sqrt{z(1\,-\,z)}}\,\frac{1}{Q^2\,\,-\,\,M^2}\,\,,
\eeq
where $\vec{\epsilon}$ is the photon polarization: for longitudinal
polarized photon we have 
$$
\epsilon_{L}\,\,=\,\,(\,\frac{q_{+}}{Q},\frac{Q}{q_{+}},\vec{0}_t\,)
$$
while for transverse polarized photons ($\vec{\epsilon}_t$)  we will use
as 
basis the  circular
polarization vectors :
$$
\epsilon_{\pm}\,\,=\,\,\frac{1}{\sqrt{2}}\,(\,0,0,1,\pm i\,)\,\,.
$$
To evaluate \eq{WFP} with $\epsilon\,=\,\epsilon_L$ we first note that
$$
q^{\mu}\,\bar u \gamma_{\mu} v \,\,=\,\,\frac{1}{2}\,(\,q_{+} \bar u
\gamma_{-} v\,\,+\,\,q_{-}\,\bar u \gamma_{+} v\,)\,\,=\,\,0
$$
and therefore
$$
\bar u \gamma_{-} v\,\,=\,\,\frac{Q^2}{q^2_{+}}\,\bar u \gamma_{+} v\,\,.
$$
We can use this identity to evaluate the wave function:
$$
\bar u \gamma \cdot \epsilon_L v\,\,=\,\,\frac{1}{2}\,\{\,\frac{q_{+}}{Q}
\bar u \gamma_{-} v\,\,+\,\,\frac{Q}{q_{+}} \bar u \gamma_{+}
v\,\}\,\,=\,\,\frac{Q}{q_{+}} \bar u \gamma_{+} v\,\,=\,\,2 Q 
\sqrt{z(1\,-\,z)} \delta_{\lambda, - \lambda'}\,\,,
$$
where  $\lambda $ is helicity of the  antiquark while $\lambda'$ is the 
helicity
of the quark ( we omitted helicities in  most of the  calculations but
put them correctly in the answer).
Coming back to \eq{WFP} one can see
that:
\beq \label{WFL}
\Psi^{\gamma^*}_L\,\,=\,\,- e Z_f \,\frac{ 4\,Q\, \delta_{\lambda,  -
\lambda'}}{Q^2\,\,+\,\,M^2}\,\,.
\eeq
For the transverse polarized photons  with helicity
$\lambda_{\gamma}\,\,=\,\,\pm 1$ one finds:
\beq \label{WFT}
\Psi^{\gamma^*}_T\,\,=\,\,- e Z_f \,\frac{\delta_{\lambda, - \lambda'}\,\{
(\,1\,-\,2z\,)\lambda_{\gamma}\,\pm\,1\,\}\,\vec{\epsilon}_{\pm}
\cdot\vec{k}_t\,\,+\,\,\delta_{\lambda,\lambda'}\,m\,\lambda_{\gamma}}
{[z(1\,-\,z]\,\,\,\,\,\,( Q^2\,\,+\,\,M^2\,)}\,\,.
\eeq

To obtain the expression for the diffractive dissociation cross section
we have to square the  amplitude and sum over all quantum numbers of the
final state ( with fixed mass $M$ ). Assuming that experimentally no
selection has been made or in other words $\Psi_{finaL}$ does not 
describe a specific state with definite angular momentum, we can sum over
helicities and obtain the following answers:
\beq \label{CRLGF}
\frac{d \sigma^{SD}_L}{d M^2 d t}\,\,=\,\,\frac{\alpha_{em}}{48 \pi^3}
\frac{Q^2 R^{exclusive}_L(M^2)}{\{\,Q^2\,\,+\,\,M^2\,\}^2}g^2_P(M^2) g^2_p
\,(\,\frac{s}{Q^2\,+\,M^2}\,)^{2 \Delta}\,\,;
\eeq
\beq \label{CRTGF}
\frac{d \sigma^{SD}_T}{d M^2 d t}\,\,=\,\,\frac{\alpha_{em}}{48 \pi^3}
\frac{M^2 R^{exclusive}_T(M^2)}{\{\,Q^2\,\,+\,\,M^2\,\}^2}g^2_P(M^2) g^2_p
\,(\,\frac{s}{Q^2\,+\,M^2}\,)^{2 \Delta}\,\,;
\eeq
where $R_{L,T}(M^2) $ is defined as the  ratio
$$
R_{L,T}(M^2)\,\,=\,\,\frac{\sigma(e^- + e^+\,\rightarrow \,hadrons\,
with\,mass\, M)}{\sigma( e^- e^+\,\rightarrow \,\mu^- \mu^+)}\,\,.
$$

 \eq{CRLGF} and \eq{CRTGF} give us the  possibility to
measure
experimentally the dependence of $g(M^2)$ on  $M$ and therefore to
determine  the value of the separation scale $M_0$. As we have discussed,
this scale separates  the long distance ( nonperturbative, ``soft" )
physics from the short distance one ( perturbative, ``hard").
It is hardly necessary  to mention how important this
 information for our understanding of the  nonperturbative QCD 
contribution in orde5r  to find a way out of the Reggeon phenomenology.

\subsection{Inclusive diffraction}
Diffractive production of a  large mass in $\gamma^* p$ scattering  in the
Reggeon approach
looks quite similar to  diffractive dissociation in hadron - hadron
collisions. We can safely apply the triple Reggeon phenomenology
and the formulae have the same form as \eq{SD}, namely
 \beq \label{SDPH}
\frac{M^2 d \sigma_{SD}(\gamma^* p)}{d M^2}\,\,=\,\, \frac{\sigma_0}{2\pi
R^2(\frac{s}{M^2})}
 \,\cdot\,( \frac{s}{M^2})^{2 \Delta}
\,\cdot\,[\,g^{\gamma^*}_P\cdot G_{3P}
(0)
\,\cdot\,
(\frac{M^2}{s_0})^{\Delta} \,+\,\,g^{\gamma^*}_R\cdot G_{PPR}(0)
 (\frac{M^2}{s_0} )^{ \alpha_R (0) - 1}\,]\,\,.
\eeq
There is   only one difference , namely, we can apply  Gribov's formula
for  the  vertex $g^{\gamma^*}_{P,R}$. Indeed,
\beq \label{GFFV}
g^{\gamma^*}_{P,R} (Q^2)\,\,=\,\,\frac{\alpha_{em}}{3\,\pi}\,\int^{M^2_0}
\frac{R(M'^2)M'^2 d M'^2}{(\,Q^2\,+\,M'^2\,)^2}
\,\,g_{P,R}(M^2)\,(\,\frac{s_0}{M'^2}\,)^{\alpha_{P,R}(0)\,-\,1}\,\,.
\eeq
We introduce explicitly the separation scale $M_0$ to demonstrate that we
trust this formula only in the region of sufficiently small masses  while
at
large masses this integral should be treated in perturbative QCD.

Notice  that the $Q^2$ dependence 
is concentrated only in the vertex $g^{\gamma^*}_{P,R}$ and, therefore,  
one of the  predictions 
of the Pomeron approach is that large mass diffraction has the
same $Q^2$ dependence as the total $\g^{*} p $ cross section. Surprisingly
the HERA data are not in contradiction with this prediction.

However, it should be stressed that the Pomeron approach can be trusted
only for rather small virtualities while at large $Q^2$  perturbative QCD
manifests itself as more powerful and reliable tool for both the DIS
total cross sections and diffractive dissociation in DIS. The challenge
for everybody who is trying to understand the photon - hadron interaction
is to find the region of applicability of pQCD and to separate  ``soft"
and `` hard" interactions. This task is impossible without a certain
knowledge of the Reggeon  phenomenology.

\section{Shadowing Corrections ( a brief outline ).} 
\subsection{Why is the Pomeron  not sufficient?}
I hope, that I have given a lot of information on the Reggeon 
phenomenology. Now, I want to touch several difficult and important
topics and, in particular, I want to explain why the Pomeron hypothesis
 is not sufficient. A full  discussion of these tropics will be
given  in the second part of my lectures, but I would like to give here a
brief review of the shadowing corrections (SC) properties to discuss the
experimental way of studying  the Pomeron structure which I will 
present in the next section.

Firstly, the  hypothesis that {\it the Pomeron gives 
the asymptotic behaviour of the scattering amplitude is not correct}.
To illustrate this fact let us consider  single diffraction
dissociation in the region of large mass $M$ ( see Fig.23c). In this
kinematic region we can apply  the  triple Pomeron formula for the
cross section. Using Eq.(76b) and the exponential parameterization for
the $t$ -
dependence of all vertices, leads to  the following expression for the
single diffraction cross section:
\beq{SDSC}
\frac{M^2 d^2 \s_{SD}}{d M^2\,d t}\,\,=\,\,( g^p_P(0) )^3 \cdot G_{PPP}
(0)\cdot e^{2 ( R^2_0 + r^2_0 + \alpha'_P \ln ( s/ M^2))t}\,
(\,\frac{s}{M^2}\,)^{2\,\Delta_P}\,(\,\frac{M^2}{s_0}\,)^{\Delta_P}\,\,,
\eeq
where $R^2_0$ and $r^2_0$ give the $t$-dependence of the Pomeron - proton
form factor and the triple Pomeron vertex respectively.
The total diffractive dissociation cross section for large masses $M$
 ( $ M\,\geq\,M_0$) is given by:
\beq \label{TSDSC}
\s_{SD}(M\,\geq\,M_0)\,\,=
\eeq
$$
\int^0_{- \infty}\,d t \int^s_{M^2_0}\,d M^2
\frac{ d^2 \s_{SD}}{d M^2\,d t}\,\,=\,\,
\frac{( g^p_P(0) )^3 \cdot G_{PPP}(0)}{2 ( R^2_0 + r^2_0 +
\alpha'_P\,\ln(s/M^2_0)}\cdot (\,\frac{s^2}{s_0 M^2_0\,})^{\Delta_P}\,\,.
$$
 One sees that $\s_{sd}\,\propto \,s^{2 \Delta_P}$ while the
total cross section is proportional to $ s^{\Delta_P}$. It means that
something is deeply wrong in our approach which can lead to a  cross
section
for  diffractive dissociation which is larger than the total cross
section.
For $\Delta_P\,>\,0$ we can take a  simpler example like the elastic cross
section or  the cross section of  diffractive dissociation in the
region of
small masses ($M \leq M_0$ ). What is important  is the fact
that the diffractive dissociation at  large masses leads to a  cross
section which increases with energy even if $\Delta_P = 0$. Indeed,
repeating all calculations in \eq{TSDSC} for $\Delta_P = 0$ we can see
that
$$
\s_{SD}\,\,\propto \,\,\ln\ln
s\,\,\gg\,\,\sigma_{tot}\,\,\propto\,\,Const\,\,.
$$  
The way out is to  take into account the
interaction between Pomerons, in particular with the triple Pomeron
vertex.  I will
postpone this problem  to the second part of the lectures. I would like to
show
here that the  interactions between Pomerons and with particles       
is a natural outcome of the parton space - time picture that we have
discussed in section 8.
\subsection{Space - time picture of the shadowing corrections.}
To understand, why we have to deal with the SC, we have to look back at
Figs.21e and 21f. When I discussed these pictures in sections 8.2 and 8.3,
I cheated  a bit or rather  I did not
stress that we had made the  assumption that only one ``wee" parton
interacts
with the  target. Now,  let  ask ourselves why only one? What is going to
happen if, let us say, two ``wee" partons interact with the  target?
Look, for example, at  Fig. 21e. If two ``wee" partons interact with the
target the coherence in two ``ladders" will be destroyed. Squaring this
amplitude with two incoherent  ``ladders", one can see that we obtain the
diagram
in which two initial hadrons interact by the  exchange of two Pomerons.
Here, I
used the second definition of the Pomeron, namely, Pomeron = ``ladder".

Therefore, the whole picture of interaction looks as follows. The fast
hadron decays into  a large  number of partons and this partonic
fluctuation
lives for a long time $ t \,\propto \,\frac{E}{\mu^2}$ ( see Fig.21d ).
During
this time each parton can create its own chain of partons and, as we have
discussed,  results  in the  production of $N$ ``wee' partons,
which can interact with the target with a standard  cross section ( $\s_0$
).
Assuming that only one ``wee" parton interacts with the target we obtained
Eq.(72), namely, $ \s^P_{tot}\,=\,\s_0 \cdot N$. This process we call the
Pomeron exchange.

 However, several ``wee"  partons can interact with  the
target.  Let us  assume that two ``wee" partons interact with
the target. The cross section will be equal to $\s^P_{tot} \cdot W$,
 where $W$ is the probability for  the second ``wee" parton  to meet and
interact with the target. As we have discussed in section 7.4 all ``wee"
partons are distributed in an  area in the transverse plane which is equal
$\pi R^2(s)$ and $R^2 \,\,\rightarrow\,\,\alpha'_P \ln(s)$. Therefore,
the cross section of the two ``wee" parton interaction is equal to
\beq \label{TWPCR}
\s^{(2)}\,\,=\,\,\s_0 \cdot N \cdot \frac{\s_0 \cdot N}{\pi R^2(s)}\,\,.
\eeq
If $\Delta_P\,>\,0$ this cross section tends to overshoot the one ``wee"
parton cross section. However, if $\Delta = 0$ $\s^{(2)}$ turns to be much
smaller than $ \s^P_{tot}$, namely, 
\beq \label{TWOPAR}
\s^{(2)}\,\,=\,\,\frac{(\,\s^P_{tot}\,)^2}{ \pi\,R^2(s)}\,\,.
\eeq 
Remember that \eq{TWOPAR} had given us the  hope that  the
Pomeron hypothesis
will  survive. However, let me recall that the problem with the triple
Pomeron interaction and with the large mass diffraction dissociation as
the first manifestation of such an interaction  remains  even 
if $\Delta_P =0$.

Now let us ask ourselves what should be the  sign for a $\s^{(2)}$
contribution
to the total cross section. A strange question, isn't it. My claim is that
the
sign should be negative. Why? Let us remember that the total cross section
is just the  probability that an  incoming particle has at least one
interaction.
Therefore, our interaction in the parton model is  the probability for the
scattering of the
flux of $N$ ``wee" partons with the target. To calculate the total cross
section we  only need  to insure   that at least one interaction
has happened.
However, we overestimate the value of the total cross section since we
assumed  that every parton out of  the total number of ``wee" partons $N$
is
able to interact with the target. Actually, the second parton cannot
interact with the target if it is just behind the first one. The
probability to find the second parton just behind the first one  is equal
to
$W$ which we have estimated. Therefore, the correct answer for   the
flux
of ``wee" partons that can interact with the target is equal to 
$$
FLUX\,\,=\,\,N\cdot (\, 1\, -\, W\,)\,\,=\,\,N\cdot(\,1\, -\,
\frac{\s^P_{tot}}{\pi R^2(s)}\,) \,\,
$$
which leads to the total cross section:
\beq \label{TCRMP}
\s_{tot}\,\,=\,\,\s^P_{tot}\,\,-\,\,\frac{(\,\s^P_{tot}\,)^2}{\pi R^2(s)}
\,\,.
\eeq
I hope that everybody recognizes the famous Glauber formula.
 
Therefore,{\it the shadowing corrections for Pomeron exchange is  the
Glauber screening of  the
flux of ``wee" partons}.

If $W\,\,\ll\,\,1$ we can restrict ourselves to the  calculation of
interactions     of two  ``wee" partons with the target. However, if $W
\,\approx\,1$ we face a complicated and challenging problem of calculating
all SC. We are far away from any solution of this problem especially in
the ``soft' interaction. Here, I am  only going to demonstrate  how
the SC works and what qualitative manifestation of the SC can be expected 
in  high energy scattering.
\subsection{The AGK cutting rules.}
 
The Abramovsky - Gribov - Kancheli  ( AGK )   cutting rules give  the
generalization of the optical
theorem for  the case of  multi -  Pomeron exchange.

Indeed, the optical theorem for one Pomeron exchange  ( $\s^{(1)}_1$ in
Fig. 32 )  decodes the Pomeron structure and says that the Pomeron
contribution to the total cross section is  closely related to the
cross section for the production of a large number of hadrons ( partons) $
< n_P>$ which in
 first approximation  are uniformly distributed in rapidity (Feynman
gas hypothesis).

In the case of a  two ``wee" parton interaction or, in other words, in the
case of  two Pomeron
exchange there are  three different contributions to the total cross
section if
we analyze them from the point of view of the type of the inelastic
events.

They  are ( see Fig.32 ):

1.  The diffractive  dissociation ( $\sigma^{(0)}_2$ ) processes  which
produces  a very small multiplicity of the particles in the final state.
A  clear signature of these processes is the fact that no particles are
produced in the large gap in rapidity ( see Fig.32 and the lego - plot for
$\s^{(0)}_2 $ typical event).

2. The production of  secondary hadrons with the same multiplicity as for
one Pomeron exchange
( see  $ \sigma^{(1)}_2$ in Fig.32 ). This event has the same
multiplicity
(  $<n_P>$) as the one Pomeron exchange ( compare $\s^{(1)}_2 $ with
$\s^{(1)}_1$ in Fig.32 ).

3.   The production of  secondary hadrons with a  multiplicity which is
twice as large as
for one Pomeron exchange ( see $\sigma^{(2)}_2$ in
Fig.32 ).

The AGK cutting rules claim \footnote{I will explain why $
\sigma^{(1)}_2$ is negative a bit later. Here, I would like only to
remind you that the cross section with the same multiplicity as for one
Pomeron exchange is impossible to separate from one Pomeron exchange. It
is clear that one Pomeron exchange gives a larger contribution than  $
\sigma^{(1)}_2$ and it results in total positive cross section with
multiplicity $< n_P >$.}: \beq \label{AGK}
\sigma^{(0)}_2\,\div\,\sigma^{(1)}_2\,\div\,\sigma^{(2)}_2\,\,=\, 
(\, 1\, )\,\div\,
(\,- 4\,)
\,
\div\,(\,2\,)
\eeq
Two important consequences follow from the AGK cutting rules:

1. The total cross section of  diffraction dissociation (DD) is equal to
 the contribution of two Pomeron exchange to the total cross section
( $\sigma^{(2P)}_{tot}$ ) with opposite sign:
$$
\sigma^{(DD)}\,\,=\,\,-\,\sigma^{(2P)}
$$

2. The two Pomeron exchange does not contribute to the total inclusive cross
 section in the central kinematic region. Indeed only two processes, $
\sigma^{(1)}$ and $\sigma^{(2)}$,  are  sources of 
particle production in
 the central region. Therefore, the total inclusive cross section due
to two Pomeron exchange is equal to
$$
\sigma_{inc}\,=\,\sigma^{(1)}\,+\,2\,\sigma^{(2)}\,=\,0
$$
The factor 2 comes from the fact that the particle can be produced from
two
 different parton showers ( see $\s^{(2)}_2$ in Fig. 32).

\begin{figure}[htbp]
\centerline{\epsfig{file=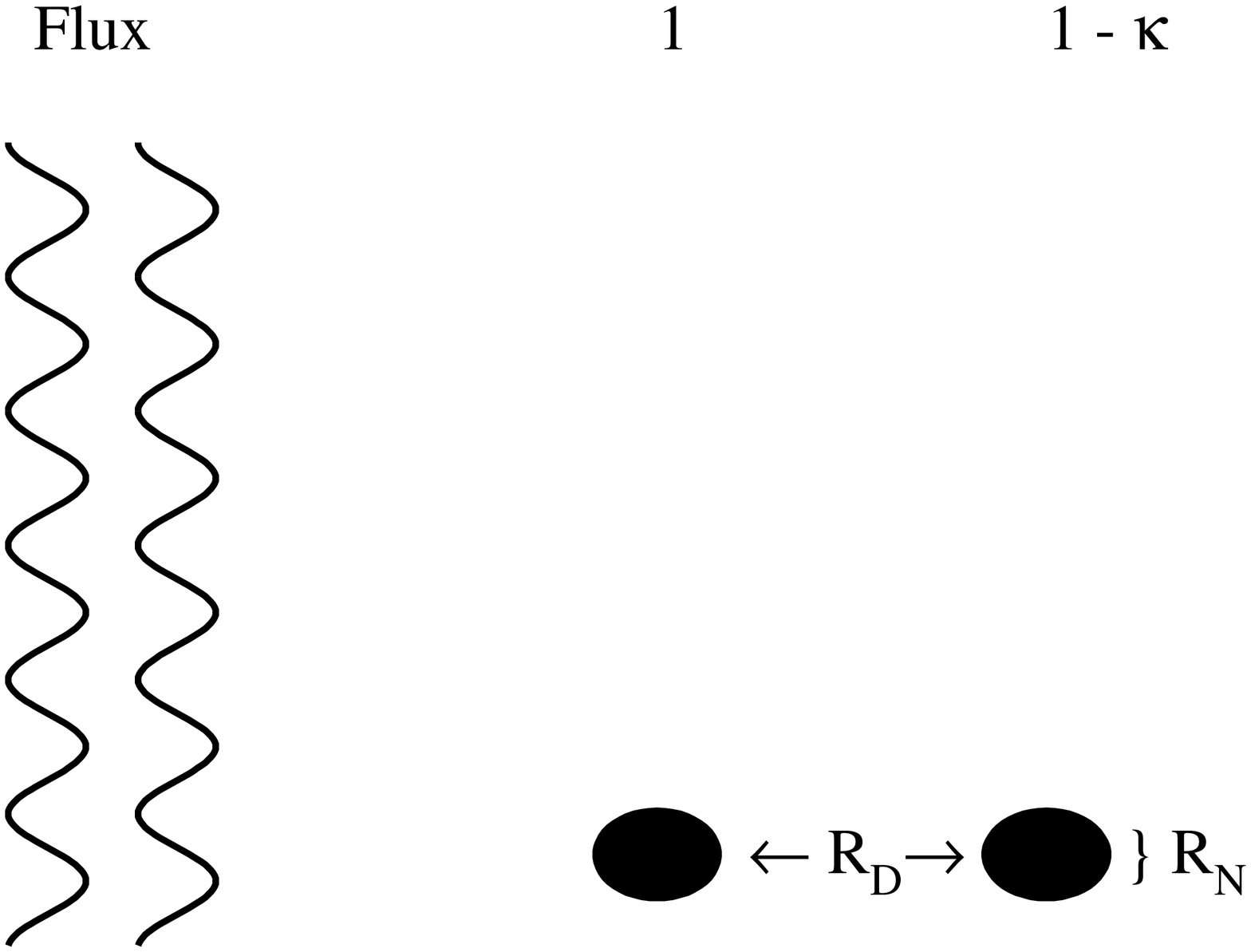, width=120mm}}
\caption{{\it Hadron - deuteron interaction.}}
\label{fig31}
\end{figure}

Now let me give you a brief proof of the AGK cutting rules for the case of
hadron - deuteron interactions ( see Fig.31).  For simplicity let us
assume
that $G_{in}(b_t)
\,=\,\kappa \,=\, Const (b_t)$ for $b_t < R_N$ and $G_{in} = 0 $ for $b_t
>
R_N$.
Than   the inelastic cross section for hadron -
 nucleon interaction is equal to
$$
\sigma^{inel}_N\,=\,\kappa \,S
$$ where S is the area of the nucleon  ( S = $\pi R^2_N$ ).
To calculate the elastic cross section we need to use the unitarity constraint
for $b_t$ ( see Eq.(20 ) which leads to
$$
\sigma^{el}_N \,=\,( \frac{\kappa}{2})^2 S
$$
Noting  that the flux of incoming particles after the first interaction
becomes
$ ( 1 - \kappa ) $ we can calculate the total inelastic interaction with the
 deutron
$$
\sigma^{inel}_D\,\,= \kappa S \,\,+\,\, \kappa ( 1 - \kappa ) S\,\,=\,\,
 2 \,\sigma^{inel}_N - \kappa^2 S
$$
Imposing the  unitarity condition the elastic cross section for
the
 interaction with deuteron  is equal to
$$
\sigma^{el}_D\,\,=\,\, (\frac{2 \kappa}{2})^2 S
$$
Note  that in the calculation of the
elastic cross section we cannot use  probabilistic argumentation,  we
have to use the unitarity constraint of Eq.(20).
Therefore the total cross section for hadron - deutron interaction can be
presented in the form:
$$
\sigma^{tot}_D\,\,= 2 \,\sigma^{tot}_N \,- \,\Delta \sigma
$$
where
$$
\Delta \sigma\,\,=\,\,\Delta \sigma^{inel} \,+\,\Delta \sigma^{el}\,=
\,-\,\frac{\kappa^2}{2}\,S\,=\,-\,\frac{ (\sigma^{inel}_N )^2}{2 \pi R^2}
$$
You will  again  recognize the  Glauber formula for hadron - deutron
interactions.

\begin{figure}[htbp]
\centerline{\epsfig{file=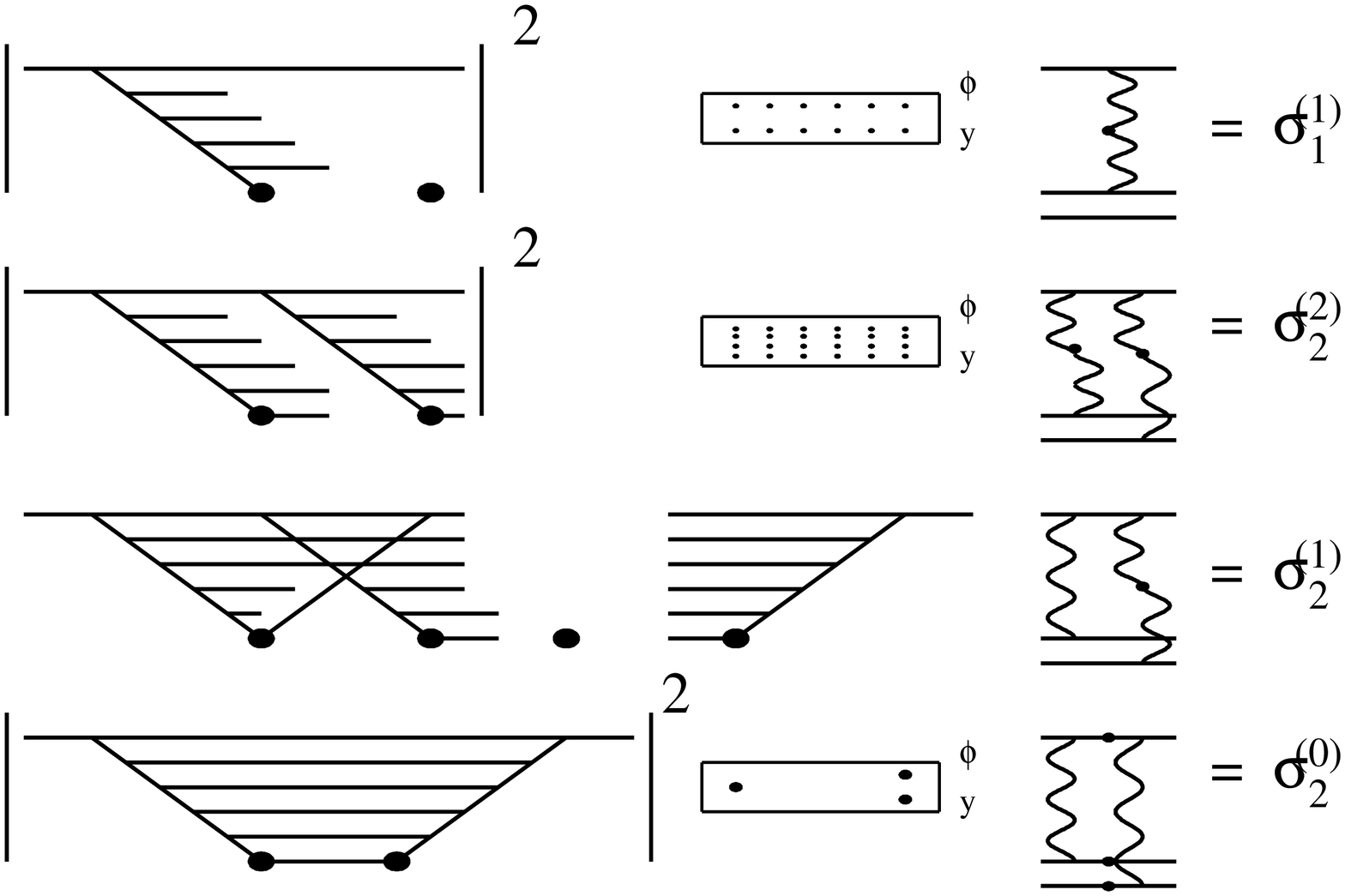, width=160mm}}
\caption{{\it The AGK cutting rules for hadron - deuteron interaction.
Dots mark cut Pomerons and particles on mass shell.}}
\label{fig32}
\end{figure}

As mentioned before there are   two sources of the inelastic cross
section which
are pictured in Fig. 32  for the case of the deutron. The cross section
for
the
inelastic process with doubled  multiplicity is easy to calculate, since 
it is
equal to the probability for  two inelastic interactions:
$$
\Delta \sigma^{(2)}_2\,\,=\,\,\kappa^2 \cdot S
$$
To calculate $\sigma^{(1)}_2$ we have to remember that
$$
\Delta
\sigma^{inel}\,\,=\,\,\sigma^{(2)}_2\,\,+\,\,\sigma^{(1)_2}\,\,=\,\,-
\kappa S
$$
Therefore $\sigma^{(1)}_2 \,=\,- 2 \kappa S$. Since  $ \Delta
\sigma^{(0)}_2\,=\,\sigma^{el}_D \,-\,2 \sigma^{el}_N$
 we get
\begin{Large}

\begin{tabular}{|l|l|l|r|}
   \hline
 n  &  0
&   $<n>_N $     &   2$<n>_N$  \\   \hline
$\Delta \sigma_{tot}$  & $ \frac{\kappa^2}{2} S $ & $ - 2 \kappa^2 S $
& $\kappa^2 S$ \\ \hline
\end{tabular}
\end{Large}
\par
Therefore we find  the AGK cutting rules for  hadron - deuteron
interactions,
since  $\sigma^{tot}_D \,=\,2 \sigma^{tot}_N$ corresponds to one Pomeron
exchange  and the correction to this simple formula just originated from the
two
Pomeron exchange in our Reggeon approach. However the above discussion, I hope,
shows you that the AGK cutting rules have more general theoretical
justification  than the
Reggeon approach. For example, they hold in QCD providing the so called
factorization theorem.

Let me give here the general formula for the AGK cutting rules.
The contribution to the total cross section from  the exchange of
$\nu$ Pomerons ( $\sigma^{\nu}_{tot}$) and   the cross sections for  the
for the production of  $\mu <n_P>$ particles ( $\sigma^{(\mu)}_{\nu}$ )
  which are generated by the
 same diagram (see Fig.32) are related\footnote{
  $<n_P>$ is the
average multiplicity in one Pomeron exchange}:

\beq
 \frac{\sigma^{(\mu)}_{\nu}}{\sigma^{\nu}_{tot}}\,|_{\mu \neq  0}\,\,=\,\,
( - 1 )^{\nu - \mu} \cdot \frac{ \nu!}{\mu! \,( \nu  - \mu)!} \cdot 2^{\nu}
\eeq
For $\mu = 0 $ the ratio is equal to 
\beq
 \frac{\sigma^{(0)}_{\nu}}{\sigma^{\nu}_{tot}}\,\,=\,\,
( - 1 )^{\nu} \cdot [\, 2^{\nu - 1}\,-\,1\,]
\eeq
These factors  for exchanging 1,2,...5 Pomerons   are given in the
following table.

\begin{tabular}{|l|r|r|r|r|r|r|}
   \hline
 $\nu \setminus \mu$  &  0 & 1 & 2 & 3 & 4& 5 \\   \hline
 1 & 0 & 1 & 0 & 0 & 0 & 0  \\ \hline
2 & 1 & - 4 & 2 & 0 & 0 & 0 \\ \hline
3 & - 3 & 12 & - 12 & 4 & 0  & 0 \\ \hline
4 & 7 & - 32 & 48 & - 32 & 8 & 0 \\ \hline
5 & - 15 & 80 & - 160 & 160 & - 80 & 16 \\ \hline
\end{tabular}
\subsection{The Eikonal model.}
The simplest way of estimating  the value of the SC is the so called 
Eikonal
model. It is easy to understand the Eikonal model  from the
general solution for  $s$-channel unitarity ( see Eq.(26)):
\beq \label{UNE}
a_{el} (s,b)\,\,=\,\,i\,\{\,1
\,\,-\,\,e^{-\,\frac{\Omega(s,b)}{2}}\,\,\}\,\,;
\eeq
\beq \label{UNI}
G_{in} (s,b)\,\,=\,\,1\,\,-\,\,e^{- \,\Omega (s,b)}\,\,.
\eeq
The weak unitarity constrains are $Im \,a_{el}(s,b)\,\,\leq\,\,1$ and
$G_{in}(s,b)\,\,\leq\,\,1$ which follow directly from \eq{UNE} and
\eq{UNI}.

At small values of opacity $\Omega$,  \, $a_{el} \,=\,\Omega/1$ and
$G_{in} =
\Omega$. Since we assume  that the Pomeron is a good approximation until
the
SC become  large, we take $\Omega$  equal to the result for  one Pomeron
exchange,
namely:
\beq \label{OMEGA}
\Omega(s,b)\,\,=\,\,\frac{\s_0 ( s = s_0)}{ \pi R^2_{ab}(s) }\cdot
(\,\frac{s}{s_0}\,)^{\Delta_P}
e^{- \,\frac{b^2}{R^2_{ab}(s)}}  \,\,,
\eeq
where $\s_0$ is the total cross section for  $a + b$ scattering at some
value of energy ($s = s_0$) where we expect  the SC to be  small.
In the exponential parameterization where the $t$ - dependence of the  
Pomeron -
hadron vertex $g^P_a (t )\,\,=\,\,exp(\,-\, R^2_{0a}|t|\,)$
$$
R^2_{ab}\,\,=\,\,4\cdot(\, R^2_{0a} \,\,+\,\,R^2_{0b}\,\,+\,\,\alpha'_P
(0)
\ln(s/s_0)\,)\,\,.
$$
Certainly,  this is an assumption which has no theoretical proof.
From the point of view of the parton model this assumption looks very
unnatural since the Eikonal model describes
the SC induced only by ``wee" partons originating  from the fastest 
parton ( hadron ). There is no reason why ``wee" partons created by the 
decay of
any  parton, not only the fastest  one, should not   interact with
the target. 
Using   $s$ - channel unitarity and our
main hypothesis that the Pomeron =  ``ladder" ( the second definition for 
the Pomeron ) one can see that the rich structure of the final state 
simplifies to two classes of events: (i)  elastic scattering  and (ii)
 inelastic particle production with  a  uniform rapidity distribution (
Feynman gas assumption). In particular, we neglected a rich structure of
the diffractive dissociation events as well as all of events with
sufficiently large rapidity gaps.

 In some sense the Eikonal model is the
direct generalization of the Feynman gas approach, namely, we take into
account the Feynman gas model for the typical inelastic event  and for 
elastic scattering since we cannot satisfy the  unitarity constraint
without including
 elastic scattering. As I have discussed the $G_{in}$ in Eq.(20) can be
treated almost classically and the probabilistic interpretation for this
term can be used. However, the first two terms in Eq.(20), namely $2\, Im
a_{el}$ and $|a_{el}|^2$, is the  quantum mechanics ( optics ) result  and
both terms must    be taken into account.  

However, the Eikonal  model can be useful in the limited range of energy
and the
physical parameter which is responsible for the accuracy of calculation
in the Eikonal model is the ratio of 
$$
\gamma\,\,=\,\,\frac{G_{PPP}(0)}{g^P_a(0)}\,=\,\frac{\s_{SD}}{2
\s_{el}}\,\,.
$$
It is interesting to note that in the whole range of  energies which are
accesible
experimentally $\gamma \,\leq 1/4$. Therefore, we can use this model for
a  first estimate of the value of SC  to find the qualitative
signature of their influence and to select an efficient  experimental way
to study SC.

First, let us fix our numerical parameters using the Fermilab data at
fixed target energy. We take $s_0 = 400GeV^2$, $\s_0$ = 40 mb,
$R^2_{0p} = 2.6 GeV^{-2}$, $\alpha'_P(0)\,=\,0.25 GeV^{-2}$ and $\Delta $
= 0.08. As you can see we take use  the  parameters of the  D-L Pomeron
but we
use the simple exponential parameterization for vertices while Donnachie
and Landshoff used a  more advanced $t$-distribution which follows from
the
additive quark model. In Fig.33 you can see the result of our calculation
for: $\Omega (s,b)$, $Im \,a_{el}(s,b)$ and $G_{in} (s,b)$ as function of
$b$ at
 $s=s_0$ and at the Tevatron energy. 
\begin{figure}[htbp]
\begin{tabular}{c}
\epsfig{file=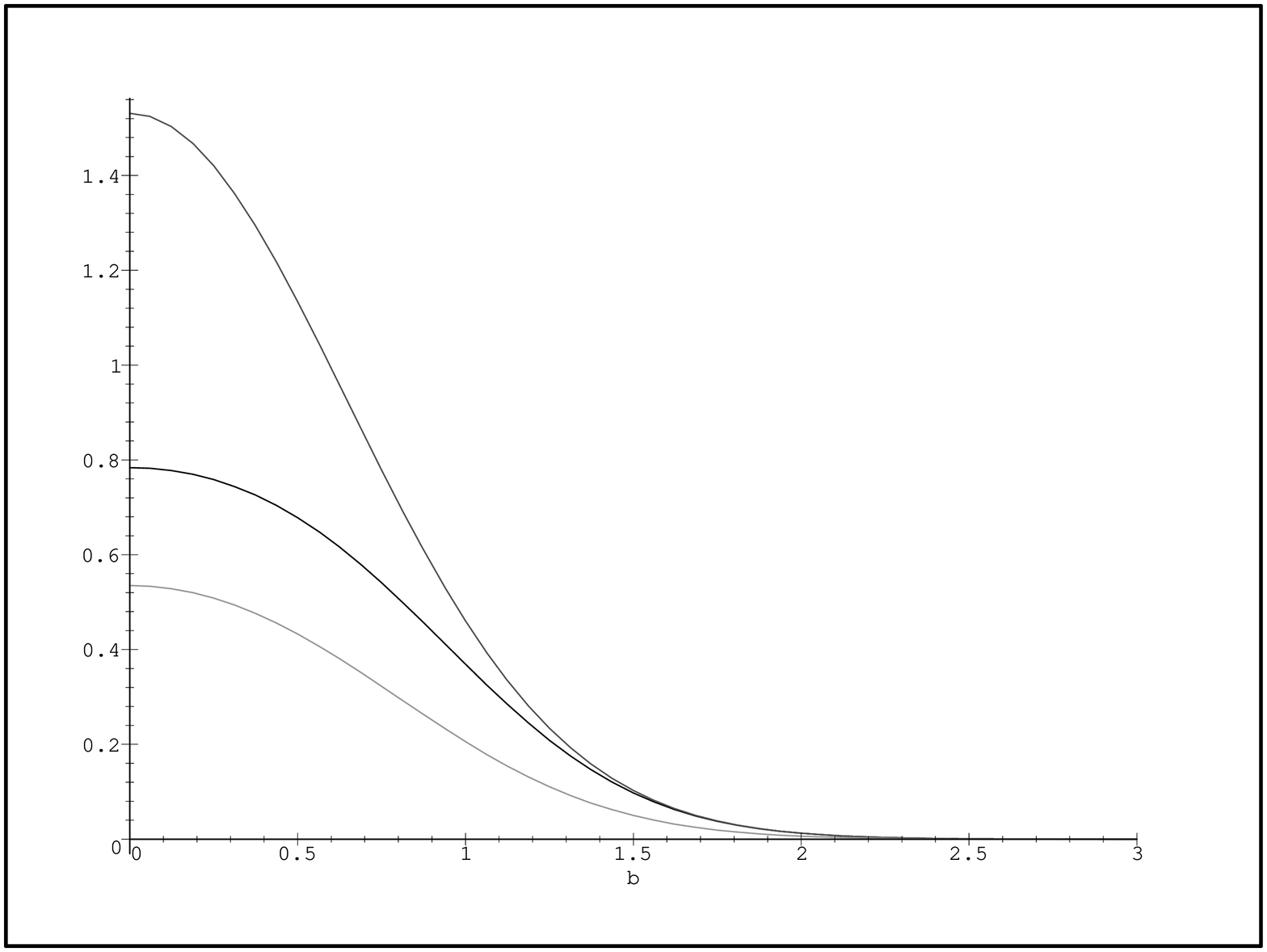,width=110mm,angle=-90} \\
Fig. 33-a\\
\epsfig{file=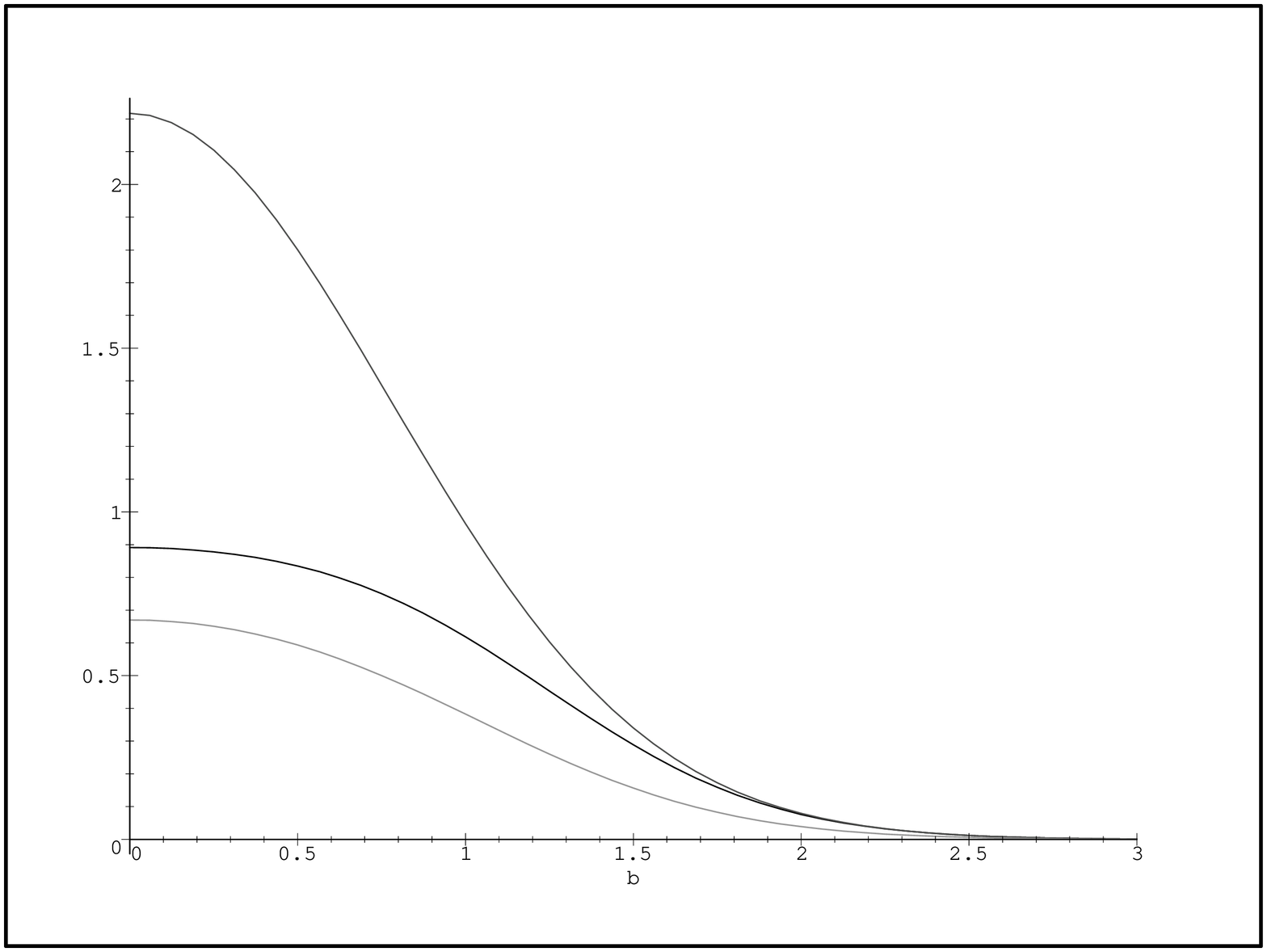,width=110mm,angle=-90}\\
 Fig. 33-b\\
\end{tabular}
\caption{{\it The behaviour of $\Omega(s,b)$ ( dashed curve),
$Im\, a_{el} (s,b)$ ( dotted curve) and $G_{in} (s,b)$ ( solid curve)
as function of impact parameter $b$( in Fm)  at $s= 400GeV^2$ ( a ) and $s
= 4
10^{6} GeV^2$ ( b ).}}
\label{fig33}
\end{figure}
\thispagestyle{empty}
For both energies $\Omega \,>\,1$ which  leads to sufficiently
big SC in $G_{in}$. However, since the SC in the  elastic amplitude
depend on
$\Omega/2$, they are much smaller. 
$G_{in}$ is close to 1:  in the intermediate energy region the 
unitarity  requirement has a simple solution, \, $a_{el}\, \approx
\,\frac{1}{2}$ which
is  obtained    $G_{in} = 1 $   is used in Eq.(20) together with
 $|a_{el}|^2 \,\ll\,Im\, a_{el}$.

The lesson  is very simple: 
{\it if we want to measure the SC we have to find a process which 
depends on $G_{in}$}. Certainly, it  is not the total cross section nor
the
elastic one. We can see from  Fig.33 that we need a
process which is  sensitive to   small impact parameters,
namely, $b\,\leq \,1 Fm$.    
 \subsection{Large Rapidity Gap processes and their survival probability.}
We call  any process a LRG process if over a   sufficiently
large
rapidity region  no hadrons are produced. 
Single and double diffractive dissociation or central diffraction are  
 examples of such processes ( see Fig.24).  Bjorken advocated
\footnote{Ingelman and Schlein (1985) and Dokshitzer, Khoze and Sjostrand
(1992) were the first who proposed to measure such processes but Bjorken
understood that LRG processes give a tool to examine the nature of the
``hard" processes and the interface them  with the ``soft" ones.} 
to study  LRG processes which have the  additional signature of a ``hard"
processes.
The simplest example of such a process is the production of two large
transverse momenta jets with LRG between them ( $\Delta y = |
y_1\,-\,y_2 |\,\gg \,1 $ ) in
back - to - back kinematics ( $ \vec{p}_{t1}\,\,\sim \,\,-\,\vec{p}_{t2}$
). Let us consider this reaction in more details to illustrate how 
Pomeron ``soft" physics enters the calculation of such a typical ``hard"
process.  I will discuss such processes in a more detailed form 
in the third part of my lectures but, I hope, that a brief discussion here
will help  to understand how important it is  to get more reliable
knowledge
on the SC.

The reaction which we consider is ( see Figs. 23-d and 24 for notations):
\beq \label{REACT}
p\,\,+\,\,p\,\,\rightarrow\,\,
\eeq
$$
 M_1\, \{\,\,hadrons\,\,+\,\,jet_1(\,x^1,\vec{p}_{t1}\,)\,\}
\,\,+\,\,[\,LRG\, (\,\Delta y\,=\,|y_1 - y_2|\,)\,]
\,\,+\,\,M_2\,\{\,\,hadrons\,\,+\,\,jet_2(\,x^2,\vec{p}_{t2}\,)\,\}\,\,,
$$
where 
$$
x^1\,\,=\,\,\frac{2 \,p_{t1}}{\sqrt{s}}\,e^{y_1}\,\,;
$$
$$
x^2\,\,=\,\,\frac{2 \,p_{t2}}{\sqrt{s}}\,e^{y_2}\,\,;
$$
and
$$
\vec{p}_{t1}\,\,\approx\,\,-\,\vec{p}_{t2}\,\,;\,\,\,\,\,\,
p_{t1}\,\approx\,p_{t2}\,\,\gg\,\,\mu\,\,.
$$
We denote by $\mu$ the  typical scale for transverse momentum  for ``soft"
 processes.

The cross section of this reaction can be described by a
factorization formula:
\beq \label{LRGCR}
f(\,\Delta y, y_1 + y_2, p_{t1},p_{t2}\,)\,\,=\,\,\frac{d \s}{d \Delta y,
d(y_1
+ y_2) d p^2_{t1} d p^2_{t2}}\,\,= 
\eeq
$$
F^{(1)}_p(x^1, p^2_{t1}) \cdot F^{(2)}_p(x^2, p^2_{t2})\cdot \s_{hard}
(p^2_{t1},x^1x^2 s )\,\,,
$$
where $F^i_p$ is probability to find a parton with $x^i$ in the proton (
deep inelastic structure function) and $\s_{hard}$ is the cross section of
the ``hard" parton - parton interaction at sufficiently large energies
( $x^1x^2 s$ ). We assume, that this ``hard" process is due to the  
exchange of a
colourless object which we call ``hard" Pomeron which will be discussed
in the later  parts of my lectures;  here we concentrate 
on the calculation of the probability for  such a  reaction.  
Reaction of \eq{REACT} can be viewed  as a statistical fluctuation of  the
typical inelastic event. The probability for such a fluctuation is small
and  is of the order
of $exp ( - < n(\Delta y) >) $ where $<  n(\Delta y) >$ is the average
hadron
multiplicity of a typical inelastic event  in the rapidity region
$\Delta y $.

Secondly, we need to multiply \eq{LRGCR} by the  so called damping
factor $< D^2 >$ . It  gives the  probability  that no partons with $x
>x^1$ from one proton and no partons with  $x > x^2$  from the other
proton 
will interact with each other inelastically. In general, such an
interaction
produces many  hadrons  in the rapidity region $\Delta y $, therefore,
they must be excluded 
in  the correct expression for the cross section. Let me point out that 
\eq{LRGCR} is correct if one believes  that the exchange of a single
Pomeron describes the ``soft" processes, an assumption which is made for
the
D-L Pomeron. Therefore, the experimental deviation from \eq{LRGCR} itself
gives  a  model independent information on whether or not the single
Pomeron
is  doing
its job.  Remember that  we need 
to calculate $< D^2 >$. We have discussed that 
\beq \label{PLRG}
P(s,b)\,\,=\,\,e^{-\,\Omega(s,b)}\,\,
\eeq
is equal to the probability that no inelastic interaction between the 
scattered hadron
has happened at energy $\sqrt{s}$ and  impact parameter $b$. Therefore,
in the Eikonal model $P(s,b)$ is just the factor that we need to multiply
\eq{LRGCR} to get a right answer:
\beq \label{RALRG}
f(\,\Delta y,y_c= y_1 + y_2, p_{t1},p_{t2}\,)\,\,=\,\,< D^2 > f
(\,\eq{LRGCR}\,)\,\,,
\eeq
where 
\beq \label{D}
< D^2 >\,\,=\,\,
 \frac{\int \,d^2 b P(s,b)\cdot f(b,\Delta  y,y_c,p_{t1},p_{t2}\,)}
{\int \,d^2 b f(b,\Delta  y,y_c,p_{t1},p_{t2}\,)}\,\,.
\eeq
Here we introduce $f(b,\Delta y)$ using the following formula:
\beq \label{CRBLRG}
f(b,\Delta y )\,\,=\,\,\int d^2 b' d^2 b" F^1_p(b'; x^1,p^2_{t1})
\cdot \s_{hard} (b' - b", x^1 x^2 s)\cdot  F^1_p(b - b';
x^2,p^2_{t2})\,\,,
\eeq
where 
$$
 F^i_p(x^i,p^2_{ti}) \,\,=\,\,\int d^2 b' F^1_p(b'; x^i,p^2_{ti})\,\,.
 $$
Actually, for the deep inelastic structure function we can prove that
the impact parameter  behaviour can be factorized out in the form
$$
F^1_p(b'; x^1,p^2_{t1})\,\,=\,\,S(b)\cdot F^1_p( x^1,p^2_{t1})\,\,,
$$
where $\int d^2 b S(b)\,=\,1$.

One   can also  see  that the  ``hard" process is  located at a
very
small value of  $b' - b" $. Just from the  uncertainty principle 
$b' - b" \,\propto\,1/p_{t1}\,\,\ll\,$ any impact parameter scale for
``soft" processes. Finally, we can use  $b'=b"$ in the integral of
\eq{CRBLRG}. 

 It is  now easy to see that the damping factor can be
calculated as:
\beq \label{DF}
< D^2 >\,\,=\,\,\int d^2 b d^2 b'  \,P(s,b)\cdot S(b')\cdot S(b - b')\,\,.
\eeq
For $S(b)$ we use the exponential parameterization,namely
\beq \label{FF}
S(b)\,\,=\,\,\frac{1}{\pi R^2_H}\,\,e^{-\,\frac{b^2}{R^2_H}}\,\,.
\eeq
Taking the integral over $b'$ we get
\beq \label{DFF}
< D^2 >\,\,=\,\,\int d b^2   \,P(s,b)\cdot \frac{1}{2  R^2_H}\,\,
e^{-\,\frac{b^2}{2 \,R^2_H}}\,\,.
\eeq
Fortunately, the value of $R^2_H$ has been measured at HERA and  is
equal to 
$$
R^2_H\,\,=\,\,8 \,\,GeV^{-2}\,\,.
$$
Of course, $R_H$ does not depend on energy. In Fig.34 one   sees the
impact parameter dependence of 
$$
\Gamma(s,b)\,\,=\,\, P(s,b)\cdot 
e^{-\,\frac{b^2}{2 \,R^2_H}}\,\,,
$$
 or, in other
words, of the profile function for the LRG process. From this figure one
 concludes  that the damping factor should be rather small since $\Gamma$
turns out to be much smaller than $G_{in}$.
 Secondly,  LRG processes in general have
 larger  impact parameters than  the average inelastic process
since the $b$ - distribution shows a dip in the region of
small $b$ and therefore, almost all  LRG processes have
$b\,\approx \,1 Fm$.

\begin{figure}[htbp]
\begin{tabular}{l l}
\epsfig{file=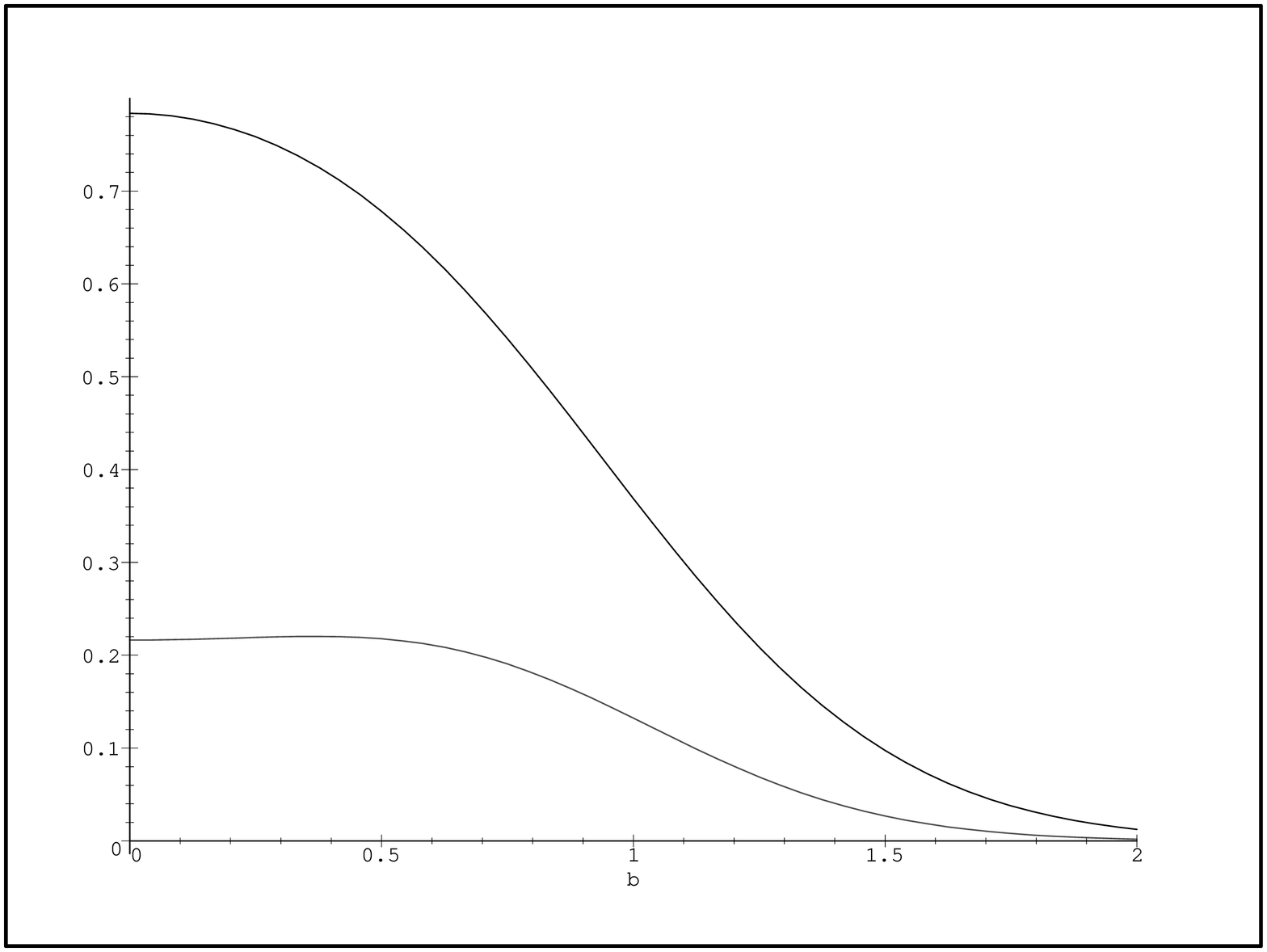,width=60mm,angle=-90} &
\epsfig{file=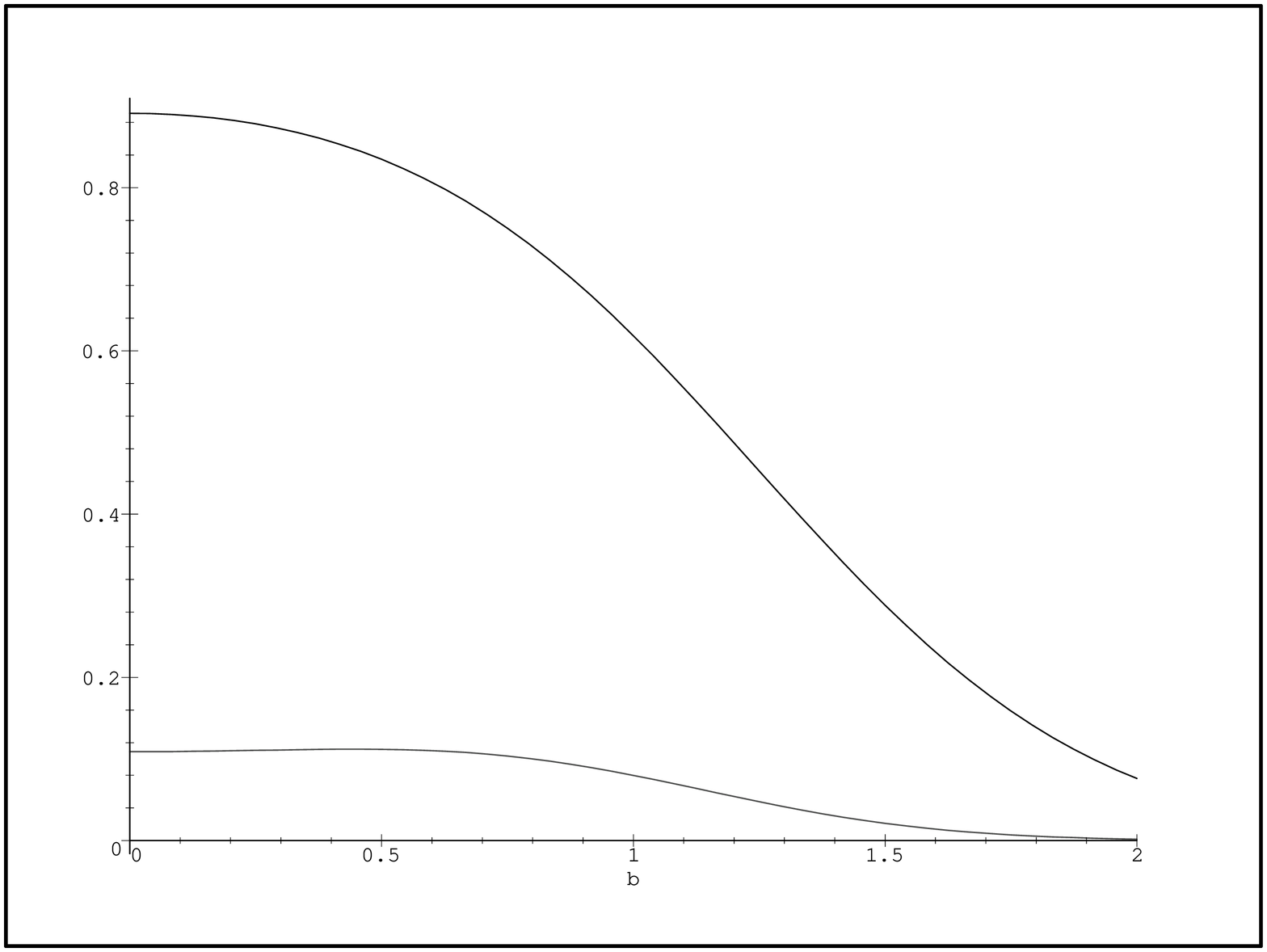,width=60mm,angle=-90}\\
 \,\,\,\,Fig. 34-a &\,\,\,\,Fig. 34-b\\
\epsfig{file=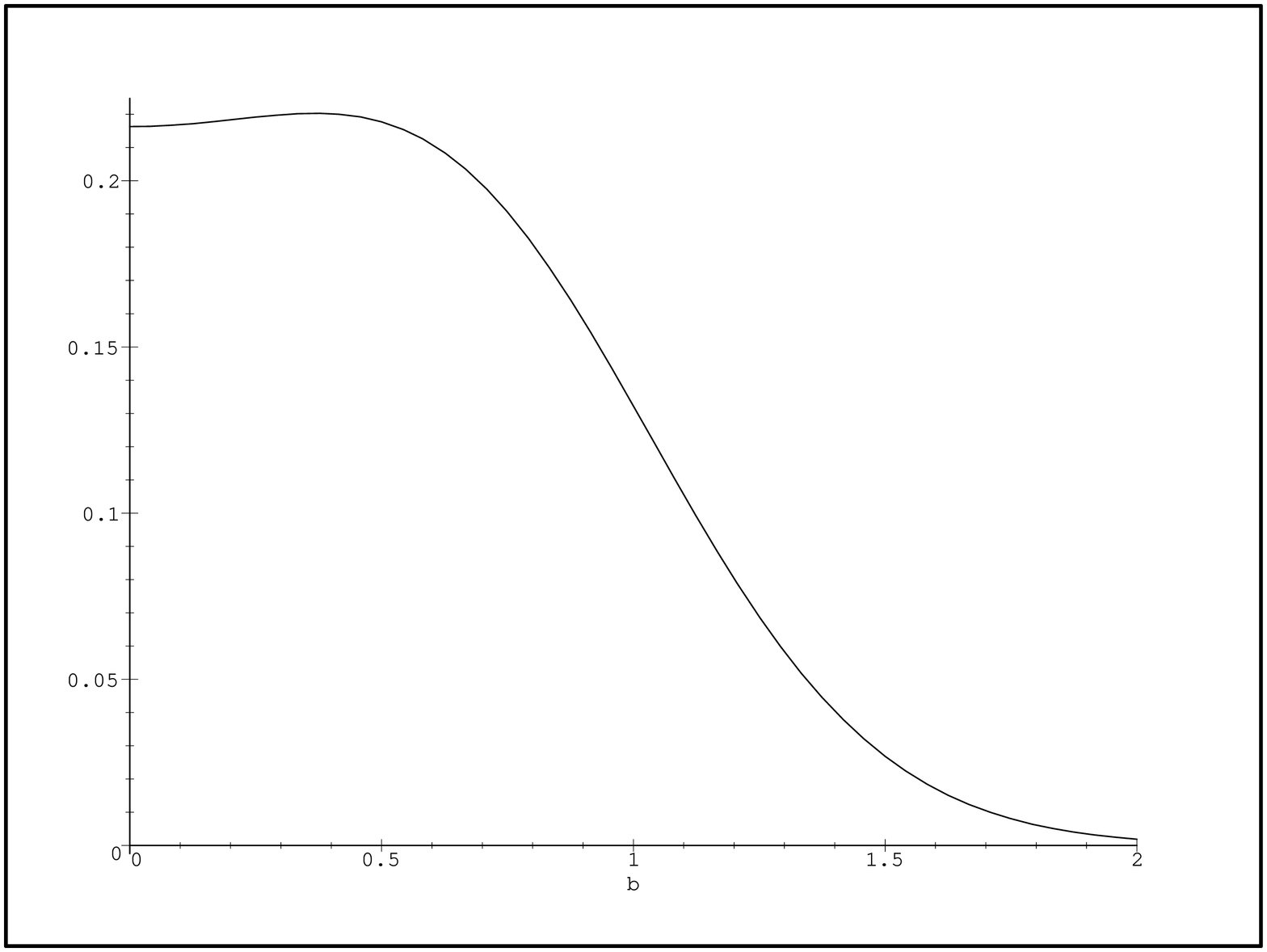,width=60mm,angle=-90} &
\epsfig{file=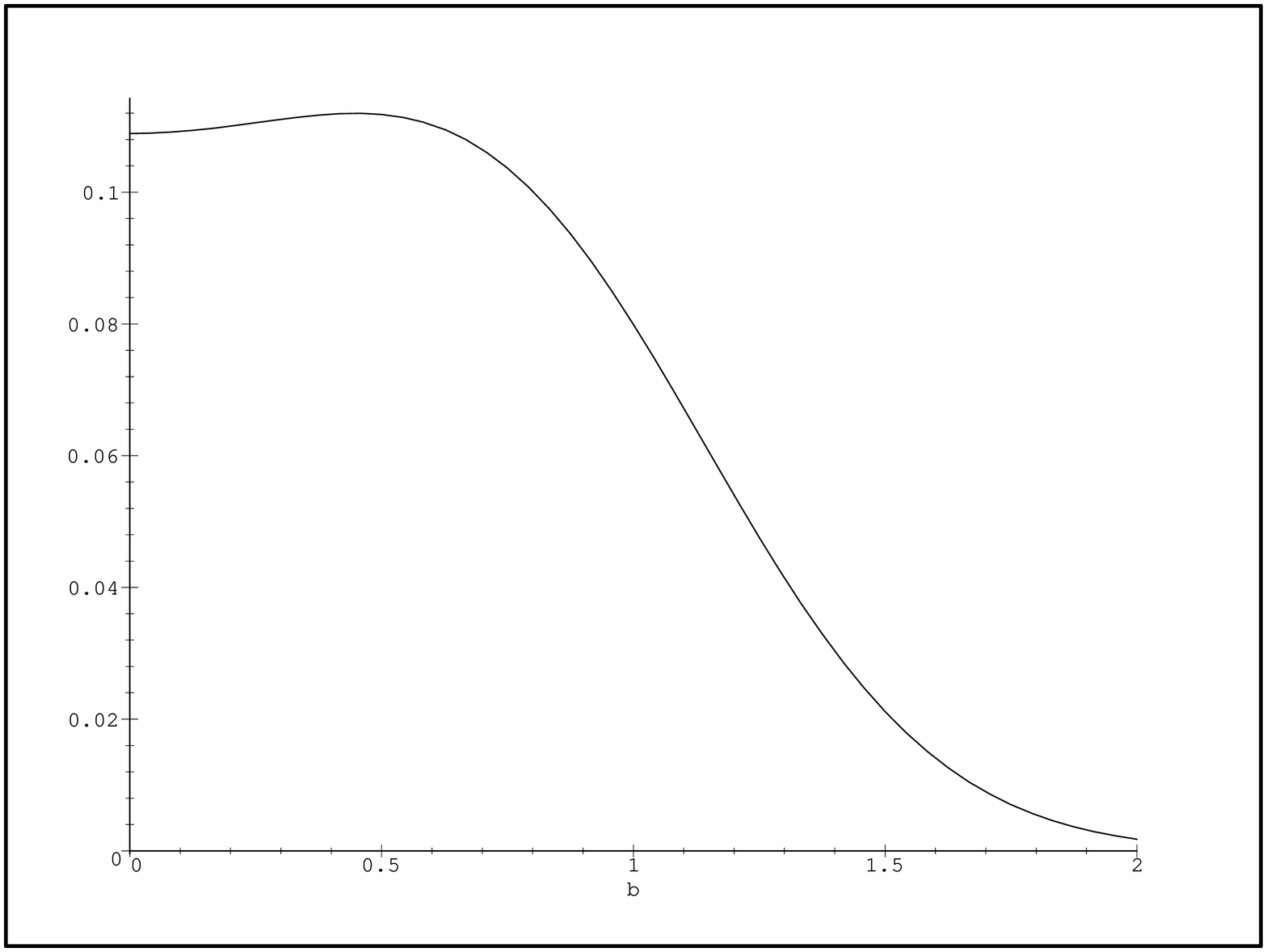,width=60mm,angle=-90}\\
 \,\,\,\,Fig. 34-c &\,\,\,\,Fig. 34-d\\
\end{tabular}
\caption{{\it The impact parameter dependence for $G_{in} (s,b)$ and
$\Gamma(s,b)$ at two values of energy $\sqrt{s}\, = \,400\, GeV$ (
Fig.34-a and
Fig. 34-c)  and $\sqrt{s}\, =\, 4 \,10^{6}\, GeV$ ( Fig. 34-b and Fig.
34-d).}}
\label{fig34}
\end{figure}

Integration  in \eq{DFF} gives:
\beq \label{DFA}
< D^2 >\,\,=\,\,a\,\cdot\,(\frac{1}{\nu(s)})^a \,\cdot\,\gamma(a, \nu(s))
\eeq
where $\gamma(a,\nu) $ is the  incomplete Gamma function
$$
\gamma(a,\nu) \,=\,\int^{\nu}_{0}\,z^{a - 1}\,e^{-z}\,d\,z\,\,
$$
and
$$
a\,\,=\,\,\frac{R^2(s)}{2 R^2_H}\,\,.
$$
The energy behaviour of the damping factor is given in Fig.35a. One can
see
that in this simple model the damping factor is about 30\% and drops by
about a factor of two
 from the Fermilab fixed target experiment energies to the
Tevatron energies.

It is interesting to note that the  single diffraction process which is
also a LRG  process   in the Eikonal model has  a similar damping factor.
The
difference is only in the definition of parameter $a$ or rather  in 
the value of $R_H$ which starts to depend on energy. It turns out that for
single diffraction
$$
a\,\,=\,\, \frac{2 R^2(s)}{(\, R^2(s ) \,+\,R^2(M^2) - R^2(s =
s_0)\,)}\,\,.
$$
In Fig.35b we plot the energy behaviour of the single diffraction cross
section in the triple Pomeron formula ( see Eq.(76) ) and after  taking
into
account $< D^2 >$. One can see that there is  a striking difference in the 
energy behaviour. In my opinion this is  direct experimental  evidence
that  the SC must be taken into account  in addition to the Pomeron
exchange.

\begin{figure}[htbp]
\begin{tabular}{ c }
\epsfig{file=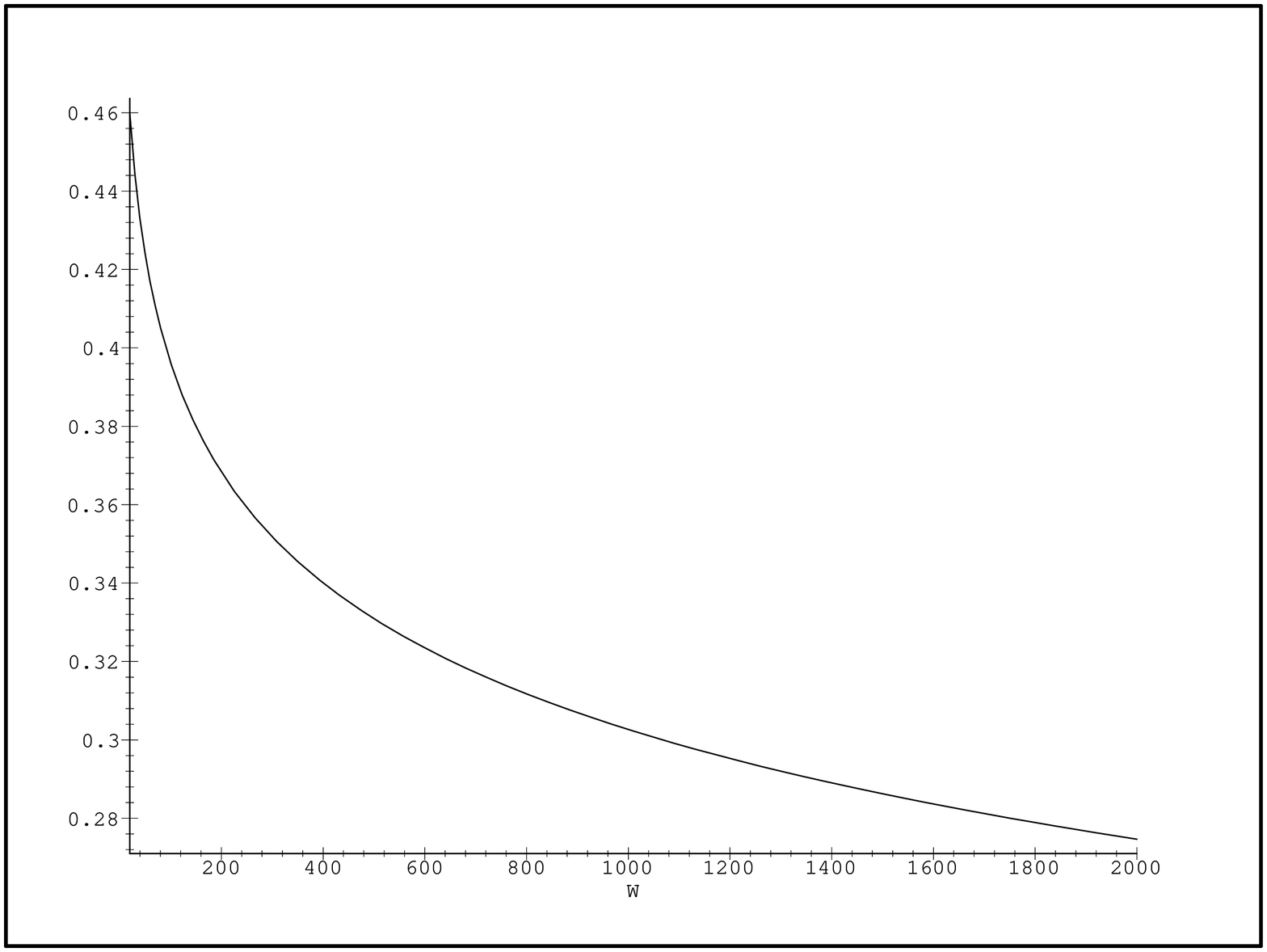,width=100mm,angle=-90} \\
Fig.35-a \\
\epsfig{file=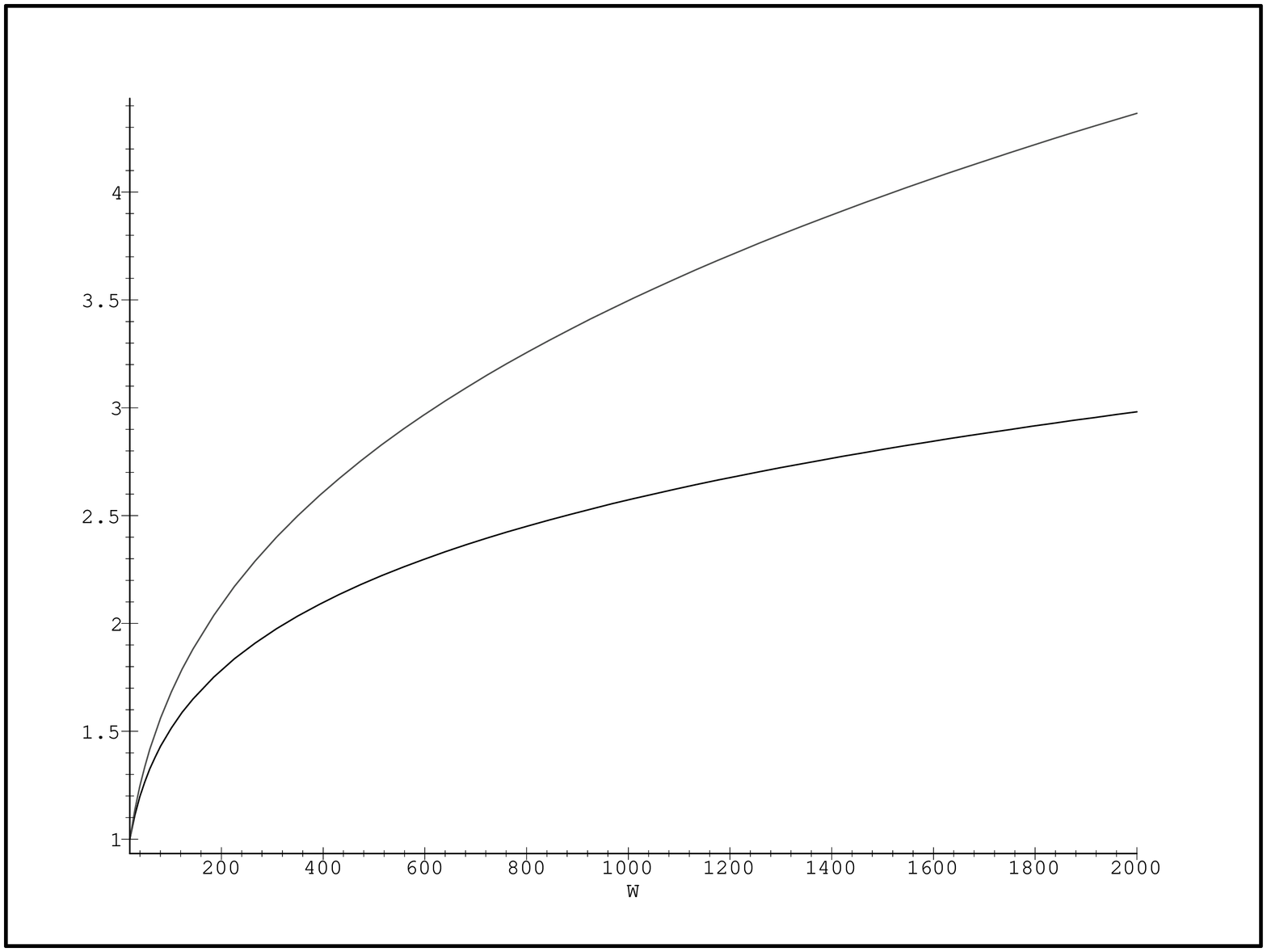,width=100mm,angle=-90}\\   
 Fig. 35-b\\
\end{tabular}
\caption{{\it The energy behaviour of the damping factor for the
``hard" processes with LRG (Fig. 35a ) and for the cross section of  
single diffraction dissociation (Fig.35b): for the triple Pomeron
formula (dashed line) and for the triple Pomeron formula multiplied by
damping
factor( solid line ), $W\,\,=\,\,\sqrt{s}$.}}
 \label{fig35} 
\end{figure} 

\subsection{$\sigma_{tot}$,\,\,\,$\sigma_{el}$ and slope $B$.}

 Using the general formulae of the Eikonal model( see Eqs.(108) - (110) )
one can easily obtain the expression for the total cross section and for
 the elastic one after integration over $b$. I hope that you will be able 
to perform this integration with $\Omega$ given by Eq.(110);  the result
is:
\beq \label{TOT}
\s_{tot}\,\,=\,\,2\,\pi\,R^2(s)\,\{\,\,ln(\Omega(s,b=0)/2)\,\,+\,\,C\,\,-\,\,
Ei( - \Omega(s,b=0)/2)\,\,\}
\eeq
and
\beq \label{ELAST}
\s_{el}\,\,=
\eeq
$$
\pi\,R^2(s)\,\{\,ln( - \Omega( s,b=0 )/4)\,+\,C\,+\,
Ei(\,- \Omega (s,b=0 )\, ) \,- \,2 Ei(\,- \Omega( s,b=0)/2 )\, )\,\}\,\,.
$$
Here, C= 0.5773 is  the Euler constant and $Ei(x) $ denotes the integral
exponent $Ei(x)\,=\,\int^x_{- \infty}\,\,\frac{e^{-t}}{t}d t$.

In Figs.36a and 36.b we plot the energy dependence of the total and
elastic cross sections. One can see that in spite of the fact that the
influence of the SC can be seen at high energy  it is not as pronounced
as in the value of the damping factor and/or in the energy dependence of
the
diffractive dissociation cross section.

For  completeness  let me present here the calculations
for the slope $B$ which  can be  defined  as
$$
B\,\,=\,\,\frac{ln \frac{d \s}{d t}}{d t}\,\,|_{t = 0}\,\,.
$$
Using the impact parameter representation one finds
\beq \label{B}
  B\,\,=\,\,\frac{\int\, b^2 \,Im\,
a_{el}(s,b)\,d^2 b}{2\,\int\,Im\,a_{el}(s,b)\,d^2 b}\,\,.
\eeq
We can rewrite \eq{B} as a ratio of two hypergeometrical functions
\beq \label{BA}
B\,\,=\,\,\frac{R^2(s)}{2}\,\cdot\,\frac{hypergeom(\,[1,1,1],[2,2,2],
-\Omega(s,b=0)/2\, )}{hypergeom(\,[1,1],[2,2], - \Omega(s,b=0)/2)\,)}\,\,.
\eeq 
Here I used the notation of Maple for the hypergeometrical functions to
give you a possibility to check the  numerics. The calculations plotted in
Fig.36c were done with the same values for the parameters of the Eikonal 
model as have been discussed before in section 12.4.
One can see that the SC induce a larger shrinkage of the diffraction  peak
($s$-dependence of the slope $B$ ) than in the simple one Pomeron exchange
approach.

The reason why I discussed these typical ``soft" processes observables is
 to show you that the SC manifest themselves differently for
different processes,  compare e.g.  Figs.36 and
Figs.35.

\begin{figure}[htbp]
\begin{tabular}{l l}
\epsfig{file=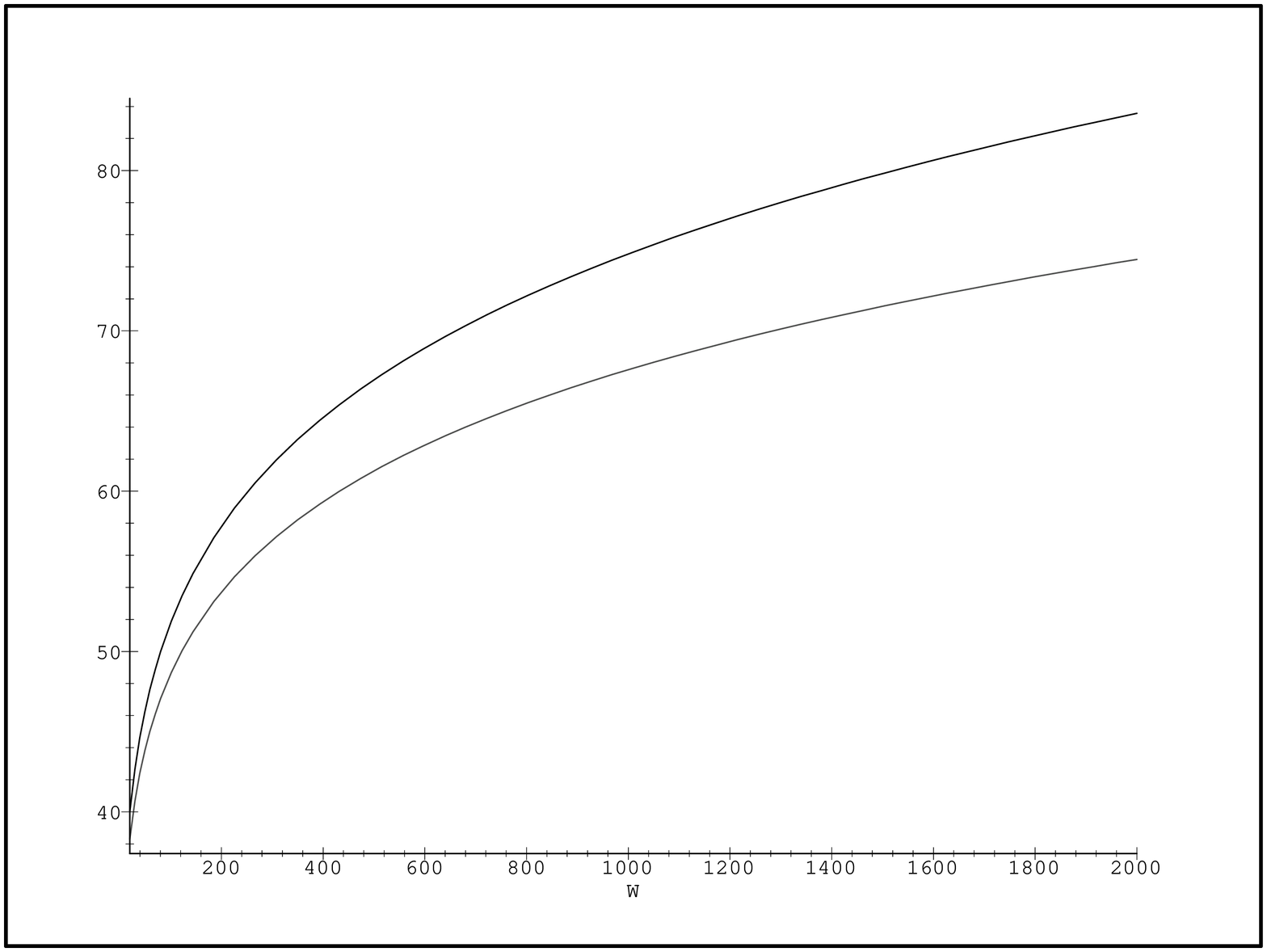,width=60mm,angle=-90} &
\epsfig{file=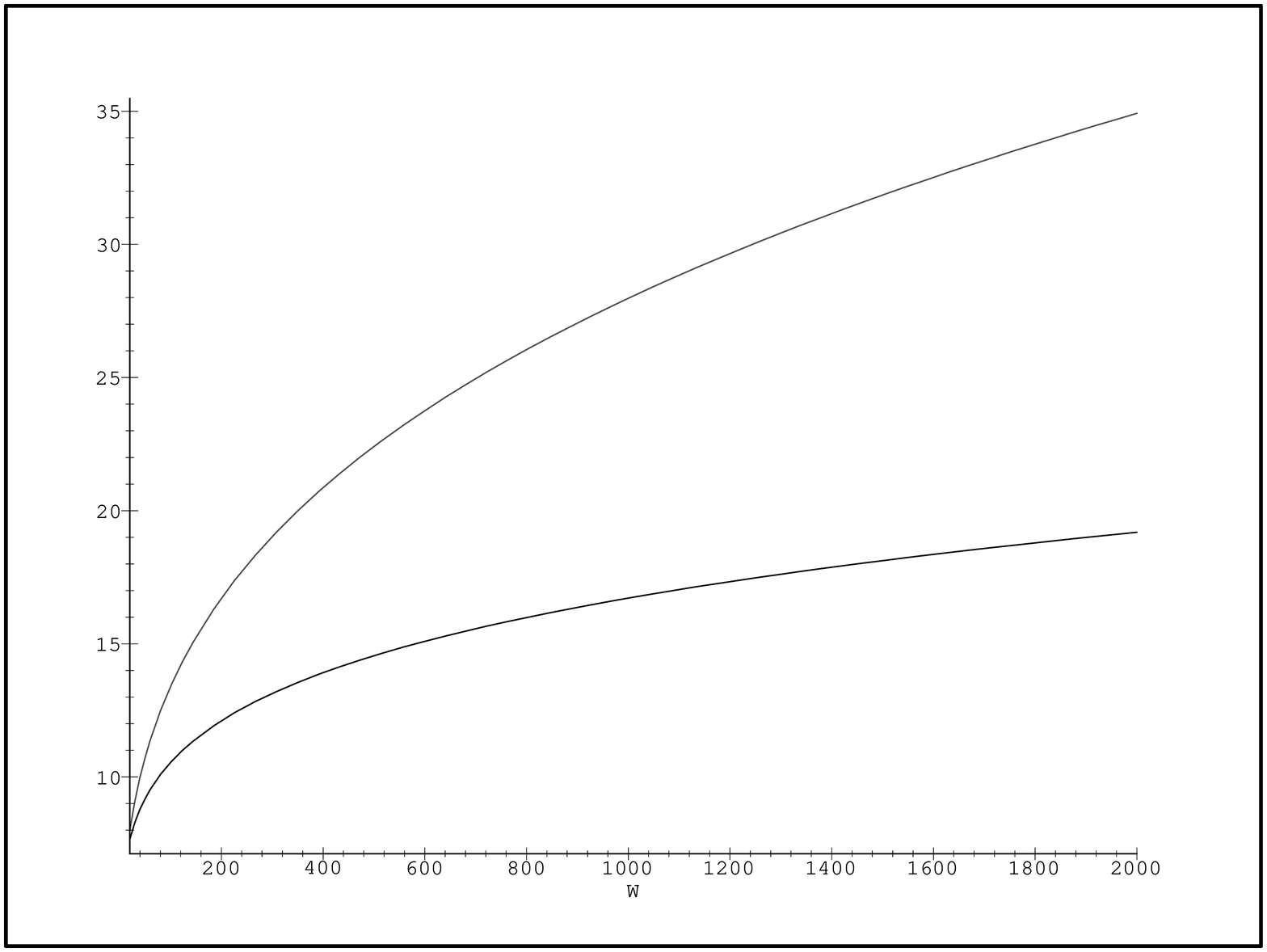,width=60mm,angle=-90}\\   
 \,\,\,\,Fig. 36-a &\,\,\,\,Fig. 36-b\\
\epsfig{file=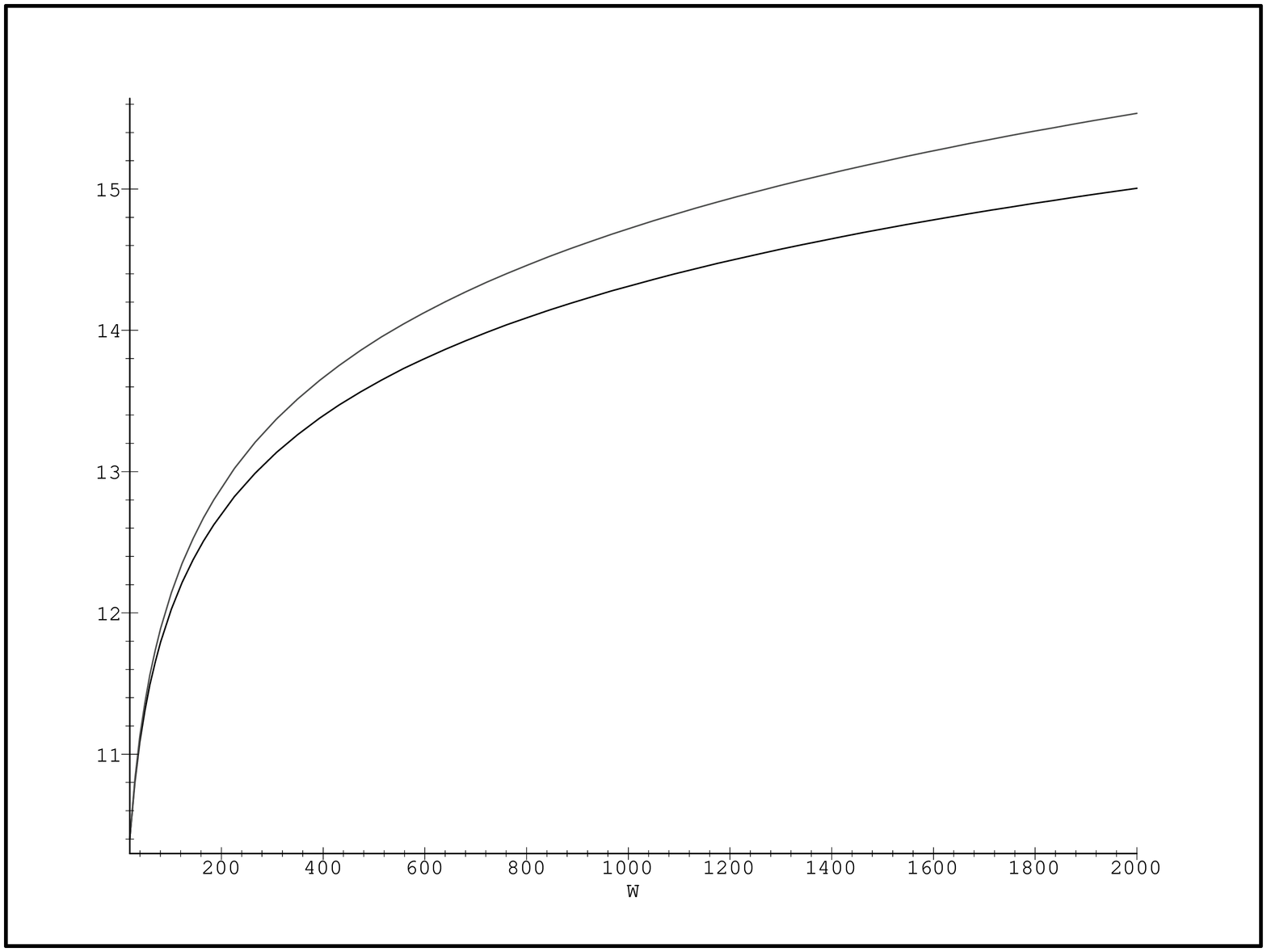,width=60mm,angle=-90} &
\\
 \,\,\,\,Fig. 36-c &\,\,\,\,\\
\end{tabular}
\caption{{\it The energy dependence of $\sigma_{tot}$, $\sigma_{el} $ and 
the slope $B$ of  the Eikonal model for the SC ( dashed curves) and for
the one Pomeron 
exchange approach ( solid curves).}}
 \label{fig36}
\end{figure} 

\subsection{Inclusive production.}

\subsubsection{Cross section.}

We have already  discussed the main result for the inclusive cross
section: { \it due to the AGK cutting rules all SC cancel  and only the
Pomeron exchange gives a nonzero  contribution} ( see Fig.37a ).
We want now to stress that according to AGK cutting rules only the SC
induced by interaction between
a parton with rapidity larger than $y$ with a parton with rapidity smaller
that $y$  give a contribution to the inclusive cross section. All
interactions between two or more partons with rapidity
larger than $y$ are possible as well as  interaction of the produced
particle with the parton cloud (see Fig.37a). In the framework of the
Eikonal model we neglect such   interactions and, therefore, only in the
Eikonal model we can prove that  the simple one Pomeron exchange diagram
for the
inclusive cross section describes the cross section for inclusive
production.

However, even in the general approach the single inclusive cross section
has
a factorization form and can be written as:
\beq \label{INFACT}
\s_{inc}(a + b \,\rightarrow\,c + X)\,\,=\,\,a\,\s^{(a + c)}_{tot}(s_1)\,
\cdot\,\s^{(b + c)}_{tot}(s_2)\,\,,
\eeq
where $a$ is constant and $s_1\, =\,x_1 s$ and $s_2\,=\,x_2 s$ with
$x_1\,=\,\frac{m_{tc}}{\sqrt{s}}\,e^{y_1}$ and
$x_2\,=\,\frac{m_{tc}}{\sqrt{s}}\,e^{- y_1}$ ($ m_{tc}\,=\,\sqrt{m^2_c +
p^2_{tc}}$ ).

In the third part of my lectures I will show  that this  factorization
 leads to the factorization theorem for  ``hard" processes.

\subsubsection{Correlations.}

We have discussed the two particle rapidity correlation for the inclusive
reaction in section 9.8 ( see Fig. 23f ) for  one Pomeron exchange.
We found  that in the correlation function $R$ given by Eq.(82) only
the contribution from the first diagram in Fig.37b survives  leading  to
the simple formula of Eq.(83).  This equation describes the so called  
`` short
range" 
rapidity correlations which vanish at large values of $\Delta y = | y_1
- y_2|$ as
\beq \label{SRCOR}
R\,\,\rightarrow|_{\Delta y\,\gg\,1}\,\,SR\,\cdot\, 
 e^{- \frac{\Delta y}{L_{cor}}}\,\,.
\eeq
A  new contribution comes from the SC and due to the same AGK cutting
rules the contributions from all diagrams cancel in the correlation
function except one, namely,
the two Pomeron exchange shown in Fig.37b. A look at this
diagram shows that  it does not depend on the rapidity difference ($\Delta
y$) and, therefore,  gives a constant contribution to the correlation
function $R$
which we call long range  rapidity  correlations. Therefore, the general
form for the rapidity correlation function $R$ (see Eq.(82) for the
definition) at large $\Delta y$ is
\beq \label{CORGEN}
R\,\,=\,\,SR \cdot\, e^{- \frac{\Delta y}{L_{cor}}}\,\,+\,\,LR
\eeq
The constant term $LR$  can be  estimated  using the AGK cutting rules.
 The cross section for  two particle production from different
Pomerons (``ladders"  in Fig.32 ) is equal to $\s^{(2)}_2
\,\,=\,\,2\,\Delta \s_{tot}$  and  $\s_{tot}\,=\,\s^P_{tot} - \Delta
\s_{tot}$, where $\s^P_{tot}$ is the total cross section in the  one
Pomeron
exchange model and $\Delta \s_{tot}$ is correction to the total cross
section  due to two Pomeron exchange. Therefore,
\beq \label{LR}
LR\,\,=\,\,2\frac{\Delta \sigma_{tot}}{\s_{tot}}\,\,.
\eeq
In the Eikonal model one   finds  $\Delta \sigma_{tot}$, expanding
Eq.(108), namely
\beq \label{EIKLR}
\Delta \s_{tot} \,\,=\,\,\frac{\pi}{2}\,\int b^2 \Omega^2(s,b)\,\,
\eeq
which gives in the parameterization of Eq.(110) for $\Omega$
\beq \label{LREIKFIN}
\Delta \s_{tot}\,\,=\,\,\frac{\s^2_0 (s=s_0)}{\pi\, 4 \,R^2(s)}\cdot
(\,\frac{s}{s_0}\,)^{2 \,\Delta_P}\,\,.
\eeq
Using \eq{LREIKFIN} and \eq{EIKLR}, we  obtain   the following
estimates for the term $LR$ in \eq{CORGEN}:
\beq \label{LREST}
RL\,\,=\,\,\,\frac{\s^2_0 (s=s_0)}{\s_{tot}\,\pi\, 4 \,R^2(s)}\cdot
(\,\frac{s}{s_0}\,)^{2 \,\Delta_P}\,\,.
\eeq
In Fig.37c the value of $LR$ term is plotted as a function of energy. For
$\s_{tot}$ we used Eq.(121). One can see that the long range rapidity
correlations are  rather essential and give positive correlation of the
order of  30\% - 50\%.

We can also write the two Pomeron contribution  to the total double
inclusive cross section ( $\s^{DP}$ ) in the form:
\beq \label{DICR}
\s^{DP}\,\,=\,\,m\,\frac{\frac{d \s}{d y_1}\,\frac{d \s}{d
y_2}}{2\,\s_{eff}}
\eeq
where $m$ is equal 1 for identical particle $1$ and $2$ while it is equal
to 2 if they are different. The Eikonal model predicts for 
$\s_{eff}\,=\,25 \,mb$ at $W = \sqrt{s} = 20 GeV$ rising  to $38 \,mb$ at
the Tevatron energies ( $W = 2000 GeV$).
We will use the form of \eq{DICR} in the next section in  discussion of
the
CDF data.

\alphfig
\begin{figure}[htbp]
\centerline{\epsfig{file=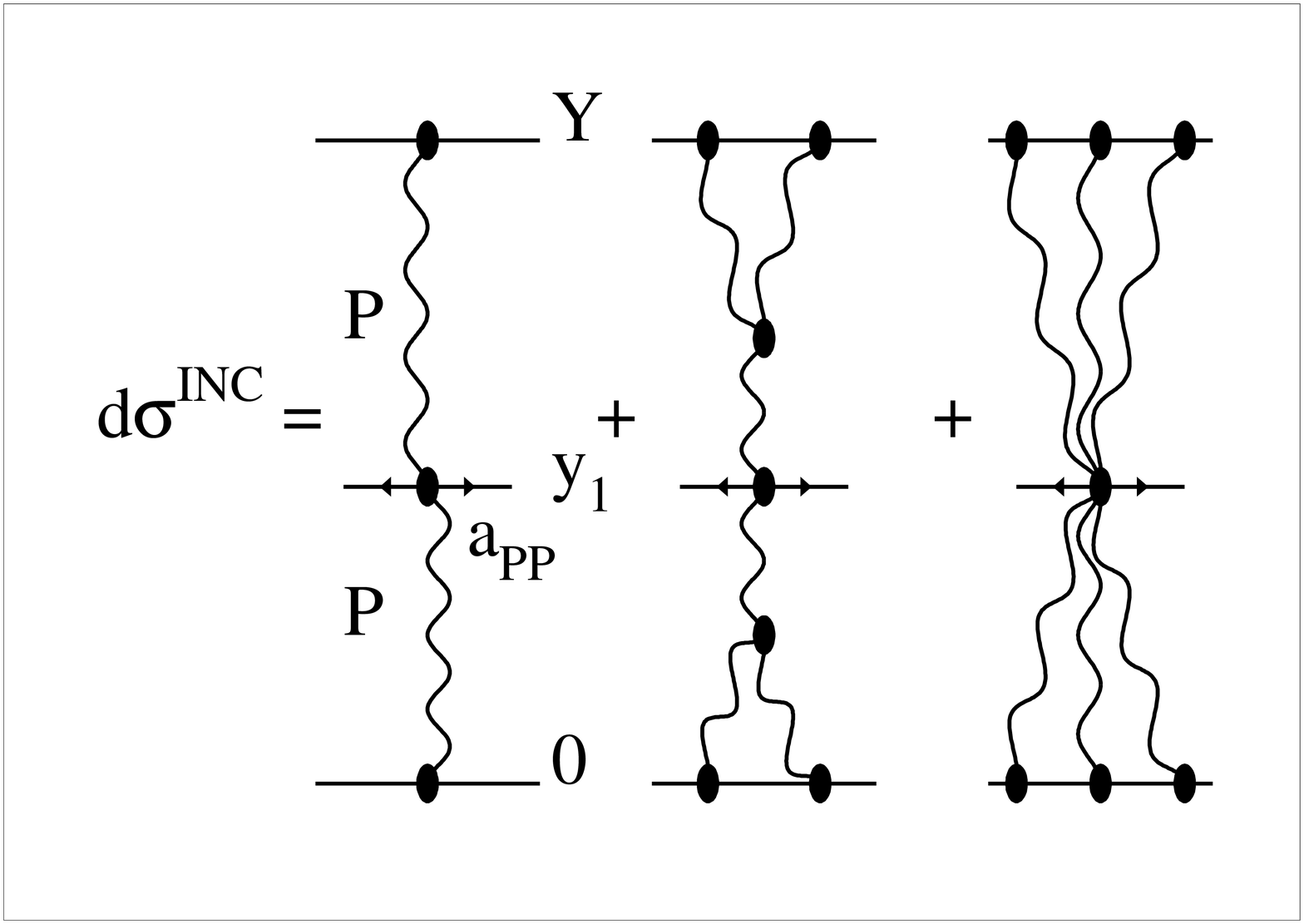,width=120mm}}
\caption{{\it Single inclusive cross section with shadowing corrections.}}
\label{fig37a}
\end{figure} 
\begin{figure}[htbp]
\centerline{\epsfig{file=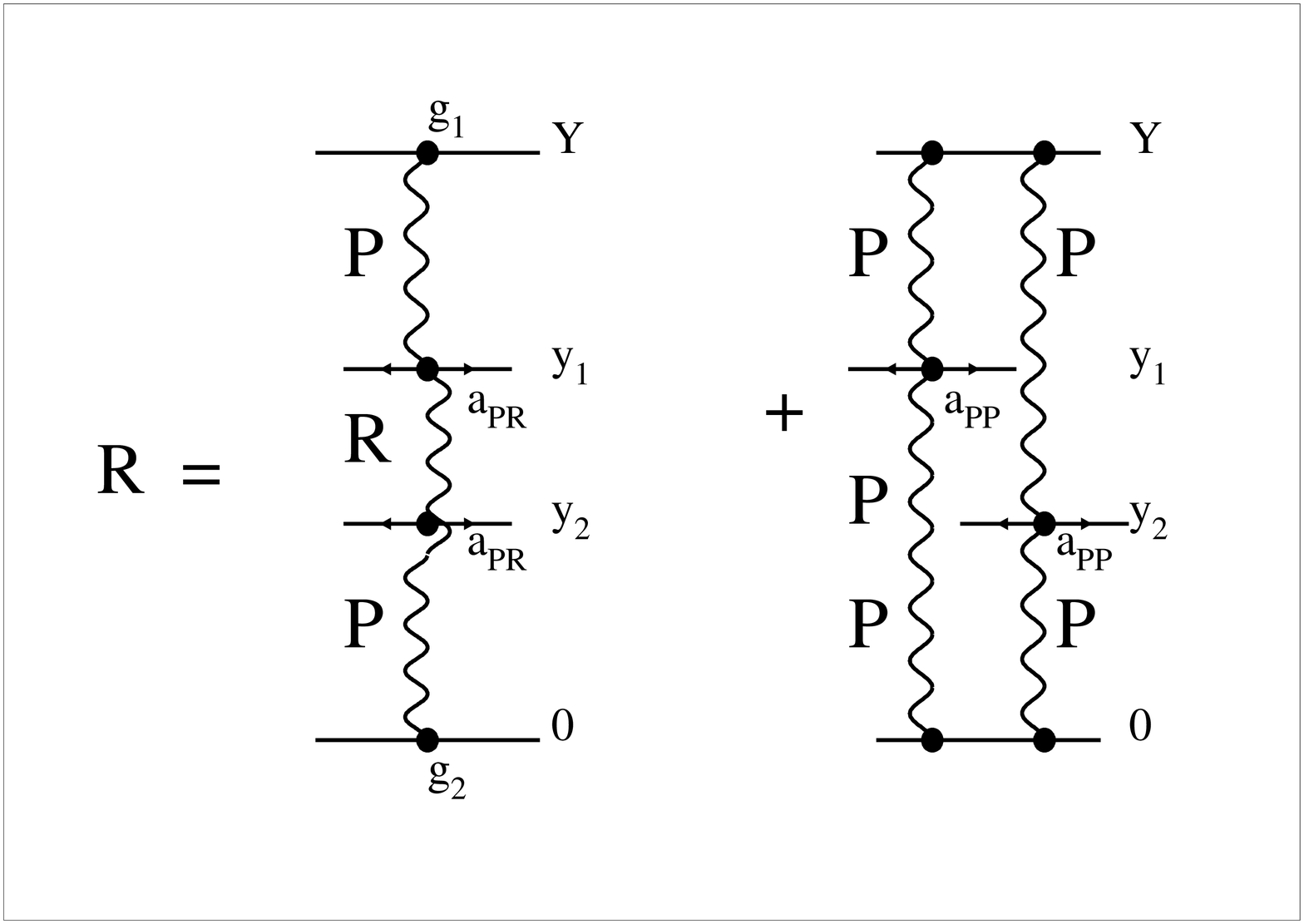,width=120mm}}
\caption{{\it Correlation function $R$ (see Eq.(82) ) with shadowing
corrections.}}
\label{fig37b}
\end{figure} 
\begin{figure}[htbp]
\centerline{\epsfig{file=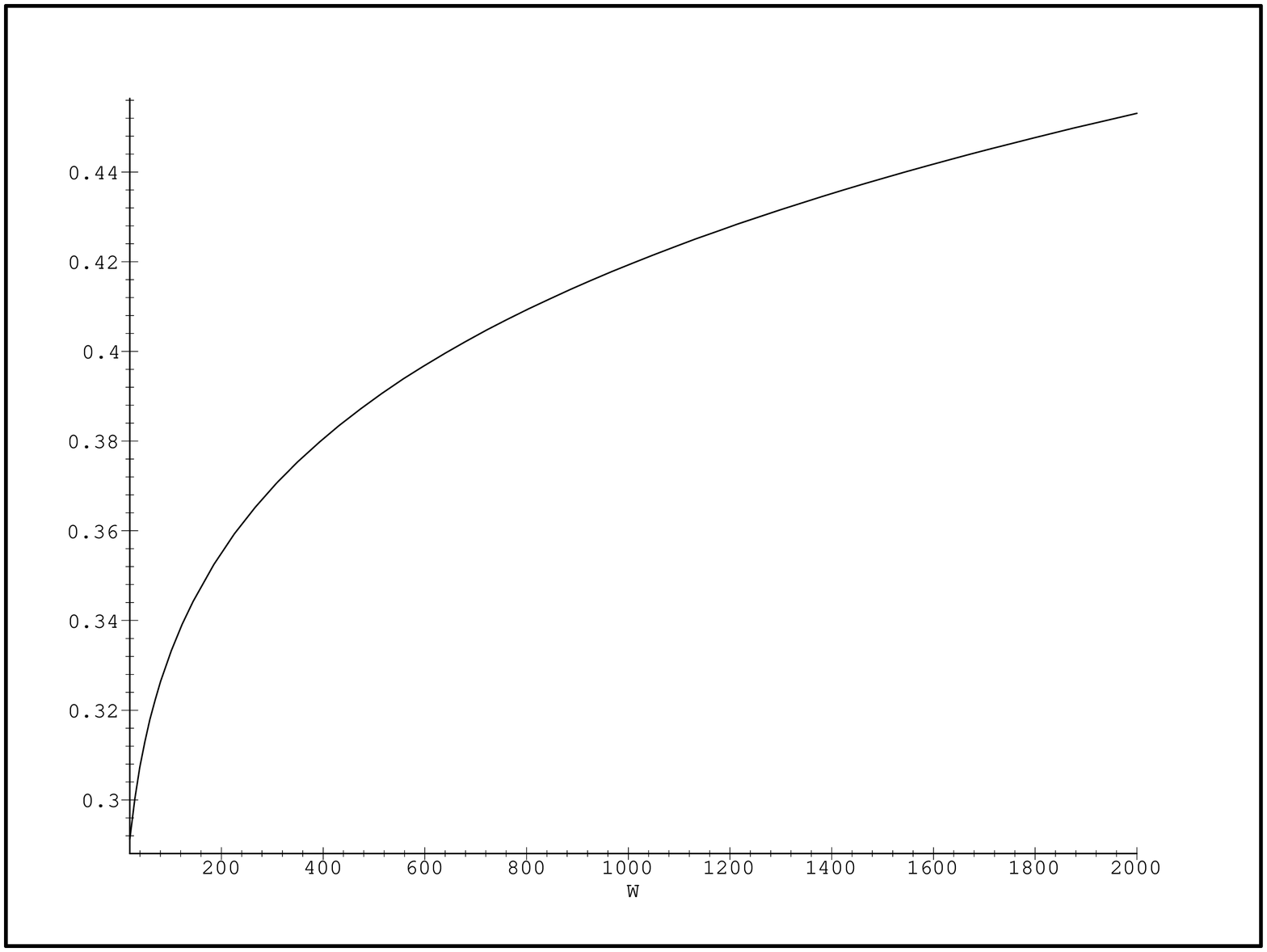,width=120mm,angle=-90}}
\caption{{\it The value of the long range rapidity correlation term (
$LR$)(see \protect\eq{CORGEN} ) in
the Eikonal model.}}
\label{fig37c}
\end{figure} 
\resetfig

\section{The Pomeron  from HERA and  Tevatron experiments.}
\subsection{ Reggeon approach strategy  for  measuring   the Pomeron
structure.}
 The
question which I am asking: { \tt what is the  correct strategy to measure
the
Pomeron structure?} I hope, that you are now in a position to  answer this
question.
 
Let me give you  my answer:

{\bf 1.}  We must measure the intercept of the  Pomeron as well as which
type  of 
inelastic processes proceed  Pomeron exchange 
 
(i) in the  inclusive processes where the most part of the
shadowing
corrections     mostly
cancel

 and 

(II) in  the short range ( in rapidity )  part of the correlation
function.

These are  the  best signatures for  one Pomeron exchange.

{\bf 2.} The traditional observables  like the total cross section, the 
elastic
cross section and the diffractive  dissociation cross sections cannot be
recommended for determining  the Pomeron structure since they
suffer from   large contribution from  shadowing
corrections. Among these three the least informative   is the diffractive
dissociation
cross section, very bad is the elastic one and bad is the total cross
section as far as the value of the SC is concerned.

{\bf 3.} The $t$ - dependence of the Pomeron trajectory and of the  
vertices
of interaction with particles and between Pomerons is very interesting
since it is closely related to the typical distances
involved in the Pomeron problem. Unfortunately, I do not know a way
of measuring  these other than  to measure the traditional observables
like the value
and the shrinkage of the $t$ - slope of elastic and diffractive cross
sections. We have discussed that these are not the best observables but
nobody knows better ones.To diminish the influence of the SC we have to
measure
all slopes at very small values of $t$.  I would like to emphasize this
point of view since this is the real justification why we need to measure
the classical  observables at a new generation of accelerators in spite of
the fact that it looks old fashioned and not very informative. At the
moment this is the only way to find out the scale of distances which are
responsible for ``soft" interactions.   

{\bf 4.} The energy behaviour of the diffractive dissociation in DIS.
For large values of  the photon virtuality ( $Q^2$) the SC  tends to be
negligible and  only  one  Pomeron  exchange contributes to this process.
However, the interpretation of this process in terms of  Pomeron
exchange at large $Q^2$ is highly questionable and demands the
experimental study of which  distances are  essential in 
diffractive  processes.

However, \,if\,  I\,  will\, ask \, you \, {\tt what}\,\,{\tt is}\,\,
{ \tt the
best} \,\,{ \tt  way} \,\,{\tt to}\,\, {\tt measure}\,\, { \tt the
shadowing}\,\,\,\,\, {\tt  corrections}, the answer will be just the  
opposite:

{\bf 1.} The best way is the diffractive dissociation processes, 
worse  elastic scattering, very bad is the total cross section and
extremely bad is the inclusive one. However, here I want to make a remark
on  diffractive dissociation in  photon - proton interaction.
As  discussed the photon looks like  a collection of hadrons with a
broad mass  spectrum. This  means that when measuring the diffractive
cross
section in photon - hadron interactions we measure mainly  elastic
scattering between  hadrons of the  appropriate masses. Therefore,
unfortunately,
in photon - hadron interaction we have not found yet the best process. The
 single photon diffraction dissociation certainly is not the best one.

{\bf 2.}
One  can improve the situation by  measuring   double
diffractive dissociation. The hadron vertex looks the same as in hadron -
hadron collisions and the double diffractive dissociation provides a tool
to measure the SC.

{\bf 3.} I am  certain that the best way of extracting the SC in
photon - hadron interactions
and one of the best in hadron - hadron collisions is the long range ( in
rapidity ) part of the correlation function.  I will show below how this 
sort of measurement by  CDF led to  a breakthrough in the  issue of
the SC.

{\bf 4.} The direct way of measuring  the value of the SC are 
LRG processes especially the ``hard" ones. The real measurement of the
damping
factor,   which as it  turns out to be rather big will provide  unique
information on the SC.

And let me ask you the third question: {\tt what is the most fundamental
problem for the Pomeron?}

As I have discussed, the Pomeron hypothesis is the  outcome mostly
of
 lack of imagination and we have no real deep argument that the high
energy asymptotic should be  Pomeron like.
However, this hypothesis works well  phenomenologically.
 My answer is that in the framework of the Pomeron hypothesis the
fundamental problem is to understand  experimentally and theoretically
what is going on with  diffractive production of  large masses.
As I have discussed, the main difficulties of the Pomeron approach are
concentrated in this problem. Theoretically,  solution to this problem
means that we understood how  the SC work  between every possible parton.

\subsection{ The Pomeron before HERA and Tevatron.}

Our  knowledge about the Pomeron before HERA and Tevatron  can be
described as Donnachie - Landshoff Pomeron \footnote{Of course, I
oversipmlify the situation. Several clouds were on the shining sky of the 
 D-L Pomeron even before HERA and Tevatron. In particular, the UA8 data
on diffractive dissociation which led the Ingelman - Schlein hypothesis
of the Pomeron structure function. Nevertheless, namely the D-L Pomeron
was a main tool to examine the `soft" high energy interaction.}. Actually,
it is not correct to call this Pomeron the  D-L Pomeron. Many
people introduced such a Pomeron  long before D-L ( K.A.Ter-Martirosian,
A.Kaidalov et al. in Moscow, A. Capella, Tranh Tran Van et al in Orsay
and  many others ) but we have to give a credit to Donnachie and
Landshoff -\,
they were the first ones  who assumed that there is nothing but  the D-L
Pomeron. We have discussed the certain theoretical difficulties with 
 the D-L Pomeron, but let us look at the  D-L Pomeron from a different
angle,namely, as the most ecanomic  way to express the experimental
data.
 Let us list the parameters of the D-L Pomeron and 
let us try to extract physics out of them.

1. $\alpha_P(0)\,\,=\,\,1.08$ or $\Delta_P\,=\,\alpha_P \,-\,1\,=\,0.08$.

This is a small number. What does  the smallness mean using
the
parton model as a guide? Firstly, let me make an essential remark about
partons. In QCD 
 partons are gluons ( massless particles with spin one ) and quarks and
antiquarks( massive with  spin 1/2 ). Going back to Eq.(60) one can see
that in the $g\phi^3$ model there is an  extra power of  $s$ in
front of  the 
formula for the cross section. Actually, in the  general case the power of
$s$ is as 
 follows : $(1/s)^{2j - 2}$ where $j$ is the spin of partons. 
It is obvious that in QCD mainly gluons  contribute to the expression for
the cross section because for them the extra power of $s$  is  just  equal
to
zero. In this particular case, the intercept of the Pomeron is equal to
$\alpha_P(0) \,\,=\,\,\Sigma (0)$ and closely related to the parton
multiplicity $( N_G )$  which is $ N_G\,\,=\,\,\Sigma(0)\ln(s)$.

Therefore, the smallness of the Pomeron intercept  means that the main
contribution to the Pomeron structure  originate from the first ``Born"
diagram = the two gluon exchange, but most likely with completely
nonperturbative gluons which carry the nonperturbative scale ($\mu$ in the
parton model). The production of gluons could be considered in a 
perturbative way  with respect to $\alpha_P(0)$.

2.  $\alpha'_P(0)\,\,=\,\,0.25 GeV^{-2}$.

Once more a small value  ( remember thayt for Reggeons typically
$\alpha'_R\,\approx\,0.7\,-\,1\,GeV^{-2}$  which , at first sight, says
that we
have a  large typical transverse momentum scale in the Pomeron. 
Let us estimate this using Eq.(60) and expanding  $\Sigma(q^2)$ as a
function
of $q^2$ at small $q^2$ which gives 

It gives:
\beq \label{TRAJPART}
\alpha_P (q^2)\,\,=\,\,\Sigma(0)\,\,-\,\, q^2 \frac{\Sigma(0)}{6
\mu^2}\,\,.
\eeq
Comparing \eq{TRAJPART} with the value for $\alpha'_P(0)$ we find that
$\mu \,\approx \,250 MeV\,\approx\,\Lambda$, where $\Lambda$ is the QCD
scale parameter.

3. The experimental slope for the $t$-dependence of the triple Pomeron
 vertex  ($r^2_0$) turned out to be  smaller than the slope of the
Pomeron proton form factor ( $R^2_0 $ ): $r^2_0\,\,\leq
\,1\,GeV^{-2}\,\,\ll \,\,R^2_0\,\,= 2.6\,GeV^{-2}$ in all so-called
triple
Reggeon fits of diffractive dissociation data.
Unfortunately, the accuracy of the data was not so good and the value of
$r^2_0$ could not be defined by the fit. Actually, the fit could be done
with $r^2_0 = 0$. This observable is very important since it gives a direct
information on the  transverse momentum scale is responsible for the
Pomeron
structure. As we have seen the value of $\alpha'_P(0)$ depends on the
unknown parton density while $r^2_0$  does not.

4. The ideas of the additive quark model that a hadron has two scales,
namely, the
hadron size $R\,\approx\,1 \,fm $ and the size of the constituent quark
$r_Q\,\approx \,0.2 - 0.4 \,fm$; is still a  working hypothesis  and it
was
included in the D-L Pomeron to the  $t$ - dependence of the Pomeron -
hadron
vertices.
 
5. Regge factorization has been checked and it works, but the number of
reactions studied  was limited.

\subsection{Pomeron at the Tevatron.}
The Tevatron data as well as the CERN data especially on diffractive
dissociation, I think, contributed a lot to our understanding of the
Pomeron structure. It is a pity that this contribution has not been
properly discussed yet. Let me tell you what I learned from the Tevatron.

1. The measurement of the inclusive cross section at Tevatron energy
allowed us to extract the intercept of the Pomeron from the energies 
behaviour of the inclusive spectra in the central rapidity region.
The result was surprising: the energy behaviour of the inclusive cross
section can be parameterized (Likhoded et al. 1989)  as 
\beq \label{INCPOM}
C_0\,+\,C_P (s/s_0)^{\Delta_P(0)}\,\,,
\eeq
with $\Delta_P(0)\,=\,0.2$  instead of the D-L value $\Delta\,
=\,0.08$ !!!. Both $C_0$ and $C_P$ are constant in the
above parameterization. In the framework of the Reggeon approach I have to
insist that the Pomeron intercept has been measured correctly only in the
inclusive cross section and to face the  problem that the D-L Pomeron is
not
a Pomeron at all.  Irritating feature of \eq{INCPOM} is the fact that one 
has to include  an additional constant $C_0$. Does this  mean that I have
to
deal with two Pomerons: the first with $\Delta_{P_1}(0) =0$ ( term $C_0$
in \eq{INCPOM} ) and the second with $\Delta_{P_2}\,=\,0.2$ ?!

2. The beautiful data on  diffractive dissociation which are shown in
Fig.26 changed the whole issue of the SC from theoretical guess to 
a  practical nessecity. I have discussed that the Eikonal model for t
SC could describe the main feature of the data but I an very certain that
data are much richer and have not been absorbed and properly discussed by
the
high energy community. You can will  find some discussions in papers and
talks of Gotsman et all (1992), Goulianos   (1993 - present), Capella et
al. (1997) , C-I Tan (1997), Schlein at al. (1997)  and others.

3. The CDF data for the  double parton cross section.
 CDF used the double inclusive cross section to measure the
contribution of the SC.  Actually,  CDF measured the process of
inclusive production of two pairs of ``hard" jets with almost compensating
transverse momenta in each pair and with almost  same values of 
rapidities. The production of two such pairs is highly supressed in the
one Pomeron model ( one parton cascade ) and can be produced only via 
double
Pomeron interaction ( double parton collisions). The Mueller diagram for
such a  process is given in Fig. 38b. We can use formula (132) for this
cross
section which predicts for   $\s_{eff}$ in Eq.(132) 
 14.5 \,$\pm$\,1.7\,$\pm$\,2.3 mb. This value is approximately half as big
as  our estimates in the Eikonal model. Therefore, these
data show that the SC exist and they are  even  bigger than the
Eikonal
model estimates. Perhaps, the problem is not in the Eikonal model but in
our picture for the hadron which was included in the Eikonal model.
Namely, if we  try to explore the two radii picture of the proton as
it is shown in Fig.38b we will get an  average radius which is a factor
of two  smaller
than what  has been used. So for me the CDF data finished the  discussion
of the
one Pomeron model, measured the value of the SC and provided  a strong
argument in favour of the two radii structure of the proton. 

\alphfig
\begin{figure}[htbp]
\centerline{\epsfig{file=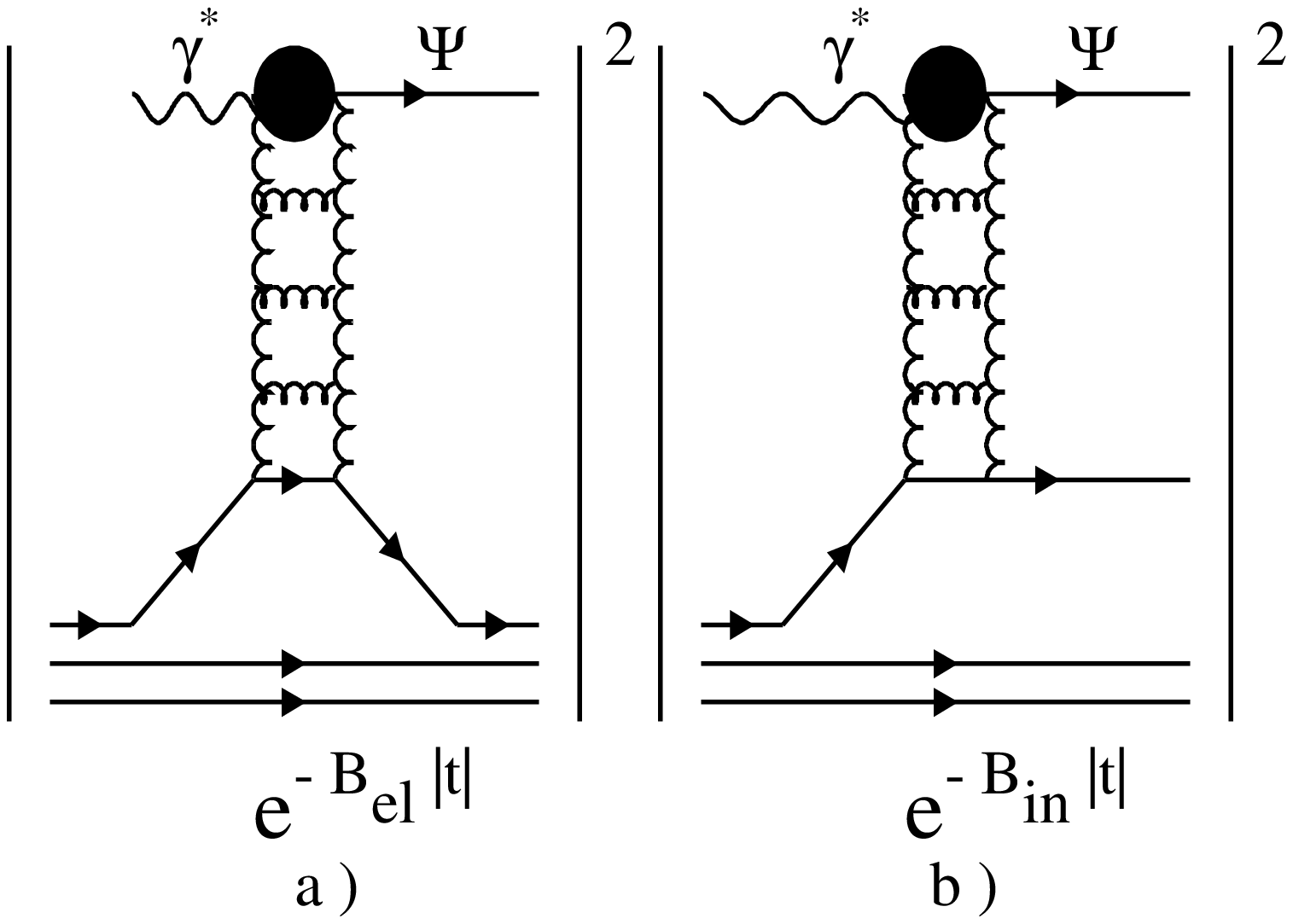,width=120mm}}
\caption{{\it The J/$\Psi$ production without (a) and with (b) proton
dissociation.}}
\label{fig38a}
\end{figure}
\begin{figure}[htbp]
\centerline{\epsfig{file=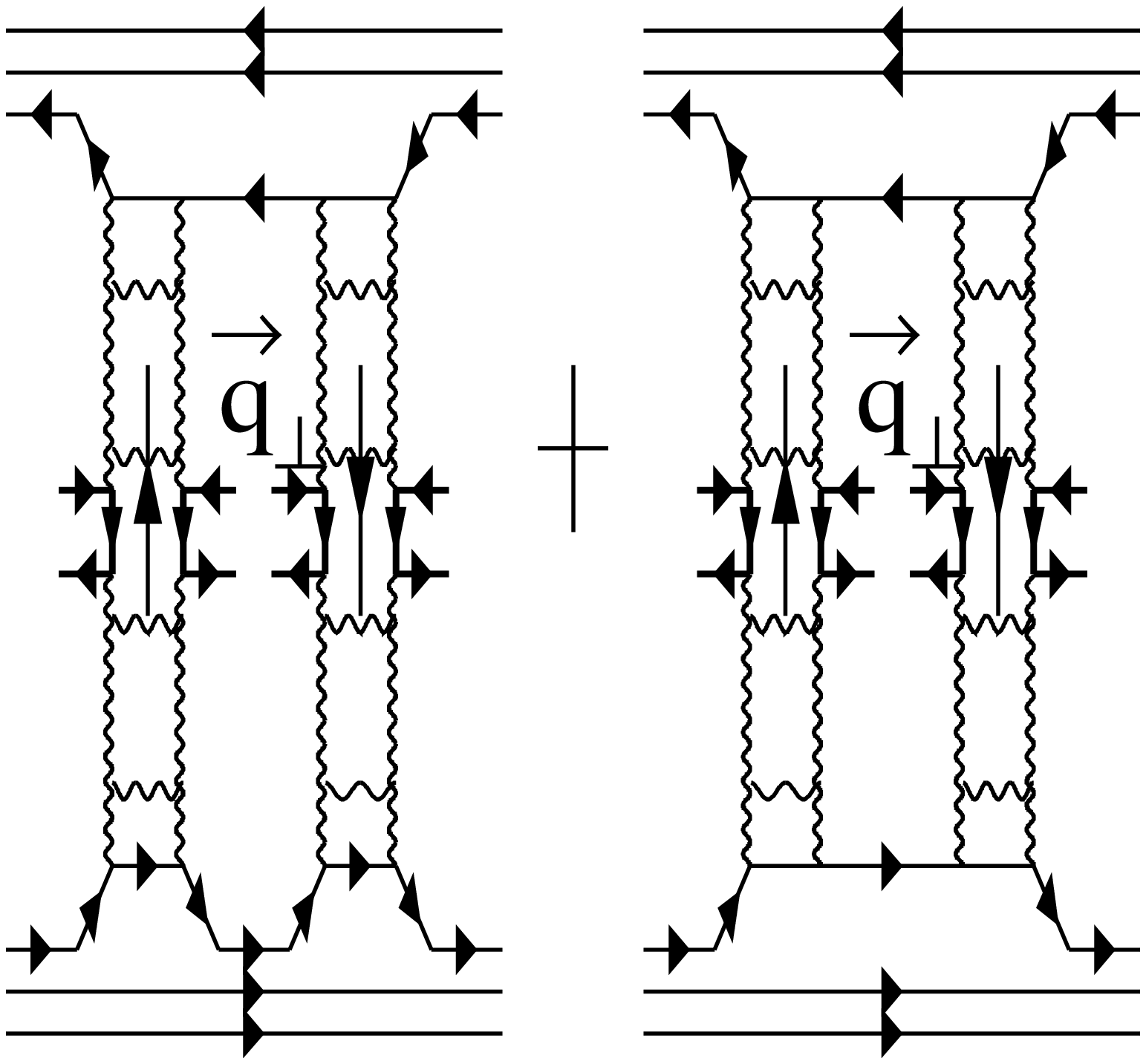,width=120mm}}
\caption{{\it Inclusive production of two pair of ``hard" jets in the
double parton scattering.}}
\label{fig38b}
\end{figure}
\begin{figure}[htbp]
\centerline{\epsfig{file=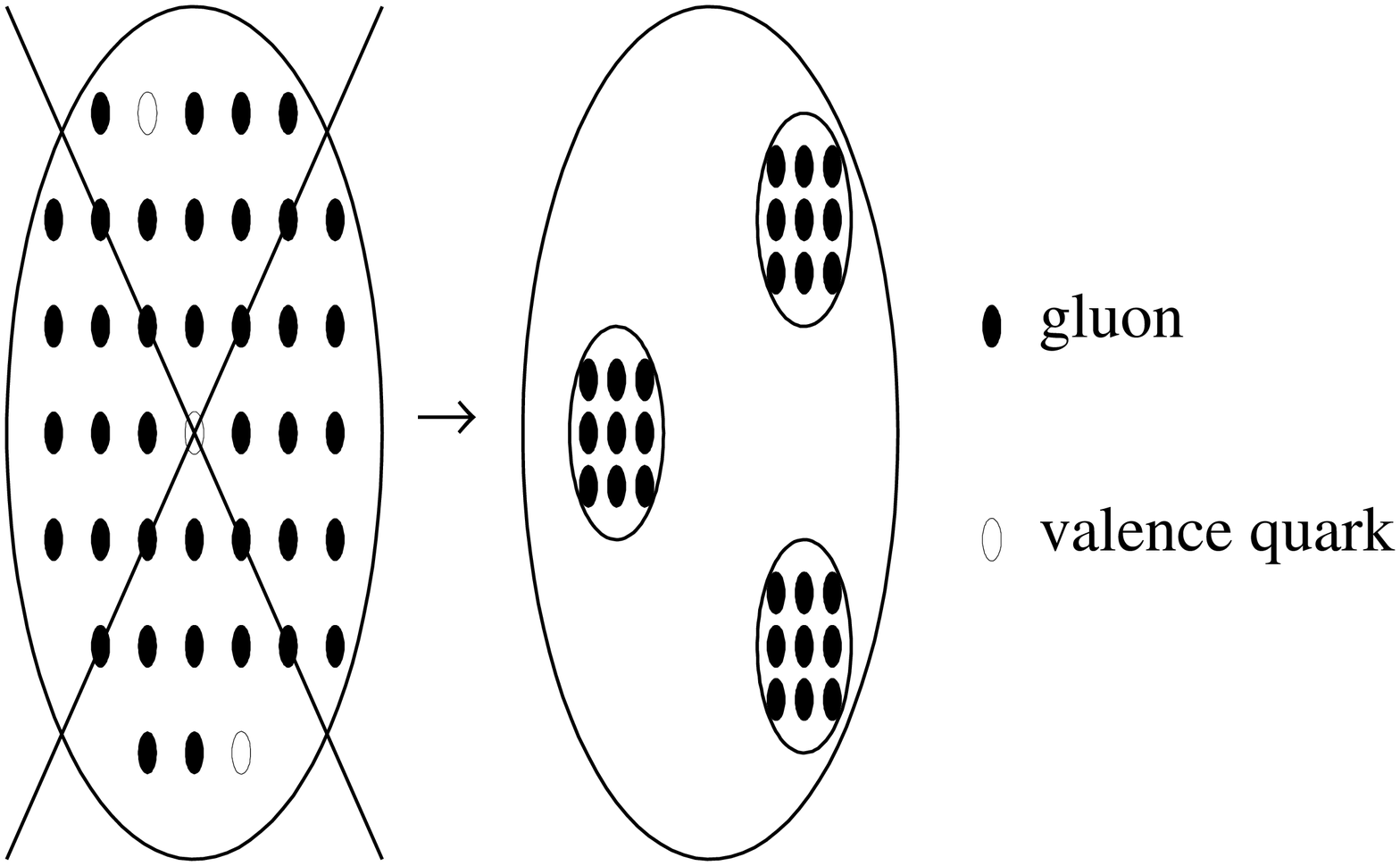,width=160mm,height=50mm}}
\caption{{\it Democratic and two radii picture of proton.}}
\label{fig38c}
\end{figure}

\begin{figure}[htbp]
\centerline{\epsfig{file=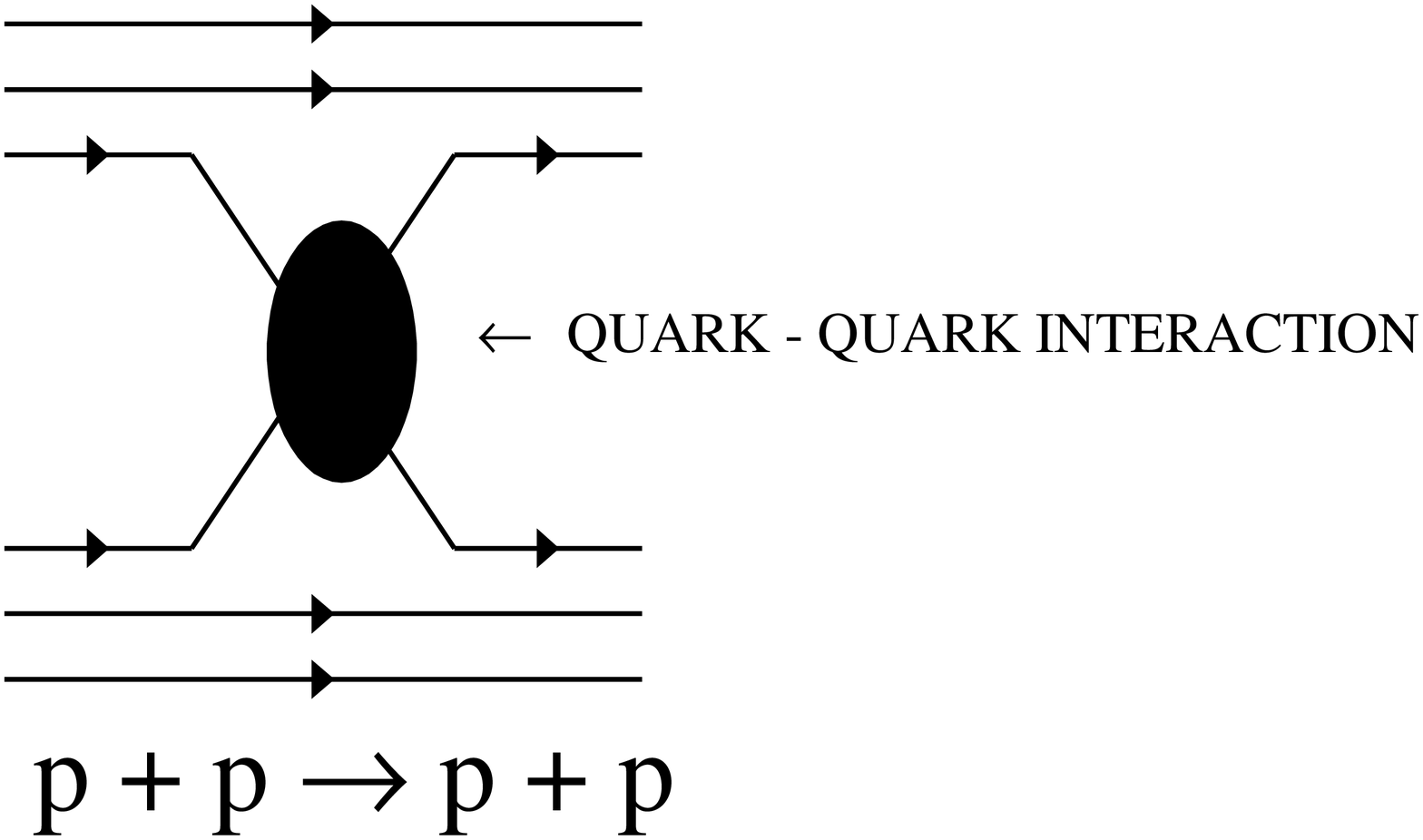,width=120mm}}
\caption{{\it Scattering in the constituent quark model.}}
\label{fig38d}
\end{figure}  
\begin{figure}[htbp]
\centerline{\epsfig{file=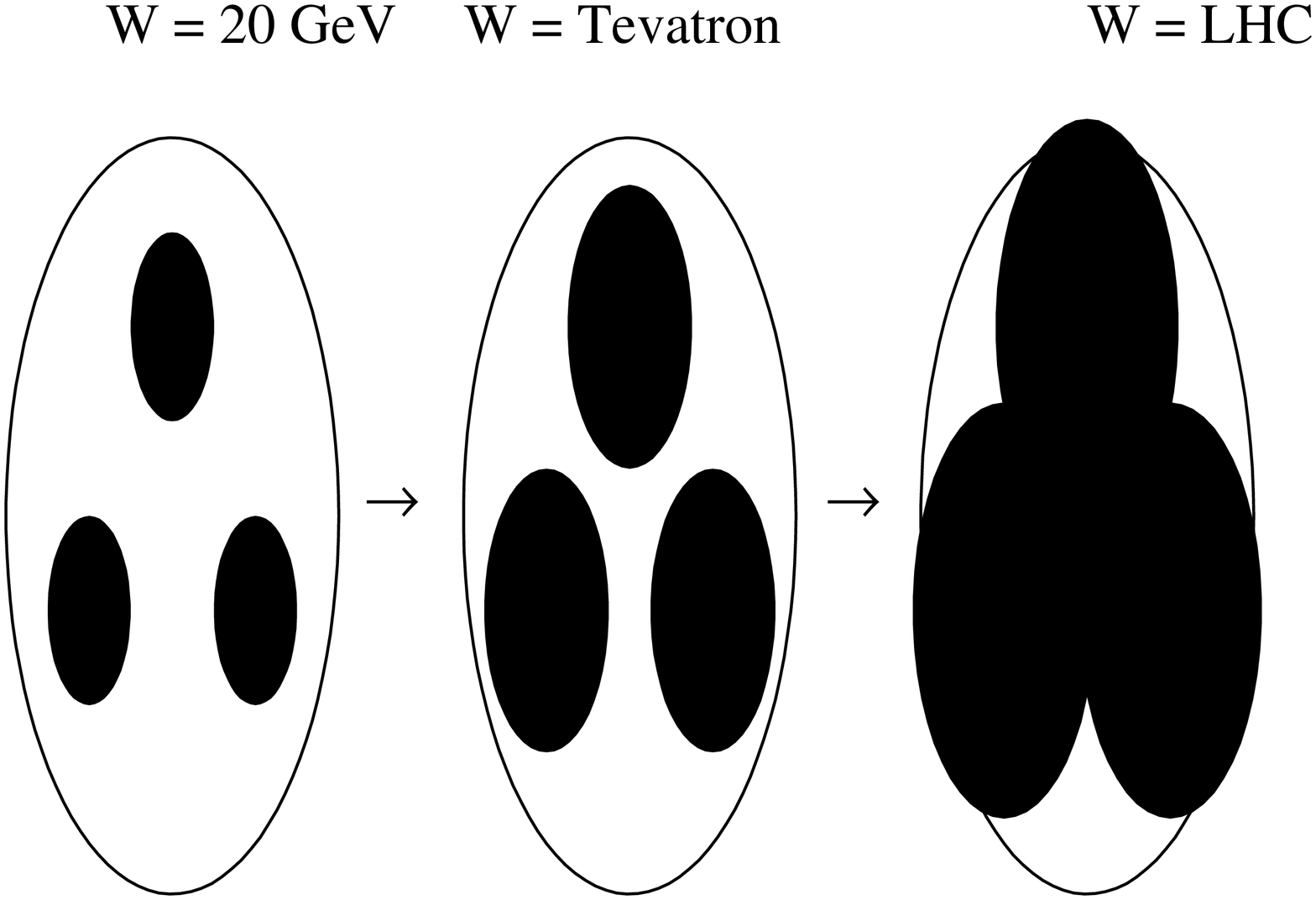,width=160mm,height=50mm}}
\caption{{\it The growth of the interaction radius and the new regime
at the LHC energies.}}
\label{fig38e}
\end{figure}

\resetfig

\subsection{Pomeron at HERA.}
1. I think that HERA did a great thing, namely, the HERA data show
explicitly
that the hadron has two ``soft" scales  or two radii. This  comes out
from the HERA data on photoproduction of  the  J/$\Psi$ meson. These data
show
two different slopes in $t$-dependence for quasi-elastic photoproduction  
and for inelastic 
photoproduction in which the proton dissociates into a  hadron system (
see
Fig.38a ): $B_{el}\,\,=\,\,4\,GeV^{-2}$ and
$B_{in}\,\,=\,\,1.66\,GeV^{-2}$, while the cross sections are about equal
for
these two  processes. Since in a such a process we can neglect the
$t$-dependence of the upper vertex we can interprete these data as
follows:

1). as  the first measurement of the radius of the  so-called Pomeron -
proton
vertex ( we also called this value the two gluon form factor in a picture
of  ``hard" processes). It turns out that this radius is smaller than
the radius given by the  electro-magnetic form factor of the proton  which
 for this vertex follows 
from the additive quark model and has been used for the D-L Pomeron.

2). the second radius which appears  can be easily interpreted in the
picture of the additive quark model (see Fig.38a) as the proper radius of
the constituent quark.

Therefore, HERA finished with a democratic picture for gluons and valence
quarks and proved  the two radii picture of the proton (see Fig. 38c),
which
could be an additive quark model. The direct observation of the second
scale in the ``soft" interaction will lead to  reconsideration of
the simple partonic estimates ( see Eq. (133) ) of the value of
$\alpha'_P(0)$.  Our conclusion that the typical scale in the
Pomeron problem is about $\Lambda$ QCD should be revised, but this is a
certain problem for the future.

2. The measurement for the energy behaviour of the diffractive
dissociation process leads to the intercept of the Pomeron
$\Delta_P(0)\,\approx\,0.2$. As I have mention there  could be different
interpretation of the experimental data , but in the framework of the
Pomeron approach this is the intercept.

3. The HERA data on the  deep inelastic structure function $F_2$  can be
described by the DGLAP evolution equation if we assume an initial
parton distribution  at $Q^2\,= \,Q^2_0\,=\,3
-5 GeV^2 $ with  the  energy behaviour
$F_2(x,Q^2_0)\,\propto\,(1/x)^{0.2}$
which corresponds to a  Pomeron intercept $\Delta_P(0)\,=\,0.2$.

4. Inclusive hadron production by real and virtual photon scattering 
 can be
described assuming a Pomeron with the intercept $\Delta_P(0) \,=\,0.2$.

Therefore, let me conclude this section with the following statement:

{\it In October 1997 following the  HERA and Tevatron data we have :

$\bullet$ The Pomeron intercept $\Delta_P(0)\,=\,0.2$;

$\bullet$ The SC are large and approximately twice as big as
the Eikonal model estimates;

$\bullet$ The two radii picture of the proton has been proven
experimentally and  the values of these two radii have been
measured.}
 
\section{Instead of conclusions.}

\subsection{My picture for the high energy interaction.}

Let me conclude my lectures sharing with you the truth that as  starting
from the
year 1977 I do not believe in the Pomeron as a specific Regge pole.
In the framework of these lectures I could not show you why and it will be 
a subject for the second part. Instead of the Pomeron approach Ryskin and
me developed a different approach which corresponds to  the following
simple
picture (see Figs.38c and 38d) with two ingredients: the constituent
quark model for hadrons and the "black" disc interaction between two
constituent quarks. Now, finally, I have the  real pleasure telling you
that
due to the results of HERA we have an experimental confirmation of this
approach. Indeed, let us estimate the value of the total cross section for
quark - quark interactions. As we have discussed in section 2.4 the cross
section is equal to
$$
\s_{tot}(q + q )\,\,=\,\,2\,\pi ( 2\, \,r^2_Q )\,\,
$$
where $r_Q$ is the quark radius, the  factor 2 emerges because two quarks
collide. Substituting $2 r^2_Q \,=\,B_{in}$ ( see Fig.38a ) we get
$\s_{tot}( q + q ) \,=\,4\,-\,5 \, mb$ which gives the correct order of
magnitude  value for the
total
proton - proton cross section $\s_{tot}(p + p) \,=\,9\,\s_{tot}(q + q)$. 

In this model the size of the constituent quarks increases with energy
(see Fig. 38e) and only at LHC energies the clouds of partons  from
different
constituent quarks will start to overlap. This is the  region where we can
expect to see new physics for  ``soft" processes.

The interesting feature of this model is  that a substantial part of
Regge
factorization relations holds  because of the simple fact that only
quarks interact.

\subsection{Apology and Acknowledgements.}

Reading back my lectures I found that I have kept my promise and have
shared with you all ideas, hopes and difficulties of the Pomeron problem.
I hope that after reading these lectures you will  know enough about
Pomerons to support or to reject this approach. 

I thought that I would not
give references in this part but one short story changed my mind. Once,
when actually I have finished my writing,  Aharon Levy called me asking
for a
help. He tried to find the book of Collins on Regge theory. Indeed, it is
the last book on the subject which summarizes our knowledge of
Reggeon theory. However, he failed to find it in any library in Hamburg,
DESY
one including. I also could not find but fortunately, I went to Paris and
found one copy in the LPTHE at Orsay. I asked the librarian to check
whether the general library  has this book. It turnes out that no and even
more no university libraries in Paris have. So, my usual excuse that you
could easily find everything that I have discussed in any book of the  
sixties
being correct is not practical. So, I decided to give a list of references
which is far from being full but at least you will be able to read 
and clarify a subject that you feel to be interesting. By the way, the end
of the story with Collins book is happy. In December, Aharon Levy (tel.
2001 at DESY ) will be the only one who will have the copy of the book.
 
 I wish to express my deep apology to  all my friends whom  I
did not cite. I hope to prepare the full  list of references including
Russian physicists
to conclude
the
last part of my lectures.

As I have mentioned I have learned everything about the Pomeron from Prof.
Gribov and I was actively involved  in this problem during the best
fifteen years of my life in the  Leningrad Nuclear Physics Institute
under the leadership and strong influence of Prof. Gribov. I cannot
express, at least in english, my deep sorry that we lost him.
These lectures are  an attempt to look at the problem with his eyes
and, unfortunately, this is the only thing  I can do now for him.

I am very grateful to my friends A. Kaidalov, O. Kancheli and M. Ryskin
for a life long discussions of the Pomeron. I am sure that a lot of ideas,
that we  discussed, have  shown  up in these lectures.
I thank E.Gotsman and U.Maor who started with me five years ago
an attempt to revise  
the whole 
issue of the Pomeron approach. Sections, devoted to the
Eikonal model, is  written on the basis of our common papers as well
as 
our common understanding that the Donnachie - Landshoff Pomeron is not the
right one. E. Gotsman calculated Fig.25 which I used here and I thank
him for this.  Travelling around the globe I met the Old Reggeon Guard:
J. Bartels, K. Boreskov, A. Capella, J.Cohen - Tannoudji,P. Collins,
A.Donnachie, A. Krzywicki,S. Matinian, A. Mueller, R. Peschanski, A.
Santoro, C-I Tan, K.A. Ter - Martirosian and A. White ( more names for
search). Most of them, as
well as me,    are doing  quite different things but all of them still
keep alive the interest for the beauty of this difficult problem. My
special thanks to them for encouraging optimism. 

These lectures would be impossible without help and support of my DESY
friends
W. Buchmueller, J. Dainton, H. Kowalski and G. Wolf. I am grateful to them
as well as to my listerners who asked many good questions which have
improved the presentation and contents. Hope for more questions and
critisism.

\section{References.}
~
{{ \Large \bf Books:}}
~
\begin{enumerate}

\item{ G.F. Chew: {\it `` S- matrix theory of strong interaction"}, Benjamin,
1962.}

\item{S.C. Frautschi:{\it `` Regge Poles and S-matrix Theory"}, Benjamin,
1963.}

\item{R. Omnes and M. Froissart: {\it ``Mandelstam Theory and Regge
Poles: An Introduction for Experimentalists"}, Benjamin, 1963.}

\item{  M. Jacob and G.F.Chew:{\it ``Strong interaction physics"},
Frontiers
in
Physics, 1964.}

\item{ R.J. Eden, P.V. Landshoff, D.I. Olive and J.C. Polkinghorne: {\it
``
The
analytic S - matrix"}, Cambridge U.P., 1966.}

\item{ R.J. Eden: {\it `` High energy collisions of elementary
particles"},
Cambridge U.P. 1967.}

\item{ V. Barger and D. Cline: {\it `` Phenomenological Theories of High
Energy Scattering: An Experimental Evaluation"}, Frontiers in Physics,
1967.}

\item{ P.D.B. Collins and E.J. Squires:{\it `` Regge Poles in Particle
Physics"},
Springer, 1968.}

\item{ A. Martin: {\it `` Scattering Theory: Unitarity, Analyticity and
Crossing"}, Lecture notes in Physics, Springer, 1969.}

\item{ R. Omnes: {\it `` Introduction to particle physics"}, Wiley, 1971.}

\item{ R. Feynman: {\it `` Photon - hadron interaction "}, Benjamin,
1972.}

\item{ D. Horn and F. Zachariasen: {\it `` Hadron Physics at Very High
Energies"}, Benjamin, 1973.}

\item{ S.J.  Lindenbaum: {\it `` Particle Interaction Physics at High
Energy"},  Clarendon Press, 1973.}

\item{ M. Perl: {\it `` High energy hadron physics"}, Wiley, 1974.}

\item{ S. Humble: {\it `` Introduction to particle production in hadron
physics"}, Academic Press,  1974.}

\item{ {\it ``Phenomenology of particles at high energy"}, eds. R.L.
Crawford and R. Jennings, Academic Press, 1974.}

\item{  P.D.B. Collins: {\it ``An introduction to Regge theory and High
energy
physics"}, Cambridge U.P., 1977.}

\item{Yu. P. Nikitin and I.L. Rosental: {\it `` Theory of Multiparticle
Production Processes"}, Harwood Academic, 1988.}

\item{{\it ``Hadronic Multiparticle Production"}, ed. P. Carruther, WS,
1988.}

\item{ J.R. Forshaw and D.A. Ross:
{\it `` QCD and the Pomeron"},
Cambridge U.P. 1997.}

\item{ { \bf The collection of the best original papers on Reggeon
approach:}
{\it ``
Regge Theory of low $p_t$ Hadronic Interaction"}, ed. L. Caneschi, North -
Holland, 1989.}

\end{enumerate}

~

{ {\Large \bf Reviews:}}

~
\begin{enumerate}

\item{ P.D.B. Collins: {\it `` Regge theory and particle physics"}, Phys.
Rep.{\bf 1C} (1970) 105. }

\item{ P. Zachariasen: {\it `` Theoretical models of diffraction
scattering"},
Phys. Rep. {\bf 2C} (1971) 1.}

\item{ D. Horn: {\it `` Many particle production"},
Phys. Rep. {\bf 4C} (1971) 1.}

\item{ F. J. Gilman: {\it `` Photoproduction and electroproduction"},
Phys. Rep. {\bf 4C} (1971) 95.}

\item{ P.V. Landshoff and J.C. Polkinghorne: {\it `` Models for hadronic
and
leptonic processes at high energy"},
Phys. Rep. {\bf 5C} (1972) 1.}

\item{S.M. Roy: {\it `` High energy theorems for strong interactions and
their comparison with experimental data"},
Phys. Rep. {\bf 5C} (1972) 125.}

\item{  J. Kogut and L. Susskind: {\it `` The parton picture of elementary
particles"},
Phys. Rep. {\bf 8C} (1973) 75.}

\item{ J. H. Schwarz: {\it `` Dual resonance theory"},
Phys. Rep. {\bf 8C} (1973) 269.}

\item{  G. Veneziano: {\it `` An introduction to dual models of strong
interaction and their physical motivation"},
Phys. Rep. {\bf 9C} (1973) 199.}

\item{ R. Slansky: {\it `` High energy hadron production and inclusive
reactions 
"}, 
Phys. Rep. {\bf 11C} (1974) 99.}

\item{ S. Mandelstam: {\it `` Dual - resonance models"}, 
Phys. Rep. {\bf 13C} (1974) 259.}

\item{ R.C. Brower, C.E. De Tar and J.H. Wers:
{\it `` Regge theory for multiparticle amplitude"},
Phys. Rep. {\bf 14C} (1974) 257.}

\item{Ya.I. Azimov,  E.M. Levin, M.G. Ryskin and V.A. Khoze:
 {\it `` What is interesting about the region of  small momentum
transfers 
 at high energies ?"},
  \, CERN-TRANS 74-8, Dec
1974. 196pp.
Translated from the Proceedings of 9th Winter School of Leningrad Nuclear
Physics Inst. on
Nuclear and Elementary Particle Physics, Feb 15-26, 1974. Leningrad,
Akademiya Nauk SSSR,
1974, pp. 5-161(translation).} 

\item{ I.M. Dremin and A.N. Dunaevskii:
{\it ``The multiperipheral cluster theory and its comparison with
experiment"},
Phys. Rep. {\bf 18C} (1975) 159.}

\item{  H.D.I. Abarbanel, J.D. Bronzan, R.L. Sugar and A.K. White:
{\it `` Reggeon field theory formulation and use"},
Phys. Rep. {\bf 21C} (1975) 119.}

\item{ G. Giacomelli:
{\it `` Total cross sections and elastic scattering at high energy"},
Phys. Rep. {\bf 23C} (1976) 123.}

\item{M. Baker and K. Ter Martirosyan:
{\it `` Gribov's Reggeon Calculus"},
Phys. Rep. {\bf 28C} (1976) 1.}

\item{ A.C. Irving and R.P. Worden:
{\it `` Regge phenomenology"},
Phys. Rep. {\bf 36C} (1977) 117.}

\item{  M. Moshe:
{\it `` Recent developments in Reggeon field theory"},
Phys. Rep. {\bf 37C} (1978) 255.}

\item{ A.B. Kaidalov:
{\it `` Diffractive production mechanisms"},
Phys. Rep. {\bf 50C} (1979) 157.}

\item{  S.N. Ganguli andd D.P. Roy:
{\it `` Regge phenomenology in inclusive reactions"},
Phys. Rep. {\bf 67C} (1980) 257.}

\item{ K. Goulianos:
{\it `` Diffraction interactions of hadron at high energy"},
Phys. Rep. {\bf 101C} (1983) 169.}

\item{ M. Kamran:
{\it `` A review of elastic hadronic scattering at high energy and small 
momentum transfer"},
Phys. Rep. {\bf 108C} (1984) 275.}

\item{ E.M. Levin and M.G. Ryskin:
{\it ``The increase  in the total cross sections for hadronic interactions
with increasing energy"}
Sov.Phys. Usp. {\bf 32} (1989) 479.}

\item{ E.M. Levin and M.G. Ryskin:
{\it `` High - energy hadron collisions in QCD"}
Phys. Rep. {\bf 189C} (1990) 267.}

\item{ A. Capella, U. Sukhatme, C-I Tan and J. Tran Thanh Van:
{\it `` Dual parton model"},
Phys. Rep. {\bf 236C} (1994) 225.}

\item{ E. Levin: {\it ``The Pomeron: yesterday, today and  tomorrow"},
 hep-ph/9503399 ,  Lectures given at 3rd Gleb Wataghin School on High
Energy
Phenomenology, Campinas,
Brazil, 11-16 Jul 1994.}

\item{ J. Stirling: {\it `` The return of the Pomeron"}
 Phys.World {\bf 7} (1994) 30.}

\item{ A.B. Kaidalov:
{\it `` Low - x physics"}, 
Survey in High Energy Physics,{\bf 9} (1996) 143.}

\end{enumerate}

 \end{document}